\documentclass[aps,twocolumn,groupedaddress,showpacs]{revtex4}

\begin{document}
\bibliographystyle{apsrev}

\title{
Quantum Orders and Symmetric Spin Liquids\\
(the original version) 
}

\author{Xiao-Gang Wen}
\homepage{http://dao.mit.edu/~wen}
\affiliation{Department of Physics, Massachusetts Institute of Technology,
Cambridge, Massachusetts 02139}

\date{June 3, 2001}

\begin{abstract}
A concept -- quantum order -- is introduced to describe a new kind of orders
that generally appear in quantum states at zero temperature. Quantum orders
that characterize universality classes of quantum states (described by {\em
complex} ground state wave-functions) is much richer then classical orders
that characterize universality classes of finite temperature classical states
(described by {\em positive} probability distribution functions).  The
Landau's theory for orders and phase transitions does not apply to quantum
orders since they cannot be described by broken symmetries and the associated
order parameters. We introduced a mathematical object -- 
projective symmetry group -- to characterize quantum
orders.  With the help of quantum orders and projective symmetry groups,
we construct hundreds of symmetric spin liquids, which have $SU(2)$, $U(1)$ or
$Z_2$ gauge structures at low energies.  We found that various spin liquids
can be divided into four classes: 
(a) Rigid spin liquid -- spinons (and all other excitations) are fully
gaped and may have bosonic, fermionic, or fractional statistics.
(b) Fermi spin liquid -- spinons are gapless and are described by a Fermi
liquid theory.
(c) Algebraic spin liquid -- spinons are gapless,
but they are not described by free fermionic/bosonic quasiparticles.  
(d) Bose spin liquid -- low lying gapless excitations are described by a
free boson theory.
The stability of those spin liquids are discussed in details.  We find that
stable 2D spin liquids exist in the first three classes (a--c).  Those
stable spin liquids occupy a finite region in phase space and represent
quantum phases. Remarkably, some of the stable quantum phases support
gapless excitations even without any spontaneous symmetry breaking.  In
particular, the gapless excitations in algebraic spin liquids interact down
to zero energy and the interaction does not open any energy gap.  We
propose that it is the quantum orders (instead of symmetries) that protect
the gapless excitations and make algebraic spin liquids and
Fermi spin liquids stable.
Since high $T_c$ superconductors are likely to be described by a gapless
spin liquid, the quantum orders and their projective symmetry group
descriptions lay the foundation for spin liquid approach to high $T_c$
superconductors. 
\end{abstract}
\pacs{73.43.Nq,  74.25.-q,  11.15.Ex}
\keywords{Quantum order, Spin liquid, High Tc superconductors, Gauge theory}

\maketitle

\tableofcontents

\section{Introduction}

Due to its long length, we would like to first outline the structure of the
paper so readers can choose to read the parts of interests.  The section
\ref{sum} summarize the main results of the paper, which also serves as a
guide of the whole paper.  The concept of quantum order is introduced in
section \ref{sec:topqun}.  A concrete mathematical description of quantum
order is described in section \ref{psg} and section \ref{Z2sec}.  Readers who
are interested in the background and motivation of quantum orders may choose
to read section \ref{sec:topqun}.  Readers who are familiar with the
slave-boson approach and just want a quick introduction to quantum orders may
choose to read sections \ref{psg} and \ref{Z2sec}.  Readers who are not
familiar with the slave-boson approach may find the review sections
\ref{revSU2} and \ref{trnANS} useful.  Readers who do not care about the
slave-boson approach but are interested in application to high $T_c$
superconductors and experimental measurements of quantum orders may choose to
read sections \ref{sec:topqun}, \ref{splqhTc}, \ref{sec:pmQO} and Fig.
\ref{specZ2l} - Fig.  \ref{spec2sU1B}, to gain some intuitive picture of
spinon dispersion and neutron scattering behavior of various spin liquids.

\subsection{Topological orders and quantum orders}
\label{sec:topqun}

Matter can have many different states, such as gas, liquid, and solid.
Understanding states of matter is the first step in understanding matter.
Physicists find matter can have much more different states than just gas,
liquid, and solid.  Even solids and liquids can appear in many different
forms and states.  With so many different states of matter, a general
theory is needed to gain a deeper understanding of states of matter.  

All the states of matter are distinguished by their internal structures or
orders.  The key step in developing the general theory for states of matter is
the realization that all the orders are associated with symmetries (or rather,
the breaking of symmetries).  Based on the relation between orders and
symmetries, Landau developed a general theory of orders and the transitions
between different orders.\cite{LL58,GL5064}
Landau's theory is so successful and one starts to have a feeling that we
understand, at in principle, all kinds of orders that matter can have.  

However, nature never stops to surprise us.  In 1982, Tsui, Stormer, and
Gossard\cite{TSG8259} discovered a new kind of state -- Fractional Quantum
Hall (FQH) liquid.\cite{L8395}  Quantum Hall liquids have many amazing
properties.  A quantum Hall liquid is more ``rigid'' than a solid (a
crystal), in the sense that a quantum Hall liquid cannot be compressed.
Thus a quantum Hall liquid has a fixed and well-defined density.  When we
measure the electron density in terms of filling factor $\nu$, we found
that all discovered quantum Hall states have such densities that the
filling factors are exactly given by some rational numbers, such as $\nu=1,
1/3, 2/3,2/5, ...$.  Knowing that FQH liquids exist only at certain magical
filling factors, one cannot help to guess that FQH liquids should have some
internal orders or ``patterns''. Different magical filling factors should
be due to those different internal ``patterns''.  However, the hypothesis
of internal ``patterns'' appears to have one difficulty -- FQH states are
liquids, and how can liquids have any internal ``patterns''?

In 1989, it was realized that the internal orders in FQH liquids (as well
as the internal orders in chiral spin liquids\cite{KL8795,WWZcsp}) are
different from any other known orders and cannot be observed and
characterized in any conventional ways.\cite{Wtop,WNtop} What is really new
(and strange) about the orders in chiral spin liquids and FQH liquids is
that they are not associated with any symmetries (or the breaking of
symmetries), and cannot be described by Landau's theory using physical
order parameters.\cite{Wrig} This kind of order is called {\it
topological order}. Topological order is a new concept and a whole new
theory was developed to describe it.\cite{Wrig,Wtoprev}

Knowing FQH liquids contain a new kind of order -- topological order,
we would like to ask why FQH liquids are so special. What is missed in
Landau's theory for states of matter so that the theory fails to capture
the topological order in FQH liquids?

When we talk about orders in FQH liquids, we are really talking about the
internal structure of FQH liquids at {\em zero} temperature. In other
words, we are talking about the internal structure of the quantum ground
state of FQH systems.  So the topological order is a property of ground
state wave-function.  The Landau's theory is developed for system at finite
temperatures where quantum effects can be ignored.  Thus one should not be
surprised that the Landau's theory does not apply to states at zero
temperature where quantum effects are important.
The very existence of topological orders 
suggests that finite-temperature orders and zero-temperature orders are
different, and zero-temperature orders contain richer structures.  We
see that what is missed by Landau's theory is simply the quantum effect.
Thus FQH liquids are not that special.  The Landau's theory and symmetry
characterization can fail for any quantum states at zero temperature.  As a
consequence, new kind of orders with no broken symmetries and local order
parameters (such as topological orders) can exist for any quantum states at
zero temperature.  Because the orders in quantum states at zero temperature
and the orders in classical states at finite temperatures are very
different, here we would like to introduce two concepts to stress their
differences:\cite{Wqos} \\
(A) {\em Quantum orders:}\footnote{A more precise definition of quantum order
is given in \Ref{Wqos}.} which describe the universality classes of
quantum ground states (\ie the  universality classes of {\em complex}
ground state wave-functions with infinity variables);\\ 
(B){\em Classical orders:} which describe the universality classes of
classical statistical states (\ie the universality classes of {\em
positive} probability distribution functions with infinity variables).\\ 
From the above definition, it is clear that the quantum orders associated with
complex functions are richer than the classical orders associated with
positive functions.  The Landau's theory is a theory for classical orders,
which suggests that classical orders may be
characterized by broken symmetries
and local order parameters.\footnote{The Landau's theory may not even be
able to describe all the classical orders. Some classical phase transitions,
such as Kosterliz-Thouless transition, do not change any symmetries.} The
existence of topological order indicates that quantum orders cannot be
completely characterized by broken symmetries and order parameters.  Thus we
need to develop a new theory to describe quantum orders.  

In a sense, the classical world described by positive probabilities is a
world with only ``black and white''. The  Landau's theory and the symmetry
principle for classical orders are color blind which
can only describe different ``shades of
grey'' in the classical world.  The quantum world described by complex wave
functions is a ``colorful'' world.  We need to use new theories, such as
the theory of topological order and the theory developed in this paper, to
describe the rich ``color'' of quantum world.

The quantum
orders in FQH liquids have a special property that all excitations above
ground state have finite energy gaps. This kind of quantum orders are called
topological orders.  In general, a topological order is defined as a quantum
order where all the excitations above ground state have finite energy gapes.

Topological orders and quantum orders are general properties of any states at
zero temperature. Non trivial topological orders not only appear in FQH
liquids, they also appear in spin liquids at zero temperature.  In fact, the
concept of topological order was first introduced in a study of spin
liquids.\cite{Wrig} FQH liquid is not even the first experimentally
observed state with non trivial topological orders.  That honor goes to
superconducting state discovered in 1911.\cite{O1122} In contrast to a common
point of view, a superconducting state cannot be characterized by broken
symmetries.  It contains non trivial topological orders,\cite{Wtopcs} and is
fundamentally different from a superfluid state.

After a long introduction, now we can state the main subject of this paper.
In this paper, we will study a new class of quantum orders where the
excitations above the ground state are gapless. We believe that the gapless
quantum orders are important in understanding high $T_c$ superconductors.  To
connect to high $T_c$ superconductors, we will study quantum orders in quantum
spin liquids on a 2D square lattice. We will concentrate on how to
characterize and classify quantum spin liquids with different quantum orders.
We introduce projective symmetry groups to help us to achieve this goal.  The
projective symmetry group can be viewed as a generalization of symmetry group
that characterize different classical orders.

\subsection{Spin-liquid approach to high $T_c$ superconductors}
\label{splqhTc}

There are many different approaches to the high $T_c$ superconductors.
Different people have different points of view on what are the key
experimental facts for the high $T_c$ superconductors.  The different choice
of the key experimental facts lead to many different approaches and
theories.  The spin liquid approach is based on a point of view that the
high $T_c$ superconductors are doped Mott
insulators.\cite{A8796,BZA8773,BA8880} (Here by Mott insulator we mean a
insulator with an odd number of electron per unit cell.)  We believe that
the most important properties of the high $T_c$ superconductors is that the
materials are insulators when the conduction band is {\it half} filled. The
charge gap obtained by the optical conductance experiments is about $2eV$,
which is much larger than the anti-ferromagnetic (AF) transition
temperature $T_{\rm AF}\sim 250$K, the superconducting transition
temperature $T_c \sim 100$K, and the spin pseudo-gap scale $\Del \sim
40$meV.\cite{MDL9641,DCN9651,MIM9541}
The insulating property is completely due to the strong correlations
present in the high $T_c$ materials.
Thus the strong correlations are expect to play very important role in
understanding high $T_c$ superconductors.  Many important properties of
high $T_c$ superconductors can be directly linked to the Mott insulator at
half filling, such as (a) the low charge density\cite{OWT8707} and superfluid 
density,\cite{HBM9399}
(b) $T_c$ being proportional to doping 
$T_c\propto x$,\cite{ULS8917,ULL9165,KE9534}
(c) the positive charge carried by the charge carrier,\cite{OWT8707} \etc.

In the spin liquid approach, the strategy is to try to understand the
properties of the high $T_c$ superconductors from the low doping limit. We
first study the spin liquid state at half filling and try to understand the
parent Mott insulator. (In this paper, by spin liquid, we mean a spin state
with translation and spin rotation symmetry.)  At half filling, the charge
excitations can be ignored due to the huge charge gap. Thus we can use a pure
spin model to describe the half filled system. After understand the spin
liquid, we try to understand the dynamics of a few doped holes in the spin
liquid states and to obtain the properties of the high $T_c$ superconductors
at low doping. One advantage of the  spin liquid approach is that experiments
(such as angle resolved photo-emission,\cite{MDL9641,DCN9651,Lo9625,HSW9665}
NMR,\cite{TRH9147}, neutron scattering,\cite{MFY0048,BSF0034,DMH0125} \etc)
suggest that underdoped cuperates have many striking and qualitatively new
properties which are very different from the well known Fermi liquids.  It is
thus easier to approve or disapprove a new theory in the underdoped regime by
studying those qualitatively new properties.

Since the properties of the doped holes (such as their statistics, spin,
effective mass, \etc) are completely determined by the spin correlation in the
parent spin liquids, thus in the spin liquid approach, each possible spin
liquid leads to a possible theory for high $T_c$ superconductors.  Using the
concept of quantum orders, we can say that possible theories for high $T_c$
superconductors in the low doping limits are classified by possible quantum
orders in spin liquids on 2D square lattice.  Thus one way to study high $T_c$
superconductors is to construct all the possible spin liquids that have the
same symmetries as those observed in high $T_c$ superconductors.  Then analyze
the physical properties of those spin liquids with dopings to see which one
actually describes the high $T_c$ superconductor. Although we cannot say that
we have constructed all the symmetric spin liquids, in this paper we have
found a way to construct a large class of symmetric spin liquids.  (Here by
symmetric spin liquids we mean spin liquids with all the lattice symmetries:
translation, rotation, parity, and the time reversal symmetries.) We also find
a way to characterize the quantum orders in those spin liquids via projective
symmetry groups.  This gives us a global picture of possible high $T_c$
theories.  We would like to mention that a particular spin liquid -- the
staggered-flux/$d$-wave state\cite{AM8874,KL8842} -- may be important for high
$T_c$ superconductors. Such a state can explain\cite{WLsu2,RWspec} the highly
unusual pseudo-gap metallic state found in underdoped
cuperates,\cite{MDL9641,DCN9651,Lo9625,HSW9665} as well as the $d$-wave
superconducting state\cite{KL8842}.

The spin liquids constructed in this paper can be divided into four class:
(a) Rigid spin liquid -- spinons are fully gaped and may have bosonic,
fermionic, or fractional statistics, (b) Fermi spin liquid -- spinons are
gapless and are described by a Fermi liquid theory, (c) Algebraic spin
liquid -- spinons are gapless,
but they are not described by free fermionic/bosonic quasiparticles.  (d)
Bose spin liquid -- low lying gapless excitations are described by a free
boson theory.  We find some of the constructed spin liquids are stable and
represent stable quantum phases, while others are unstable at low energies
due to long range interactions caused by gauge fluctuations.  The algebraic
spin liquids and Fermi spin liquids are interesting since they can be
stable despite their gapless excitations.  Those gapless excitations are
not protected by symmetries.  This is particularly striking for algebraic
spin liquids since their gapless excitations interact down to zero energy
and the states are still stable.  We propose that it is the quantum orders
that protect the gapless excitations and ensure the stability of the
algebraic spin liquids and Fermi spin liquids.

We would like to point out that both stable and unstable spin liquids may
be important for understanding high $T_c$ superconductors. Although at zero
temperature high $T_c$ superconductors are always described stable quantum
states, some important states of high $T_c$ superconductors, such as the
pseudo-gap metallic state for underdoped samples, are observed only at
finite temperatures. Such finite temperature states may correspond to
(doped) unstable spin liquids, such as staggered flux state.
Thus even unstable spin liquids can be useful in understanding finite
temperature metallic states.

There are many different approach to spin liquids. In addition to the
slave-boson approach,\cite{BZA8773,BA8880,AM8874,KL8842,SHF8868,%
AZH8845,DFM8826,WWZcsp,Wsrvb,LN9221,MF9400,WLsu2} spin liquids has been
studied using slave-fermion/$\si$-model
approach,\cite{AA8816,RS8994,RS9173,S9277,WSC9794,SP0114} quantum dimer
model,\cite{KRS8765,RK8876,IL8941,FK9025,MS0181} and various numerical
methods.\cite{CJZ9062,MLB9964,CBP0104,KI0148} In particular, the numerical
results and recent experimental results\cite{CTT0135} strongly support the
existence of quantum spin liquids in some frustrated systems.  A 3D quantum
orbital liquid was also proposed to exist in $L_aT_iO_3$.\cite{KM0050}

However, I must point out that there is no generally accepted numerical
results yet that prove the existence of spin liquids with odd number of
electron per unit cell for spin-1/2 systems, despite intensive search in last
ten years.  But it is my faith that spin liquids (with odd number of electron
per unit cell) exist in spin-1/2 systems.  For more general systems, spin
liquids do exist. Read and Sachdev\cite{RS9173} found stable spin liquids in a
$Sp(N)$ model in large $N$ limit. The spin-1/2 model studied in this paper can
be easily generalized to $SU(N)$ model with $N/2$ fermions per
site.\cite{AM8874,Wlight} In the large $N$ limit, one can easily construct
various Hamiltonians whose ground states realize various $U(1)$ and $Z_2$ spin
liquids constructed in this paper.\cite{Wtoap} The quantum orders in those
large-$N$ spin liquids can be described by the methods introduced in this
paper. Thus, despite the uncertainty about the existence of spin-1/2 spin
liquids, the methods and the results presented in this paper are not about
(possibly) non-existing ``ghost states''. Those methods and results apply, at
least, to certain large-$N$ systems. In short, non-trivial quantum orders
exist in theory. We just need to find them in nature. (In fact, our vacuum is
likely to be a state with a non-trivial quantum order, due to the fact that
light exists.\cite{Wlight}) Knowing the existence of spin liquids in large-$N$
systems, it is not such a big leap to go one step further to speculate that
spin liquids exist for spin-1/2 systems.

\subsection{ Spin-charge separation in (doped) spin liquids}
\label{sec:spnchg}
  
Spin-charge separation and the associated gauge theory in spin liquids (and
in doped spin liquids) are very important concepts in our attempt to
understand the properties of high $T_c$
superconductors.\cite{A8796,BZA8773,BA8880,IL8988,LN9221} However, the exact
meaning of spin-charge separation is different for different researchers.
The term ``spin-charge separation'' has at lease in two different
interpretations.  In the first interpretation, the term means that it is
better to introduce separate spinons (a neutral spin-1/2 excitation) and
holons (a spinless excitation with unit charge) to understand the dynamical
properties of high $T_c$ superconductors, instead of using the original
electrons. However, there may be long range interaction (possibly, even
confining interactions at long distance) between the spinons and holons,
and the spinons and holons may not be well defined quasiparticles.  We will
call this interpretation pseudo spin-charge separation.  The algebraic spin
liquids have the pseudo spin-charge separation. The essence of the pseudo
spin-charge separation is not that spin and charge separate.  {\em The
pseudo spin-charge separation is simply another way to say that the gapless
excitations cannot be described by free fermions or bosons.} In the second
interpretation, the term ``spin-charge separation'' means that there are
only {\em short ranged} interactions between the spinons and holons.  The
spinons and holons are well defined quasiparticles at least in the dilute
limit or at low energies.  We will call the second interpretation the true
spin-charge separation.  The rigid spin liquids and the Fermi spin liquids
have the true spin-charge separation.

Electron operator is not a good starting point to describe states
with pseudo spin-charge separation or true spin-charge separation. To study
those states, we usually rewrite the electron operator as a product of
several other operators.  Those operators are called parton operators.
(The spinon operator and the holon operator are examples of parton
operators).  We then construct mean-field state in the enlarged Hilbert
space of partons.  The gauge structure can be determined as the most
general transformations between the partons that leave the electron operator
unchanged.\cite{Wpc}  After identifying the gauge structure, we can project the
mean-field state onto the physical (\ie the gauge invariant) Hilbert space and
obtain a strongly correlated electron state.  This procedure in its general
form is called
projective construction. It is a generalization of the slave-boson
approach.\cite{BZA8773,BA8880,AZH8845,DFM8826,Wsrvb,MF9400,WLsu2} The general 
projective construction and the related gauge structure has been
discussed in detail for quantum Hall states.\cite{Wpc} Now we see a third
(but technical) meaning of spin-charge separation: to construct a strongly
correlated electron state, we need to use partons and projective
construction. The resulting effective theory naturally contains a gauge
structure.

Although, it is not clear which interpretation of spin-charge separation
actually applies to high $T_c$ superconductors, the possibility of true
spin-charge separation in an electron system is very interesting.  The first
concrete example of true spin-charge separation in 2D is given by the chiral
spin liquid state,\cite{KL8795,WWZcsp} where the gauge interaction between the
spinons and holons becomes short-ranged due to a Chern-Simons term. The
Chern-Simons term breaks time reversal symmetry and gives the spinons and
holons a fractional statistics. Later in 1991, it was realized that there is
another way to make the gauge interaction short-ranged through the
Anderson-Higgs mechanism.\cite{RS9173,Wsrvb}  This led to a mean-field
theory\cite{Wsrvb,MF9400} of the short-ranged Resonating Valence Bound (RVB)
state\cite{KRS8765,RK8876} conjectured earlier.  We will call such a state
$Z_2$ spin liquid state, to stress the {\em unconfined} $Z_2$ gauge field that
appears in the {\em low energy} effective theory of those spin liquids.  (See
remarks at the end of this section. We also note that the $Z_2$ spin liquids
studied in \Ref{RS9173} all break the $90^\circ$ rotation symmetry and are
different from the short-ranged RVB state studied
\Ref{KRS8765,RK8876,Wsrvb,MF9400}.) Since the $Z_2$ gauge fluctuations are
weak and are not confining, the spinons and holons have only short ranged
interactions in the $Z_2$ spin liquid state. The $Z_2$ spin liquid state also
contains a $Z_2$ vortex-like excitation.\cite{RC8933,Wsrvb} The spinons and
holons can be bosons or fermions depending on if they are bound with the $Z_2$
vortex.

Recently, the true spin-charge separation, the $Z_2$ gauge structure and
the $Z_2$ vortex excitations were also proposed in a study of quantum
disordered superconducting state in a continuum model\cite{BFN9833} and in
a $Z_2$ slave-boson approach\cite{SF0050}.  The resulting liquid state
(which was named nodal liquid)  has all the novel properties of $Z_2$ spin
liquid state such as the $Z_2$ gauge structure and the $Z_2$ vortex
excitation (which was named vison).  From the point of view of universality
class, the nodal liquid is one kind of $Z_2$ spin liquids.  However, 
the particular 
$Z_2$ spin liquid studied in \Ref{Wsrvb,MF9400} and the nodal liquid
are two different $Z_2$ spin liquids,
despite they have the same symmetry.  The spinons in the first $Z_2$ spin
liquid have a finite energy gap while the spinons in the nodal liquid are
gapless and have a Dirac-like dispersion.  In this paper, we will use the
projective construction to obtain more general spin liquids.  We find
that one can construct hundreds of different $Z_2$ spin liquids.  Some $Z_2$
spin liquids have finite energy gaps, while others are gapless.  Among
those gapless $Z_2$ spin liquids, some have finite Fermi surfaces while
others have only Fermi points. The spinons near the Fermi points can have
linear $E(\v k) \propto |\v k|$ or quadratic $E(\v k) \propto \v k^2$
dispersions.  We find there are more than one $Z_2$ spin liquids whose
spinons have a massless Dirac-like dispersion.  Those $Z_2$ spin liquids
have the same symmetry but different quantum orders. Their ansatz are
give by \Eq{Z2lCs}, \Eq{Z2lE}, \Eq{Z2lG}, \etc. 

Both chiral spin liquid and $Z_2$ spin liquid states are Mott insulators
with one electron per unit cell if not doped. Their internal structures are
characterized by a new kind of order -- topological order, if they are
gapped or if the gapless sector decouples.  Topological order is not
related to any symmetries and has no (local) order parameters.  Thus, the
topological order is robust against all perturbations that can break any
symmetries (including random perturbations that break translation
symmetry).\cite{Wrig,Wtoprev} (This point was also emphasized in
\Ref{SF0192} recently.) Even though there are no order parameters to
characterize them, the topological orders can be characterized by other
measurable quantum numbers, such as ground state degeneracy in compact
space as proposed in \Ref{Wrig,Wtoprev}.  Recently, \Ref{SF0192} introduced
a very clever experiment to test the ground state degeneracy associated
with the non-trivial topological orders.  In addition to ground state
degeneracy, there are other practical ways to detect topological orders.
For example, the excitations on top of a topologically ordered state can be
defects of the under lying topological order, which usually leads to
unusual statistics for those excitations. Measuring the statistics of those
excitations also allow us to measure topological orders.

The concept of topological order and quantum order are very important in
understanding quantum spin liquids (or any other strongly correlated
quantum liquids). In this paper we are going to construct hundreds of
different spin liquids. Those spin liquids all have the same symmetry. To
understand those spin liquids, we need to first learn how to characterize
those spin liquids.  Those states break no symmetries and hence have no
order parameters. One would get into a wrong track if trying to find an
order parameter to characterize the spin liquids. We need to use a
completely new way, such as topological orders and quantum orders, to
characterize those states. 

In addition to the above $Z_2$ spin liquids, in this paper we will also
study many other spin liquids with different low energy gauge structures,
such as $U(1)$ and $SU(2)$ gauge structures. We will use the terms $Z_2$
spin liquids, $U(1)$ spin liquids, and $SU(2)$ spin liquids to describe
them. We would like to stress that $Z_2$, $U(1)$, and $SU(2)$ here are
gauge groups that appear in the low energy effective theories of those spin
liquids. They should not be confused with the $Z_2$, $U(1)$, and $SU(2)$
gauge group in slave-boson approach or other theories of the projective
construction. The latter are high energy gauge groups. The high energy
gauge groups have nothing to do with the low energy gauge groups.
A high energy $Z_2$ gauge theory (or a $Z_2$ slave-boson approach) can have
a low energy effective theory that contains $SU(2)$, $U(1)$ or $Z_2$
gauge fluctuations.  Even the $t$-$J$ model, which has no gauge structure
at lattice scale, can have a low energy effective theory that contains
$SU(2)$, $U(1)$ or $Z_2$ gauge fluctuations.  The spin liquids studied in
this paper all contain some kind of low energy gauge fluctuations. Despite
their different low energy gauge groups, all those spin liquids can be
constructed from any one of $SU(2)$, $U(1)$, or $Z_2$ slave-boson
approaches.  After all, all those slave-boson approaches describe the same
$t$-$J$ model and are equivalent to each other.  In short, the high energy
gauge group is related to the way in which we write down the Hamiltonian,
while the low energy gauge group is a property of ground state.  
Thus we should not regard $Z_2$ spin liquids as the spin liquids
constructed using $Z_2$ slave-boson approach.  A $Z_2$ spin liquid can be
constructed from the $U(1)$ or $SU(2)$ slave-boson approaches as well.  A
precise mathematical definition of the low energy gauge group will be given
in section \ref{psg}.

\subsection{Organization}

In this paper we will use the method outlined in \Ref{Wsrvb,MF9400} to study
gauge structures in various spin liquid states.  In section \ref{revSU2} we
review $SU(2)$ mean-field theory of spin liquids.  In section \ref{trnANS}, we
construct simple symmetric spin liquids using translationally invariant
ansatz.  In section \ref{qois}, projective symmetry group is introduced to
characterize quantum orders in spin liquids.  In section \ref{simp}, we study
the transition between different symmetric spin liquids, using the results
obtained in appendix B, where we find a way to construct all the symmetric spin
liquids in the neighborhood of some well known spin liquids. We also study the
spinon spectrum to gain some intuitive understanding on the properties of the
spin liquids.  Using the relation between two-spinon spectrum and quantum
order, we propose, in section \ref{sec:pmQO}, a practical way to use neutron
scattering to measure quantum orders.  We study the stability of Fermi spin
liquids and algebraic spin liquids in section \ref{stb}. We find that both
Fermi spin liquids and algebraic spin liquids can exist as zero temperature
phases. This is particularly striking for algebraic spin liquids since their
gapless excitations interacts even at lowest energies and there are no free
fermionic/bosonic quasiparticle excitations at low energies.  We show how
quantum order can protect gapless excitations.  Appendix A contains a more
detailed discussion on projective symmetry group, and a classification of
$Z_2$, $U(1)$ and $SU(2)$ spin liquids using the projective symmetry group.
Section \ref{sum} summarizes the main results of the paper.

\section{Projective construction of 2D spin liquids --
a review of SU(2) slave-boson approach }
\label{revSU2}

In this section, we are going to use projective construction to
construct 2D spin liquids.  We are going to review a particular projective
construction, namely the $SU(2)$ slave-boson
approach.\cite{BZA8773,BA8880,AZH8845,DFM8826,Wsrvb,MF9400,WLsu2} The gauge
structure discovered by Baskaran and Anderson\cite{BA8880} in the
slave-boson approach plays a crucial role in our understanding of strongly
correlated spin liquids.

We will concentrate
on the spin liquid states of a pure spin-1/2 model on a 2D square lattice
\begin{equation}
 H_{spin} =\sum_{<\v i \v j>} J_{\v i \v j}\v S_{\v i}\cdot \v S_{\v j} + ...
\label{sl2.1}
\end{equation}
where the summation is over different links (\ie $\<\v i \v j\>$ and $\<\v j
\v i\>$ are regarded as the same) and $...$ represents possible terms which
contain three or more spin operators.
Those terms are needed in order for many exotic spin liquid
states introduced in this paper to become the ground state.
To obtain the mean-field ground state of the spin liquids, we introduce
fermionic parton operators 
$f_{\v i\al}, \ \al=1,2$ which carries spin $1/2$ and no charge.
The spin operator $\v S_{\v i}$ is represented as
\begin{equation} 
\v S_{\v i}=\frac12 f^\dag_{\v i\al}\v \si_{\al\bt}f_{\v i\bt}
\label{sl2.2}
\end{equation}
In terms of the fermion operators the Hamiltonian \Eq{sl2.1} can be
rewritten as
\begin{equation} H=\sum_{\< \v i \v j\>} -\frac{1}{2} J_{\v i \v j}
\left( 
 f^\dag_{\v i\al}f_{\v j\al}f^\dag_{\v j\bt}f_{\v i\bt}
+\frac12 f^\dag_{\v i\al}f_{\v i\al}f^\dag_{\v j\bt}f_{\v j\bt}
\right)
\label{sl2.3}
\end{equation}
Here we have used $\v \si_{\al\bt} \cdot \v \si_{\al'\bt'} = 2 \del_{\al\bt'}
\del_{\al'\bt} - \del_{\al\bt} \del_{\al'\bt'} $.  We also added proper
constant terms $\sum_{\v i} f^\dag_{\v i\al}f_{\v i\al}$ and $\sum_{\<\v i \v
j\>} f^\dag_{\v i\al}f_{\v i\al}f^\dag_{\v j\bt}f_{\v j\bt}$ to get the above
form.  Notice that the Hilbert space of \Eq{sl2.3} is generated by the parton
operators $f_\al$ and is larger than that of
\Eq{sl2.1}.  The equivalence between \Eq{sl2.1} and \Eq{sl2.3} is valid only
in the subspace where there is exactly one fermion per site. Therefore
to use \Eq{sl2.3} to describe the spin state we need to impose the
constraint\cite{BZA8773,BA8880}
\begin{equation} f^\dag_{\v i\al}f_{\v i\al}=1,\ \ \
 f_{\v i\al} f_{\v i\bt}\eps_{\al\bt} =0
\label{sl2.4}\end{equation}
The second constraint is actually a consequence of the first one.

A mean-field ground state at ``zeroth'' order is obtained by
making the following
approximations. First we replace constraint \Eq{sl2.4} by its
ground-state average
\begin{equation} 
\< f^\dag_{\v i\al}f_{\v i\al}\> =1,\ \ \ 
\< f_{\v i\al} f_{\v i\bt}\eps_{\al\bt}\> =0
\label{sl2.5}\end{equation}
Such a constraint can be enforced by including a {\it site dependent} and
time independent
Lagrangian multiplier: $a_0^{l}(i)( f^\dag_{\v i\al}f_{\v i\al}-1)$, $l=1,2,3$,
 in the Hamiltonian.
At the zeroth order we ignore the fluctuations (\ie the time dependence)
of $a_0^{l}$. If we included the
fluctuations of $a_0^{l}$, the constraint \Eq{sl2.5} would become the
original constraint \Eq{sl2.4}.\cite{BZA8773,BA8880,AZH8845,DFM8826}
Second we replace the operators $f^\dag_{\v i\al}f_{\v j\bt}$ and
$f_{\v i\al} f_{\v i\bt}$
by their ground-state expectations value 
\begin{align}
\eta_{\v i \v j} \eps_{\al\bt}=&  - 2 \< f_{\v i \alpha}~f_{\v j \bt} \> ,
& \eta_{\v i \v j} =& \eta_{\v j \v i}
\nonumber \\
\chi_{\v i \v j}\del_{\al\bt}=& 
2 \< f_{\v i\al}^{\dag} f_{\v j \bt} \>,
& \chi_{\v i \v j} =&  \chi_{\v j \v i}^\dag
\label{sl2a.4}
\end{align}
again ignoring their fluctuations.
In this way we obtain the zeroth order mean-field Hamiltonian:
\begin{align}
\label{sl2.6}
& H_{mean} \nonumber\\
=&  \sum_{\<\v i\v j\>} -\frac{3}{8} J_{\v i \v j} 
\left[(\chi_{\v j \v i}f^\dag_{\v i\al}f_{\v j\al}  + \eta_{\v i \v j}
f^\dag_{ i \alpha} f^\dag_{ j \bt}~\eps_{\alpha \bt} + h.c)
\right. \nonumber\\
& \ \ \ \ \ \ \ \left. - |\chi_{\v i \v j}|^2  -|\eta_{\v i \v j}|^2 \right]
\\
& +\sum_{\v i} \left[ a_0^3(f_{\v i\al}^\dag f_{\v i\al} -1)+[(a_0^1+i a_0^2)
 f_{\v i\al} f_{\v i\bt}\eps_{\al\bt}+h.c.]\right] \nonumber 
\end{align}
$\chi_{\v i \v j}$ and $\eta_{\v i \v j}$
in \Eq{sl2.6} must satisfy the self consistency condition \Eq{sl2a.4}
and the site dependent fields $a_0^{l}(i)$ are chosen such that \Eq{sl2.5}
is satisfied by the mean-field ground state. 
Such $\chi_{\v i \v j}$, $\eta_{\v i \v j}$ and $a_0^l$ give us a
mean-field solution.
The fluctuations in $\chi_{\v i \v j}$, $\eta_{\v i \v j}$ and $a_0^l(i)$ 
describe the collective excitations above the mean-field ground state.

The Hamiltonian \Eq{sl2.6} and the constraints \Eq{sl2.4} have
a local $SU(2)$ symmetry.\cite{AZH8845,DFM8826}  The local
$SU(2)$ symmetry becomes explicit if we introduce doublet
\begin{equation}
\psi = \bpm \psi_1 \\ \psi_2 \epm = \bpm f_\up \\ f_\down^\dag \epm 
\label{sl2a.6}
\end{equation}
and matrix
\begin{equation}
U_{\v i \v j }= \bpm \chi_{\v i \v j}^\dag &\eta_{\v i \v j} \\
               \eta_{\v i \v j}^\dag & -\chi_{\v i \v j}
                                        \epm
=U_{\v j \v i}^\dag
\label{sl2a.7}
\end{equation}
Using \Eq{sl2a.6} and \Eq{sl2a.7} we can rewrite \Eq{sl2.5} and \Eq{sl2.6} as:
\begin{equation}
\vev{\psi_{\v i}^\dag \tau^l \psi_{\v i}} = 0
\end{equation}
\begin{align}
H_{mean} =& \sum_{\<\v i \v j\>} \frac{3}{8}
 J_{\v i \v j} \left[ 
\frac12 \Tr (U_{\v i \v j}^\dag~U_{\v i \v j}) -(\psi^\dag_ i
U_{\v i \v j}
\psi_ j +~h.c.)\right] \nonumber\\ 
& +\sum_{\v i}  a_0^l \psi_{\v i}^\dag \tau^l \psi_{\v i} 
\label{sl2a.8}
\end{align}
where $\tau^l, \ l=1,2,3$ are the Pauli matrices.
{}From \Eq{sl2a.8} we can see clearly that the Hamiltonian is invariant
under a local $SU(2)$ transformation $W(\v i)$:
\begin{eqnarray}
\psi_{\v i} &\to & W(\v i)~\psi_{\v i}  \nonumber \\
U_{\v i \v j} &\to &W(\v i)~U_{\v i \v j}~W^\dag(\v j)    
\label{sl2a.9}
\end{eqnarray}
The $SU(2)$ gauge structure is originated from \Eq{sl2.2}.  The $SU(2)$ is the
most general transformation between the partons that leave the physical
spin operator unchanged. Thus once we write down the parton expression of
the spin operator \Eq{sl2.2}, the gauge structure of the theory is
determined.\cite{Wpc} (The $SU(2)$ gauge structure discussed here is a high
energy gauge structure.)

We note that both components of $\psi$ carry spin-up. Thus the
spin-rotation symmetry is not explicit in our formalism and it is hard to
tell if \Eq{sl2a.8} describes a spin-rotation invariant state or not. In
fact, for a general $U_{\v i \v j}$ satisfying $U_{\v i \v j}=U_{\v j \v
i}^\dag$, \Eq{sl2a.8} may not describe a spin-rotation invariant state.
However, if $U_{\v i \v j}$ has a form
\begin{eqnarray}
\label{UuW}
U_{\v i \v j} &=& i \rho_{\v i \v j} W_{\v i \v j}, \nonumber\\
\rho_{\v i \v j} &=& \hbox{real number}, \nonumber\\
W_{\v i \v j} &\in& SU(2),
\end{eqnarray}
then \Eq{sl2a.8} will describe a spin-rotation invariant state.  This is
because the above $U_{\v i \v j}$ can be rewritten in a form \Eq{sl2a.7}.
In this case \Eq{sl2a.8} can be rewritten as \Eq{sl2.6} where the
spin-rotation invariance is explicit.

To obtain the mean-field theory, we have enlarged the Hilbert space.
Because of this, the mean-field theory is not even qualitatively correct.
Let $|\Psi_{mean}^{(U_{\v i \v j})}\>$ be the ground state of the
Hamiltonian \Eq{sl2a.8} with energy $E(U_{\v i\v j},a^l_{\v i}\tau^l)$.
It is clear that the mean-field ground state is not
even a valid wave-function for the spin system since it may not have one
fermion per site.  Thus it is very important to include fluctuations of
$a_0^{l}$ to enforce one-fermion-per-site constraint. With this
understanding, we may obtain a valid wave-function of the spin system
$\Psi_{spin}(\{\al_{\v i}\})$ by projecting the mean-field state to the
subspace of one-fermion-per-site:
\begin{equation}
\Psi_{spin}(\{\al_{\v i}\}) = \<0|\prod_{\v i} f_{\v i\al_{\v i}}
|\Psi_{mean}^{(U_{\v i \v j})}\> .
\label{PsiPsi}
\end{equation}

Now the local $SU(2)$ transformation \Eq{sl2a.9} can have a very physical
meaning: $|\Psi_{mean}^{(U_{\v i \v j})}\>$ and $|\Psi_{mean}^{(W(\v
i)U_{\v i \v j}W^\dag(\v j))}\>$ give rise to the same spin wave-function
after projection:
\begin{equation}
\label{UUp}
  \<0|\prod_{\v i} f_{\v i\al_{\v i}} |\Psi_{mean}^{(U_{\v i \v j})}\> 
= \<0|\prod_{\v i} f_{\v i\al_{\v i}} |\Psi_{mean}^{(W(\v i)U_{\v i \v j}
W^\dag(\v j))}\>
\end{equation}
Thus $U_{\v i \v j}$ and $U'_{\v i \v j}=W(\v i)U_{\v i \v j}W^\dag(\v j)$
are just two different labels which label the {\em same physical state}.
Within the mean-field theory, a local $SU(2)$ transformation changes a
mean-field state  $|\Psi_{mean}^{(U_{\v i \v j})}\>$ to a different
mean-field state  $|\Psi_{mean}^{(U'_{\v i \v j})}\>$. If the two
mean-field states always have the same physical properties, the system has
a local $SU(2)$ {\em symmetry}. However, after projection, the physical
spin quantum state described by wave-function $ \Psi_{spin}(\{\al_{\v
i}\})$ is invariant under the local $SU(2)$ transformation. A  local
$SU(2)$ transformation just transforms one label, $U_{\v i \v j}$, of a
physical spin state to another label, $U'_{\v i \v j}$, which labels the
exactly the same physical state. Thus after projection, local $SU(2)$
transformations become gauge transformations.  The fact that $U_{\v i \v
j}$ and $U'_{\v i \v j}$ label the same physical spin state creates a
interesting situation when we consider the fluctuations of $U_{\v i \v j}$
around a mean-field solution -- some fluctuations of $U_{\v i \v j}$ do not
change the physical state and are unphysical. Those fluctuations are called
the pure gauge fluctuations.

The above discussion also indicates that in order for the mean-field theory
to make any sense, we must at least include the $SU(2)$ gauge (or other
gauge) fluctuations described by $a_0^l$ and $W_{\v i \v j}$ in \Eq{UuW},
so that the $SU(2)$ gauge structure of the mean-field theory is revealed
and the physical spin state is obtained.  We will include the gauge
fluctuations to the zeroth-order mean-field theory. The new theory will be
called the first order mean-field theory. It is this first order mean-field
theory that represents a proper low energy effective theory of the spin
liquid.

Here, we would like make a remark about ``gauge symmetry'' and ``gauge
symmetry breaking''. We see that two ansatz $U_{\v i \v j}$ and $U'_{\v i
\v j}=W(\v i)U_{\v i \v j}W^\dag(\v j)$ have the same physical properties.
This property is usually  called the ``gauge symmetry''.  However, from the
above discussion, we see that the ``gauge symmetry'' is {\em not} a
symmetry. A symmetry is about two {\em different} states having the same
properties. $U_{\v i \v j}$ and $U'_{\v i \v j}$ are just two labels that
label the same state, and the same state always have the same properties.
We do not usually call the same state having the same properties a
symmetry.  Because the same state alway have the same properties, the
``gauge symmetry'' can never be broken. It is very misleading to call the
Anderson-Higgs mechanism ``gauge symmetry breaking''.  With this
understanding, we see that a superconductor is fundamentally different from
a superfluid. A superfluid is characterized by $U(1)$ symmetry breaking,
while a superconductor has no symmetry breaking once we include the
dynamical electromagnetic gauge fluctuations.  A superconductor is actually
the first topologically ordered state observed in experiments,\cite{Wtopcs}
which has no symmetry breaking, no long range order, and no (local) order
parameter.  However, when the speed of light $c=\infty$, a superconductor
becomes similar to a superfluid and is characterized by $U(1)$ symmetry
breaking.  

The relation between the mean-field state and the physical spin wave function
\Eq{PsiPsi} allows us to construct transformation of the physical spin
wave-function from the mean-field ansatz.  For example the mean-field state
$|\Psi_{mean}^{(U^\prime_{\v i \v j})}\>$ with $U^\prime_{\v i \v j} = U_{\v
i-\v l, \v j-\v l}$ produces a physical spin wave-function which is translated
by a distance $\v l$ from the physical spin wave-function produced by
$|\Psi_{mean}^{(U_{\v i \v j})}\>$.  The physical state is translationally
symmetric if and only if the translated ansatz $U^\prime_{\v i \v j}$ and the
original ansatz $U_{\v i \v j}$ are gauge equivalent (it does not require
$U^\prime_{\v i \v j}= U_{\v i \v j}$). We see that the gauge structure can
complicates our analysis of symmetries, since the physical spin wave-function
$ \Psi_{spin}(\{\al_{\v i}\}) $ may has more symmetries than the mean-field
state $|\Psi_{mean}^{(U_{\v i \v j})}\>$ before projection.

Let us discuss time reversal symmetry 
in more detail. 
A quantum system described by
\begin{equation}
 i\hbar \prt_t \Psi(t) = H\Psi(t)
\end{equation}
has a time reversal symmetry if $\Psi(t)$ satisfying the equation of motion
implies that $\Psi^*(-t)$ also satisfying the equation of motion. This
requires that $H=H^*$. We see that, for time reversal symmetric system, if
$\Psi$ is an eigenstate, then $\Psi^*$ will be an eigenstate with the same
energy.  

For our system, the time reversal symmetry means that if the
mean-field wave function $\Psi_{mean}^{(U_{\v i\v j}, a^l_{\v i}\tau^l)}$ is a
mean-field ground state wave function for ansatz $(U_{\v i\v j},a^l_{\v
i}\tau^l)$, then $\left(\Psi_{mean}^{(U_{\v i\v j},a^l_{\v
i}\tau^l)}\right)^*$ will be the mean-field ground state wave function
for ansatz $(U^*_{\v i\v j},a^l_{\v i}(\tau^l)^*)$.  That is
\begin{equation}
\left(\Psi_{mean}^{(U_{\v i\v j},a^l_{\v i}\tau^l)}\right)^*
= \Psi_{mean}^{(U^*_{\v i\v j},a^l_{\v i}(\tau^l)^*)}
\end{equation}
For a system with time reversal symmetry, the mean-field energy
$E(U_{\v i\v j},a^l_{\v i}\tau^l)$ satisfies
\begin{equation}
E(U_{\v i\v j},a^l_{\v i}\tau^l)
=
E(U^*_{\v i\v j},a^l_{\v i}(\tau^l)^*)
\end{equation}
Thus if an ansatz
$(U_{\v i\v j}, a_{\v i}^l\tau^l)$ 
is a mean-field solution, then
$(U^*_{\v i\v j}, a_{\v i}^l(\tau^l)^*)$ is also a mean-field solution with
the same mean-field energy. 

From the above discussion, we see that
under the time reversal transformation, 
the ansatz 
transforms as
\begin{eqnarray}
 U_{\v i \v j} &\to & U_{\v i \v j}^\prime =  
 (-i\tau^2) U_{\v i \v j}^* (i\tau^2) = -U_{\v i\v j} , \nonumber\\
a_{\v i}^l\tau^l &\to & a_{\v i}^{\prime l}\tau^l =
 (-i\tau^2) (a_{\v i}^l\tau^l)^* (i\tau^2)= - a_{\v i}^l\tau^l  .
\end{eqnarray}
Note here we have included an additional $SU(2)$ gauge transformation $W_{\v
i} = -i\tau^2$.  We also note that under the time reversal transformation, the
loop operator transforms as $P_C=e^{i\th+i\th^l\tau^l} \to (-i\tau^2) P_C^*
(i\tau^2)= e^{-i\th+i\th^l\tau^l}$. We see that the $U(1)$ flux changes the
sign while the $SU(2)$ flux is not changed.

Before ending this review section, we would like to point out that the
mean-field ansatz of the spin liquids $U_{\v i \v j}$ can be divided into
two classes: unfrustrated ansatz where $U_{\v i \v j}$ only link an even
lattice site to an odd lattice site and frustrated ansatz where $U_{\v i \v
j}$ are nonzero between two even sites and/or two odd sites.
An unfrustrated ansatz has only pure $SU(2)$ flux through each plaquette,
while an frustrated ansatz has $U(1)$ flux of multiple of $\pi/2$ through
some plaquettes in addition to the $SU(2)$ flux.

\section{Spin liquids from translationally invariant ansatz} \label{trnANS}

In this section, we will study many simple examples of spin
liquids and their ansatz. Through those simple examples, we gain some
understandings on what kind of spin liquids are possible. Those
understandings help us to develop the characterization and classification
of spin liquids using projective symmetry group.

Using the above $SU(2)$ projective construction, one can construct many
spin liquid states.  To limit ourselves, we will concentrate on spin
liquids with translation and $90^{\circ}$ rotation symmetries.  Although a
mean-field ansatz with translation and rotation invariance always generate
a spin liquid with translation and rotation symmetries, a mean-field ansatz
without those invariances can also generate a spin liquid with those
symmetries.\footnote{ In this paper we will distinguish the invariance of
an ansatz and the symmetry of an ansatz. We say an ansatz has a translation
invariance when the ansatz itself does not change under translation.  We
say an ansatz has a translation symmetry when the physical spin wave
function obtained from the ansatz has a translation symmetry.} Because of
this, it is quite difficult to construct all the translation and rotation
symmetric spin liquids.  In this section we will consider a simpler
problem.  We will limit ourselves to spin liquids generated from
translationally invariant ansatz:
\begin{equation}
U_{\v i+\v l, \v j +\v l} = U_{\v i\v j},\ \ \ \
a_0^l(\v i) = a_0^l
\end{equation}
In this case, we only need to find the conditions under which the above
ansatz can give rise to a rotationally symmetric spin liquid.  First let us
introduce $u_{\v i\v j}$:
\begin{equation}
\frac{3}{8}J_{\v i\v j}U_{\v i\v j} =u_{\v i\v j}
\end{equation}
For translationally invariant ansatz, we can introduce a short-hand notation:
\begin{equation}
u_{\v i\v j} = u_{-\v i +\v j}^\mu \tau^\mu \equiv u_{-\v i +\v j}
\end{equation}
where $u_{\v l}^{1,2,3}$ are real, $u_{\v l}^0$ is imaginary, 
$\tau^0$ is the identity matrix and
$\tau^{1,2,3}$ are the Pauli matrices.  The fermion spectrum is determined by
Hamiltonian
\begin{align}
 H =& -\sum_{\<\v i\v j\>} \left( 
\psi^\dag_{\v i} u_{\v j-\v i} \psi_{\v j} + h.c. 
\right) 
 +\sum_{\v i}  \psi^\dag_{\v i} a_0^l \tau^l \psi_{\v i}
\end{align}
In $\v k$-space we have
\begin{equation}
 H = -\sum_{\v k} \psi^\dag_{\v k} (u^\mu(\v k)-a_0^\mu) \tau^\mu
\psi_{\v k}
\end{equation}
where $\mu = 0,1,2,3$,
\begin{equation}
 u^\mu(\v k) = \sum_{\v l} u^\mu_{\v l}e^{i\v l \cdot \v k},
\end{equation}
$a_0^0=0$, and $N$ is the total number of site.  The fermion spectrum has
two branches and is given by
\begin{align}
 E_{\pm}(\v k) = & u^0(\v k)\pm E_0(\v k)  \nonumber\\
 E_0(\v k) = & \sqrt{\sum_l (u^l(\v k)-a_0^l)^2}
\end{align}
The constraints can be obtained from $\frac{\prt E_{ground}}{\prt a_0^l}=0$
and have a form
\begin{align}
 & N \<\psi_{\v i}^\dag \tau^l \psi_{\v i}\> \nonumber\\
=&
 \sum_{\v k, E_-(\v k) <0} \frac{u^l(\v k) - a_0^l}{E_0(\v k)} 
-\sum_{\v k, E_+(\v k) <0} \frac{u^l(\v k) - a_0^l}{E_0(\v k)} 
= 0
\label{constraintK}
\end{align}
which allow us to determine $a_0^l$, $l=1,2,3$.  It is interesting to see
that if $u^0_{\v i}=0$ and the ansatz is unfrustrated, then we can simply
choose $a_0^l=0$ to satisfy the mean-field constraints (since $u^\mu(\v k)
= -u^\mu(\v k+(\pi,\pi))$ for unfrustrated ansatz).  Such ansatz always
have time reversal symmetry. This is because $U_{\v i\v j}$ and $- U_{\v
i\v j}$ are gauge equivalent for unfrustrated ansatz.

Now let us study some simple examples. First let us assume that only the
nearest neighbor coupling $u_{\hat{\v x}}$ and $u_{\hat{\v y}}$ are
non-zero.  In order for the ansatz to describe a rotationally symmetric
state, the rotated ansatz must be gauge equivalent to the original ansatz.
One can easily check that the following ansatz has the rotation symmetry 
\begin{eqnarray}
a_0^l&=& 0\nonumber\\
u_{\hat{\v x}} &=& \chi \tau^3 +\eta \tau^1  \nonumber\\
u_{\hat{\v y}} &=& \chi \tau^3 -\eta \tau^1
\label{u-chi-eta}
\end{eqnarray}
since the $90^\circ$ rotation followed by a gauge transformation $W_{\v
i}=i\tau^3$ leave the ansatz unchanged.  The above ansatz also has the time
reversal symmetry, since time reversal transformation $u_{\v i\v j} \to
-u_{\v i\v j}$ followed by a gauge transformation $W_{\v i}=i\tau^2$ leave
the ansatz unchanged.

To understand the gauge fluctuations around the above mean-field state, we
note that the mean-field ansatz may generate non-trivial $SU(2)$ flux through
plaquettes. Those flux may break $SU(2)$ gauge structure down to $U(1)$ or
$Z_2$ gauge structures as discussed in \Ref{Wsrvb,MF9400}. In particular, the
dynamics of the gauge fluctuations in the break down from $SU(2)$ to $Z_2$ has
been discussed in detail in \Ref{MF9400}.  According to \Ref{Wsrvb,MF9400},
the $SU(2)$ flux plays a role of Higgs fields.  A non-trivial $SU(2)$ flux
correspond to a condensation of Higgs fields which can break the gauge
structure and give $SU(2)$ and/or $U(1)$ gauge boson a mass. Thus to
understand the dynamics of the  gauge fluctuations, we need to find the
$SU(2)$ flux.  

The $SU(2)$ flux is defined for loops with a base point. The loop starts and
ends at the base point.  For example, we can consider the following two
loops $C_{1,2}$ with the same base point $\v i$: $C_1 =   \v i \to \v i
+\hat{\v x} \to \v i +\hat{\v x} +\hat{\v y} \to \v i +\hat{\v y} \to \v i
$ and $C_2$ is the 90$^\circ$ rotation of $C_1$: $ C_2 = \v i \to \v i
+\hat{\v y} \to \v i -\hat{\v x} +\hat{\v y} \to \v i -\hat{\v x} \to \v i
$.  The $SU(2)$ flux for the two loops is defined as
\begin{align}
& P_{C_1} \equiv 
u_{\v i,\v i+\hat{\v y}}
u_{\v i+\hat{\v y},\v i+\hat{\v x}+\hat{\v y}}
u_{\v i+\hat{\v x}+\hat{\v y},\v i+\hat{\v x}}
u_{\v i+\hat{\v x},\v i} =
u_{\hat{\v y}}^\dag   
u_{\hat{\v x}}^\dag 
u_{\hat{\v y}} 
u_{\hat{\v x}}
\nonumber\\
& P_{C_2} \equiv 
u_{\v i,\v i-\hat{\v x}}
u_{\v i-\hat{\v x},\v i-\hat{\v x}+\hat{\v y}}
u_{\v i-\hat{\v x}+\hat{\v y},\v i+\hat{\v y}}
u_{\v i+\hat{\v y},\v i} =
u_{\hat{\v x}}
u_{\hat{\v y}}^\dag 
u_{\hat{\v x}}^\dag 
u_{\hat{\v y}} 
\label{PC12}
\end{align}

As discussed in \Ref{Wsrvb,MF9400}, if the $SU(2)$ flux $P_C$ for all loops are
trivial: $P_C\propto \tau^0$, then the $SU(2)$ gauge structure is unbroken.
This is the case when $\chi=\eta$ or when $\eta=0$ in the above ansatz
\Eq{u-chi-eta}.  The spinon in the spin liquid described by $\eta=0$ has a
large Fermi surface.  We will call this state $SU(2)$-gapless state (This
state was called uniform RVB state  in literature). The state with
$\chi=\eta$ has gapless spinons only at isolated $\v k$ points. We will
call such a state $SU(2)$-linear state to stress the linear dispersion
$E\propto |\v k|$ near the Fermi points.  (Such a state was called the
$\pi$-flux state  in literature).  The low energy effective theory for the
$SU(2)$-linear state is described by massless Dirac fermions (the spinons)
coupled to a $SU(2)$ gauge field.

After proper gauge transformations, the $SU(2)$-gapless ansatz can be
rewritten as
\begin{eqnarray}
\label{SU2gl}
u_{\hat{\v x}} &=& i\chi \nonumber\\
u_{\hat{\v y}} &=& i\chi 
\end{eqnarray}
and the $SU(2)$-linear ansatz as
\begin{eqnarray}
\label{SU2lA}
u_{\v i,\v i+\hat{\v x}} &=& i\chi \nonumber\\
u_{\v i,\v i+\hat{\v y}} &=& i(-)^{i_x} \chi 
\end{eqnarray}
In these form, the $SU(2)$ gauge structure is explicit since $u_{\v i \v
j}\propto i\tau^0$.  Here we would also like to mention that under the
projective-symmetry-group classification, the $SU(2)$-gapless ansatz
\Eq{SU2gl} is labeled by SU2A$n 0$ and the $SU(2)$-linear ansatz
\Eq{SU2lA} by SU2B$n0$ (see \Eq{SU2ASU2B}).

When $\chi \neq \eta \neq 0$, The flux $P_C$ is non trivial. However, $P_C$
commute with $P_{C'}$ as long as the two loops $C$ and $C'$ have the same
base point. In this case the $SU(2)$ gauge structure is broken down to a
$U(1)$ gauge structure.\cite{Wsrvb,MF9400} The gapless spinon still only appear at
isolated $\v k$ points.  We will call such a state $U(1)$-linear state.
(This state was called staggered flux state and/or $d$-wave pairing state
in literature.) After a proper gauge transformation, the $U(1)$-linear
state can also be described by the ansatz
\begin{eqnarray}
\label{U1lA}
&& u_{\v i,\v i+\hat{\v x}} = i\chi  -(-)^{\v i} \eta\tau^3 \nonumber\\
&& u_{\v i,\v i+\hat{\v y}} = i\chi  +(-)^{\v i} \eta\tau^3 
\end{eqnarray}
where the $U(1)$ gauge structure is explicit.  Under the
projective-symmetry-group classification, such a state is labeled by
U1C$n0 1n$ (see \Eq{PSGU1lin} and \ref{U1SU2sec}).  The
low energy effective theory is described by massless Dirac fermions (the
spinons) coupled to a $U(1)$ gauge field.

The above results are all known before. In the following we are going to
study a new class of translation and rotation symmetric ansatz, which has a
form
\begin{align}
\label{Z2glA}
 a_0^l =& 0 \nonumber\\
 u_{\hat{\v x}} =& i\eta \tau^0 -\chi (\tau^3-\tau^1) \nonumber\\
 u_{\hat{\v y}} =& i\eta \tau^0 -\chi (\tau^3+\tau^1)
\end{align}
with $\chi$ and $\eta$ non-zero.  The above ansatz describes the
$SU(2)$-gapless spin liquid if $\chi=0$, and the $SU(2)$-linear spin liquid
if $\eta=0$. 

After a $90^\circ$ rotation $R_{90}$, the above ansatz becomes
\begin{eqnarray}
u_{\hat{\v x}} &=& -i\eta \tau^0 -\chi (\tau^3+\tau^1) \nonumber\\
u_{\hat{\v y}} &=&  i\eta \tau^0 -\chi (\tau^3-\tau^1)
\end{eqnarray}
The rotated ansatz is gauge equivalent to the original ansatz
under the gauge transformation $G_{R_{90}}(\v
i)=(-)^{i_x} (1-i\tau^2)/\sqrt{2}$.  After a parity $x\to -x$
transformation $P_x$, \Eq{Z2glA} becomes
\begin{eqnarray}
u_{\hat{\v x}} = -i\eta \tau^0 -\chi (\tau^3 -\tau^1)  \nonumber\\
u_{\hat{\v y}} = i\eta \tau^0 -\chi (\tau^3 + \tau^1)
\end{eqnarray}
which is gauge equivalent to the original ansatz under the gauge
transformation $G_{P_x}(\v i)=(-)^{i_x} i(\tau^3+\tau^1)/\sqrt{2} $.  
Under time reversal transformation $T$, \Eq{Z2glA} is changed to
\begin{eqnarray}
u_{\hat{\v x}} = -i\eta \tau^0 +\chi (\tau^3 -\tau^1) \nonumber\\
u_{\hat{\v y}} = -i\eta \tau^0 +\chi (\tau^3 +\tau^1)
\end{eqnarray}
which is again gauge equivalent to the original ansatz under the gauge
transformation $G_T(\v i)=(-)^{\v i}$.  (In fact any ansatz which only has
links between two non-overlapping sublattices (\ie the unfrustrated ansatz) is
time reversal symmetric if $a_0^l=0$ .) To summarize the ansatz \Eq{Z2glA} is
invariant under the rotation $R_{90}$, parity $P_x$, and time reversal
transformation $T$, \emph{followed by the following gauge transformations}
\begin{align}
G_{R_{90}}(\v i)=& (-)^{i_x} (1-i\tau^2)/\sqrt{2}  \nonumber\\
G_{P_x}(\v i)=& (-)^{i_x} i(\tau^3+\tau^1)/\sqrt{2} \nonumber\\
G_T(\v i)=& (-)^{\v i}
\end{align}
Thus the ansatz \Eq{Z2glA}
describes a spin liquid which translation, rotation, parity and time
reversal symmetries.

Using the time reversal symmetry we can show that the vanishing $a_0^l$
in our ansatz \Eq{Z2glA} indeed satisfy the constraint \Eq{constraintK}.
This is because $a_0^l \to -a_0^l$ under the time reversal transformation.
Thus $\frac{\prt E_{mean}}{\prt a_0^l} = 0$ when $a^l_0=0$
for any time reversal symmetric
ansatz, including the ansatz \Eq{Z2glA}.

The spinon spectrum is given by (see Fig. \ref{specZ2glq}a)
\begin{equation}
 E_\pm = 2\eta (\sin(k_x)+\sin(k_y))\pm 
2|\chi| \sqrt{ 2\cos^2(k_x)+2\cos^2(k_y)}
\end{equation}
The spinons have two Fermi points and two small Fermi pockets (for small
$\eta$).  The $SU(2)$ flux is non-trivial. Further more $P_{C_1}$ and
$P_{C_2}$ do not commute. Thus the $SU(2)$ gauge structure is broken down to a
$Z_2$ gauge structure by the $SU(2)$ flux $P_{C_1}$ and
$P_{C_2}$.\cite{Wsrvb,MF9400} We will call the spin liquid described by
\Eq{Z2glA} $Z2$-gapless spin liquid.  The low energy effective theory is
described by massless Dirac fermions and fermions with small Fermi surfaces,
coupled to a $Z_2$ gauge field. Since the $Z_2$ gauge interaction is
irrelevant at low energies, the spinons are {\it free} fermions at low
energies and we have a true spin-charge separation in the $Z_2$-gapless spin
liquid.  The $Z_2$-gapless spin liquid is one of the $Z_2$ spin liquids
classified in appendix A.  Its projective symmetry group is labeled by
Z2A$\tau^{13}_-\tau^{1\bar3}_+\tau^3\tau^0_-$ or equivalently by Z2A$x2(12)n$
(see section \ref{Z2sec} and \Eq{Z2Alb}).

Now let us include longer links. First we still limit ourselves to
unfrustrated ansatz. An interesting ansatz is given by
\begin{eqnarray}
a_0^l&=& 0 \nonumber\\
u_{\hat{\v x}} &=& \chi \tau^3 +\eta \tau^1  \nonumber\\
u_{\hat{\v y}} &=& \chi \tau^3 -\eta \tau^1  \nonumber\\
u_{2\hat{\v x}+\hat{\v y}} &=& \la \tau^2  \nonumber\\
u_{-\hat{\v x}+2\hat{\v y}} &=& -\la \tau^2  \nonumber\\
u_{2\hat{\v x}-\hat{\v y}} &=& \la \tau^2  \nonumber\\
u_{\hat{\v x}+2\hat{\v y}} &=& -\la \tau^2 
\label{Z2lE}
\end{eqnarray}
By definition, the ansatz is invariant under translation and parity $x\to
-x$. After a $90^\circ$ rotation, the ansatz is changed to
\begin{eqnarray}
u_{\hat{\v x}} &=& -\chi \tau^3 -\eta \tau^1  \nonumber\\
u_{\hat{\v y}} &=& -\chi \tau^3 +\eta \tau^1  \nonumber\\
u_{2\hat{\v x}+\hat{\v y}} &=& -\la \tau^2  \nonumber\\
u_{-\hat{\v x}+2\hat{\v y}} &=& +\la \tau^2  \nonumber\\
u_{2\hat{\v x}-\hat{\v y}} &=& -\la \tau^2  \nonumber\\
u_{\hat{\v x}+2\hat{\v y}} &=& +\la \tau^2 
\end{eqnarray}
which is gauge equivalent to \Eq{Z2lE} under the gauge transformation
$G_{R_{90}}(\v i)=i\tau_3$. 
Thus the ansatz describe a spin liquid with translation,
rotation, parity and the time reversal symmetries.  The spinon spectrum is
given by (see Fig. \ref{specZ2l}c)
\begin{eqnarray}
 E_\pm &=& \pm \sqrt{
\eps_1(\v k)^2 + 
\eps_2(\v k)^2 + 
\eps_3(\v k)^2 } \nonumber\\
\eps_1 &=& -2\chi (\cos(k_x)+\cos(k_y)) \nonumber\\
\eps_2 &=& -2\eta (\cos(k_x)-\cos(k_y)) \nonumber\\
\eps_3 &=& -2\la [\cos(2k_x+k_y)+\cos(2k_x-k_y) \nonumber\\
&& -\cos(k_x-2k_y)-\cos(k_x+2k_y)]
\end{eqnarray}
Thus the spinons are gapless only at four $\v k$ points
$(\pm \pi/2, \pm \pi/2)$. We also find that $P_{C_3}$ and $P_{C_4}$ do not
commute, where the loops
$ C_3 = \v i 
\to \v i +\hat{\v x} 
\to \v i +2 \hat{\v x} 
\to \v i +2\hat{\v x} + \hat{\v y} 
\to \v i $ and
$ C_4 = \v i 
\to \v i +\hat{\v y} 
\to \v i +2 \hat{\v y} 
\to \v i +2\hat{\v y} - \hat{\v x} 
\to \v i $.
Thus the $SU(2)$ flux $P_{C_3}$ and $P_{C_4}$ break the $SU(2)$ gauge
structure down to a $Z_2$ gauge structure.  The spin liquid described by
\Eq{Z2lE} will be called the $Z_2$-linear spin liquid.  The low energy
effective theory is described by massless Dirac fermions coupled to a
$Z_2$ gauge field. Again the $Z_2$ coupling is irrelevant and the spinons
are free fermions at low energies. We have a true spin-charge separation.
According to the classification scheme summarized in section
\ref{Z2sec}, the above $Z_2$-linear spin liquid is labeled by Z2A$003n$.

Next let us discuss frustrated ansatz.  A simple $Z_2$ spin liquid can be
obtained from the following frustrated ansatz
\begin{align}
\label{Z2lCs}
a_0^3& \neq 0,\ \ \ a^{1,2}_0=0  \nonumber\\
u_{\hat{\v x}} =&  \chi \tau^3 +\eta \tau^1  \nonumber\\
u_{\hat{\v y}} =&  \chi \tau^3 -\eta \tau^1  \nonumber\\
u_{\hat{\v x}+\hat{\v y}} =&   \ga \tau^3 \nonumber\\
u_{-\hat{\v x}+\hat{\v y}} =&   \ga \tau^3 
\end{align}
The ansatz has translation, rotation, parity, and the time reversal symmetries.
When $a^3_0\neq 0$, $\chi \neq \pm \eta$ and $\chi\eta \neq 0$,
$a^l_0\tau^l$ does not commute with the loop operators. Thus the ansatz
breaks the $SU(2)$ gauge structure to a $Z_2$ gauge structure.  The spinon
spectrum is given by (see Fig. \ref{specZ2l}a)
\begin{align}
 E_\pm =& \pm \sqrt{\eps^2(\v k) + \Del^2(\v k)}  \nonumber\\
 \eps(\v k) =&  2\chi (\cos(k_x)+\cos(k_y)) + a_0^3  \nonumber\\
             &  2\ga (\cos(k_x+k_y)+\cos(k_x-k_y))   \nonumber\\
 \Del(\v k) =&  2\eta (\cos(k_x)-\cos(k_y)) + a_0^3  
\end{align}
which is gapless only at four $\v k$ points with a linear dispersion.  Thus
the spin liquid described by \Eq{Z2lCs} is a $Z_2$-linear spin liquid,
which has a true spin-charge separation.  The $Z_2$-linear spin liquid is
described by the projective symmetry group Z2A$0032$ or equivalently
Z2A$0013$. (see section \ref{Z2sec}.) From the above two examples of
$Z_2$-linear spin liquids, we find that it is possible to obtain true
spin-charge separation with massless Dirac points (or nodes) within a pure
spin model without the charge fluctuations. We also find that there are
more than one way to do it. 

A well known frustrated ansatz is the ansatz for the chiral spin
liquid\cite{WWZcsp}
\begin{eqnarray}
\label{chiralspin}
u_{\hat{\v x}} &=& -\chi \tau^3 -\chi \tau^1 \nonumber\\
u_{\hat{\v y}} &=& -\chi \tau^3 +\chi \tau^1 \nonumber\\
u_{\hat{\v x}+\hat{\v y}} &=& \eta \tau^2 \nonumber\\
u_{-\hat{\v x}+\hat{\v y}} &=& -\eta \tau^2 \nonumber\\
a_0^l &=& 0
\end{eqnarray}
The chiral spin liquid breaks the time reversal and parity symmetries.  The
$SU(2)$ gauge  structure is unbroken.\cite{Wsrvb} The low energy effective
theory is an $SU(2)$ Chern-Simons theory (of level 1).  The spinons are
gaped and have a semionic statistics.\cite{KL8795,WWZcsp} The third
interesting frustrated ansatz is given in \Ref{Wsrvb,MF9400}
\begin{align}
u_{\hat{\v x} }=&  u_{ \hat{\v y} }= -\chi \tau^3 \nonumber \\
u_{ \hat{\v x} + \hat{\v y}} =&  \eta\tau^1 +\la\tau^2 \nonumber \\
u_{-\hat{\v x} + \hat{\v y}} =&  \eta\tau^1 -\la\tau^2 \nonumber\\
a_0^{2,3} =&  0, \ \ \ a_0^1 \neq 0
\label{Z2gA}
\end{align}
This ansatz has translation, rotation, parity and the time reversal symmetries.
The spinons are fully gaped and the $SU(2)$ gauge structure is broken down
to $Z_2$ gauge structure. We may call such a state $Z_2$-gapped spin liquid
(it was called sRVB state in \Ref{Wsrvb,MF9400}).  It is described by the
projective symmetry group Z2A$xx0z$.  Both the chiral spin
liquid and the $Z_2$-gapped spin liquid  have true spin-charge separation.

\section{Quantum orders in symmetric spin liquids}
\label{qois}

\subsection{Quantum orders and projective symmetry groups}
\label{psg}

We have seen that there can be many different spin liquids with the {\em same}
symmetries.  The stability analysis in section \ref{stb} shows that many of
those spin liquids occupy a finite region in phase space and represent stable
quantum phases. So here we are facing a similar situation as in quantum Hall
effect: there are many distinct quantum phases not separated by symmetries and
order parameters. The quantum Hall liquids have finite energy gaps and are
rigid states. The concept of topological order was introduced to describe the
internal order of those rigid states. Here we can also use the topological
order to describe the internal orders of rigid spin liquids.  However, we also
have many other stable quantum spin liquids that have gapless excitations. 

To describe internal orders in gapless quantum spin liquids (as well as gapped
spin liquids), we have introduced a new concept -- quantum order -- that
describes the internal orders in any quantum phases.  
The key point in introducing quantum orders is that quantum phases, in
general, cannot be completely characterized by broken symmetries and local
order parameters. This point is illustrated by quantum Hall states and by the
stable spin liquids constructed in this paper.
However, to make the concept of quantum order useful, we need to find concrete
mathematical characterizations the quantum orders.  Since quantum orders are
not described by symmetries and order parameters, we need to find a completely
new way to characterize them. Here we would like to propose to use Projective
Symmetry Group to characterize quantum (or topological) orders in quantum spin
liquids.  
The projective
symmetry group is motivated from the following observation. 
Although ansatz for different symmetric spin liquids all have the same
symmetry, the ansatz are invariant under transformations followed by
\emph{different} gauge transformations. We can use those different gauge
transformations to distinguish different spin liquids with the same symmetry.
In the following, we will introduce projective symmetry group in a general and
formal setting.

We know that to find quantum numbers that characterize a phase is to find the
universal properties of the phase. For classical systems, we know that
symmetry is a universal property of a phase and we can use symmetry to
characterize different classical phases. To find universal properties of
quantum phases we need to find universal properties of many-body wave
functions.  This is too hard.  Here we want to simplify the problem by
limiting ourselves to a subclass of many-body wave functions which can be
described by ansatz $(u_{\v i\v j}, a_0^{l}\tau^l)$ via \Eq{PsiPsi}.  Instead
of looking for the universal properties of many-body wave functions, we try to
find the universal properties of ansatz $(u_{\v i\v j}, a_0^{l}\tau^l)$.
Certainly, one may object that the universal properties of the ansatz (or the
subclass of wave functions) may not be the universal properties of spin
quantum phase. This is indeed the case for some ansatz. However, if the
mean-field state described by ansatz $(u_{\v i\v j}, a_0^{l}\tau^l)$ is stable
against fluctuations (\ie the fluctuations around the mean-field state do not
cause any infrared divergence), then the mean-field state faithfully describes
a spin quantum state and the universal properties of the ansatz will be the
universal properties of the correspond spin quantum phase. This completes the
link between the properties of ansatz and properties of physical spin liquids.
Motivated by the Landau's theory for classical orders, here we whould like to
propose that the invariance group (or the ``symmetry'' group) of an ansatz is
a universal property of the ansatz.  Such a group will be called the
projective symmetry group (PSG). We will show that PSG can be used to
characterize quantum orders in quantum spin liquids.

Let us give a detailed definition of PSG. A PSG is a property of an ansatz.  
It is formed by all the transformations that keep the ansatz unchanged.
Each transformation (or each element in the PSG) can be written as a
combination of a symmetry transformation $U$ (such as translation) and a gauge
transformation $G_U$.
The invariance of the ansatz under its PSG can be expressed as
\begin{align}
\label{GUU}
 G_U U (u_{\v i\v j}) =&  u_{\v i\v j}  \nonumber\\
 U (u_{\v i\v j})  \equiv & u_{U(\v i), U(\v j)}  \nonumber\\
 G_U( u_{\v i\v j})  \equiv & G_U(\v i) u_{\v i\v j}  G_U^\dag(\v j) \nonumber\\
 G_U(\v i) \in & SU(2)
\end{align}
for each $G_U U \in PSG$.

Every PSG contains a special subgroup,
which will be called invariant gauge group (IGG).  
IGG (denoted by $\cG$) for an ansatz is formed by
all the gauge transformations that leave the ansatz unchanged:
\begin{equation}
\label{cGdef}
 \cG = \{ W_{\v i} | W_{\v i} u_{\v i\v j}W_{\v j}^\dag = u_{\v i\v j},
 W_{\v i}\in SU(2)\}
\end{equation}
If we want to relate IGG to a symmetry transformation, then the associated
transformation is simply an identity transformation. 

If IGG is non-trivial, then for a fixed symmetry transformation $U$, there are
can be many gauge transformations $G_U$ that leave the ansatz unchanged.  
If $G_U U$ is in the PSG of $u_{\v i\v j}$, $GG_U U$ will also be in the PSG
iff $G \in \cG$.  Thus for each symmetry transformation $U$, the different
choices of $G_U$ have a one to one correspondence with the elements in IGG.
From the above definition, we see that the PSG, the IGG, and the symmetry
group (SG) of an ansatz are related:
\begin{equation}
 SG = PSG/IGG
\end{equation}
This relation tells us that a PSG is a projective representation or an
extension of the symmetry group.\footnote{In his unpublished study of
quantum antiferromagnetism with a symmetry group of large rank,
Wiegmenn\cite{W91} constructed a gauge theory which realizes a double valued
magnetic space group. The double valued magnetic space group extends the space
group and is a special case of projective symmetry group.}
(In section \ref{genconPSG} we will introduce a closely related but different
definition of PSG.  To distinguish the two definitions, we will call the PSG
defined above invariant PSG and the PSG defined in section \ref{genconPSG}
algebraic PSG.)

Certainly the PSG's for two gauge equivalent ansatz $u_{\v i\v j}$ and $W(\v
i) u_{\v i\v j}W^\dag(\v j)$ are related.  From $ W G_U U (u_{\v i\v j}) =
W(u_{\v i\v j}) $, where $W(u_{\v i\v j}) \equiv W(\v i) u_{\v i\v j}W^\dag(\v
j)$, we find $ W G_U U W^{-1} W(u_{\v i\v j}) = W G_U  W_U^{-1} U W(u_{\v i\v
j}) = W(u_{\v i\v j}) $, where $W_U \equiv UWU^{-1}$ is given by $ W_U(\v i) =
W(U(\v i))$.  Thus if $G_UU$ is in the PSG of ansatz $u_{\v i\v j}$, then
$(WG_UW_U) U $ is in the PSG of gauge transformed ansatz $W(\v i) u_{\v i\v
j}W^\dag(\v j)$.  We see that the gauge transformation $G_U$ associated with
the symmetry transformation $U$ is changed in the following way
\begin{equation}
\label{WGUWU}
G_U(\v i) \to W(\v i)G_U(\v i) W^\dag(U(\v i))
\end{equation}
after a gauge transformation $W(\v i)$.

Since PSG is a property of an ansatz, we can group all the ansatz sharing the
same PSG together to form a class. We claim that such a class is formed by one
or several universality classes that correspond to quantum phases. (A more
detailed discussion of this important point is given in section \ref{qostb}.)
It is in this sense we say that quantum orders are characterized by PSG's. 

We know that a classical order can be described by its symmetry properties.
Mathematically, we say that a classical order is characterized by its symmetry
group.  Using projective symmetry group to describe a quantum order,
conceptually, is similar to using symmetry group to describe a classical
order.  The symmetry description of a classical order is very useful since it
allows us to obtain many universal properties, such as the number of
Nambu-Goldstone modes, without knowing the details of the system.  Similarly,
knowing the PSG of a quantum order also allows us to obtain low energy
properties of a quantum system without knowing its details.  As an example, we
will discuss a particular kind of the low energy fluctuations -- the gauge
fluctuations -- in a quantum state. We will show that the low energy gauge
fluctuations can be determined completely from the PSG. In fact the gauge
group of the low energy gauge fluctuations is nothing but the IGG of the
ansatz.

To see this, let us assume that, as an example, an IGG $\cG$ contains a $U(1)$
subgroup which is formed by the following constant gauge transformations
\begin{equation}
 \{ W_{\v i} = e^{i\th \tau^3} | \th \in [0,2\pi) \} \subset \cG
\end{equation}
Now we consider the following type of fluctuations around the mean-field
solution $\bar u_{\v i\v j}$: $u_{\v i\v j}=\bar u_{\v i\v j}e^{i a^3_{\v i\v
j}\tau^3}$.  Since $\bar u_{\v i\v j}$ is invariant under the constant gauge
transformation $ e^{i\th \tau^3}$, a spatial dependent gauge transformation
$e^{i\th_i \tau^3}$ will transform the fluctuation $a^3_{\v i\v j}$ to $\t
a^3_{\v i\v j}=a^3_{\v i\v j} +\th_{\v i}-\th_{\v j}$.  This means that
$a^3_{\v i\v j}$ and $\t a^3_{\v i\v j}$ label the same physical state and
$a^3_{\v i\v j}$ correspond to gauge fluctuations.  The energy of the
fluctuations has a gauge invariance $E(\{ a^3_{\v i\v j} \}) = E( \{ \t
a^3_{\v i\v j} \}) $. We see that the mass term of the gauge field, $(a^3_{\v
i\v j})^2$, is not allowed and the $U(1)$ gauge fluctuations described by
$a^3_{\v i\v j}$ will appear at low energies.

If the $U(1)$ subgroup of $\cG$ is formed by 
spatial dependent gauge transformations
\begin{equation}
 \{ W_{\v i} = e^{i\th \v n_{\v i}\cdot \v \tau} | 
 \th \in [0,2\pi), |\v n_{\v i}|=1 \} \subset \cG ,
\end{equation}
we can always use a $SU(2)$ gauge transformation to rotate $\v n_{\v i}$
to the $\hat{\v z}$ direction on every site and reduce the problem to the one
discussed above. Thus, regardless if the gauge transformations in IGG have
spatial dependence or not, the gauge group for low energy gauge fluctuations
is always given by $\cG$.

We would like to remark that some times low energy gauge fluctuations not only
appear near $\v k=0$, but also appear near some other $\v k$ points.  In this
case, we will have several low energy gauge fields, one for each $\v k$
points.  Examples of this phenomenon are given by some ansatz of $SU(2)$
slave-boson theory discussed in section \ref{mpdo}, which have an $SU(2)\times
SU(2)$ gauge structures at low energies.  We see that the low energy gauge
structure $SU(2)\times SU(2)$ can even be larger than the high energy gauge
structure $SU(2)$. Even for this complicated case where low energy gauge
fluctuations appear around different $\v k$ points, IGG still correctly
describes the low energy gauge structure of the corresponding ansatz.  If IGG
contains gauge transformations that are independent of spatial coordinates,
then such transformations correspond to the gauge group for gapless gauge
fluctuations near $\v k=0$. If IGG contains gauge transformations that depend
on spatial coordinates, then those transformations correspond to the gauge
group for gapless gauge fluctuations near non-zero $\v k$. Thus IGG gives us a
unified treatment of all low energy gauge fluctuations, regardless their
momenta.

In this paper, we have used the terms $Z_2$ spin liquids, $U(1)$ spin liquids,
$SU(2)$ spin liquids, and $SU(2)\times SU(2)$ spin liquids in many places.
Now we can have a precise definition of those low energy $Z_2$, $U(1)$,
$SU(2)$, and $SU(2)\times SU(2)$ gauge groups.  Those low energy gauge groups
are nothing but the IGG of the corresponding ansatz. They have nothing to do
with the high energy gauge groups that appear in the $SU(2)$, $U(1)$, or $Z_2$
slave-boson approaches.  We also used the terms $Z_2$ gauge structure, $U(1)$
gauge structure, and $SU(2)$ gauge structure of a mean-field state.  Their
precise mathematical meaning is again the IGG of the corresponding ansatz. When we
say a $U(1)$ gauge structure is broken down to a $Z_2$ gauge structure, we
mean that an ansatz is changed in such a way that its IGG  is changed from
$U(1)$ to $Z_2$ group.

\subsection{Classification of symmetric $Z_2$ spin liquids}
\label{Z2sec}

As an application of PSG characterization of quantum orders in spin
liquids, we would like to classify the PSG's associated with translation
transformations assuming the IGG $\cG=Z_2$.  Such a classification leads to
a classification of translation symmetric $Z_2$ spin liquids.  

When $\cG=Z_2$, it contains two elements -- gauge transformations $G_1$ and
$G_2$:
\begin{align}
 \cG =& \{G_1,G_2\} \nonumber\\
G_1(\v i)=& \tau^0,\ \ \
G_2(\v i)= -\tau^0  .
\end{align}
Let us assume that a $Z_2$ spin liquid has a translation symmetry.  The PSG
associated with the translation group is generated by four elements $\pm G_x
T_x$, $\pm G_y T_x$ where
\begin{equation}
 T_x (u_{\v i\v j}) = u_{\v i-\hat{\v x}, \v j-\hat{\v x}}, \ \ \
 T_y (u_{\v i\v j}) = u_{\v i-\hat{\v y}, \v j-\hat{\v y}} .
\end{equation}
Due to the translation symmetry of the ansatz, we can choose a gauge in
which all the loop operators of the ansatz are translation invariant. That
is $P_{C_1}=P_{C_2}$ if the two loops $C_1$ and $C_2$ are related by a
translation.  We will call such a gauge uniform gauge.

Under transformation $G_xT_x$, a loop operator $P_{C}$ based at $\v i$
transforms as $P_C\to G_x(\v i^\prime) P_{T_xC}G_x^\dag(\v i^\prime)= G_x(\v
i^\prime) P_{C}G_x^\dag(\v i^\prime)$ where $\v i^\prime= T_x \v i$ is the
base point of the translated loop $T_x(C)$.  We see that translation
invariance of $P_C$ in the uniform gauge requires
\begin{equation}
 G_x(\v i)=\pm \tau^0,\ \ \
 G_y(\v i)=\pm \tau^0 .
\end{equation}
since different loop operators based at the same base point do not commute
for $Z_2$ spin liquids.  We note that the gauge transformations of form
$W(\v i)=\pm \tau^0$ do not change the translation invariant property of
the loop operators.  Thus we can use such gauge transformations to further
simplify $G_{x,y}$ through \Eq{WGUWU}.  First we can choose a gauge to make 
\begin{equation}
\label{Gy0}
G_y(\v i)=\tau^0. 
\end{equation}
We note that a gauge transformation satisfying $W(\v i) = W(i_x)$ does not
change the condition $G_y(\v i)=\tau^0$. We can use such kind of gauge
transformations to make 
\begin{equation}
\label{Gx0}
G_x(i_x, i_y=0) =\tau^0 .
\end{equation}

Since the translations in $x$- and $y$-direction commute,  $G_{x,y}$ must
satisfy (for any ansatz, $Z_2$ or not $Z_2$)
\begin{align}
\label{GxGyT1}
& G_x T_x G_y T_y (G_x T_x)^{-1} (G_y T_y)^{-1} =   \nonumber\\
& G_x T_x G_y T_y T_x^{-1} G_x^{-1} T_y^{-1} G_y^{-1} \in \cG.
\end{align}
That means
\begin{align}
& G_x(\v i)G_y(\v i -\hat{\v x}) G_x^{-1}(\v i -\hat{\v y}) G_y(\v i)^{-1}
\in \cG
\label{genGxGy1}
\end{align}

For $Z_2$ spin liquids, \Eq{genGxGy1} reduces to
\begin{align}
& G_x(\v i) G_x^{-1}(\v i -\hat{\v y}) =+\tau^0
\label{Z2GxGy1}
\end{align}
or
\begin{align}
& G_x(\v i) G_x^{-1}(\v i -\hat{\v y}) =-\tau^0
\label{Z2GxGy2}
\end{align}
When combined with \Eq{Gy0} and \Eq{Gx0}, we find that there are only two
gauge inequivalent extensions of the translation group when
IGG is $\cG=Z_2$. The two PSG's are given by
\begin{align}
\label{GZ2t1}
 G_x(\v i) =&  \tau^0, & G_y(\v i) =&  \tau^0  
\end{align}
and
\begin{align}
\label{GZ2t2}
 G_x(\v i) =&  (-)^{i_y} \tau^0, & G_y(\v i) =&  \tau^0  
\end{align}
Thus, under PSG classification, there are only two types of $Z_2$ spin
liquids if they have {\em only} the translation symmetry and no other
symmetries.  The ansatz that satisfy \Eq{GZ2t1} have a form
\begin{align}
\label{Z2t1}
 u_{\v i, \v i+\v m} =& u_{\v m}
\end{align}
and the ones that satisfy \Eq{GZ2t2} have a form
\begin{align}
\label{Z2t2}
 u_{\v i, \v i+\v m} =& (-)^{m_yi_x} u_{\v m}
\end{align}

Through the above example, we see that PSG is a very powerful tool. It can
lead to a complete classification of (mean-field) spin liquids with
prescribed symmetries and low energy gauge structures.

In the above, we have studied $Z_2$ spin liquids which have {\em only} the
translation symmetry and no other symmetries.  We find there are only two
types of such spin liquids.  However, if spin liquids have more symmetries,
then they can have much more types.  In the appendix A, we will give a
classification of symmetric $Z_2$ spin liquids using PSG.  Here we use the
term symmetric spin liquid to refer to a spin liquid with the translation
symmetry $T_{x,y}$, the time reversal symmetry $T$: $u_{\v i\v j} \to
-u_{\v i\v j}$, and the three parity symmetries $P_x$:
$(i_x,i_y)\to(-i_x,i_y)$, $P_y$: $(i_x,i_y)\to(i_x,-i_y)$,  and $P_{xy}$:
$(i_x,i_y)\to(i_y,i_x)$.  The three parity symmetries also imply the
$90^\circ$ rotation symmetry.  In the appendix A, we find that there are
272 different extensions of the symmetry group $\{ T_{x,y},
P_{x,y,xy}, T\}$ if IGG $\cG=Z_2$.  Those PSG's are generated by $( G_xT_x,
G_yT_y, G_TT, G_{P_x}P_x, G_{P_y}P_y, G_{P_{xy}}P_{xy})$.  The PSG's can be
divided into two classes. The first class is given by
\begin{align}
\label{Z2sum1}
 G_x(\v i) =&  \tau^0, 
& 
G_y(\v i) =&  \tau^0  \nonumber \\
 G_{P_x}(\v i) =&  \eta_{xpx}^{i_x}\eta_{xpy}^{i_y}  g_{P_x} 
&
 G_{P_y}(\v i) =&  \eta_{xpy}^{i_x}\eta_{xpx}^{i_y}  g_{P_y} 
\nonumber\\
 G_{P_{xy}}(\v i) =&  g_{P_{xy}} 
& 
 G_T(\v i) =&  \eta_t^{\v i} g_T
\end{align}
and the second class by
\begin{align}
\label{Z2sum2}
 G_x(\v i) =&  (-)^{i_y}\tau^0, 
& 
G_y(\v i) =&  \tau^0  \nonumber \\
 G_{P_x}(\v i) =&  \eta_{xpx}^{i_x}\eta_{xpy}^{i_y}  g_{P_x} 
&
 G_{P_y}(\v i) =&  \eta_{xpy}^{i_x}\eta_{xpx}^{i_y}  g_{P_y} 
\nonumber\\
 G_{P_{xy}}(\v i) =&  (-)^{i_xi_y} g_{P_{xy}} 
& 
 G_T(\v i) =&  \eta_t^{\v i} g_T
\end{align}
Here the three $\eta$'s can independently take two values $\pm 1$.  $g$'s
have 17 different choices which are given by \Eq{gggg1} - \Eq{gggg17} in
the appendix A.  Thus there are $2\times 17 \times 2^3 = 272$ different PSG's.
They can potentially lead to 272 different types of
symmetric $Z_2$ spin liquids on 2D square lattice.  

To label the 272 PSG's, we propose the following scheme:
\begin{align}
\label{Z2Alb}
Z2A(g_{px})_{\eta_{xpx}}(g_{py})_{\eta_{xpy}}g_{pxy}(g_t)_{\eta_t} , \\
\label{Z2Blb}
Z2B(g_{px})_{\eta_{xpx}}(g_{py})_{\eta_{xpy}}g_{pxy}(g_t)_{\eta_t} .
\end{align}
The label Z2A$...$ correspond to the case \Eq{Z2sum1}, and the label
Z2B$...$ correspond to the case \Eq{Z2sum2}.  A typical label will looks
like Z2A$\tau^1_+\tau^2_-\tau^{12}\tau^3_-$.  We will also use an
abbreviated notation.  An abbreviated notation is obtained by replacing
$(\tau^0,\tau^1,\tau^2,\tau^3)$ or $(\tau^0_+,\tau^1_+,\tau^2_+,\tau^3_+)$
by $(0,1,2,3)$ and $(\tau^0_-,\tau^1_-,\tau^2_-,\tau^3_-)$ by $(n,x,y,z)$.
For example, Z2A$\tau^1_+\tau^0_-\tau^{12}\tau^3_-$ can be abbreviated as
Z2A$1n(12)z$.

Those $272$ different $Z_2$ PSG's, strictly speaking, are the so called
algebraic PSG's.  The algebraic PSG's are defined as 
extensions of the symmetry group. They can be calculated through the
algebraic relations listed in section \ref{genconPSG}. The algebraic PSG's
are different from the invariant PSG's which are defined as a collection of
all transformations that leave an ansatz $u_{\v i\v j}$ invariant.  Although
an invariant PSG must be an algebraic PSG, an algebraic PSG may not be an
invariant PSG.  This is because certain algebraic PSG's have the following
properties: any ansatz $u_{\v i\v j}$ that is invariant under an algebraic PSG
may actually be invariant under a larger PSG.  In this case the original
algebraic PSG cannot be an invariant PSG of the ansatz.  The invariant PSG of
the ansatz is really given by the larger PSG.  If we limit ourselves to the
spin liquids constructed through the ansatz $u_{\v i\v j}$, then we should
drop the algebraic PSG's are not invariant PSG's.  This is
because those algebraic PSG's do not characterize mean-field spin liquids.  

We find that among the 272 algebraic $Z_2$ PSG's, at least 76 of them are
not invariant PSG's. Thus the 272 algebraic $Z_2$ PSG's can at most lead to
196 possible $Z_2$ spin liquids.  
Since some of the mean-field spin liquid states may not survive the quantum
fluctuations, the number of physical $Z_2$ spin liquids is even smaller.
However, for the physical spin liquids that can be obtained through the
mean-field states, the PSG's do offer a characterization of the quantum
orders in those spin liquids.

\subsection{Classification of symmetric $U(1)$ and $SU(2)$ spin liquids}
\label{U1SU2sec}

In addition to the $Z_2$ symmetric spin liquids studied above, there can be 
symmetric spin liquids whose low energy gauge structure is $U(1)$ or
$SU(2)$. Such $U(1)$ and $SU(2)$ symmetric
spin liquids (at mean-field level) are classified by $U(1)$ and $SU(2)$
symmetric PSG's.  The $U(1)$ and $SU(2)$ symmetric PSG's are
calculated in the appendix A.  In the following we just summarize the results.

We find that the PSG's that characterize mean-field symmetric $U(1)$ spin liquids
can be divided into four types: U1A, U1B, U1C and U1$^m_n$.  There are 24 type
U1A PSG's:
\begin{align}
\label{U1Apsg1}
 G_x =& g_3(\th_x), \ \ \ \ G_y = g_3(\th_y), \nonumber\\
 G_{P_x} =& \eta_{ypx}^{i_y} g_3(\th_{px}),\ \ \ \
 G_{P_y} = \eta_{ypx}^{i_x} g_3(\th_{py})  \nonumber\\
 G_{P_{xy}} =& g_3(\th_{pxy}), \ \ g_3(\th_{pxy})i\tau^1
\nonumber\\
 G_T =& \eta_{t}^{\v i} g_3(\th_t)|_{\eta_t=-1},\ \
        \eta_{t}^{\v i} g_3(\th_t)i\tau^1
\end{align}
and
\begin{align}
\label{U1Apsg2}
 G_x =& g_3(\th_x), \ \ \ \ G_y = g_3(\th_y), \nonumber\\
 G_{P_x} =& \eta_{xpx}^{i_x} g_3(\th_{px})i\tau^1,\ \ \ \
 G_{P_y} = \eta_{xpx}^{i_y} g_3(\th_{py})i\tau^1  \nonumber\\
 G_{P_{xy}} =& g_3(\th_{pxy}),\ g_3(\th_{pxy})i\tau^1  \nonumber \\
 G_T =& \eta_{t}^{\v i} g_3(\th_t)|_{\eta_t=-1},\ \ 
        \eta_{t}^{\v i} g_3(\th_t)i\tau^1
\end{align}
where 
\begin{equation}
g_a(\th)\equiv e^{i\th\tau^a}.
\end{equation}
We will use U1A$a_{\eta_{xpx}}b_{\eta_{ypx}}cd_{\eta_t}$ to label the 24
PSG's. $a,\ b,\ c,\ d$ are associated with $G_{P_x}$, $G_{P_y}$,
$G_{P_{xy}}$, $G_T$ respectively. They are equal to $\tau^1$ if the
corresponding $G$ contains a $\tau^1$ and equal to $\tau^0$ otherwise.  A
typical notation looks like U1A$\tau^1_-\tau^1\tau^0\tau^1_-$ which can be
abbreviated as  U1A$x10x$.

There are also 24 type U1B PSG's:
\begin{align}
\label{U1Bpsg1}
 G_x =& (-)^{i_y}g_3(\th_x), \ \ \ \ G_y = g_3(\th_y),
 \nonumber\\
 G_{P_x} =& \eta_{ypx}^{i_y} g_3(\th_{px}),\ \ \ \
 G_{P_y} = \eta_{ypx}^{i_x} g_3(\th_{py})  \nonumber\\
(-)^{i_xi_y} G_{P_{xy}} =& g_3(\th_{pxy}), \ \ g_3(\th_{pxy})i\tau^1 
\nonumber \\
 G_T =& \eta_{t}^{\v i} g_3(\th_t)|_{\eta_t=-1},\ \
        \eta_{t}^{\v i} g_3(\th_t)i\tau^1
\end{align}
and
\begin{align}
\label{U1Bpsg2}
 G_x =& (-)^{i_y}g_3(\th_x), \ \ \ \ G_y = g_3(\th_y),
 \nonumber\\
 G_{P_x} =& \eta_{xpx}^{i_x} g_3(\th_{px})i\tau^1,\ \ \ \
 G_{P_y} = \eta_{xpx}^{i_y} g_3(\th_{py})i\tau^1  \nonumber\\
(-)^{i_xi_y} G_{P_{xy}} =& g_3(\th_{pxy}), \ \ g_3(\th_{pxy})i\tau^1 
\nonumber \\
 G_T =& \eta_{t}^{\v i} g_3(\th_t)|_{\eta_t=-1},\ \ 
        \eta_{t}^{\v i} g_3(\th_t)i\tau^1
\end{align}
We will use U1B$a_{\eta_{xpx}}b_{\eta_{ypx}}cd_{\eta_t}$ to label the 24
PSG's. 

The 60 type U1C PSG's are given by
\begin{align}
\label{PSGU1Ca}
 G_x =&  g_3(\th_x)i\tau^1,\ \ \ G_y =  g_3(\th_y)i\tau^1,
 \nonumber\\
 G_{P_x} =& \eta_{xpx}^{i_x}  \eta_{ypx}^{i_y}   g_3(\th_{px}),\ \ \ \
 G_{P_y} = \eta_{ypx}^{i_x}   \eta_{xpx}^{i_y}   g_3(\th_{py})  \nonumber\\
 G_{P_{xy}} =
 & \eta_{pxy}^{i_x}g_3( \eta_{pxy}^{\v i}\frac{\pi}{4} + \th_{pxy} ), \nonumber\\
G_T = &\eta_t^{\v i} g_3(\th_t)|_{\eta_t=-1}, \ \ 
       \eta_{pxy}^{i_x} g_3(\th_t)i\tau^1
\end{align}

\begin{align}
\label{PSGU1Cb}
 G_x =&  g_3(\th_x)i\tau^1,\ \ \ G_y =  g_3(\th_y)i\tau^1,
 \nonumber\\
 G_{P_x} =&\eta_{xpx}^{i_x}                   g_3(\th_{px})i\tau^1,\ \ \ \
 G_{P_y} = \eta_{xpx}^{i_y} \eta_{pxy}^{\v i} g_3(\th_{py})i\tau^1  
\nonumber\\
 G_{P_{xy}} =
 & \eta_{pxy}^{i_x}g_3( \eta_{pxy}^{\v i}\frac{\pi}{4} + \th_{pxy} ), \nonumber\\
G_T = &\eta_t^{\v i} g_3(\th_t)|_{\eta_t=-1}, \ \ 
       \eta_{pxy}^{i_x} \eta_t^{\v i} g_3(\th_t)i\tau^1
\end{align}

\begin{align}
\label{PSGU1Cc1}
 G_x =&  g_3(\th_x)i\tau^1,\ \ \ G_y =  g_3(\th_y)i\tau^1,
 \nonumber\\
 G_{P_x} =& \eta_{xpx}^{i_x}  \eta_{ypx}^{i_y}   g_3(\th_{px}),\ \ \ \
 G_{P_y} = \eta_{ypx}^{i_x}   \eta_{xpx}^{i_y}   g_3(\th_{py})  \nonumber\\
 G_{P_{xy}} =
 & g_3( \th_{pxy} )i\tau^1  \nonumber\\
G_T = &\eta_t^{\v i} g_3(\th_t)|_{\eta_t=-1}
\end{align}

\begin{align}
\label{PSGU1Cc2}
 G_x =&  g_3(\th_x)i\tau^1,\ \ \ G_y =  g_3(\th_y)i\tau^1,
 \nonumber\\
 G_{P_x} =& \eta_{xpx}^{i_x}  \eta_{ypx}^{i_y}   g_3(\th_{px}),\ \ \ \
 G_{P_y} = \eta_{ypx}^{i_x}   \eta_{xpx}^{i_y}   g_3(\th_{py})  \nonumber\\
 G_{P_{xy}} =
 & g_3( \eta_{pxy}^{\v i}\frac{\pi}{4} + \th_{pxy} )i\tau^1  \nonumber\\
G_T = & \eta_{pxy}^{i_x} \eta_t^{\v i} g_3(\th_t)i\tau^1
\end{align}

\begin{align}
\label{PSGU1Cd}
 G_x =&  g_3(\th_x)i\tau^1,\ \ \ G_y =  g_3(\th_y)i\tau^1,
 \nonumber\\
 G_{P_x} =&\eta_{xpx}^{i_x}                   g_3(\th_{px})i\tau^1,\ \ \ \
 G_{P_y} = \eta_{xpx}^{i_y} \eta_{pxy}^{\v i} g_3(\th_{py})i\tau^1  
\nonumber\\
 G_{P_{xy}} =
 & g_3( \eta_{pxy}^{\v i} \frac{\pi}{4} + \th_{pxy} )i\tau^1 \nonumber\\
G_T = &\eta_t^{\v i} g_3(\th_t)|_{\eta_t=-1}, \ \ 
       \eta_t^{\v i} \eta_{pxy}^{i_x} g_3(\th_t)i\tau^1
\end{align}
which will be labeled by
U1C$a_{\eta_{xpx}}b_{\eta_{ypx}}c_{\eta_{pxy}}d_{\eta_t}$.

The type U1$^m_n$ PSG's have not been classified. However, we do know that
for each rational number $m/n \in (0,1)$, there exist at least one mean-field
symmetric spin liquid, which is described by the ansatz
\begin{align}
\label{U1mnS1}
u_{\v i,\v i+\hat{\v x}} = \chi \tau^3,\ \ \
u_{\v i,\v i+\hat{\v y}} = \chi g_3( \frac{m\pi}{n} i_x)\tau^3
\end{align}
It has $\pi m/n$ flux per plaquette.  Thus there are infinite many
type U1$^m_n$ spin liquids.

We would like to point out that the above 108 U1[A,B,C] PSG's are
algebraic PSG's. They are only a subset of all possible algebraic $U(1)$
PSG's.  However, they do contain all the invariant $U(1)$ PSG's of type U1A,
U1B and U1C.  We find 46 of the 108 PSG's are also invariant PSG's.  Thus
there are 46 different mean-field $U(1)$ spin liquids of type U1A, U1B and
U1C.  Their ansatz and labels are given by \Eq{U1Aan1}, \Eq{U1Aan2},
\Eq{U1Ban1}, \Eq{U1Ban2}, and \Eq{U1Clb2} -- \Eq{U1Can10}.

To classify symmetric $SU(2)$ spin liquids, we find 8 different $SU(2)$
PSG's which are given by
\begin{align}
\label{SU2Apsg}
 G_x(\v i) =& g_x,\ \ \  G_y(\v i) = g_y
 \nonumber\\
 G_{P_x}(\v i) =& \eta_{xpx}^{i_x}\eta_{xpy}^{i_y}  g_{P_x} , \ \ \
 G_{P_y}(\v i) = \eta_{xpy}^{i_x}\eta_{xpx}^{i_y}  g_{P_y} 
\nonumber\\
 G_{P_{xy}}(\v i) =& g_{P_{xy}}, \ \ \  
 G_T(\v i) = (-)^{\v i} g_T
\end{align}
and
\begin{align}
\label{SU2Bpsg}
 G_x(\v i) =& (-)^{i_y} g_x,\ \ \  G_y(\v i) = g_y
 \nonumber\\
 G_{P_x}(\v i) =& \eta_{xpx}^{i_x}\eta_{xpy}^{i_y}  g_{P_x} , \ \ \
 G_{P_y}(\v i) = \eta_{xpy}^{i_x}\eta_{xpx}^{i_y}  g_{P_y} 
\nonumber\\
 G_{P_{xy}}(\v i) =& (-)^{i_xi_y}  g_{P_{xy}} , \ \ \
 G_T(\v i) = (-)^{\v i} g_T
\end{align}
where $g$'s are in $SU(2)$.
We would like to use the following two notations 
\begin{align}
\label{SU2ASU2B}
& \text{SU2A}\tau^0_{\eta_{xpx}}\tau^0_{\eta_{xpy}}  \nonumber\\
& \text{SU2B}\tau^0_{\eta_{xpx}}\tau^0_{\eta_{xpy}}
\end{align}
to denote the above 8 PSG's.  SU2A$\tau^0_{\eta_{xpx}}\tau^0_{\eta_{xpy}}$
is for \Eq{SU2Apsg} and SU2B$\tau^0_{\eta_{xpx}}\tau^0_{\eta_{xpy}}$ for
\Eq{SU2Bpsg}.  
We find only 4 of the 8 $SU(2)$
PSG's, SU2A$[n0,0n]$ and SU2B$[n0,0n]$, leads to $SU(2)$
symmetric spin liquids. 
The SU2A$n0$ state is the uniform RVB state and 
the SU2B$n0$ state is the $\pi$-flux state.
The other two $SU(2)$ spin liquids are given by
SU2A$0n$: 
\begin{align}
\label{SU2A0b0}
 u_{\v i,\v i+2\hat{\v x}+\hat{\v y}} =& + i\chi \tau^0 \nonumber\\
 u_{\v i,\v i-2\hat{\v x}+\hat{\v y}} =& - i\chi \tau^0 \nonumber\\
 u_{\v i,\v i+\hat{\v x}+2\hat{\v y}} =& + i\chi \tau^0 \nonumber\\
 u_{\v i,\v i-\hat{\v x}+2\hat{\v y}} =& + i\chi \tau^0 
\end{align}
and SU2B$0n$: 
\begin{align}
\label{SU2B0b0}
 u_{\v i,\v i+2\hat{\v x}+\hat{\v y}} =& + i(-)^{i_x}\chi \tau^0 \nonumber\\
 u_{\v i,\v i-2\hat{\v x}+\hat{\v y}} =& - i(-)^{i_x}\chi \tau^0 \nonumber\\
 u_{\v i,\v i+\hat{\v x}+2\hat{\v y}} =& + i\chi \tau^0 \nonumber\\
 u_{\v i,\v i-\hat{\v x}+2\hat{\v y}} =& + i\chi \tau^0 
\end{align}

The above results give us a classification of symmetric $U(1)$ and $SU(2)$
spin liquids at mean-field level.  If a mean-field state is stable against
fluctuations, it will correspond to a physical $U(1)$ or $SU(2)$ symmetric
spin liquids.  In this way the $U(1)$ and the $SU(2)$ PSG's also provide an
description of some physical spin liquids.

\section{Continuous transitions and spinon spectra in
 symmetric spin liquids } \label{simp}

\subsection{Continuous phase transitions without symmetry breaking}

After classifying mean-field symmetric spin liquids, we would like to know how
those symmetric spin liquids are related to each other.  In particular, we
would like to know which spin liquids can change into
each other through a {\em continuous} phase transition.
This problem is studied in detail in
appendix B, where we study the symmetric spin liquids in 
the neighborhood of some important symmetric spin
liquids.  After lengthy calculations, we found all the mean-field symmetric
spin liquids around the $Z_2$-linear state Z2A$001n$ in \Eq{Z2lE}, the
$U(1)$-linear state U1C$n01n$ in \Eq{U1lA}, the $SU(2)$-gapless state SU2A$n0$
in \Eq{SU2gl}, and the $SU(2)$-linear state SU2B$n0$ in \Eq{SU2lA}.
Those ansatz are given by \Eq{Z2lg1} for the $Z_2$-linear state, by
\Eq{U1glin}, \Eq{Z2U1g1}, \Eq{Z2U1g2}, \Eq{Z2U1g3}, and \Eq{Z2U1g4} for the
$U(1)$-linear state, by \Eq{SU2glg}, \Eq{U1SU2gl1} -- \Eq{U1SU2gl6}, and
\Eq{Z2SU2gl1} -- \Eq{Z2SU2gl8} for the $SU(2)$-gapless state, and by
\Eq{SU2lin}, \Eq{U1SU2lin1} -- \Eq{U1SU2lin6}, and \Eq{Z2SU2lin1} --
\Eq{Z2SU2lin7}
for the $SU(2)$-linear state.  
According to the above results, we find that, at the mean-field level, the
$U(1)$-linear spin liquid U1C$n01n$ can continuously change into
8 different $Z_2$ spin liquids, the $SU(2)$-gapless spin liquid SU2A$n0$
can continuously change into 12 $U(1)$ spin liquids and 52 $Z_2$
spin liquids, and the $SU(2)$-linear spin liquid SU2B$n0$ can
continuously change into 12 $U(1)$ spin liquids and 58 $Z_2$ spin liquids.

We would like to stress that the above results on the continuous
transitions are valid only at mean-field level. Some of the mean-field results
survive the quantum fluctuations while others do not. One need to do a case by
case study to see which mean-field results can be valid beyond the mean-field
theory.  In \Ref{MF9400}, a mean-field transition between a $SU(2)\times
SU(2)$-linear spin liquid and a $Z_2$-gapped spin liquid was studied. In
particular the effects of quantum fluctuations were discussed.

We would also like to point out that all the above spin liquids have the same
symmetry. Thus the continuous transitions between them, if exist, represent a
new class of continuous transitions which do not change any
symmetries.\cite{Wctpt}

\subsection{Symmetric spin liquids around the $U(1)$-linear spin liquid U1C$n01n$}

The $SU(2)$-linear state SU2B$n0$ (the $\pi$-flux state), the $U(1)$-linear
state U1C$n01n$ (the staggered-flux/$d$-wave state), and the $SU(2)$-gapless
state SU2A$n0$ (the uniform RVB state), are closely related to high $T_c$
superconductors.  They reproduce the observed electron spectra function for
undoped, underdoped, and overdoped samples respectively.  However,
theoretically, those spin liquids are unstable at low energies due to the
$U(1)$ or $SU(2)$ gauge fluctuations. Those states may change into more stable
spin liquids in their neighborhood.  In the next a few subsections, we are
going to study those more stable spin liquids.  Since there are still many
different spin liquids involved, we will only present some simplified results
by limiting the length of non-zero links.  Those spin liquids with short links
should be more stable for simple spin Hamiltonians.  The length of a link
between $\v i$ and $\v j$ is defined as $|i_x-j_x|+|i_y-j_y|$.  By studying
the spinon dispersion in those mean-field states, we can understand some basic
physical properties of those spin liquids, such as their stability against the
gauge fluctuations and the qualitative behaviors of spin correlations which
can be measured by neutron scattering.  Those results allow us to identify
them, if those spin liquids exist in certain samples or appear in numerical
calculations. We would like to point out that we will only study symmetric
spin liquids here. The above three unstable spin liquids may also change into
some other states that break certain symmetries. Such symmetry breaking
transitions actually have been observed in high $T_c$ superconductors (such as
the transitions to antiferromagnetic state, $d$-wave superconducting state,
and stripe state).

First, let us consider the spin liquids around the $U(1)$-linear state
U1C$n01n$.  In the neighborhood of the U1C$n01n$ ansatz \Eq{U1lA}, there are 8
classes of symmetric ansatz \Eq{Z2U1g1}, \Eq{Z2U1g2} \Eq{Z2U1g3}, and
\Eq{Z2U1g4} that break the $U(1)$ gauge structure down to a $Z_2$ gauge 
structure.
The first one is labeled by Z2A$0013$ and takes the following form
\begin{align}
\label{Z2lC}
 u_{\v i,\v i+\hat{\v x}} =& \chi \tau^1 - \eta\tau^2 \nonumber\\
 u_{\v i,\v i+\hat{\v y}} =& \chi \tau^1 + \eta\tau^2 \nonumber\\
 u_{\v i,\v i+\hat{\v x}+\hat{\v y}} =& + \ga_1 \tau^1 \nonumber\\
 u_{\v i,\v i-\hat{\v x}+\hat{\v y}} =& + \ga_1 \tau^1 \nonumber\\
 u_{\v i,\v i+2\hat{\v x}} =& \ga_2\tau^1 + \la_2 \tau^2 \nonumber\\
 u_{\v i,\v i+2\hat{\v y}} =& \ga_2\tau^1 - \la_2 \tau^2 
\nonumber\\
 a^1_0 & \neq 0,\ \ \ a^{2,3}_0=0
\end{align}
It has the same quantum order as that in the ansatz \Eq{Z2lCs}.  
The label Z2A$0013$ tells us the PSG that characterizes the spin liquid.

The second ansatz is labeled by Z2A$zz13$:
\begin{align}
\label{Z2lA}
u_{\v i,\v i+\hat{\v x}} =& \chi \tau^1 - \eta\tau^2 \nonumber\\
u_{\v i,\v i+\hat{\v y}} =& \chi \tau^1 + \eta\tau^2 \nonumber\\
 u_{\v i,\v i+\hat{\v x}+\hat{\v y}} =& - \ga_1 \tau^1 \nonumber\\
 u_{\v i,\v i-\hat{\v x}+\hat{\v y}} =& + \ga_1 \tau^1 \nonumber\\
 u_{\v i,\v i+2\hat{\v x}} =& 
 u_{\v i,\v i+2\hat{\v y}} = 0
 \nonumber\\
 a^{1,2,3}_0 & =0
\end{align}
The third one is labeled by Z2A$001n$ (or equivalently Z2A$003n$):
\begin{align}
a_0^l =& 0 \nonumber\\
u_{\v i,\v i+\hat{\v x}}  =& \chi \tau^1 +\eta \tau^2  \nonumber\\
u_{\v i,\v i+\hat{\v y}}  =& \chi \tau^1 -\eta \tau^2  \nonumber\\
u_{\v i,\v i+2\hat{\v x}+\hat{\v y}}  =& \la \tau^3  \nonumber\\
u_{\v i,\v i-\hat{\v x}+2\hat{\v y}}  =& -\la \tau^3  \nonumber\\
u_{\v i,\v i+2\hat{\v x}-\hat{\v y}}  =& \la \tau^3  \nonumber\\
u_{\v i,\v i+\hat{\v x}+2\hat{\v y}}  =& -\la \tau^3 
\label{Z2lEa}
\end{align}
Such a spin liquid has the same quantum order as \Eq{Z2lE}.
The fourth one is labeled by Z2A$zz1n$:
\begin{align}
a_0^l =& 0 \nonumber\\
u_{\v i,\v i+\hat{\v x}}  =& 
\chi \tau^1 +\eta \tau^2  
\nonumber\\
u_{\v i,\v i+\hat{\v y}}  =& \chi \tau^1 -\eta \tau^2  \nonumber\\
u_{\v i,\v i+2\hat{\v x}+\hat{\v y}}  =& 
\chi_1 \tau^1 +\eta_1 \tau^2  
+\la \tau^3  
\nonumber\\
u_{\v i,\v i-\hat{\v x}+2\hat{\v y}}  =& 
\chi_1 \tau^1 -\eta_1 \tau^2  
+\la \tau^3  
\nonumber\\
u_{\v i,\v i+2\hat{\v x}-\hat{\v y}}  =& 
\chi_1 \tau^1 +\eta_1 \tau^2  
-\la \tau^3  
\nonumber\\
u_{\v i,\v i+\hat{\v x}+2\hat{\v y}}  =& 
\chi_1 \tau^1 -\eta_1 \tau^2  
-\la \tau^3  
\label{Z2lG}
\end{align}

The above four ansatz have translation invariance.  The next four $Z_2$
ansatz do not have translation invariance. (But they still describe
translation symmetric spin liquids after the projection.) Those $Z_2$ spin
liquids are Z2B$0013$:
\begin{align}
\label{Z2lF}
 u_{\v i,\v i+\hat{\v x}} =& \chi \tau^1 - \eta\tau^2 \nonumber\\
 u_{\v i,\v i+\hat{\v y}} =& (-)^{i_x}( \chi \tau^1 + \eta\tau^2) \nonumber\\
 u_{\v i,\v i+2\hat{\v x}} =& -\ga_2\tau^1 + \la_2 \tau^2 \nonumber\\
 u_{\v i,\v i+2\hat{\v y}} =& -\ga_2\tau^1 - \la_2 \tau^2 
\nonumber\\
 a^1_0 & \neq 0,\ \ \ a^{2,3}_0=0 ,
\end{align}
Z2B$zz13$:
\begin{align}
\label{Z2BlA}
u_{\v i,\v i+\hat{\v x}} =& \chi \tau^1 - \eta\tau^2 \nonumber\\
u_{\v i,\v i+\hat{\v y}} =& (-)^{i_x} (\chi \tau^1 + \eta\tau^2) \nonumber\\
u_{\v i,\v i+2\hat{\v x}+2\hat{\v y}} =& -\ga_1 \tau^1 
\nonumber\\
u_{\v i,\v i-2\hat{\v x}+2\hat{\v y}} =& \ga_1 \tau^1  
 \nonumber\\
 a^{1,2,3}_0 & =0 ,
\end{align}
Z2B$001n$:
\begin{align}
u_{\hat{\v i,\v i+\v x}}  =& \chi \tau^1 +\eta \tau^2  \nonumber\\
u_{\hat{\v i,\v i+\v y}}  =& (-)^{i_x} (\chi \tau^1 -\eta \tau^2)  \nonumber\\
u_{\hat{\v i,\v i+2\v x+\v y}}  =& (-)^{i_x} \la \tau^3  \nonumber\\
u_{\hat{\v i,\v i-\v x+2\v y}}  =& -\la \tau^3  \nonumber\\
u_{\hat{\v i,\v i+2\v x-\v y}}  =& (-)^{i_x} \la \tau^3  \nonumber\\
u_{\hat{\v i,\v i+\v x+2\v y}}  =& -\la \tau^3 \nonumber\\
a_0^l =& 0  ,
\label{Z2BlE}
\end{align}
and Z2B$zz1n$:
\begin{align}
u_{\hat{\v x}}  =& 
\chi \tau^1 +\eta \tau^2  
\nonumber\\
u_{\hat{\v y}}  =& (-)^{i_x} (\chi \tau^1 -\eta \tau^2)  \nonumber\\
u_{\hat{2\v x+\v y}}  =& 
(-)^{i_x} (\chi_1 \tau^1 +\eta_1 \tau^2  
+\la  \tau^3  )
\nonumber\\
u_{\hat{-\v x+2\v y}}  =& 
\chi_1 \tau^1 -\eta_1 \tau^2  
+\la \tau^3  
\nonumber\\
u_{\hat{2\v x-\v y}}  =& 
(-)^{i_x} (\chi_1 \tau^1 +\eta_1 \tau^2  
-\la \tau^3  )
\nonumber\\
u_{\hat{\v x+2\v y}}  =& 
\chi_1 \tau^1 -\eta_1 \tau^2  
-\la \tau^3    \nonumber\\
a_0^l =& 0  .
\label{Z2BlG}
\end{align}

\begin{figure}[tb]
\centerline{
\hfil 
\includegraphics[height=1.3in]{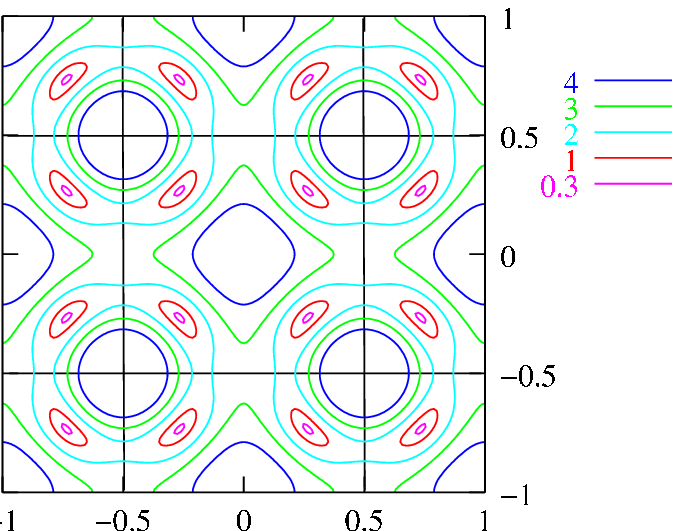}
\hfil  
\includegraphics[height=1.3in]{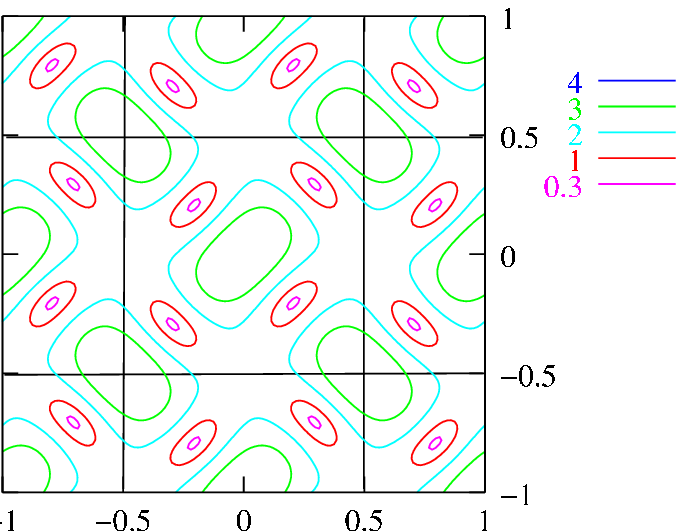} 
\hfil  
}
\centerline{ (a) \hspace{4cm} (b)\hspace{0.5cm} }
\centerline{
\hfil 
\includegraphics[height=1.3in]{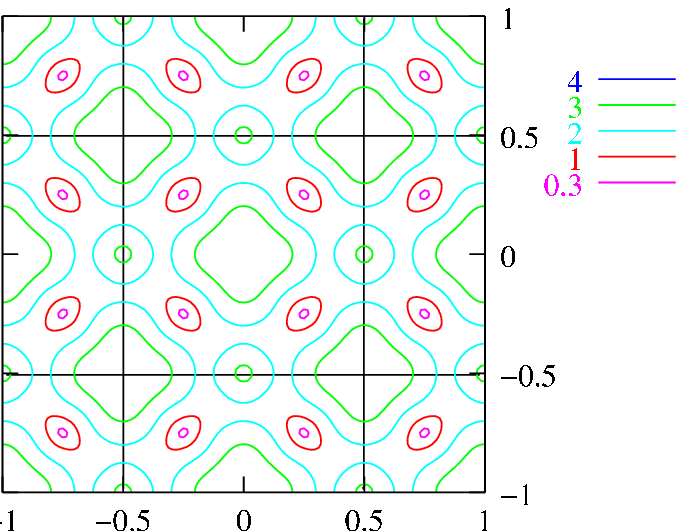} 
\hfil  
\includegraphics[height=1.3in]{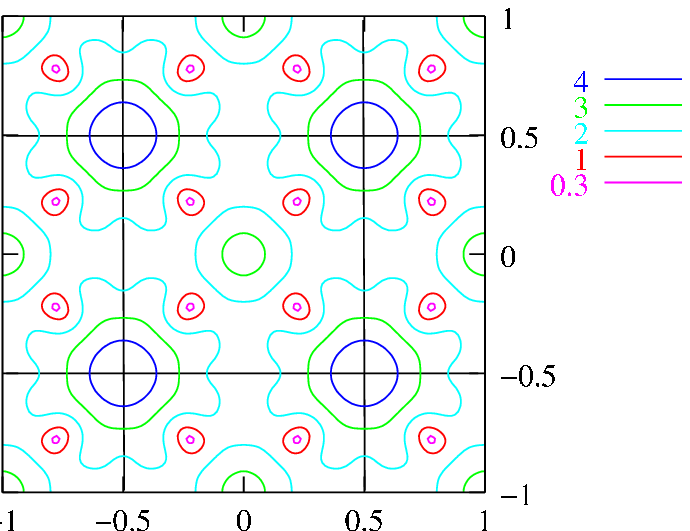}
\hfil  
}
\centerline{ (c) \hspace{4cm} (d)\hspace{0.5cm} }
\caption{
Contour plot of the spinon dispersion $E_+(\v k)$ as a function
of $(k_x/2\pi, k_y/2\pi)$ for the $Z_2$-linear spin liquids.
(a) is for the Z2A$0013$ state in \Eq{Z2lC},
(b) for the Z2A$zz13$ state in \Eq{Z2lA},
(c) for the Z2A$001n$ state in \Eq{Z2lEa},
(d) for the Z2A$zz1n$ state in \Eq{Z2lG}.
}
\label{specZ2l}
\end{figure}

The spinons are gapless at four isolated points with a linear dispersion for
the first four $Z_2$ spin liquids \Eq{Z2lC}, \Eq{Z2lA}, \Eq{Z2lEa}, and
\Eq{Z2lG}. (See Fig. \ref{specZ2l})
Therefore the four ansatz describe symmetric $Z_2$-linear spin liquids.  The
single spinon dispersion for the second $Z_2$ spin liquid Z2A$zz13$ is quite
interesting. It has the $90^\circ$ rotation symmetry around $\v k=(0,\pi)$ and
the parity symmetry about $\v k = (0,0)$.  One very important thing to notice
is that the spinon dispersions for the four $Z_2$-linear spin liquids,
\Eq{Z2lC}, \Eq{Z2lA}, \Eq{Z2lEa}, and \Eq{Z2lG} have some qualitative
differences between them.  Those differences can be used to physically measure
quantum orders (see section \ref{sec:pmQO}).

\begin{figure}[tb]
\centerline{
\hfil 
\includegraphics[height=1.3in]{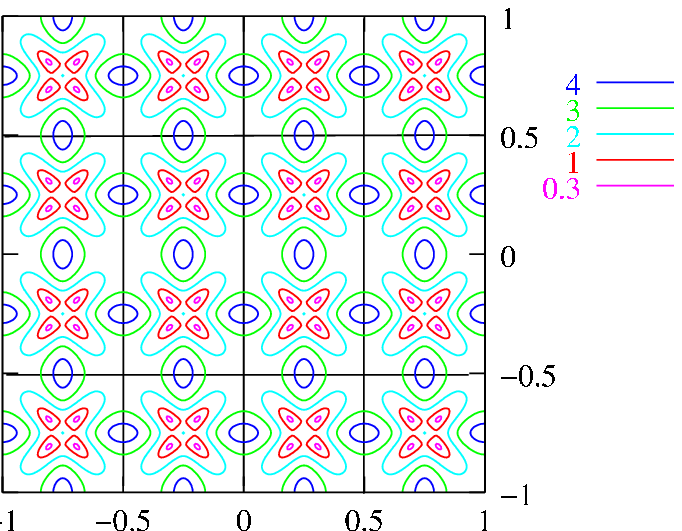}
\hfil  
\includegraphics[height=1.3in]{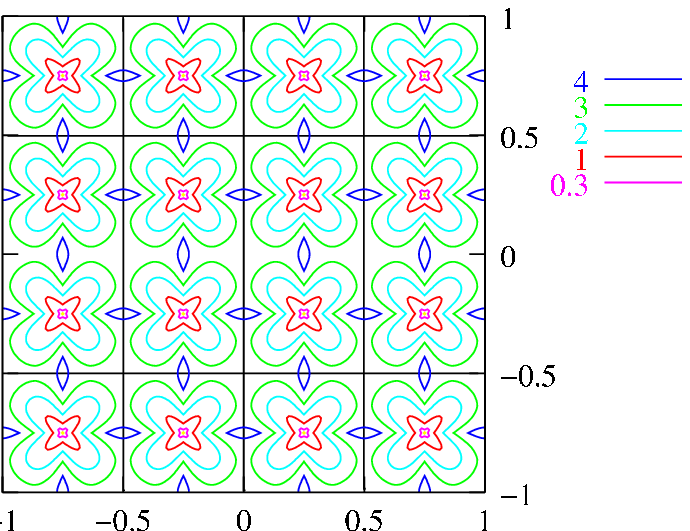} 
\hfil  
}
\centerline{ (a) \hspace{4cm} (b)\hspace{0.5cm} }
\centerline{
\hfil 
\includegraphics[height=1.3in]{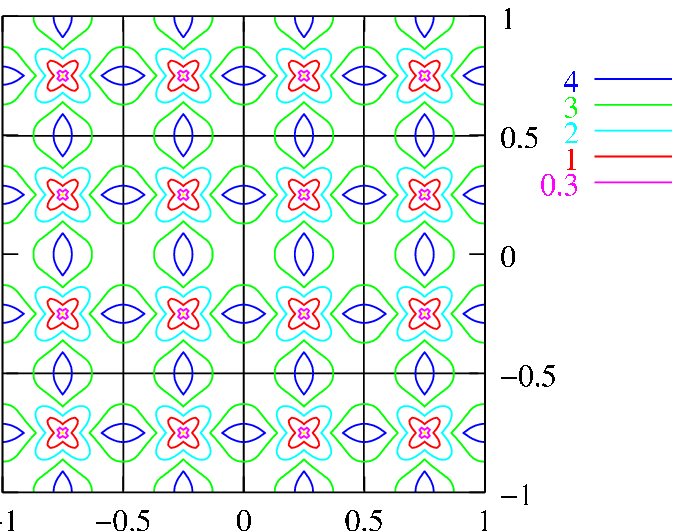} 
\hfil  
\includegraphics[height=1.3in]{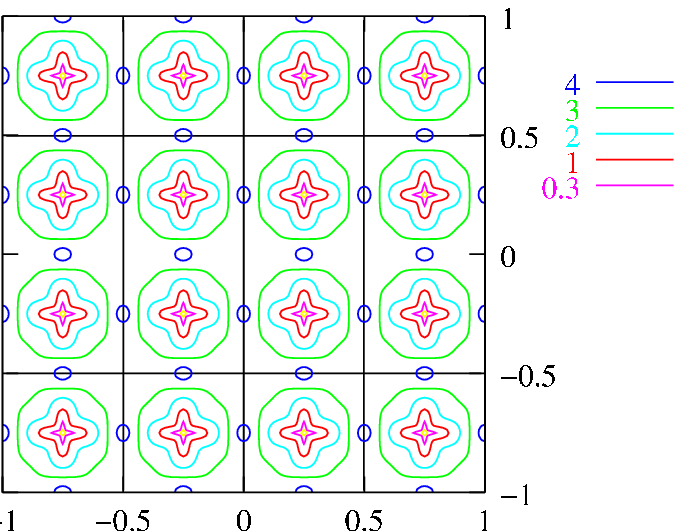}
\hfil  
}
\centerline{ (c) \hspace{4cm} (d)\hspace{0.5cm} }
\caption{
Contour plot of the spinon dispersion min$(E_1(\v k), E_2(\v k))$ 
as a function of $(k_x/2\pi, k_y/2\pi)$ for the $Z_2$-linear states.
(a) is for the Z2B$0013$ state in \Eq{Z2lF},
(b) for the Z2B$zz13$ state in \Eq{Z2BlA},
(c) for the Z2B$001n$ state in \Eq{Z2BlE},
(d) for the Z2B$zz1n$ state in \Eq{Z2BlG}.
}
\label{specZ2Bl}
\end{figure}

Next let us consider the ansatz Z2B$0013$ in
\Eq{Z2lF}.  The spinon spectrum for ansatz \Eq{Z2lF} is determined by
\begin{align}
 H =& -2\chi \cos(k_x) \Ga_0 - 2\eta \cos(k_x) \Ga_2 \nonumber\\
    & -2\chi \cos(k_y) \Ga_1 + 2\eta \cos(k_y) \Ga_3 + \la \Ga_4
\end{align}
where $k_x \in (0,\pi)$, $k_y \in (-\pi,\pi)$ and
\begin{align}
 \Ga_0 =& \tau^1\otimes \tau^3, &
 \Ga_1 =& \tau^1\otimes \tau^1, \nonumber\\
 \Ga_2 =& \tau^2\otimes \tau^3, &
 \Ga_3 =& \tau^2\otimes \tau^1, \nonumber\\
 \Ga_4 =& \tau^1\otimes \tau^0.
\end{align}
assuming $\ga_{1,2}=\la_2 =0$.  The four bands of spinon dispersion have a
form $\pm E_1(\v k)$, $\pm E_2(\v k)$.  We find the spinon spectrum vanishes
at 8 isolated points near $\v k =(\pi/2, \pm \pi/2)$. (See Fig.
\ref{specZ2Bl}a.) Thus the state Z2B$0013$ is a $Z_2$-linear spin liquid.

Knowing the translation symmetry of the above $Z_2$-linear spin liquid, it
seems strange to find that the spinon spectrum is defined only on half of
the lattice Brillouin zone. However, this is not inconsistent with translation
symmetry since the single spinon excitation is not physical. Only two-spinon
excitations correspond to physical excitations and their spectrum
should be defined on the full Brillouin zone. Now the problem is that how to
obtain two-spinon spectrum defined on the full Brillouin zone from the
single-spinon spectrum defined on half of the Brillouin zone.  Let $|\v k,1\>$
and $|\v k,2\>$ be the two eigenstates of single spinon with positive
energies $E_1(\v k)$ and $E_2(\v k)$ (here $k_x \in (-\pi/2,\pi/2)$  and
$k_y \in (-\pi,\pi)$).  The translation by $\hat{\v x}$ (followed by a
gauge transformation) change $|\v k,1\>$ and $|\v k,2\>$ to the other two
eigenstates with the same energies:
\begin{align}
|\v k,1\> \to & |\v k+\pi \hat{\v y},1\>  \nonumber\\
|\v k,2\> \to & |\v k+\pi \hat{\v y},2\>  
\end{align}
Now we see that momentum and the energy of two-spinon states
$|\v k_1,\al_1\>|\v k_2,\al_2\> \pm 
|\v k_1+\pi\hat{\v y},\al_1\>|\v k_2+\pi\hat{\v y},\al_2\> $
are given by
\begin{align}
\label{Z22spinon}
 E_{2spinon} = & E_{\al_1}(\v k_1) + E_{\al_2}(\v k_2)  \nonumber\\
 \v k = & \v k_1+\v k_2,\ \ \v k_1+\v k_2 + \pi \hat{\v x}
\end{align}
\Eq{Z22spinon} allows us to construct two-spinon spectrum from single-spinon
spectrum.

Now let us consider the ansatz Z2B$zz13$ in
\Eq{Z2BlA}.  The spinon spectrum for ansatz \Eq{Z2BlA} is determined by
\begin{align}
 H =& -2\chi \cos(k_x) \Ga_0 - 2\eta \cos(k_x) \Ga_2 \nonumber\\
    & -2\chi \cos(k_y) \Ga_1 + 2\eta \cos(k_y) \Ga_3  \\
    & -2\ga_1 \cos(2k_x+2k_y) \Ga_4 + 2\ga_1 \cos(2k_x-2k_y) \Ga_4
\nonumber 
\end{align}
where $k_x \in (0,\pi)$, $k_y \in (-\pi,\pi)$ and
\begin{align}
 \Ga_0 =& \tau^1\otimes \tau^3, &
 \Ga_1 =& \tau^1\otimes \tau^1, \nonumber\\
 \Ga_2 =& \tau^2\otimes \tau^3, &
 \Ga_3 =& \tau^2\otimes \tau^1, \nonumber\\
 \Ga_4 =& \tau^1\otimes \tau^0.
\end{align}
We find the spinon spectrum to vanish at 2 isolated points $\v k =(\pi/2, \pm
\pi/2)$. (See Fig. \ref{specZ2Bl}b.) The state Z2B$zz13$ is a $Z_2$-linear
spin liquid.

The spinon spectrum for the ansatz Z2B$001n$ in 
\Eq{Z2BlE} is determined by
\begin{align}
 H =& -2\chi \cos(k_x) \Ga_0 - 2\eta \cos(k_x) \Ga_2 \nonumber\\
    & -2\chi \cos(k_y) \Ga_1 + 2\eta \cos(k_y) \Ga_3  \nonumber\\
    & +2\la (\cos(k_x+2k_y)+\cos(-k_x+2k_y)) \Ga_4  \nonumber\\
    &  -2\la (\cos(2k_x+k_y)+\cos(2k_x-k_y)) \Ga_5
\end{align}
where $k_x \in (0,\pi)$, $k_y \in (-\pi,\pi)$ and
\begin{align}
 \Ga_0 =& \tau^1\otimes \tau^3, &
 \Ga_1 =& \tau^1\otimes \tau^1, \nonumber\\
 \Ga_2 =& \tau^2\otimes \tau^3, &
 \Ga_3 =& \tau^2\otimes \tau^1, \nonumber\\
 \Ga_4 =& \tau^3\otimes \tau^3, &
 \Ga_5 =& \tau^3\otimes \tau^1.
\end{align}
The spinon spectrum vanishes at 2 isolated points $\v k =(\pi/2, \pm \pi/2)$.
(See Fig. \ref{specZ2Bl}c.) The state Z2B$001n$ is also a $Z_2$-linear spin
liquid.

The spinon spectrum for the ansatz Z2B$zz1n$ in 
\Eq{Z2BlG} can be obtained from
\begin{align}
 H =& -2\chi \cos(k_x) \Ga_0 - 2\eta \cos(k_x) \Ga_2 \nonumber\\
    & -2\chi \cos(k_y) \Ga_1 + 2\eta \cos(k_y) \Ga_3  \nonumber\\
    & +2\la (\cos(k_x+2k_y)-\cos(-k_x+2k_y)) \Ga_4 \nonumber\\
    & -2\la (\cos(2k_x+k_y)-\cos(2k_x-k_y)) \Ga_5
\end{align}
where $k_x \in (0,\pi)$, $k_y \in (-\pi,\pi)$ and
\begin{align}
 \Ga_0 =& \tau^1\otimes \tau^3, &
 \Ga_1 =& \tau^1\otimes \tau^1, \nonumber\\
 \Ga_2 =& \tau^2\otimes \tau^3, &
 \Ga_3 =& \tau^2\otimes \tau^1, \nonumber\\
 \Ga_4 =& \tau^3\otimes \tau^3, &
 \Ga_5 =& \tau^3\otimes \tau^1.
\end{align}
We have also assumed that $\chi_1=\eta_1=0$.  The spinon spectrum vanishes at
2 isolated points $\v k =(\pi/2, \pm \pi/2)$. (See Fig. \ref{specZ2Bl}d.) The
state Z2B$zz1n$ is again a $Z_2$-linear spin liquid.

\subsection{Symmetric spin liquids around the $SU(2)$-gapless spin liquid SU2A$n0$}

There are many types of symmetric ansatz in the neighborhood of the
$SU(2)$-gapless state \Eq{SU2gl}.  Let us first consider the 12 classes of
symmetric $U(1)$ spin liquids around the $SU(2)$-gapless state \Eq{U1SU2gl1}
-- \Eq{U1SU2gl6}.  Here we just present the simple cases where $u_{\v i\v j}$
are non-zero only for links with length $\leq 2$.  Among the 12 classes of
symmetric ansatz, We find that 5 classes actually give us the $SU(2)$-gapless
spin liquid when the link length is $\leq 2$.  The other 7 symmetric $U(1)$
spin liquids are given bellow.  

{}From \Eq{U1SU2gl1tr} we get
\begin{align}
\label{U1SU2gl1sb}
  u_{\v i,\v i+\hat{\v x}} =& \chi \tau^1  -  \eta \tau^2
 &
  u_{\v i,\v i+\hat{\v y}} =& \chi \tau^1  +  \eta \tau^2
  \nonumber\\
  a^{1,2,3}_0 =& 0 \nonumber\\
  G_x =& G_y = \tau^0,  
  &
  G_{P_x} =& G_{P_y} = \tau^0, \nonumber\\
  G_{P_{xy}} =&  i\tau^1, 
  & 
  G_T =& (-)^{\v i}\tau^0
\end{align}
In the above, we have also listed the gauge transformations $G_{x,y}$,
$G_{P_x,P_y,P_{xy}}$ and $G_T$ associated translation, parity and time
reversal transformations. Those  gauge transformations define the 
PSG that characterizes the $U(1)$ spin liquid.  In section
\ref{U1SU2sec}, we have introduced a notation U1C$n01n$ to label 
the PSG and its associated ansatz.  In the following, we will list
ansatz together with their labels and the associated gauge transformations.  

{}From \Eq{U1SU2gl2tr} we get U1C$n00x$ state
\begin{align}
\label{U1lB}
  u_{\v i,\v i+\hat{\v x}} =& \chi \tau^1  
 &
  u_{\v i,\v i+\hat{\v y}} =& \chi \tau^1  
 \nonumber\\
  u_{\v i,\v i+\hat{\v x}+\hat{\v y}} =&  \eta_1 \tau^3
 &
  u_{\v i,\v i-\hat{\v x}+\hat{\v y}} =&  \eta_1 \tau^3
 \nonumber\\
  u_{\v i,\v i+2\hat{\v x}} =& \eta_2 \tau^3
 &
  u_{\v i,\v i+2\hat{\v y}} =& \eta_2 \tau^3
  \nonumber\\
  a^3_0 =& \eta_3, & a^{1,2}_0 =& 0 \nonumber\\
  G_x =& G_y = \tau^0,  
  &
  G_{P_x} =& G_{P_y} = \tau^0, \nonumber\\
  G_{P_{xy}} =&  \tau^0, 
  & 
  G_T =& i\tau^2
\end{align}
U1C$n01x$ state
\begin{align}
\label{U1glA}
  u_{\v i,\v i+\hat{\v x}} =& \chi \tau^1  
 &
  u_{\v i,\v i+\hat{\v y}} =& \chi \tau^1  
 \nonumber\\
  u_{\v i,\v i+2\hat{\v x}} =& -\eta_2 \tau^3
 &
  u_{\v i,\v i+2\hat{\v y}} =& \eta_2 \tau^3
  \nonumber\\
  a^{1,2,3}_0 =& 0 \nonumber\\
  G_x =& G_y = \tau^0,  
  &
  G_{P_x} =& G_{P_y} = \tau^0, \nonumber\\
  G_{P_{xy}} =&  i\tau^1, 
  & 
  G_T =& i\tau^2
\end{align}
and U1C$x10x$ state
\begin{align}
\label{U1qA}
  u_{\v i,\v i+\hat{\v x}} =& \chi \tau^1  
 &
  u_{\v i,\v i+\hat{\v y}} =& \chi \tau^1  
 \nonumber\\
  u_{\v i,\v i+\hat{\v x}+\hat{\v y}} =&  -\eta \tau^3
 &
  u_{\v i,\v i-\hat{\v x}+\hat{\v y}} =&  \eta \tau^3
  \nonumber\\
  a^{1,2,3}_0 =& 0 \nonumber\\
  G_x =& G_y = \tau^0,  
  &
  G_{P_x} =& G_{P_y} = i\tau^1, \nonumber\\
  G_{P_{xy}} =&  \tau^0, 
  & 
  G_T =& i\tau^2
\end{align}

From \Eq{U1SU2gl5} we get
U1A$0001$ state
\begin{align}
\label{U1glB}
  u_{\v i,\v i+\hat{\v x}} =& i\chi \tau^0  
 &
  u_{\v i,\v i+\hat{\v y}} =& i\chi \tau^0  
 \nonumber\\
  u_{\v i,\v i+\hat{\v x}+\hat{\v y}} =&  -\eta_1 \tau^3
 &
  u_{\v i,\v i-\hat{\v x}+\hat{\v y}} =&  \eta_1 \tau^3
 \nonumber\\
  u_{\v i,\v i+2\hat{\v x}} =& \eta_2 \tau^3
 &
  u_{\v i,\v i+2\hat{\v y}} =& \eta_2 \tau^3
  \nonumber\\
  a^{1,2,3}_0 =& 0 \nonumber\\
  G_x =& G_y = \tau^0,  
  &
  (-)^{i_x}G_{P_x} =& (-)^{i_x}G_{P_y} = \tau^0, \nonumber\\
  G_{P_{xy}} =&  \tau^0, 
  & 
  G_T =& i(-)^{\v i}\tau^1
\end{align}
and
U1A$0011$ state
\begin{align}
\label{U1glC}
  u_{\v i,\v i+\hat{\v x}} =& i\chi \tau^0  
 &
  u_{\v i,\v i+\hat{\v y}} =& i\chi \tau^0  
 \nonumber\\
  u_{\v i,\v i+2\hat{\v x}} =&  -\eta_2 \tau^3
 &
  u_{\v i,\v i+2\hat{\v y}} =&  \eta_2 \tau^3
  \nonumber\\
  a^{1,2,3}_0 =& 0 \nonumber\\
  G_x =& G_y = \tau^0,  
  &
  (-)^{i_x}G_{P_x} =& (-)^{i_x}G_{P_y} = \tau^0, \nonumber\\
  G_{P_{xy}} =&  i\tau^1, 
  & 
  G_T =& i(-)^{\v i}\tau^1
\end{align}

From \Eq{U1SU2gl6} we get, for $G_{P_{xy}}=g_3(\th_{pxy})$,
U1A$x10x$ state
\begin{align}
\label{U1glD}
  u_{\v i,\v i+\hat{\v x}} =& i\chi \tau^0  
 &
  u_{\v i,\v i+\hat{\v y}} =& i\chi \tau^0  
 \nonumber\\
  u_{\v i,\v i+\hat{\v x}+\hat{\v y}} =&  \eta \tau^3
 &
  u_{\v i,\v i-\hat{\v x}+\hat{\v y}} =&  \eta \tau^3
  \nonumber\\
  a^{1,2,3}_0 =& 0 \nonumber\\
  G_x =& G_y = \tau^0,  
  &
  (-)^{i_x}G_{P_x} =& (-)^{i_x}G_{P_y} = i\tau^1, \nonumber\\
  G_{P_{xy}} =&  \tau^0, 
  & 
  G_T =& i(-)^{\v i}\tau^1
\end{align}

\begin{figure}[tb]
\centerline{
\hfil 
\includegraphics[height=1.3in]{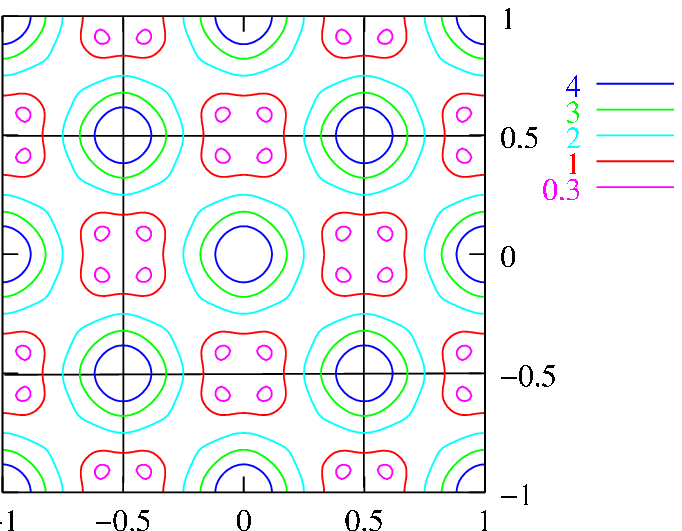} 
\includegraphics[height=1.3in]{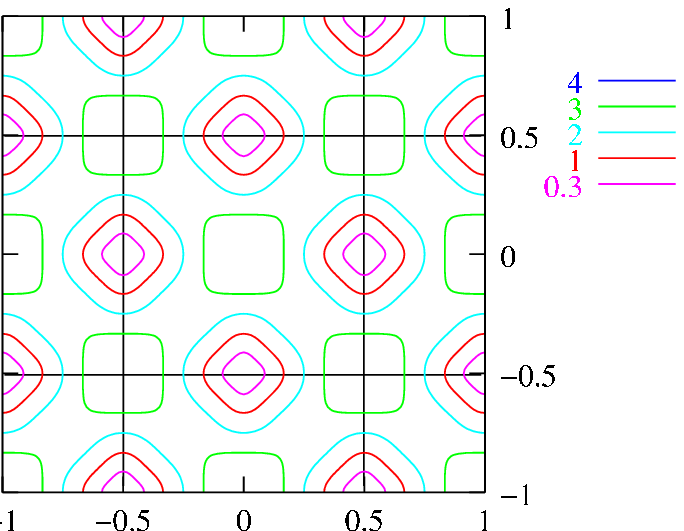}
\hfil  
}
\centerline{ (a) \hspace{4cm} (b)\hspace{0.5cm} }
\caption{
Contour plot of the spinon dispersion $E_+(\v k)$ as a function
of $(k_x/2\pi, k_y/2\pi)$ for (a) the $U(1)$-linear state
U1C$n00x$ in \Eq{U1lB}, and (b) the $U(1)$-quadratic state 
U1C$x10x$ in \Eq{U1qA}.  In the $U(1)$-quadratic state,
the spinon energy vanishes as $\Del \v k^2$ near two points 
$\v k =(\pi, 0), (0,\pi)$.
}
\label{specU1lq}
\end{figure}

\begin{figure}[tb]
\centerline{
\hfil 
\includegraphics[height=1.3in]{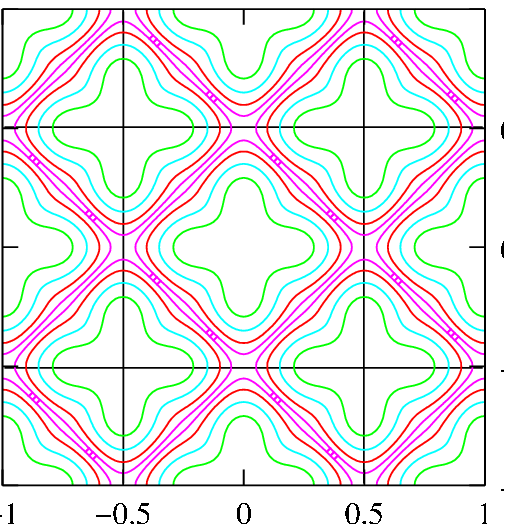} 
\includegraphics[height=1.3in]{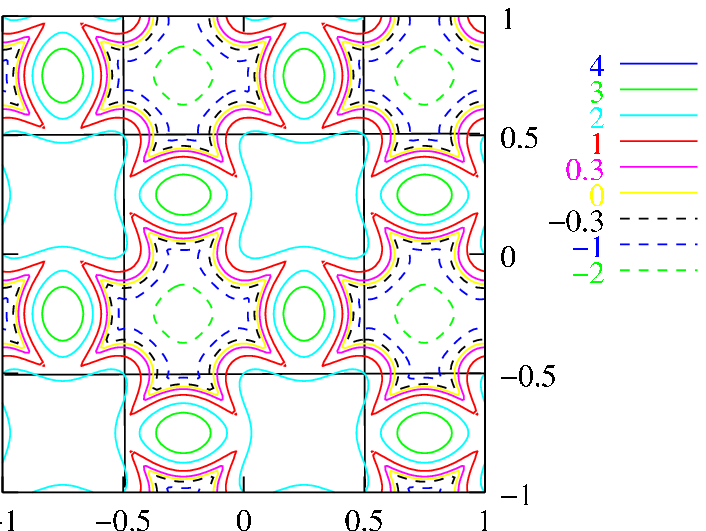}
\hfil  
}
\centerline{ (a) \hspace{4cm} (b)\hspace{0.5cm} }
\centerline{
\hfil 
\includegraphics[height=1.3in]{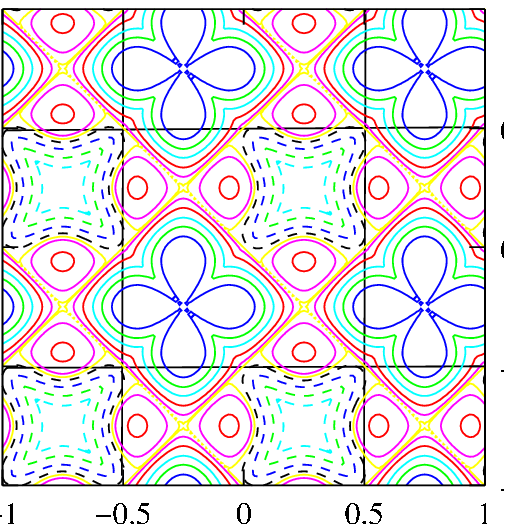} 
\includegraphics[height=1.3in]{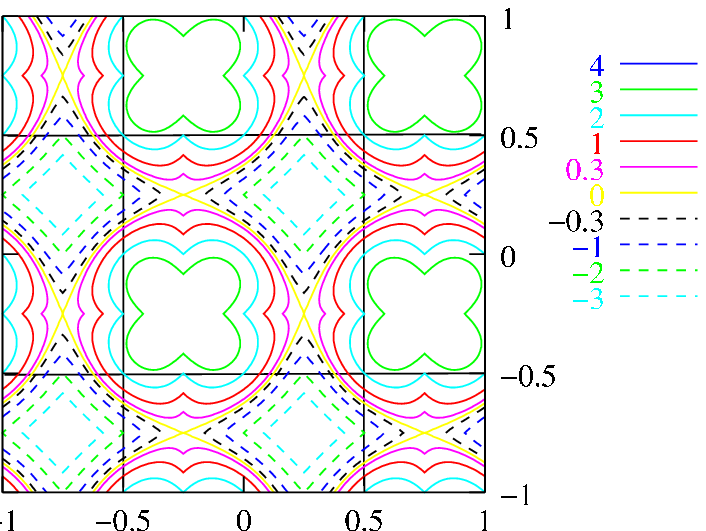}
\hfil  
}
\centerline{ (c) \hspace{4cm} (d)\hspace{0.5cm} }
\caption{
Contour plot of the spinon dispersion $E_+(\v k)$ as a function
of $(k_x/2\pi, k_y/2\pi)$ for the $U(1)$-gapless states.
(a) is for the U1C$n01x$ state \Eq{U1glA},
(b) for the U1A$0001$ state \Eq{U1glB},
(c) for the U1A$0011$ state \Eq{U1glC}, and
(d) for the U1A$x10x$ state \Eq{U1glD}.
}
\label{specU1gl}
\end{figure}

\Eq{U1SU2gl1sb} is the U1C$n01n$ $U(1)$-linear state (the staggered flux
state) studied in the last section.  After examining the spinon dispersion, we
find that the U1C$n00x$ state in \Eq{U1lB} can be a $U(1)$-linear or a
$U(1)$-gapped state depending on the value of $a^3_0$.  If it is a
$U(1)$-linear state, it will have 8 isolated Fermi points (see Fig.
\ref{specU1lq}a).  The U1C$n01x$ state in \Eq{U1glA} is a $U(1)$-gapless state
(see Fig.  \ref{specU1gl}a).  The U1C$x10x$ state in \Eq{U1qA} has two Fermi
points at $\v k_1=(\pi,0)$ and $\v k_2 = (0,\pi)$.  (see Fig.
\ref{specU1lq}b).  However, the spinon energy has a form $E(\v k)\propto (\v k
-\v k_{1,2})^2$ near $\v k_1$ and $\v k_2$.  Thus we call the U1C$x10x$ spin
liquid \Eq{U1qA} a $U(1)$-quadratic state.  The U1A$0001$ state in \Eq{U1glB},
the U1A$0011$ state in \Eq{U1glC}, and the U1A$x10x$ state in \Eq{U1glD} are
$U(1)$-gapless states (see Fig.  \ref{specU1gl}).  Again the spinon
dispersions for the $U(1)$ spin liquids have some qualitative differences
between each other, which can be used to detect different quantum orders in
those $U(1)$ spin liquids.

We next consider the 52 classes of symmetric $Z_2$ spin liquids around the
$SU(2)$-gapless state \Eq{Z2SU2gl1} -- \Eq{Z2SU2gl8}.  Here we just present
the simplest case where $u_{\v i\v j}$ are non-zero only for links with length
$\leq 1$.  We find that 48 out of 52 classes of ansatz describe $U(1)$ or
$SU(2)$ spin liquids when the link's length is $\leq 1$.  In the following we
discuss the 4 remaining $Z_2$ ansatz.

We obtain one
$Z_2$ spin liquid Z2A$x2(12)n$ from
\Eq{Z2SU2gl2}.  It is described by \Eq{Z2glA}.  
{}From \Eq{Z2SU2gl4a}, we
obtain a $Z_2$ spin liquid Z2A$0013$.  It is
described by \Eq{Z2lC} or \Eq{Z2lCs}.  {}From \Eq{Z2SU2gl5}, we obtain a
$Z_2$ spin liquid  Z2B$y1(12)n$ (note Z2B$y1(12)n$ 
is gauge equivalent to Z2B$x2(12)n$ ):
\begin{align}
\label{Z2qA}
  u_{\v i,\v i+\hat{\v x}} =& i\chi \tau^0  + \eta_1 \tau^1 
 \nonumber\\
  u_{\v i,\v i+\hat{\v y}} =& (-)^{i_x} (i\chi \tau^0  + \eta_1 \tau^2)
 \nonumber\\
 a^{1,2,3}_0 =& 0  
\end{align}
{}From \Eq{Z2SU2gl8}, we obtain a $Z_2$ spin liquid
Z2B$0013$, 
which is described by \Eq{Z2lF}.

\begin{figure}[tb]
\centerline{
\hfil 
\includegraphics[height=1.3in]{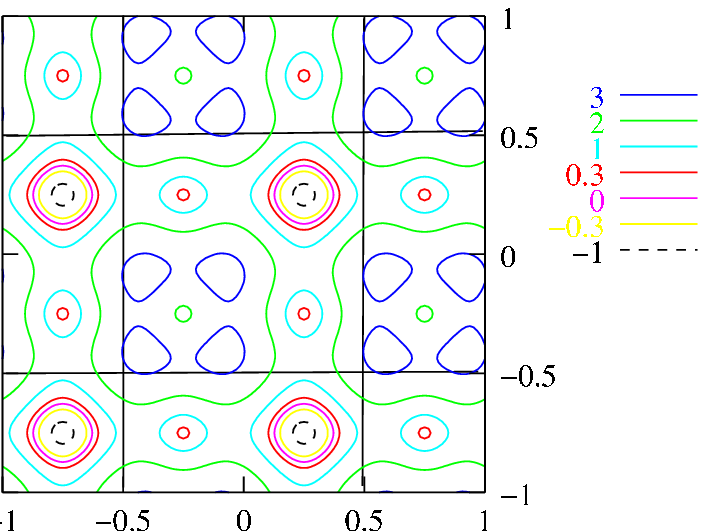} 
\includegraphics[height=1.3in]{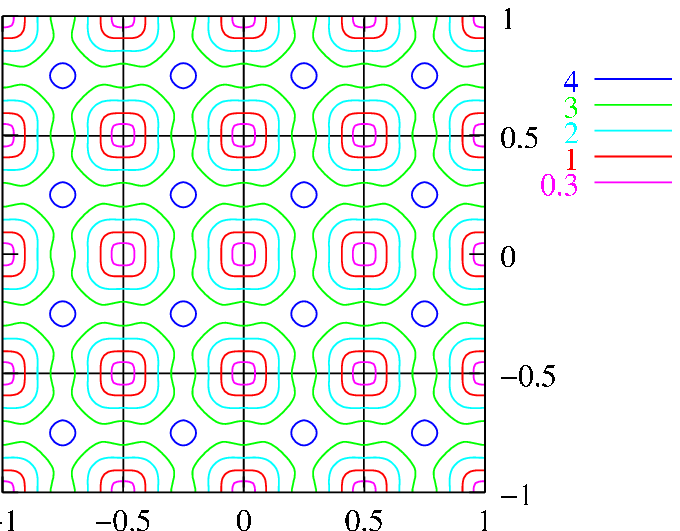}
\hfil  
}
\centerline{ (a) \hspace{4cm} (b)\hspace{0.5cm} }
\caption{
Contour plot of the spinon dispersion $E_+(\v k)$ as a function
of $(k_x/2\pi, k_y/2\pi)$ for the $Z_2$ spin liquids.
(a) is for $Z_2$-gapless state 
Z2A$x2(12)n$ in \Eq{Z2glA},
and 
(b) is for $Z_2$-quadratic state 
Z2B$x2(12)n$ in \Eq{Z2qA}.
Despite the lack of rotation and parity symmetries in the single 
spinon dispersion in (a), the two-spinon spectrum does have those 
symmetries.
}
\label{specZ2glq}
\end{figure}

The ansatz Z2B$x2(12)n$
in \Eq{Z2qA} is a new $Z_2$ spin liquid.
The spinon spectrum for ansatz \Eq{Z2qA} is determined by
\begin{align}
 H =& -2\chi \sin(k_x) \Ga_0 + 2\eta \cos(k_x) \Ga_2 \nonumber\\
    & -2\chi \sin(k_y) \Ga_1 + 2\eta \cos(k_y) \Ga_3
\end{align}
where $k_x \in (-\pi/2,\pi/2)$, $k_y \in (-\pi,\pi)$ and
\begin{align}
 \Ga_0 =& \tau^0\otimes \tau^3, &
 \Ga_2 =& \tau^1\otimes \tau^3, \nonumber\\
 \Ga_1 =& \tau^0\otimes \tau^1, &
 \Ga_3 =& \tau^2\otimes \tau^1  .
\end{align}
The spinon spectrum can be calculated exactly and its four branches take a
form $\pm E_1(\v k)$ and $\pm E_2(\v k)$.  The spinon energy vanishes at
two isolated points $\v k=(0,0), (0,\pi)$. Near $\v k =0$ the low energy
spectrum is given by (see Fig. \ref{specZ2glq}b)
\begin{equation}
E=\pm \eta^{-1} \sqrt{(\chi^2+\eta^2)^2( k_x^2-k_y^2)^2 + 4 \chi^4 k_x^2k_y^2}
\end{equation}
It is interesting to see that the energy does not vanish linearly as $\v
k\to 0$, instead it vanishes like $\v k^2$.

We find that the loop operators for the following loops
$ 
\v i\to
\v i+\hat{\v x} \to
\v i+\hat{\v x} +\hat{\v y} \to
\v i+\hat{\v y} \to
\v i
$
and
$ 
\v i\to
\v i+\hat{\v y} \to
\v i-\hat{\v x} +\hat{\v y} \to
\v i-\hat{\v x} \to
\v i
$
do not commute as long as both $\chi$ and $\eta$ are non-zero.  Thus the
spin liquid described by \Eq{Z2qA} indeed has a $Z_2$ gauge
structure.  We will call such a state $Z_2$-quadratic spin liquid
to stress the $E\propto \v k^2$ dispersion.  Such a state cannot be
constructed from translation invariant ansatz, and it is the reason
why we missed this state in the last section.  The two-spinon
spectrum is still related to the one-spinon spectrum through
\Eq{Z22spinon}.

\subsection{Symmetric spin liquids around the $SU(2)$-linear spin liquid SU2B$n0$}

Last, we consider symmetric states in the neighborhood of the
$SU(2)$-linear state \Eq{SU2lA}.  The PSG's for those
symmetric states can be obtained through the mapping \Eq{newPSG} from the
PSG's of symmetric spin liquids around the
$SU(2)$-gapless spin liquid.  Here we will only consider the 12 classes of
symmetric $U(1)$ spin liquids around the $SU(2)$-linear state given by
\Eq{U1SU2lin1} -- \Eq{U1SU2lin6}.  We will just present the simple cases
where $u_{\v i\v j}$ are non-zero only for links with length $\leq 2$.  We
find that 7 of 12 classes of ansatz actually give us $SU(2)$-gapless spin
liquids when the link length is $\leq 2$.  Thus we only obtain the
following 5 symmetric $U(1)$ spin liquids.

From \Eq{U1SU2lin1tr} we get
U1C$n01n$ ansatz
\begin{align}
\label{U1B1A}
  u_{\v i,\v i+\hat{\v x}} =&  \chi \tau^1  -  \eta \tau^2
 &
  u_{\v i,\v i+\hat{\v y}} =&  \chi \tau^1  +  \eta \tau^2
  \nonumber\\
  a^{1,2,3}_0 =&  0 \nonumber\\
  G_x =&  G_y = \tau^0,  
  &
  G_{P_x} =&  G_{P_y} = \tau^0, \nonumber\\
  G_{P_{xy}} =& i\tau^1, 
  & 
  G_T =& (-)^{\v i}\tau^0
\end{align}
which has the same quantum order as in the $U(1)$-linear state \Eq{U1lA}
(the staggered-flux state).

From \Eq{U1SU2lin2tr} we get
U1C$n0x1$ ansatz
\begin{align}
\label{U1B2A}
  u_{\v i,\v i+\hat{\v x}} =&  \chi \tau^2  
 &
  u_{\v i,\v i+\hat{\v y}} =&  \chi \tau^1  
 \nonumber\\
  u_{\v i,\v i+2\hat{\v x}} =&  -\eta \tau^3
 &
  u_{\v i,\v i+2\hat{\v y}} =&  \eta \tau^3
  \nonumber\\
  a^{1,2,3}_0 =&  0 \nonumber\\
  G_x =&  G_y = \tau^0,  
&
  G_{P_x} =&  G_{P_y} = \tau^0, \nonumber\\
  G_{P_{xy}} =&   i\tau^{12}, 
  & 
  G_T =& (-)^{i_y}i\tau^1
\end{align}
and
U1C$n0n1$ ansatz
\begin{align}
\label{U1B2B}
  u_{\v i,\v i+\hat{\v x}} =&  \chi \tau^2  
 ,\ \ \ \
  u_{\v i,\v i+\hat{\v y}} = \chi \tau^1  
 \nonumber\\
  u_{\v i,\v i+2\hat{\v x}} =&  \eta \tau^3
 ,\ \ \ \
  u_{\v i,\v i+2\hat{\v y}} = \eta \tau^3
  \nonumber\\
  a^3_0 =&  \eta_1,\ \ \ \  a^{1,2}_0 = 0 
\nonumber\\
  G_x =&  G_y = \tau^0, \ \ \ \ 
  G_T =(-)^{i_y}i\tau^1
  \nonumber\\
  G_{P_x} =&  G_{P_y} = \tau^0, \nonumber\\
  G_{P_{xy}} =&   (-)^{i_xi_y} g_3( ((-)^{i_x}-(-)^{i_y})\pi/4)  . 
\end{align}
From \Eq{U1SU2lin5} we get
U1B$0001$ ansatz
\begin{align}
\label{U1B5A}
  u_{\v i,\v i+\hat{\v x}} =&  i\chi \tau^0  
 &
  u_{\v i,\v i+\hat{\v y}} =&  i(-)^{i_x} \chi \tau^0  
 \nonumber\\
  u_{\v i,\v i+2\hat{\v x}} =&  \eta \tau^3
 &
  u_{\v i,\v i+2\hat{\v y}} =&  \eta \tau^3
  \nonumber\\
  a^3_0 =&  \eta_1, & a^{1,2}_0 =&  0 \nonumber\\
  (-)^{i_y} G_x =&  G_y = \tau^0,  
  &
  (-)^{i_x}G_{P_x} =&  (-)^{i_x}G_{P_y} = \tau^0, \nonumber\\
  G_{P_{xy}} =&  (-)^{i_xi_y} \tau^0, 
  & 
  G_T =& i(-)^{\v i}\tau^1
\end{align}
and
U1B$0011$ ansatz
\begin{align}
\label{U1B5B}
  u_{\v i,\v i+\hat{\v x}} =&  i\chi \tau^0  
 &
  u_{\v i,\v i+\hat{\v y}} =&  i(-)^{i_x}\chi \tau^0  
 \nonumber\\
  u_{\v i,\v i+2\hat{\v x}} =&   -\eta \tau^3
 &
  u_{\v i,\v i+2\hat{\v y}} =&   \eta \tau^3
  \nonumber\\
  a^{1,2,3}_0 =&  0 \nonumber\\
  (-)^{i_y} G_x =&  G_y = \tau^0,  
  &
  (-)^{i_x}G_{P_x} =&  (-)^{i_x}G_{P_y} = \tau^0, \nonumber\\
  G_{P_{xy}} =&  i(-)^{i_xi_y} \tau^1, 
  & 
  G_T =& i(-)^{\v i}\tau^1
\end{align}

\begin{figure}[tb]
\centerline{
\hfil 
\includegraphics[height=1.3in]{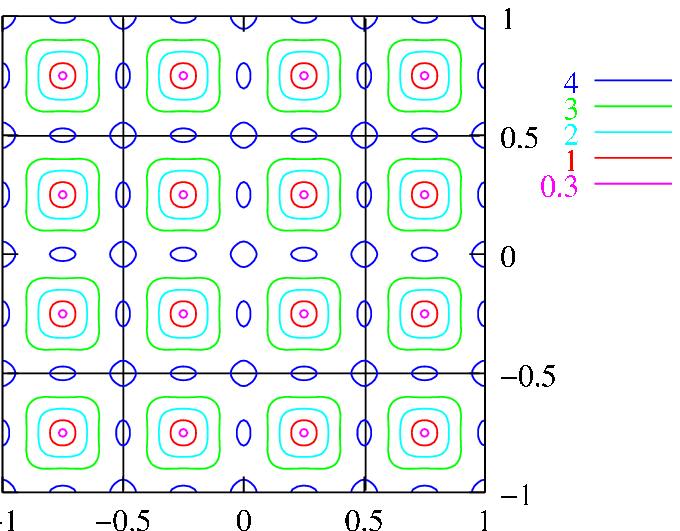}
\includegraphics[height=1.3in]{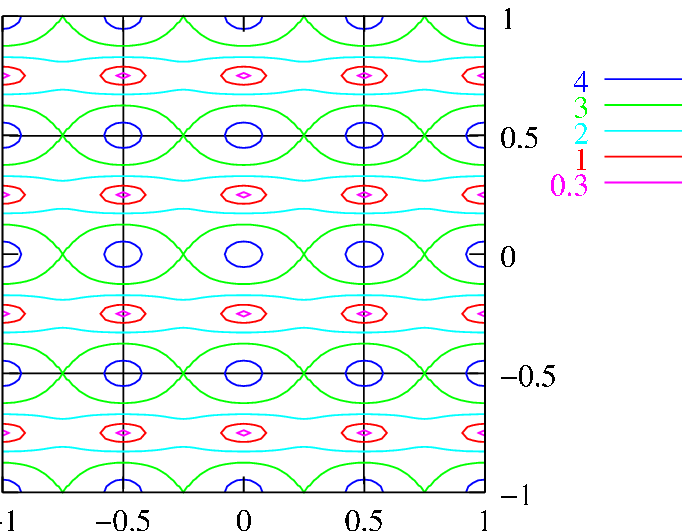}
\hfil 
}
\centerline{ (a) \hspace{4cm} (b)\hspace{0.5cm} }
\caption{
Contour plot of the spinon dispersion $E_+(\v k)$ 
as a function of $(k_x/2\pi, k_y/2\pi)$ for (a) the 
$U(1)$-linear 
state
U1C$n0x1$ in
\Eq{U1B2A} and (b) the $U(1)$-linear state \Eq{u1l90b}.
}
\label{specU1B2}
\end{figure}

\begin{figure}[tb]
\centerline{
\hfil 
\includegraphics[height=1.3in]{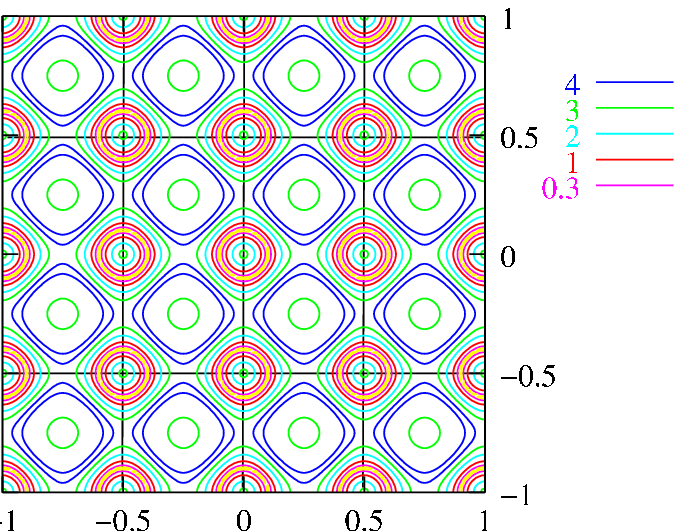}
\includegraphics[height=1.3in]{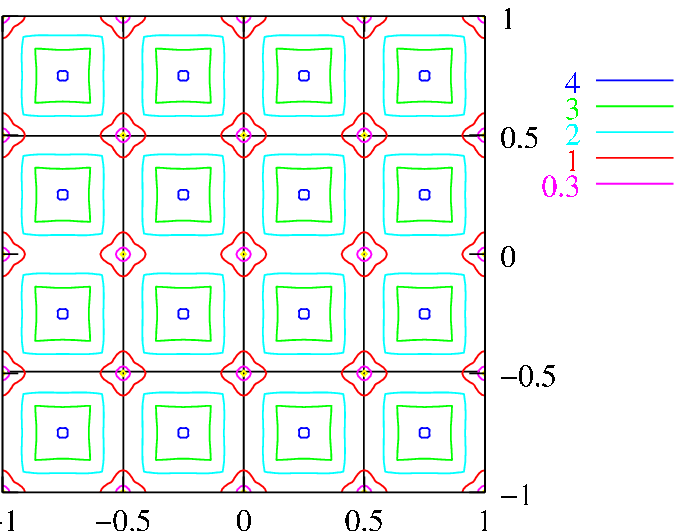} 
\hfil  
}
\centerline{ (a) \hspace{4cm} (b)\hspace{0.5cm} }
\caption{
Contour plot of the spinon dispersion min$(E_1(\v k), E_2(\v k))$ 
as a function of $(k_x/2\pi, k_y/2\pi)$ for the $U(1)$ spin liquid states.
(a) is for the $U(1)$-gapless state U1B$0001$ in \Eq{U1B5A} and
(b) is for the $U(1)$-linear state U1B$0011$ in \Eq{U1B5B}.
}
\label{specU1B}
\end{figure}

Now let us discuss spinon dispersions in the above $U(1)$ spin liquids.
The spinon in the U1C$n0x1$ state \Eq{U1B2A} has 4 linear nodes at
$(\pm\pi/2,\pm\pi/2)$.  Thus U1C$n0x1$ state is a $U(1)$-linear 
spin liquid.  The U1C$n0n1$
state \Eq{U1B2B} has fully gapped spinons and is a $U(1)$-gapped spin liquid.

The four spinon bands in the U1B$0001$ state \Eq{U1B5A} are given by
(see Fig. \ref{specU1B}a)
\begin{align}
 \pm 2 \chi \sqrt{ \sin^2(k_x) + \sin^2(k_y) } 
 \pm (2\eta \cos(2k_x)+2\eta\cos(2k_y) + \eta_1)  
\end{align}
We find that the U1B$0001$ state is a $U(1)$-gapless spin liquid.
The four spinon bands in the U1B$0011$ state \Eq{U1B5B} are given by
(see Fig. \ref{specU1B}b)
\begin{align}
 \pm 2 \chi \sqrt{ \sin^2(k_x) + \sin^2(k_y) } 
 \pm 2\eta (\cos(2k_x)-\cos(2k_y))  
\end{align}
Hence, the U1B$0011$ state is a $U(1)$-linear spin liquid.

To summarize we list all the spin liquids discussed so far in the following
table:\\
\centerline{
\begin{tabular}{|r|l|}
\hline
$Z_2$-gapped & Z2A$xx0z$ \\
\hline
$Z_2$-linear & Z2A$0013$, Z2A$zz13$, Z2A$001n$ \\
             & Z2A$zz1n$, Z2B$0013$, Z2B$zz13$ \\
	     & Z2B$001n$, Z2B$zz1n$  \\
\hline
$Z_2$-quadratic & Z2B$x2(12)n$ \\
\hline
$Z_2$-gapless & Z2A$x2(12)n$ \\
\hline
$U(1)$-gapped & U1C$n00x$ \\
\hline
$U(1)$-linear & U1B$0011$, U1C$n00x$, U1C$n01n$ \\
              & U1C$n0x1$ \\
\hline
$U(1)$-quadratic & U1C$x10x$ \\
\hline
$U(1)$-gapless & U1A$0001$, U1A$0011$, U1A$x10x$ \\
               & U1B$0001$, U1C$n01x$ \\
\hline
$SU(2)$-linear & SU2B$n0$ \\
\hline
$SU(2)$-gapless & SU2A$n0$ \\
\hline
\end{tabular}
}

\section{Mean-field phase diagram of $J_1$-$J_2$ model}
\label{mpdo}

\begin{figure}[tb]
\centerline{
\hfil 
\includegraphics[width=3.0in]{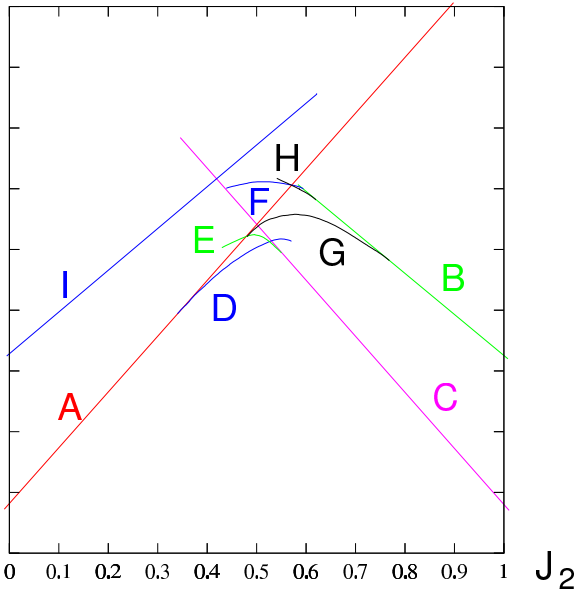}
\hfil  
}
\caption{
The mean-field energies for various phases in a $J_1$-$J_2$ spin system.
(A) the $\pi$-flux state  (the $SU(2)$-linear state SU2B$n0$). 
(B) the $SU(2)\times SU(2)$-gapless state in \Eq{su2su2gl}. 
(C) the $SU(2)\times SU(2)$-linear state in \Eq{su2su2}. 
(D) the chiral spin state (an $SU(2)$-gapped state). 
(E) the $U(1)$-linear state \Eq{u1l90b} which breaks $90^\circ$ 
rotation symmetry. 
(F) the $U(1)$-gapped state U1C$n00x$ in \Eq{U1lB}. 
(G) the $Z_2$-linear state Z2A$zz13$ in \Eq{Z2lA}.
(H) the $Z_2$-linear state Z2A$0013$ in \Eq{Z2lC}.
(I) the uniform RVB state  (the $SU(2)$-gapless state SU2A$n0$). 
}
\label{phaseJ12}
\end{figure}

To see which of the $Z_2$, $U(1)$, and $SU(2)$ spin liquids discussed in the
last section have low ground energies and may appear in real high $T_c$
superconductors, we calculate the mean-field energy of a large class of
\emph{translation invariant}
ansatz.  In Fig. \ref{phaseJ12}, we present the resulting mean-field phase
diagram for a $J_1$-$J_2$ spin system. Here $J_1$ is the nearest-neighbor spin
coupling and $J_2$ is the next-nearest-neighbor spin coupling. We have fixed
$J_1+J_2=1$.  The $y$-axis is the mean-field energy per site
(multiplied by a factor $8/3$).  The phase (A) is the $\pi$-flux state (the
SU2B$n0$ $SU(2)$-linear state) \Eq{SU2lA}.  The phase (B) is a state with two
independent uniform RVB states on the diagonal links. It has $SU(2)\times
SU(2)$ gauge fluctuations at low energies and will be called an $SU(2)\times
SU(2)$-gapless state.  Its ansatz is given by
\begin{align}
\label{su2su2gl}
u_{\v i,\v i+\hat{\v x}+\hat{\v y}} =&   \chi \tau^3   \nonumber\\
u_{\v i,\v i+\hat{\v x}-\hat{\v y}} =&   \chi \tau^3   \nonumber\\
a^l_0 =& 0
\end{align}
The phase (C) is a state with two independent $\pi$-flux states on the
diagonal links. It has $SU(2)\times SU(2)$ gauge fluctuations at low
energies and will be called an $SU(2)\times SU(2)$-linear state. Its ansatz
is given by
\begin{align}
\label{su2su2}
u_{\v i,\v i+\hat{\v x}+\hat{\v y}} =&   \chi (\tau^3  + \tau^1) \nonumber\\
u_{\v i,\v i+\hat{\v x}-\hat{\v y}} =&   \chi (\tau^3  - \tau^1) \nonumber\\
a^l_0 =& 0
\end{align}

The phase (D) is the chiral spin state \Eq{chiralspin}.  The phase (E) is
described by an ansatz
\begin{align}
\label{u1l90b}
u_{\v i,\v i+\hat{\v x}+\hat{\v y}} =& \chi_1 \tau^1 + \chi_2 \tau^2 \nonumber\\
u_{\v i,\v i+\hat{\v x}-\hat{\v y}} =& \chi_1 \tau^1 - \chi_2 \tau^2 \nonumber\\
u_{\v i,\v i+\hat{\v y}} =&   \eta \tau^3  \nonumber\\
a^l_0 =& 0
\end{align}
which break the $90^\circ$ rotation symmetry and is a $U(1)$-linear state
(see Fig. \ref{specU1B2}b).  The phase (F) is described by the U1C$n00x$
ansatz in \Eq{U1lB}.  The  U1C$n00x$ state can be a $U(1)$-linear or a
$U(1)$-gapped state.
The state for phase (F) turns out to be a $U(1)$-gapped state.  The phase
(G) is described by the Z2A$zz13$ ansatz in \Eq{Z2lA} which is a
$Z_2$-linear state.  The phase (H) is described by the Z2A$0013$ ansatz in
\Eq{Z2lC} and is also a $Z_2$-linear state.  The phase (I) is the uniform
RVB state (the $SU(2)$-gapless state SU2A$n0$ \Eq{SU2gl}).

From Fig. \ref{phaseJ12}, we see continuous phase transitions (at mean-field
level) between the following pairs of phases: (A,D), (A,G), (B,G), (C,E), and
(B,H). The three continuous transitions (B,G), (B,H) and (A,G) do not change
any symmetries.  We also note that the $SU(2)$ gauge structure in the phase
(A) breaks down to $Z_2$ in the continuous transition from the phase (A) to the
phase (G).  The $SU(2)\times SU(2)$ gauge structure in the phase (B) breaks
down to $Z_2$ in the two transitions (B,G) and (B,H).

\section{Physical measurements of quantum orders}
\label{sec:pmQO}

\begin{figure}[tb]
\centerline{
\hfil 
\includegraphics[height=1.3in]{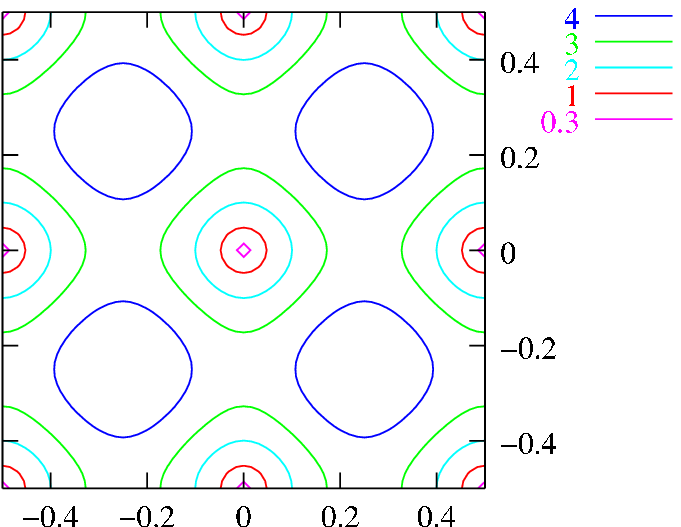} 
\includegraphics[height=1.3in]{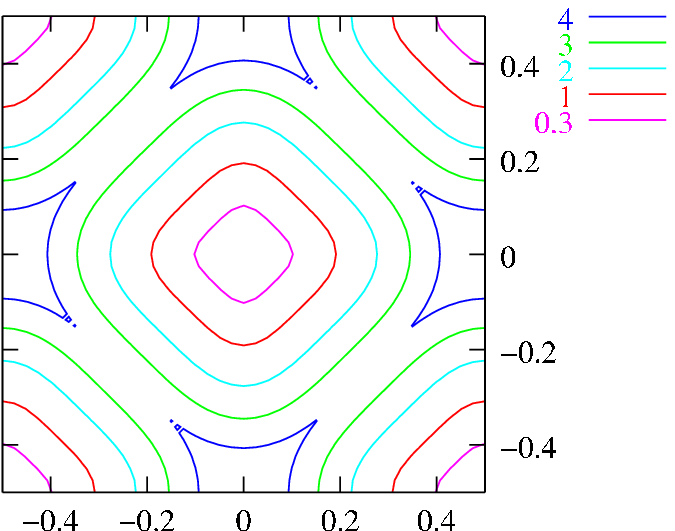}
\hfil  
}
\centerline{ (a) \hspace{4cm} (b)\hspace{0.5cm} }
\caption{
Contour plot of the dispersion for spin-1 excitation, $E_{2s}(\v k)$,
as a function
of $(k_x/2\pi, k_y/2\pi)$ for 
(a) the $SU(2)$-linear spin liquid SU2B$n0$ in 
\Eq{SU2lA} (the $\pi$-flux phase)
and (b) the $U(1)$-quadratic spin liquid U1C$x10x$ in \Eq{U1qA}.
}
\label{spec2sSU2lU1q}
\end{figure}

After characterizing the quantum orders using PSG mathematically, we would
like to ask how to measure quantum orders in experiments.  The quantum orders
in gapped states are related to the topological orders. The measurement of
topological orders are discussed in \Ref{Wrig,Wtoprev,SF0192}.  The quantum
order in a state with gapless excitations can be measured, in general, by the
dynamical properties of gapless excitation.  However, not all dynamical
properties are universal. Thus we need to identify the universal properties of
gapless excitations, before using them to characterize and measure quantum
orders. The PSG characterization of quantum orders allows us obtain
those universal properties. We simply need to identify the common properties
of gapless excitations that are shared by all the ansatz with the same PSG.

\begin{figure}[tb]
\centerline{
\hfil 
\includegraphics[height=1.3in]{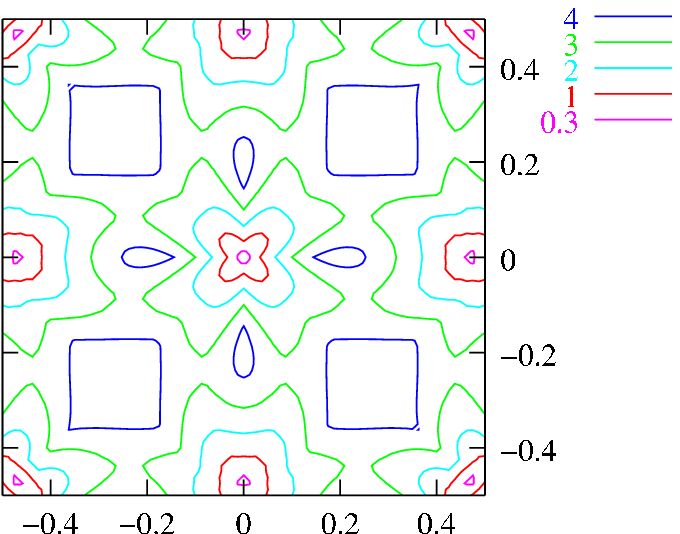} 
\includegraphics[height=1.3in]{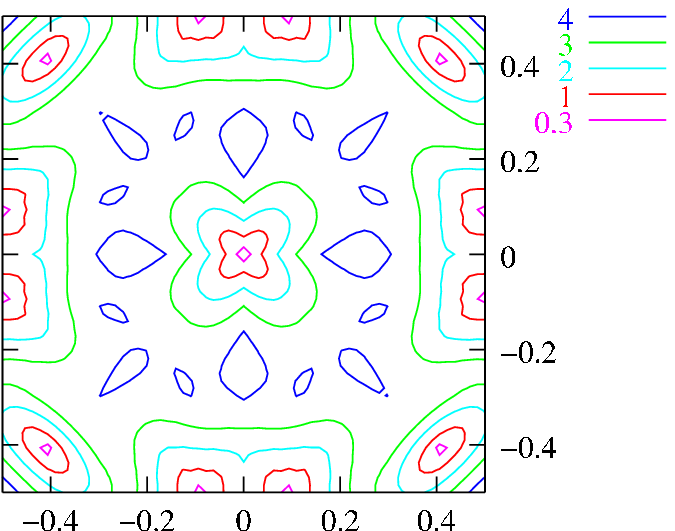}
\hfil  
}
\centerline{ (a) \hspace{4cm} (b)\hspace{0.5cm} }
\centerline{
\hfil 
\includegraphics[height=1.3in]{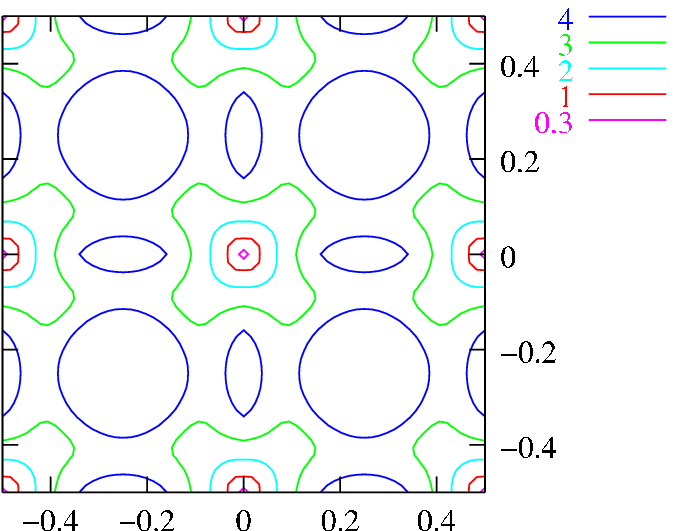} 
\includegraphics[height=1.3in]{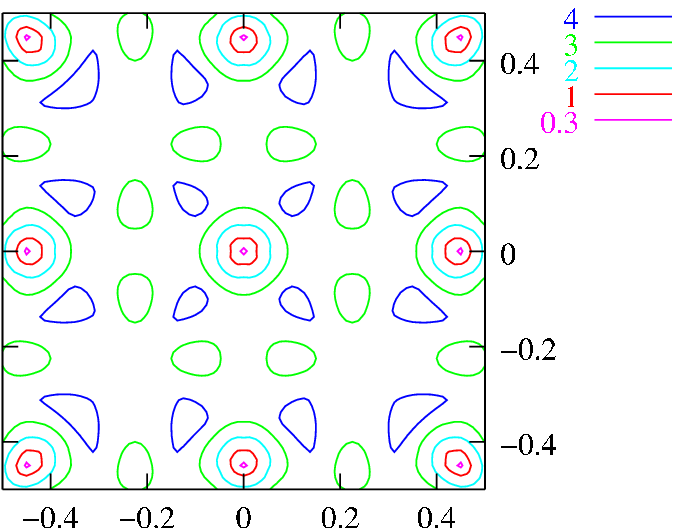} 
\hfil  
}
\centerline{ (c) \hspace{4cm} (d)\hspace{0.5cm} }
\caption{
Contour plot of $E_{2s}(\v k)$ as a 
function
of $(k_x/2\pi, k_y/2\pi)$ for the $Z_2$-linear spin liquids.
(a) is for the Z2A$0013$ state in \Eq{Z2lC},
(b) for the Z2A$zz13$ state in \Eq{Z2lA},
(c) for the Z2A$001n$ state in \Eq{Z2lEa},
(d) for the Z2A$zz1n$ state in \Eq{Z2lG}.
}
\label{spec2sZ2}
\end{figure}

\begin{figure}[tb]
\centerline{
\hfil 
\includegraphics[height=1.3in]{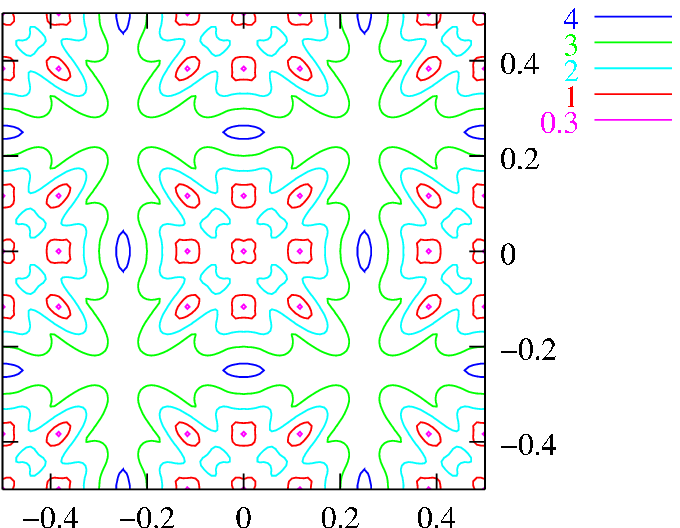} 
\includegraphics[height=1.3in]{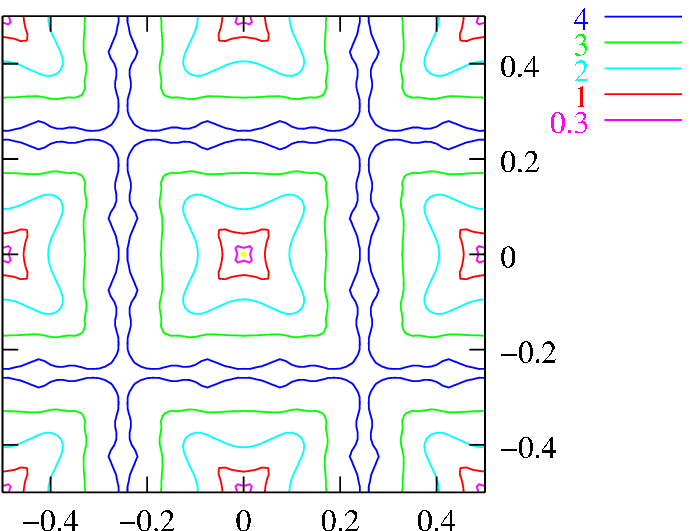}
\hfil  
}
\centerline{ (a) \hspace{4cm} (b)\hspace{0.5cm} }
\centerline{
\hfil 
\includegraphics[height=1.3in]{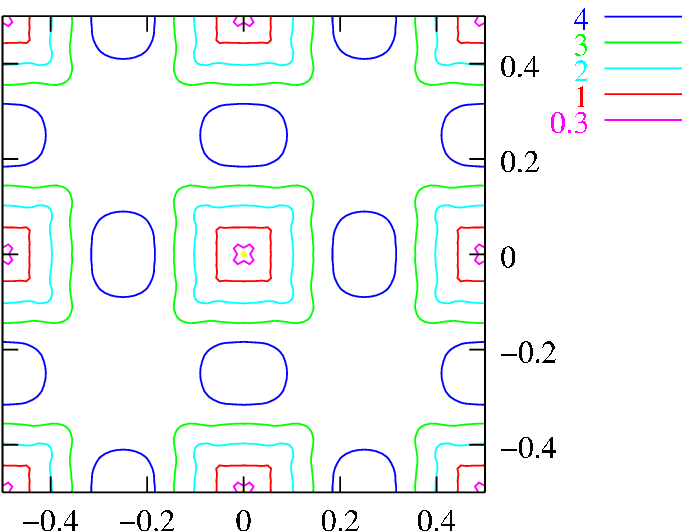} 
\includegraphics[height=1.3in]{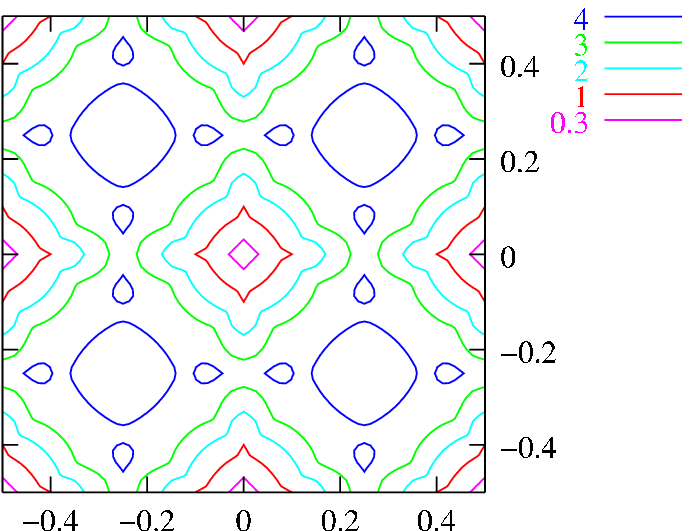} 
\hfil  
}
\centerline{ (c) \hspace{4cm} (d)\hspace{0.5cm} }
\caption{
Contour plot of $E_{2s}(\v k)$ as a 
function
of $(k_x/2\pi, k_y/2\pi)$ for the $Z_2$-linear spin liquids.
(a) is for the Z2B$0013$ state in \Eq{Z2lF},
(b) for the Z2B$zz13$ state in \Eq{Z2BlA},
(c) for the Z2B$001n$ state in \Eq{Z2BlE},
(d) for the Z2B$zz1n$ state in \Eq{Z2BlG}.
}
\label{spec2sZ2B}
\end{figure}

\begin{figure}[tb]
\centerline{
\hfil
\includegraphics[height=1.3in]{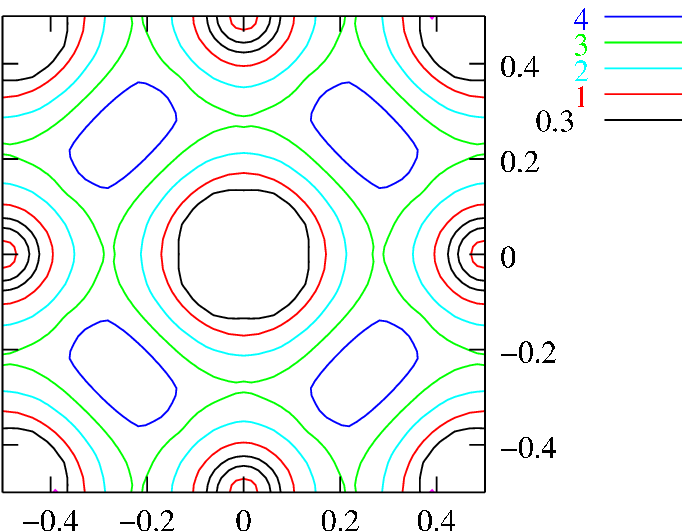}
\includegraphics[height=1.3in]{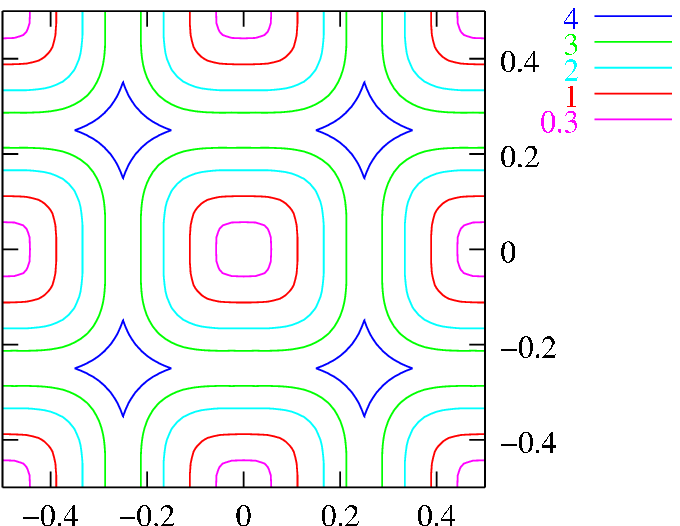}
\hfil
}
\centerline{ (a) \hspace{4cm} (b)\hspace{0.5cm} }
\caption{
Contour plot of $E_{2s}(\v k)$ as a function of $(k_x/2\pi, k_y/2\pi)$ 
for
(a) the $Z_2$-gapless state 
Z2A$x2(12)n$ in \Eq{Z2glA},
and 
(b) the $Z_2$-quadratic state 
Z2B$x2(12)n$ in \Eq{Z2qA}.
}
\label{spec2sZ2gl}
\end{figure}

\begin{figure}[tb]
\centerline{
\hfil 
\includegraphics[height=1.3in]{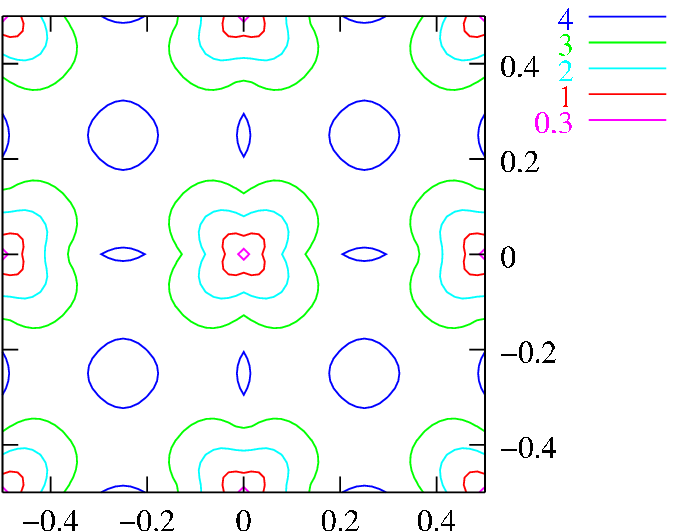} 
\includegraphics[height=1.3in]{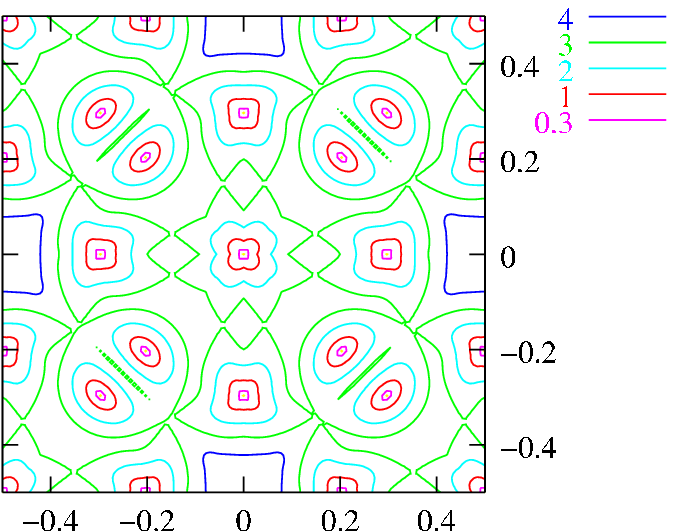}
\hfil  
}
\centerline{ (a) \hspace{4cm} (b)\hspace{0.5cm} }
\caption{
Contour plot of $E_{2s}(\v k)$ as a function
of $(k_x/2\pi, k_y/2\pi)$ for two $U(1)$-linear spin liquids.
(a) is for the U1C$n01n$ state \Eq{U1lA} (the staggered flux phase),
and (b) for the U1C$n00x$ state \Eq{U1lB} in the gapless phase. 
}
\label{spec2sU1l}
\end{figure}

\begin{figure}[tb]
\centerline{
\includegraphics[height=1.3in]{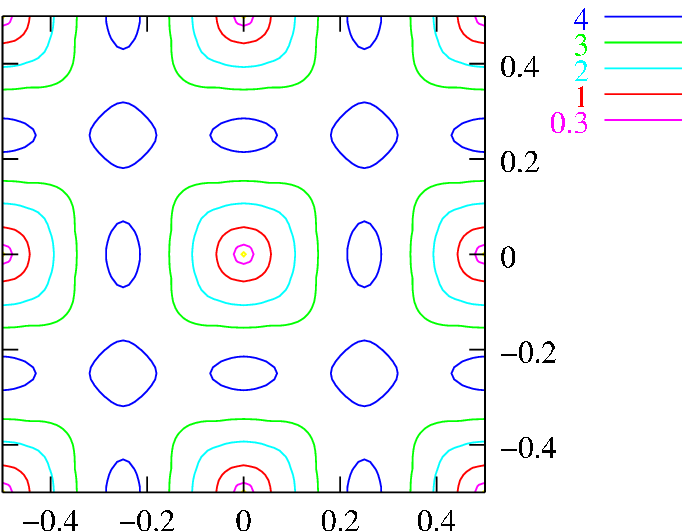}
\includegraphics[height=1.3in]{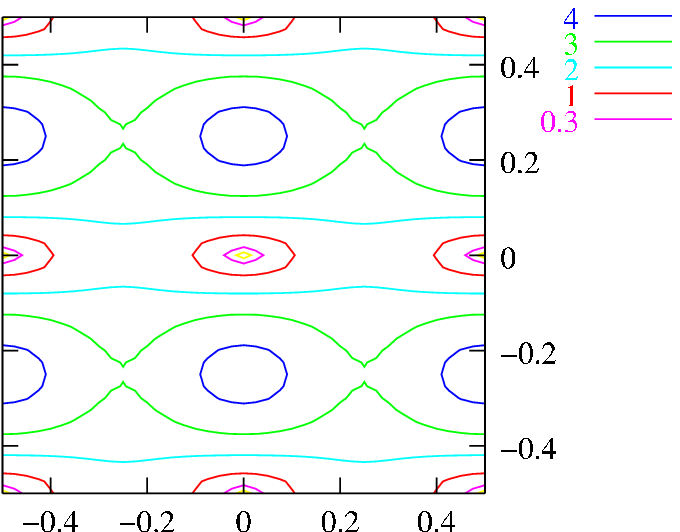}
\hfil 
}
\centerline{ (a) \hspace{4cm} (b)\hspace{0.5cm} }
\caption{
Contour plot of the two-spinon dispersion $E_{2s}(\v k)$ 
as a function of $(k_x/2\pi, k_y/2\pi)$ for (a) the $U(1)$-linear
spin liquid state U1C$n0x1$ in \Eq{U1B2A} and 
(b) the $U(1)$-linear spin liquid \Eq{u1l90b}.
}
\label{spec2sU1B2}
\end{figure}

\begin{figure}[tb]
\centerline{
\hfil 
\includegraphics[height=1.3in]{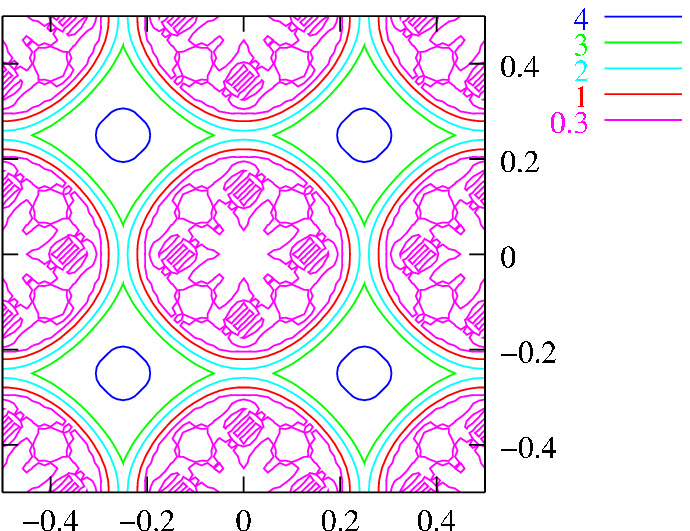}
\includegraphics[height=1.3in]{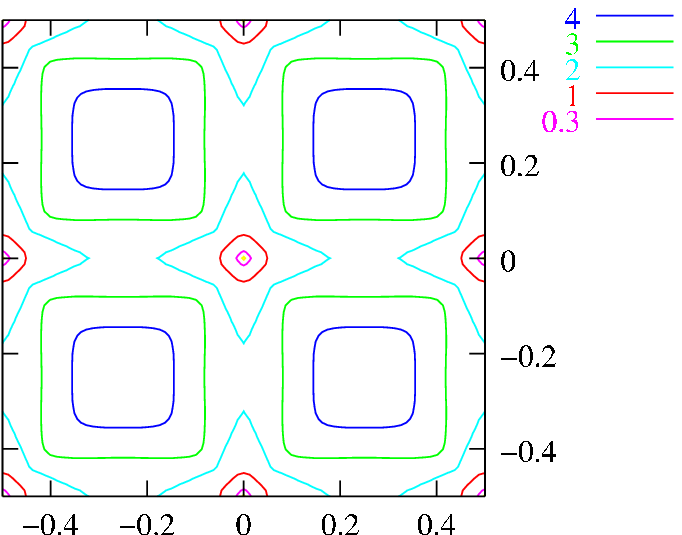} 
\hfil  
}
\centerline{ (a) \hspace{4cm} (b)\hspace{0.5cm} }
\caption{
Contour plot of the two-spinon dispersion $E_{2s}(\v k)$ 
as a function of $(k_x/2\pi, k_y/2\pi)$ for the $U(1)$ spin liquid states.
(a) is for the $U(1)$-gapless state U1B$0001$ in \Eq{U1B5A} and
(b) is for the $U(1)$-linear state U1B$0011$ in \Eq{U1B5B}.
}
\label{spec2sU1B}
\end{figure}

To demonstrate the above idea, we would like to study the spectrum of
two-spinon excitations. We note that spinons can only be created in pairs.
Thus the one-spinon spectrum is not physical. We also note that the two-spinon
spectrum include spin-1 excitations which can be measured in experiments.
At a given momentum, the
two-spinon spectrum is distributed in one or several ranges of energy. Let
$E_{2s}(\v k)$ be the lower edge of the two-spinon spectrum at momentum $\v
k$.  In the mean-field theory, the two-spinon spectrum can be constructed
from the one-spinon dispersion
\begin{equation}
 E_{\text{2-spinon}}(\v k) = E_{\text{1-spinon}}(\v q) +E_{\text{1-spinon}}(\v k-\v q)
\end{equation}
In Fig. \ref{spec2sSU2lU1q} -- \ref{spec2sU1B} we present mean-field
$E_{2s}$ for some simple spin liquids. If the mean-field state is stable
against the gauge fluctuations, we expect the mean-field $E_{2s}$ should
qualitatively agrees with the real $E_{2s}$.

Among our examples, there are eight $Z_2$-linear spin liquids (see Fig.
\ref{spec2sZ2} and Fig. \ref{spec2sZ2B}).  We see that some of those eight
different $Z_2$-linear spin liquids (or eight different quantum orders)
have different number of gapless points.  The gapless points of some spin
liquids are pinned at position $\v k = (\pi,\pi)$ and/or $\v k = (\pi, 0),
(0,\pi)$. By measuring the low energy spin excitations (say using neutron
scattering), we can distinguish those $Z_2$ spin liquids.  We note that all
the two-spinon spectra have rotation and parity symmetries around $\v k=0$.
This is expected.  Since the two-spinon spectra are physical, they should
have all the symmetries the spin liquids have.  

We also have four $U(1)$-linear spin liquids. Some of them can be
distinguished by their different numbers of gapless points.  It is
interesting to note that all the $U(1)$ spin liquids discussed here have a
gapless point in the two-spinon spectrum pinned at position $\v k =
(\pi,\pi)$.  The $U(1)$-linear spin liquids are also different from the
$Z_2$-linear spin liquids in that the spin-spin correlations have different
decay exponents once the $U(1)$ gauge fluctuations are included.  
We also see that $E_{2s}$ has a quadratic form $E_{2s}
\propto \v k^2$ for the $U(1)$-quadratic spin liquid. $E_{2s}$ vanishes in
two finite regions in $\v k$-space for the $Z_2$-gapless spin liquids.

Neutron scattering experiments probe the two-spinon sector. Thus low energy
neutron scattering allows us to measure quantum orders in high $T_c$
superconductors.

Let us discuss the $U(1)$ linear state U1C$n01n$ (the staggered-flux state) in
more detail. The U1C$n01n$ state is proposed to describe the pseudo-gap metallic
state in underdoped high $T_c$ superconductors.\cite{WLsu2,RWspec} The
U1C$n01n$ state naturally explains the spin pseudo-gap in the underdoped
metallic state. As an algebraic spin liquid, the  U1C$n01n$ state also explain
the Luttinger-like electron spectral function\cite{RWspec} and the enhancement
of the $(\pi,\pi)$ spin fluctuations\cite{RWspin} in the pseudo-gap state.  From
Fig. \ref{spec2sU1l}a, we see that gapless points of the spin-1 excitations in
the U1C$n01n$ state are always at $\v k=(\pi,\pi)$, $(0,0)$, $(\pi,0)$ and
$(0,\pi)$.  The equal energy contour for the edge of the spin-1 continuum has
a shape of two overlapped ellipses at all the four $\v k$ points. Also the
energy contours are not perpendicular to the zone boundary. All those are the
universal properties of the U1C$n01n$ state. Measuring those properties in
neutron scattering experiments will allow us to determine if the pseudo-gap
metallic state is described by the U1C$n01n$ (the staggered-flux) state or
not.

We have seen that at low energies, the U1C$n01n$ state is unstable due to the
instanton effect. Thus the U1C$n01n$ state has to change into some other
states, such as the 8 $Z_2$ spin liquids discussed in section \ref{simp} or
some other states not discussed in this paper.
From Fig. \ref{spec2sZ2}a,
we see that the transition from the U1C$n01n$ state
to the $Z_2$-linear state Z2A$0013$ can be detected by neutron scattering
if one observe the splitting of the node at $(\pi,\pi)$ into four
nodes at $(\pi\pm\del,\pi\pm\del)$ and the splitting of the nodes at $(\pi,0)$
and $(0,\pi)$ into two nodes at $(\pi\pm\del,0)$ and $(0,\pi\pm\del)$.
From Fig. \ref{spec2sZ2}b,
we see that, for the transition from the U1C$n01n$ state
to the $Z_2$-linear state Z2A$zz13$, the node at $(\pi,\pi)$ still splits
into four nodes at $(\pi\pm\del,\pi\pm\del)$. However, the nodes at $(\pi,0)$
and $(0,\pi)$ split differently
into two nodes at $(\pi,\pm\del)$ and $(\pm\del,\pi)$.
We can also study the transition from the U1C$n01n$ state to other 6 $Z_2$
spin liquids. We  find the spectrum of spin-1 excitations all change in certain
characteristic ways. Thus by measuring the spin-1 excitation spectrum and its
evolution, we not
only can detect a quantum transition that do not change any symmetries, we can
also tell which transition is happening.

The neutron scattering on high $T_c$ superconductor indeed showed a
splitting of the scattering peak at $(\pi,\pi)$ into four peaks at
$(\pi\pm\del,\pi)$,  $(\pi,\pi\pm\del)$
\cite{YMK8886,BEH8968,CAM9191,FKM9713,BFR9739,YLK9865,FBS0073,DMH0125} or into
two peaks at  $(\pi,\pi)\to (\pi+\del,\pi-\del), (\pi-\del,\pi+\del)$
\cite{WBK0099,MFY0048} as we lower the energy.  This is consistent with our
belief that the U1C$n01n$ state is unstable at low energies. However, it is
still unclear if we can identify the position of the neutron scattering peak
as the position of the node in the spin-1 spectrum.  If we do identify the
scattering peak as the node, then non of the 8 $Z_2$ spin liquids in the
neighborhood of the U1C$n01n$ state can explain the splitting pattern
$(\pi\pm\del,\pi)$,  $(\pi,\pi\pm\del)$. This will imply that the U1C$n01n$
state change into another state not studied in this paper.  This example
illustrates that detailed neutron scattering experiments are powerful tools in
detecting quantum orders and studying new transitions between quantum orders
that may not change any symmetries.

\section{Four classes of spin liquids and their stability} \label{stb}

We have concentrated on the mean-field states of spin liquids and presented
many examples of mean-field ansatz for symmetric spin liquids. In order for
those mean-field states to represent real physical spin liquids, we need to
include the gauge fluctuations. We also need to show that the inclusion of the
gauge fluctuations does not destabilize the mean-field states at low energies.
This requires that (a) the gauge interaction is not too strong and (b) the
gauge interaction is not a relevant perturbation. (The gauge interaction,
however, can be a marginal perturbation.) The requirement (a) can be satisfied
through large $N$ limit and/or adjustment of short-range spin couplings in the
spin Hamiltonian, if necessary. Here we will mainly consider the requirement
(b).  We find that, at least in certain large $N$ limits, many (but not all)
mean-field states do correspond to real quantum spin liquids which are stable
at low energies.  In this case, the characterization of the mean-field states
by PSG's correspond to the characterization of real quantum spin liquids. 

All spin liquids (with odd number of electron per unit cell) studied so far
can be divided into four classes. In the following we will study each
classes in turn.

\subsection{Rigid spin liquid}

In rigid spin liquids, by definition, the spinons and all other
excitations are fully gaped. The gapped gauge field only induces short
range interaction between spinons due to Chern-Simons terms or
Anderson-Higgs mechanism. By definition, the rigid spin liquids are locally
stable and self consistent. 
The rigid spin liquids are characterized by topological orders and they
have the true spin-charge separation.  The low energy effective theories
for rigid spin liquids are topological field theories.  The $Z_2$-gapped
spin liquid and chiral spin liquid are examples of rigid spin liquids.

\subsection{Bose spin liquid}

The $U(1)$-gapped spin liquid discussed in the last section is not a rigid
spin liquid. It is a Bose spin liquid.  Although the spinon excitations are
gapped, the $U(1)$ gauge fluctuations are gapless in the $U(1)$-gapped spin
liquid.  The dynamics of the gapless $U(1)$ gauge fluctuations are
described by low energy effective theory
\begin{equation}
 \cL = \frac{1}{2g} (f_{\mu\nu})^2
\end{equation}
where $f_{\mu\nu}$ is the field strength of the $U(1)$ gauge field.
However, in 1+2 dimension and after including the instanton effect, the
$U(1)$ gauge fluctuations will gain an energy gap.\cite{P7729} The
properties of the resulting quantum state remain to be an open problem.

\subsection{Fermi spin liquid}
\label{fsl}

The Fermi spin liquids have gapless excitations that are described by spin
1/2 fermions. Those gapless excitations have only short range interactions
between them. 
The $Z_2$-linear, $Z_2$-quadratic and the $Z_2$-gapless spin liquid
discussed above are examples of the Fermi spin liquids. 

The spinons have a massless Dirac dispersion in $Z_2$-linear spin liquids.
Thus $Z_2$-linear spin liquids are locally stable since short range
interactions between massless Dirac fermions are irrelevant at 1+2
dimensions.  We would like to point out that the massless Dirac dispersion
of the $Z_2$-linear spin liquids are protected by the PSG (or the quantum
order).  That is
any perturbations around, for example, the $Z_2$-linear ansatz \Eq{Z2lE}
cannot destroy the massless Dirac dispersion as long as the PSG are
not changed by the perturbations.  To understand this result, we start with the most general form
of symmetric perturbations \Eq{Z2lg1} around the $Z_2$-linear ansatz
\Eq{Z2lE}.  We find that such perturbations vanish in the momentum space at
$\v k = (\pm \pi/2, \pm \pi/2)$.  The translation, parity, and the time
reversal symmetries do not allow any mass terms or chemical potential
terms.  Thus the $Z_2$-linear spin liquid is a phase that occupy a finite
region in the phase space (at $T=0$). \emph{One does not need any fine tuning of
coupling constants and $u_{\v i\v j}$ to get massless Dirac spectrum.}

Now let us consider the stability of the $Z_2$-quadratic spin liquid
\Eq{Z2qA}.  The spinons have a gapless quadratic dispersion in the
$Z_2$-quadratic spin liquid.  The gapless quadratic dispersion of the
$Z_2$-quadratic spin liquid is also protected by the symmetries.  The most
general form of symmetric perturbations around the $Z_2$-quadratic ansatz
\Eq{Z2qA} is given by \Eq{Z2SU2gl5} (Z2B$x2(12)n$).
In the momentum space, the most general symmetric
$Z_2$-quadratic ansatz give rise to the following Hamiltonian
(after considering the $90^\circ$ rotation symmetry)
\begin{align}
 H =& -2\sum\chi_{mn} [\sin(nk_x-mk_y) \Ga_0+\sin(mk_x+nk_y) \Ga_1] \nonumber\\
    & +2\sum\eta_{mn} [\cos(nk_x-mk_y) \Ga_2+\cos(mk_x+nk_y) \Ga_3] \nonumber\\
    & +2\sum\la_{mn}  [\cos(nk_x-mk_y) \Ga_4+\cos(mk_x+nk_y) \Ga_5]
\end{align}
where
\begin{align}
 \Ga_0 =&  \tau^0\otimes \tau^3, &
 \Ga_2 =&  \tau^1\otimes \tau^3, \nonumber\\
 \Ga_1 =&  \tau^0\otimes \tau^1, &
 \Ga_3 =&  \tau^2\otimes \tau^1, \nonumber\\
 \Ga_4 =& -\tau^2\otimes \tau^3, &
 \Ga_5 =&  \tau^1\otimes \tau^1  .
\end{align}
and the summation is over $m=$even, $n=$odd.
We find that the spinon dispersion still vanish at $\v k = (0,0), (0,\pi)$ and
the energy still satisfy $E\propto \v k^2$.
The translation, parity, and the time reversal symmetric
perturbations do not change the qualitative behavior of the low energy spinon
dispersion.
Thus, at mean-field level,
the $Z_2$-quadratic  spin liquid is a phase that occupy a finite
region in the phase space (at $T=0$). One does not need any fine tuning of
coupling constants to get gapless quadratic dispersion of the spinons.
However, unlike the $Z_2$-linear spin liquid, the short
range four-fermion interactions between the gapless spinons in the
$Z_2$-quadratic state are marginal at 1+2
dimensions. Further studies are needed to understand the dynamical stability
of the $Z_2$-quadratic spin liquid beyond the mean-field level. 

The $Z_2$-gapless spin liquid is as stable as Fermi liquid in 1+2 dimensions.
Again we expect $Z_2$-gapless spin liquid to be a phase that occupy a finite
region in the phase space, at least at mean-field level.

\subsection{Algebraic spin liquid}
\label{asl}

$U(1)$-linear spin liquids are examples of algebraic spin liquids. Their
low lying excitations are described by massless Dirac fermions coupled to
$U(1)$ gauge field.  Although the massless Dirac fermions are protected by
quantum orders, the gauge couplings
remain large at low energies. Thus the low lying excitations in the
$U(1)$-linear spin liquids are not described by free fermions. This makes
the discussion on the stability of those states much more difficult.

Here we would like to concentrate on the $U(1)$-linear spin liquid
U1C$n01n$ in \Eq{U1lA}.  The spinons have a massless Dirac dispersion in
the $U(1)$-linear spin liquid. First we would like to know if the  massless
Dirac dispersion is generic property of the $U(1)$-linear spin liquid, \ie
if the massless Dirac dispersion is a property shared by all the spin
liquids that have the same quantum order as that in \Eq{U1lA}.  The most
general perturbations around the $U(1)$-linear ansatz \Eq{U1lA} are given
by \Eq{U1glin}, if the perturbations respect translation, parity, and the
time reversal symmetries, and if the perturbation do not break the $U(1)$
gauge structure.  Since $\del u^3_{\v m}= \del u^0_{\v m}= 0$ for $\v
m=even$, their contributions in the momentum space vanish at $\v k=(0,0)$
and $\v k=(0,\pi)$. The spinon energy also vanish at those points for the
ansatz \Eq{U1lA}.  Thus the massless Dirac dispersion is protected by the
symmetries and the $U(1)$ gauge structure in the $U(1)$-linear spin liquid
\Eq{U1lA}.  In other words, the massless Dirac dispersion is protected by
the quantum order in the  $U(1)$-linear spin liquid.

Next we consider if the symmetries and the $U(1)$ gauge structure in the
$U(1)$-linear spin liquid can be broken spontaneously due to
interactions/fluctuations at low energy.  The low energy effective theory
is described  by Lagrangian (in imaginary time)
\begin{equation}
 \cL= \sum_{a,\mu} \psi^\dag_a\ga^0[v_{\mu,a} 
\ga^\mu(\prt_\mu + i a_\mu)] \psi_a
\end{equation}
where $\mu=0,1,2$, $a=1,2$, $\ga^\mu$ are $4\times 4$ $\ga$-matrices,
$v^0_a =1$, and $(v_{1,a}, v_{2,a})$ are velocities for $a^{\th}$ fermion
in $x$ and $y$ directions.  We make a large $N$ generalization of the above
effective theory and allow $a=1,2,..,N$. Our first concern is about whether
the self energy from the gauge interaction can generate any
mass/chemical-potential term, due to infrared divergence.  It turns out
that, in the $1/N$ expansion, the gauge fluctuations represent an exact
marginal perturbation that does not generate any mass/chemical=potential
term.\cite{AN9021} Instead the gauge interaction changes the quantum fixed
point described by free massless Dirac fermions to a new quantum fixed
point which has no free fermionic excitations at low
energies.\cite{AN9021,RWspec} The new quantum fixed point has gapless
excitations and correlation functions all have algebraic decay.  Such a
quantum fixed point was called algebraic spin liquid.\cite{RWspec}
Actually, it is easy to understand why the gauge fluctuations represent an
exact marginal perturbation.  This is because the conserved current that
couple to the gauge potential cannot have any anomalous dimensions. Thus if
the gauge interaction is marginal at first order, then it is marginal at
all orders.  Gauge interaction as an exact marginal perturbation is also
supported by the following results.  The gauge invariant Green's function
of $\psi$ is found to be gapless after coupling to gauge field, to all
orders in $1/N$ expansion.\cite{AN9021}  Recently it was argued that the
$U(1)$ gauge interaction do not generate any mass perturbatively even when
$N$ is as small as $2$.\cite{ACS9903} 

Now let us discuss other possible instabilities. First we would like consider
a possible instability that change the $U(1)$-linear state to the
$Z_2$-linear state. To study such an instability we add a charge-2 Higgs
field to our effective theory
\begin{align}
 \cL =& \psi^\dag_a\ga^0[\ga^\mu(\prt_\mu + i a_\mu)] \psi_a \\
  &    + |(\prt_0 - 2 i a_0) \phi|^2  
      + v^2 |(\prt_i - 2 i a_i) \phi|^2  + V(\phi) \nonumber 
\end{align}
where $V(\phi)$ has its minimum at $\phi=0$ and 
we have assumed $v_{1,a}=v_{2,a} = 1$ for simplicity.
(Note that $\phi$ corresponds
to $\la$ in \Eq{Z2lE}. It is a non-zero $\la$ that break the $U(1)$
gauge structure down to the $Z_2$ gauge structure.) If after integrating out
$\psi$ and $a_\mu$, the resulting effective potential $V_{eff}(|\phi|)$ has its
minimum at a non zero $\phi$, then  the $U(1)$-linear state has a
instability towards the $Z_2$-linear state.

To calculate $V_{eff}(|\phi|)$, we first integrate out $\psi$ and get
\begin{align}
 \cL =& \frac12 a_\mu \pi_{\mu\nu} a_\nu  \\
  &    + |(\prt_0 - 2 i a_0) \phi|^2  
      + v^2 |(\prt_i - 2 i a_i) \phi|^2  + V(\phi)   \nonumber 
\end{align}
where
\begin{equation}
 \pi_{\mu\nu} = \frac{N}{8}(p^2)^{-1/2}(p^2\del_{\mu\nu}-p_\mu p_\nu)
\end{equation}
Now the effective potential $V_{eff}(|\phi|)$ can be obtained by integrating
out $a_\mu$ (in the $a_0=0$ gauge) and the phase $\th$ of the $\phi$ field,
$\phi=\rho e^{i\th}$:\footnote{We need to integrate out the phase of the $\phi$
field to get a gauge invariant result.}
\begin{align}
& V_{eff}(\phi) - V(\phi) \\
=&\int_0^\infty \frac{d\om}{\pi} \int \frac{d^2k}{(2\pi)^2}
\frac12 \Im \left( 
\ln[-\cK^\perp(i\om)] +
\ln[-\cK^{||}(i\om)] \right)  \nonumber\\
=& \int_0^\infty \frac{d\om}{\pi} \int \frac{d^2k}{(2\pi)^2}
\Im \ln( -\frac{N}{8}(k^2-\om^2-0^+)^{1/2} 
- 4|\phi|^2 )  \nonumber 
\end{align}
where
\begin{eqnarray}
 \cK^{\perp}  &=& \frac{N}{8}(\om^2+k^2)^{1/2}
+ 4|\phi|^2  \\
 \cK^{||}  &=& \frac{N}{8}(\om^2+k^2)^{-1/2}\om^2+ 
4|\phi|^2\frac{\om^2}{\om^2+v^2 k^2}  \nonumber\\
&=&
\frac{\om^2}{\om^2+v^2 k^2} \left( \frac{N}{8}(\om^2+k^2)^{1/2}
+ 4|\phi|^2 \right)
\nonumber 
\end{eqnarray}
We find that $V_{eff}=V-C_1|\phi|^6\ln|\phi|$ where $C_1$ is a constant.  Now
it is clear that the gapless gauge fluctuations cannot shift the minimum of
$V$ from $\phi=0$ and the $U(1)$-linear state is stable against
spontaneously changing into the $Z_2$-linear state.

So far we only considered the effects of perturbative fluctuations. The
non-perturbative instanton effects can also cause instability of the
algebraic spin liquid. The instanton effects have been discussed in
\Ref{IL8988} for the case $v^1_a=v^2_a$. It was found that the instanton
effects represent a relevant perturbation which can destabilize the
algebraic spin liquid when $N < 24$.  In the following, we will generalize
the analysis of \Ref{IL8988} to $v^1_a\neq v^2_a$ case.  First we rewrite
\begin{eqnarray}
 S &=& \int \frac{d^3 k}{(2\pi)^3}
\frac12 a_\mu(-k) \pi_{\mu\nu} a_\nu(k)  \nonumber\\
 &=&\int \frac{d^3 k}{(2\pi)^3}
 \frac12 f_\mu(-k) K_{\mu\nu} f_\nu(k)  
\label{fKf}
\end{eqnarray}
where 
\begin{equation}
 f_\mu = \eps_{\mu\nu\la}\prt_\nu a_\la
\end{equation}
When $\pi_{\mu\nu} = k^2\del_{\mu\nu} - k_\mu k_\nu$, we find 
$K_{\mu\nu} = \del_{\mu\nu}$.
When $\pi_{\mu\nu} = (k^2\del_{\mu\nu} - k_\mu k_\nu)/\sqrt{k^2}$, we may
assume $K_{\mu\nu} = \del_{\mu\nu}/\sqrt{k^2}$.
When $v_{1,a}\neq v_{2,a}$ we have
\begin{align}
 (K_{\mu\nu}) =
\sum_a \frac18 & (\om^2+v_{1,a}^2 k_1^2 + v_{2,a}^2k_2^2)^{-1/2}\times \nonumber\\
&
\bpm
v_{1,a} v_{2,a} &0 &0 \\
0& v_{2,a}/v_{1,a} & 0  \\
0&  0&  v_{1,a}/v_{2,a} 
\epm 
\end{align}
The instanton field $f_\mu$ minimize the action \Eq{fKf} and satisfies
\begin{equation}
 K_{\mu\nu} f_\nu = c(k) k_\mu
\end{equation}
where $c(k)$ is chosen such that $k_\mu f_\mu = 2i\pi$
We find that
\begin{align}
 f_0 =&  \frac{8c \om }{
 \sum_a  (\om^2+v_{1,a}^2 k_1^2 + v_{2,a}^2k_2^2)^{-1/2} v_{1,a} v_{2,a} }
\nonumber\\
 f_1 =&  \frac{8c k_1}{
 \sum_a  (\om^2+v_{1,a}^2 k_1^2 + v_{2,a}^2k_2^2)^{-1/2} v_{2,a} / v_{1,a} }
\nonumber\\
 f_2 =&  \frac{8c k_2}{
 \sum_a  (\om^2+v_{1,a}^2 k_1^2 + v_{2,a}^2k_2^2)^{-1/2} v_{1,a} / v_{2,a} }
\end{align}
and
\begin{align}
 c =& 2i\pi \left(
\frac{8 \om^2}{
 \sum_a  (\om^2+v_{1,a}^2 k_1^2 + v_{2,a}^2k_2^2)^{-1/2} v_{1,a} v_{2,a} }
\right.  \nonumber\\
&\ 
+\frac{8k_1^2}{
 \sum_a  (\om^2+v_{1,a}^2 k_1^2 + v_{2,a}^2k_2^2)^{-1/2} v_{2,a}/ v_{1,a} }
\nonumber\\
&\ \left.
+\frac{8k_2^2}{
 \sum_a  (\om^2+v_{1,a}^2 k_1^2 + v_{2,a}^2k_2^2)^{-1/2} v_{1,a}/ v_{2,a} }
\right)^{-1}
\end{align}
Using the above solution, we can calculate the action for a single instanton,
which has a form 
\begin{equation}
 S_{inst} = \frac{N}{2} \al(v_2/v_1) \ln(L)
\end{equation}
where $L$ is the size of the system and we have assumed that $N/2$ fermions
have velocity $(v_x,v_y)=(v_1,v_2)$ and the other $N/2$ fermions have
velocity $(v_x,v_y)=(v_2,v_1)$.  We find $\al(1)=1/4+O(1/N)$ and
$\al(0.003)=3.+O(1/N)$.  When $\frac{N}{2} \al(v_2/v_1) > 3$, the instanton
effect is irrelevant.  We see that even for the case $N=2$, the instanton
effect can be irrelevant for small enough $v_2/v_1$.  Therefore, the
algebraic spin liquid exists and can be stable.  

It has been proposed that the pseudo-gap metallic state in underdoped high
$T_c$ superconductors is described by the (doped) staggered flux state (the
$U(1)$-linear state U1C$n01n$) {\em which contains a long range $U(1)$ gauge
interaction}.\cite{WLsu2,RWspec} From the above result, we see that, for
realistic $v_2/v_1\sim 0.1$ in high $T_c$ superconductors, the U1C$n01n$
spin liquid is unstable at low energies. However, this does not mean that
we cannot not use the algebraic spin liquid U1C$n01n$ to describe the
pseudo-gap metallic state.  It simply means that, at low temperatures, the
algebraic spin liquid will change into other stable quantum states, such as
superconducting state or antiferromagnetic state\cite{KL9930} as observed
in experiments.

The unstable algebraic spin liquid can be viewed as an unstable quantum fixed
point. Thus the algebraic-spin-liquid approach to the pseudo-gap metallic state
in underdoped samples looks similar to the quantum-critical-point
approach\cite{CS9369,SP9313}.  However, there is an important distinction
between the two approaches. The quantum-critical-point approach assumes a
nearby continuous phase transition that changes symmetries and strong
fluctuations of local order parameters that cause the criticality.  The
algebraic-spin-liquid approach does not require any nearby symmetry breaking
state and there is no local order parameter to fluctuate.

\subsection{Quantum order and the stability of spin liquids}
\label{qostb}

After introducing quantum orders and PSG, we can have a deeper discussion
on the stability of mean-field states.  The existence of the algebraic spin
liquid is a very striking phenomenon, since gapless excitations interact
down to zero energy and cannot be described by free fermions or free
bosons. According to a conventional wisdom, if bosons/fermions interact at
low energies, the interaction will open an energy gap for those low lying
excitations. This implies that a system can either has free
bosonic/fermionic excitations at low energies or has no low energy
excitations at all. According to the discussion in section \ref{stb}, such
a conventional wisdom is incorrect.  But it nevertheless rises an important
question: what protects gapless excitations (in particular when they
interact at all energy scales).  There should be a ``reason'' or
``principle'' for the existence of the gapless excitations. Here we would
like to propose that {\em it is the quantum order that protects the gapless
excitations.} We would like to stress that gapless excitations in the Fermi
spin liquids and in the algebraic spin liquids exist even without any
spontaneous symmetry breaking and they are not protected by symmetries.
The existence of gapless excitations without symmetry breaking is a truly
remarkable feature of quantum ordered states.  In addition to the gapless
Nambu-Goldstone modes from spontaneous continuous symmetry breaking, quantum
orders offer another origin for gapless excitations.

We have seen from several examples discussed in section \ref{simp}
that the quantum order (or the PSG) not only protect
the zero energy gap,  it also protects certain qualitative properties of the
low energy excitations.  Those properties include the linear, quadratic, or
gapless dispersions, the $\v k$ locations where the 2-spinon energy 
$E_{2s}(\v k)$ vanishes, \etc.  

Since quantum order is a generic property for any quantum state at zero
temperature, we expect that the existence of interacting gapless excitations is
also a generic property of quantum state.  We see that algebraic state is a
norm. It is the Fermi liquid state that is special. 

In the following, we would like to argue that the PSG can be a stable 
(or universal) property of a quantum state. It is robust against perturbative
fluctuations. Thus, the PSG, as a universal property, can be used to
characterize a quantum phase. From the examples discussed in sections
\ref{fsl} and \ref{asl}, we see that PSG protects gapless excitations.
Thus, the stability of PSG also imply the stability of gapless excitations.

We know that a mean-field spin liquid state is
characterized by $U_{\v i\v j} = \<\psi_{\v i}\psi^\dag_{\v j}\>$.  If we
include perturbative fluctuations around the mean-field state, we expect
$U_{\v i\v j}$ to receive perturbative corrections $\del U_{\v i\v j}$.
Here we would like to argue that the perturbative fluctuations can only
change $U_{\v i\v j}$ in such a way that $U_{\v i\v j}$ and $U_{\v i\v
j}+\del U_{\v i\v j}$ have the same PSG.  

First we would like to note the following well know facts: the perturbative
fluctuations cannot change the symmetries and the gauge structures. For
example, if $U_{\v i\v j}$ and the Hamiltonian have a symmetry, then $\del
U_{\v i\v j}$ generated by perturbative fluctuations will have the same
symmetry.  Similarly, the perturbative fluctuations cannot generate $\del
U_{\v i\v j}$ that, for example, break a $U(1)$ gauge structure down to a
$Z_2$ gauge structure.  

Since both the gauge structure (described by the
IGG) and the symmetry are part of the PSG, it is reasonable to generalize
the above observation by saying that not only the IGG and the symmetry in
the PSG cannot be changed, the whole PSG cannot be changed by the
perturbative fluctuations.  In fact, the mean-field Hamiltonian and the
mean-field ground state are invariant under the transformations in the PSG.
Thus in a perturbative calculation around a mean-field state, the
transformations in the PSG behave just like symmetry transformations.
Therefore, the perturbative fluctuations can only generate $\del U_{\v i\v
j}$ that are invariant under the transformations in the PSG.

Since the perturbative fluctuations (by definition) do not change the
phase, $U_{\v i\v j}$ and $U_{\v i\v j}+\del U_{\v i\v j}$ describe the
same phase.  In other words, we can group $U_{\v i\v j}$ into classes
(which are called universality classes) such that $U_{\v i\v j}$ in each
class are connected by the the perturbative fluctuations and describe the
same phase.  We see that if the above argument is true then the
universality classes are classified by the PSG's (or quantum orders).

We would like to point out that we have assumed the perturbative
fluctuations to have no infrared divergence in the above discussion.
The infrared divergence implies  the perturbative
fluctuations to be relevant perturbations, which cause phase transitions.  

\section{Relation to previously constructed spin liquids}

Since the discovery of high $T_c$ superconductor in 1987, many spin liquids
were constructed. After  classifying and constructing a large class of spin
liquids, we would like to understand the relation between the previously
constructed spin liquids and spin liquids constructed in this paper.

Anderson, Baskaran, and Zou\cite{A8796,BZA8773,BA8880} first used the slave
boson approach to construct uniform RVB state. The uniform RVB state is a
symmetric spin liquid which has all the symmetries of the lattice. It is a
$SU(2)$-gapless state characterized by the PSG SU2A$n0$.  Later two more spin
liquids were constructed using the same $U(1)$ slave boson approach. One is
the $\pi$-flux phase and the other is the staggered-flux/$d$-wave
state.\cite{AM8874,K8864,KL8842} The $\pi$-flux phase is a $SU(2)$-linear
symmetric spin liquid characterized by PSG SU2B$n0$. The
staggered-flux/$d$-wave state is a $U(1)$-linear symmetric  spin liquid
characterized by PSG U1C$n01n$. The U1C$n01n$ state is found to be the
mean-field ground for underdoped samples.  Upon doping the  U1C$n01n$ state
becomes a metal with a pseudo-gap at high temperatures and a $d$-wave
superconductor at low temperatures.  

It is amazing to see that the slave boson approach, which is regarded as a
very unreliable approach, predicted the $d$-wave superconducting state 5 years
before its experimental confirmation.\cite{HBM9399,SDW9353,BO9430,KTS9525}
Maybe predicting the $d$-wave superconductor is not a big deal. After all, the
$d$-wave superconductor is a commonly known state and the paramagnon
approach\cite{SLH8690,SLH8794} predicted $d$-wave superconductor before the
slave boson approach. However, what is really a big deal is that the slave
boson approach also predicted the pseudo-gap metal which is a completely new
state. It is very rare in condensed matter physics to predict a new state of
matter before experiments. It is also interesting to see that not many people
believe in the slave boson approach despite such a success.

The above $U(1)$ and $SU(2)$ spin liquids are likely to be unstable at low
energies and may not appear as the ground states of spin systems. The first
known stable spin liquid is the chiral spin liquid.\cite{KL8795,WWZcsp}.  It
has a true spin-charge separation. The spinons and holons carry fractional
statistics. Such a state breaks the time reversal and parity symmetries and is
a $SU(2)$-gapped state. The $SU(2)$ gauge fluctuations in the chiral spin
state does not cause any instability since the gauge fluctuations are
suppressed and become massive due to the Chern-Simons term.  Due to the broken
time reversal and parity symmetries, the chiral spin state does not fit within
our classification scheme.

Spin liquids can also be constructed using the slave-fermion/$\si$-model
approach.\cite{AA8816,RS8994} Some gapped spin liquids were constructed using
this approach.\cite{RS9173,S9277} Those states turn out to be $Z_2$ spin
liquids.  But they are not symmetric spin liquids since the $90^\circ$
rotation symmetry is broken.  Thus they do not fit within our classification
scheme.  Later, a $Z_2$-gapped symmetric spin liquid was constructed using the
$SU(2)$ slave-boson approach (or the $SU(2)$ projective
construction).\cite{Wsrvb} The PSG for such a state is Z2A$xx0z$. Recently,
another $Z_2$ state was constructed using slave-boson
approach.\cite{BFN9833,SF0050} It is a $Z_2$-linear symmetric spin liquid. Its
PSG is given by Z2A$0013$.  New $Z_2$ spin liquids were also obtained recently
using the slave-fermion/$\si$-model approach.\cite{SP0114} It appears that
most of those states break certain symmetries and are not symmetric spin
liquids.  We would like to mention that $Z_2$ spin liquids have a nice
property that they are stable at low energies and can appear as the ground
states of spin systems.

Many spin liquids were also obtained in quantum dimer
model,\cite{KRS8765,RK8876,IL8941,FK9025,MS0181} and in various numerical
approach.\cite{CJZ9062,MLB9964,CBP0104,KI0148} It is hard to compare those
states with the spin liquids constructed here.  This is because either the
spectrum of spin-1 excitations was not calculated or the model has a very
different symmetry than the model discussed here.  We need to generalized our
classification to models with different symmetries so that we can have a
direct comparison with those interesting results and with the non-symmetric
spin liquids obtained in the  slave-fermion/$\si$-model approach.  In quantum
dimer model and in numerical approach, we usually know the explicit form of
ground state wave function. However, at moment, we do not know how to obtain
PSG from ground wave function. Thus, knowing the explicit ground state wave
function does not help us to obtain PSG. We see that it is important to
understand the relation between the ground state wave function and PSG so that
we can understand quantum order in the states obtained in numerical
calculations.

\section{Summary of the main results}
\label{sum}

In the following we will list the main results obtained in this paper.
The summary also serves as a guide of the whole paper.

(1) A concept of quantum order is introduced.  The quantum order describes the
orders in zero-temperature quantum states.  The opposite of quantum order --
classical order  describes the orders in finite-temperature classical states.
Mathematically, the quantum order characterizes universality classes of complex
ground state wave-functions. It is richer then the classical order that
characterizes the universality classes of positive distribution functions.
Quantum orders cannot be completely described by symmetries and order
parameters. Landau's theory of orders and phase transitions does not apply to
quantum orders.  (See section \ref{sec:topqun})

(2) Projective symmetry group is introduced to describe different quantum
orders. It is argued that PSG is a universal property of a quantum phase. PSG
extends the symmetry group description of classical orders and can
distinguishes different quantum orders with the same symmetries. (See section
\ref{psg} and \ref{qostb})

(3) As an application of the PSG description of quantum phases, we 
propose the following principle that govern the continuous phase
transition between quantum phases.  Let $PSG_1$ and $PSG_2$ be the PSG's of
the two quantum phases on the two sides of a transition, and $PSG_{cr}$
be the PSG that describes the quantum critical state. Then $PSG_1 \subseteq
PSG_{cr}$ and $PSG_2 \subseteq PSG_{cr}$.  We note that the two quantum
phases may have the same symmetry and continuous quantum phase transitions
are possible between quantum phases with same symmetry.\cite{Wctpt} The
continuous transitions between different mean-field symmetric spin liquids
are discussed in section \ref{simp} and
appendix B which demonstrate the above principle.
However, for continuous transitions between mean-field states, we have an
additional condition $PSG_1=PSG_{cr}$ or $PSG_2=PSG_{cr}$.  

(4) With the help of PSG, we find that, within the $SU(2)$ mean-field
slave-boson approach, there are 4 symmetric $SU(2)$ spin liquids and infinite
many
symmetric $U(1)$ spin liquids. There are at least 103 and at most 196
symmetric $Z_2$ spin liquids.  Those symmetric spin liquids have translation,
rotation, parity and the time reversal symmetries.  
Although the classifications are done for the mean-field states, they apply to
real physical spin liquids if the corresponding mean-field states turn out to
be stable against fluctuations.  (See section \ref{qois} and appendix A)

(5) The stability of mean-field spin liquid states are discussed in detail.
We find many gapless mean-field spin liquids to be stable against quantum
fluctuations. They can be stable even in the presence of long range gauge
interactions. In that case the mean-field spin liquid states become
algebraic spin liquids where the gapless excitations interact down to zero
energy. (See section \ref{stb})

(6) The existence of algebraic spin liquids is a striking phenomenon since
there is no spontaneous broken symmetry to protect the gapless excitations.
There should be a ``principle'' that prevents the interacting gapless
excitations from opening an energy gap and makes the algebraic spin liquids
stable.  We propose that quantum order is such a principle. To support our
idea, we showed that just like the symmetry group of a classical state
determines the gapless Nambu-Goldstone modes, the PSG of a quantum state 
determines the structure of gapless excitations. The gauge group of the low
energy gauge fluctuations is given by the IGG, a subgroup of the PSG.  The PSG
also protects massless Dirac fermions from gaining a mass due to radiative
corrections.  We see that the stabilities of algebraic spin liquids and Fermi
spin liquids are protected by their PSG's.  The existence of gapless
excitations (the gauge bosons and gapless fermions) without symmetry breaking
is a truly remarkable feature of quantum ordered states.  The gapless gauge
and fermion excitations are originated from the quantum orders, just like the
phonons are originated from translation symmetry breaking.  (See sections
\ref{fsl}, \ref{asl}, \ref{qostb} and discussions below \Eq{WGUWU})

(7) Many $Z_2$ spin liquids are constructed.  Their low energy excitations
are described by free fermions.  Some $Z_2$ spin liquids have gapless
excitations and others have finite energy gap. For those gapless $Z_2$ spin
liquids some have Fermi surface while others have only Fermi points. The
spinon dispersion near the Fermi points can be linear $E\propto|\v k|$
(which gives us $Z_2$-linear spin liquids) or quadratic $E\propto\v k^2$
(which gives us $Z_2$-quadratic spin liquids).  In particular, we find
there can be many $Z_2$-linear spin liquids with different quantum orders.
All those different $Z_2$-linear spin liquids have nodal spinon
excitations.  (See section \ref{trnANS}, \ref{simp}, and appendix B)

(8) Many $U(1)$ spin liquids are constructed.  Some $U(1)$ spin liquids
have gapless excitations near isolated Fermi point with a linear dispersion.
Those $U(1)$ linear states can be stable against quantum
fluctuations.  Due to long range $U(1)$ gauge fluctuations, the gapless
excitations interact at low energies. The $U(1)$-linear spin liquids can be
concrete realizations of algebraic spin liquids.\cite{RWspec,RWspin}  (See
section \ref{trnANS}, \ref{simp}, and appendix B)

(9) Spin liquids with the same symmetry and different quantum orders can
have continuous phase transitions between them. Those phase transitions are
very similar to the continuous topological phase transitions between
quantum Hall states.\cite{WW9301,CFW9349,RG0067,Wctpt} We find that, at
mean-field level, the U1C$n01n$ spin liquid in \Eq{U1lA} (the staggered
flux phase) can continuously change into 8 different symmetric $Z_2$ spin
liquids.  The SU2A$n0$ spin liquid in \Eq{SU2gl} (the uniform RVB state)
can continuously change into 12 symmetric $U(1)$ spin liquids and 52
symmetric $Z_2$ spin liquids.  The SU2B$n0$ spin liquid in \Eq{SU2lA} (the
$\pi$-flux phase) can continuously change into 12 symmetric $U(1)$ spin
liquids and 58 symmetric $Z_2$ spin liquids.  (See appendix B)

(10) We show that spectrum of spin-1 excitations (\ie the two-spinon
spectrum), which can be probed in neutron scattering experiments, can be used
to measure quantum orders.  The gapless points of the spin-1 excitations in
the U1C$n01n$ (the staggered-flux) state are always at $\v k=(\pi,\pi)$,
$(0,0)$, $(\pi,0)$ and $(0,\pi)$.  In the pseudo-gap metallic phase of
underdoped high $T_c$ superconductors, the observed splitting of the neutron
scattering peak $(\pi,\pi)\to (\pi\pm\del,\pi), (\pi,\pi\pm\del)$
\cite{YMK8886,BEH8968,CAM9191,FKM9713,BFR9739,YLK9865,FBS0073,DMH0125} or
$(\pi,\pi)\to (\pi+\del,\pi-\del), (\pi-\del,\pi+\del)$ \cite{WBK0099,MFY0048}
at low energies indicates a transition of the U1C$n01n$ state into a state
with a different quantum order, {\em if we can indeed identify the scattering
peak as the gapless node}.  Non of the 8 symmetric $Z_2$ spin liquids in the
neighborhood of the the U1C$n01n$ state can explain the splitting pattern.
Thus we might need to construct a new low energy state to explain the
splitting.  This illustrates that detailed neutron scattering experiments are
powerful tools in detecting quantum orders and studying transitions between
quantum orders.  (See section \ref{sec:pmQO}.)

(11) The mean-field phase diagram Fig. \ref{phaseJ12} for a $J_1$-$J_2$
spin system is calculated. (Only translation symmetric states are considered.)
 We find four mean-field ground states as we
change $J_2/J_1$: the $\pi$-flux state  (the SU2A$n0$ state), the
chiral spin state (an $SU(2)$-gapped state), the $U(1)$-linear state
\Eq{u1l90b} which breaks $90^\circ$ rotation symmetry, and the
$SU(2)\times SU(2)$-linear state \Eq{su2su2}.  We also find several locally
stable mean-field states: 
the $U(1)$-gapped state U1C$n00x$ in \Eq{U1lB} and two
$Z_2$-linear states Z2A$zz13$ in \Eq{Z2lA} and Z2A$0013$ in
\Eq{Z2lC}.  Those spin liquids have a better chance to appear in
underdoped high $T_c$ superconductors. The Z2A$0013$ $Z_2$-linear state
has a spinon dispersion very similar to electron dispersion observed in
underdoped samples.  The spinon dispersion in the Z2A$zz13$
$Z_2$-linear state may also be consistent with electron dispersion in
underdoped samples. We note that the two-spinon spectrum for the two
$Z_2$-linear states have some qualitative differences (see Fig.
\ref{spec2sZ2}a and Fig.  \ref{spec2sZ2}b and note the positions of the
nodes). Thus we can use neutron scattering to distinguish the two states.
(See section \ref{mpdo}.)

Next we list some remarks/comments that may clarify certain confusing points
and help to avoid possible misunderstanding.

(A) Gauge structure is simply a redundant labeling of quantum states.  The
``gauge symmetries'' (referring different labels of same physical state give
rise to the same result) are not symmetries and can never be broken. (See
the discussion below \Eq{UUp})

(B) The gauge structures referred in this paper (such as in $Z_2$, $U(1)$,
or $SU(2)$ spin liquids) are ``low energy'' gauge structures. They are
different from the ``high energy'' gauge structure that appear in $Z_2$,
$U(1)$, and $SU(2)$ slave-boson approaches.  The ``low energy'' gauge
structures are properties of the quantum orders in the ground state of a
spin system.  The ``high energy'' gauge structure is a particular way of
writing down the Hamiltonian of spin systems.  The two kinds of gauge
structures have nothing to do with each other. (See discussions at the end
of section \ref{sec:spnchg} and at the end of section \ref{psg})

(C) There are (at least) two different interpretations of spin-charge
separation.  The first interpretation (pseudo spin-charge separation)
simply means that the low energy excitations cannot be described by
electron-like quasiparticles.  The second interpretation (true spin-charge
separation) means the existence of free spin-1/2 neutral quasiparticles
and spin-0 charged quasiparticles.  In this paper both interpretations are
used. The algebraic spin liquids have a pseudo spin-charge separation. The
$Z_2$ and chiral spin liquids have a true spin-charge separation. 
(See section \ref{sec:spnchg})

(D) Although in this paper we stress that quantum orders can be
characterized by the PSG's, we need to point out that the PSG's do not
completely characterize quantum orders. Two different quantum orders may be
characterized by the same PSG. As an example, we have seen that the ansatz
\Eq{U1lB} can be a $U(1)$-linear state or a $U(1)$-gapped state depending
on the values of parameters in the ansatz.  Both states are described by
the same PSG U1C$n00x$.  Thus the PSG can not distinguish the
different quantum orders carried by the $U(1)$-linear state and the
$U(1)$-gapped state.

(E) The unstable spin liquids can be important in understanding the finite
temperature states in high $T_c$ superconductors.  The pseudo-gap metallic
state in underdoped samples is likely to be described by the unstable
U1C$n01n$ algebraic spin liquid (the staggered flux state) which
contains a long range $U(1)$ gauge interaction.\cite{WLsu2,RWspec} (See
discussions at the end of section \ref{stb}.)

(F) Although we have been concentrated on the characterization of stable
quantum states, quantum order and the PSG characterization can also be used to
describe the internal order of quantum critical states.  Here we define 
``quantum critical states'' as states that appear at the continuous phase
transition points between two states with different symmetries or between two
states with different quantum orders (but the same symmetry). We would like to
point out that ``quantum critical states'' thus defined are more general
than ``quantum critical points''. ``Quantum critical points'', by
definition, are the continuous phase transition points between two states with
different symmetries. The distinction is important.
``Quantum critical points'' are associated with broken symmetries and order
parameters. Thus the low energy excitations at ``quantum critical points''
come from the strong fluctuations of order parameters. While
``quantum critical states'' may not be related to broken symmetries and order
parameters. In that case it is impossible to relate the gapless fluctuations
in a ``quantum critical state'' to fluctuations of an order parameter.
The unstable spin liquids mentioned in (E) can be more general
quantum critical states.  Since some finite temperature phases in high $T_c$
superconductors may be described by quantum critical states or stable
algebraic spin liquids, their characterization through quantum order and PSG's
is useful for describing those finite temperature phases.

(G) In this paper, we only studied quantum orders and topological orders at
zero temperature. However, we would like point out that topological orders and
quantum orders may also apply to finite temperature systems. Quantum effect
can be important even at finite temperatures. In \Ref{Wtopcs}, a dimension
index (DI) is introduced to characterize the robustness of the ground state
degeneracy of a topologically ordered state. We find that if DI$\leq 1$
topological orders cannot exist at finite temperature. However, if DI$>1$,
topological order can exist at finite temperatures and one expect a
finite-temperature phase transition without any change of symmetry.
Topological orders in FQH states have DI=1, and they cannot exist at finite
temperatures.  The topological order in 3D superconductors has DI=2. Such a
topological order can exist at finite temperatures, and we have a continuous
finite-temperature superconductor-metal transition that do not change any
symmetry.

Although we mainly discussed quantum orders in 2D spin systems, the concept of
quantum order is not limited to 2D spin systems. The concept applies to any
quantum systems in any dimensions.  Actually, a superconductor is the simplest
example of a state with non trivial quantum order if the dynamical
electromagnetic fluctuations are included. A superconductor breaks no
symmetries and cannot be characterized by order parameters. An $s$-wave and a
$d$-wave superconductors, having the same symmetry, are distinguished only by
their different quantum orders. The gapless excitations in a $d$-wave
superconductor are not produced by broken symmetries, but by quantum orders.
We see that a superconductor has many properties characteristic of quantum
ordered states, and it is a quantum ordered state. The quantum orders in the
superconducting states can also be characterized using PSG's.  The IGG
$\cG=Z_2$ if the superconducting state is caused by electron-pair
condensation, and the IGG $\cG=Z_4$ if the superconducting state is caused by
four-electron-cluster condensation.  The different quantum orders in an
$s$-wave and a $d$-wave superconductors can be distinguished by their
different PSG's. The ansatz of the $s$-wave superconductor is invariant under
the $90^\circ$ rotation, while the ansatz of the $d$-wave superconductor is
invariant under the $90^\circ$ rotation followed by gauge transformations
$c_{\v i} \to \pm e^{i\pi/2} c_{\v i}$.

It would be interesting to study quantum orders in 3D systems. In particular,
it is interesting to find out the quantum order that describes the physical
vacuum that we all live in. The existence of light -- a massless excitation --
without any sign of spontaneous symmetry breaking suggests that our vacuum
contains a non-trivial quantum order that protect the massless photons.  Thus
quantum order provides an origin of light.\cite{Wlight}

I would like to thank P. A. Lee, J. Moore, E. Fradkin, W. Rantner, T. Senthil
for many helpful discussions.  This work is supported by NSF Grant No.
DMR--97--14198 and by NSF-MRSEC Grant No. DMR--98--08941.

\appendix

\section{Classification of projective symmetry groups}

\subsection{General conditions on projective symmetry groups}
\label{genconPSG}

The transformations in a symmetry group satisfy various algebraic 
relations so that they form a group. Those algebraic relations leads to
conditions on the elements of the PSG. Solving those conditions for a 
given symmetry group and a given IGG allows us to
find possible extensions of the symmetry group, or in another
word, to find possible PSG's associated with the symmetry group.
In section \ref{psg}, we have seen that the relation
$T_xT_yT_x^{-1}T_y^{-1}=1$ between translations in $x$- and $y$-directions
leads to condition
\begin{align}
\label{GxGyT}
& G_x T_x G_y T_y (G_x T_x)^{-1} (G_y T_y)^{-1} =   \nonumber\\
& G_x T_x G_y T_y T_x^{-1} G_x^{-1} T_y^{-1} G_y^{-1} \in \cG; 
\end{align}
or
\begin{align}
& G_x(\v i)G_y(\v i -\hat{\v x}) G_x^{-1}(\v i -\hat{\v y}) G_y(\v i)^{-1}
\in \cG
\label{genGxGy}
\end{align}
on elements $G_x T_x$ and $ G_y T_y$ of the PSG.  Here $\cG$ is the IGG.
This condition allows us to determine that there are only two
different extensions (given by \Eq{GZ2t1} and \Eq{GZ2t2}) for the
translation group generated by $T_x$ and $T_y$, if $\cG=Z_2$.

However, a bigger symmetry groups can have many more extensions.
In the following we are going to consider PSG's for the symmetry group
generated by two translations $T_{x,y}$, three parity transformations
$P_{x,y,xy}$, and the time reversal transformation $T$.  Since translations
and the time reversal transformation commute we have,
\begin{align}
 & (G_xT_x)^{-1} (G_T T)^{-1} G_xT_x G_T T \in \cG  \nonumber\\
 & (G_yT_y)^{-1} (G_T T)^{-1} G_yT_y G_T T \in \cG 
\end{align}
which reduces to the following two conditions on $G_{x,y}(\v i)$ and $G_T(\v
i)$
\begin{align}
\label{genGxyT}
&G_x^{-1}(\v i)G_T^{-1}(\v i) G_x(\v i) G_T(\v i-\hat{\v x}) \in \cG
\nonumber\\
&G_y^{-1}(\v i)G_T^{-1}(\v i) G_y(\v i) G_T(\v i-\hat{\v y}) \in \cG
\end{align}
Since $T^{-1}P_x^{-1}TP_x=1$, $T^{-1}P_y^{-1}TP_y=1$, and 
$T^{-1}P_{xy}^{-1}TP_{xy}=1$, one can also show that
\begin{align}
\label{genTPxyxy}
G_T^{-1}(P_x(\v i)) G_{P_x}^{-1}(\v i) G_T(\v i) G_{P_x}(\v i) \in & \cG
\nonumber\\
G_T^{-1}(P_y(\v i)) G_{P_y}^{-1}(\v i) G_T(\v i) G_{P_y}(\v i) \in & \cG
\nonumber\\
G_T^{-1}(P_{xy}(\v i)) G_{P_{xy}}^{-1}(\v i) G_T(\v i) G_{P_{xy}}(\v i)\in & \cG
\end{align}
From the relation between the translations and the parity transformations,
$
T_x P_x^{-1} T_x P_x = 
T_y^{-1} P_x^{-1} T_y P_x = 
T_y P_y^{-1} T_y P_y = 
T_x^{-1} P_y^{-1} T_x P_y = 
T_y^{-1} P_{xy}^{-1} T_x P_{xy} = 
T_x^{-1} P_{xy}^{-1} T_y P_{xy} = 1
$,
we find that
\begin{align}
  (G_xT_x) (G_{P_x} P_x)^{-1} G_xT_x G_{P_x} P_x \in & \cG  \nonumber\\
  (G_yT_y)^{-1} (G_{P_x} P_x)^{-1} G_yT_y G_{P_x} P_x \in & \cG  
\end{align}
\begin{align}
  (G_yT_y) (G_{P_y} P_y)^{-1} G_yT_y G_{P_y} P_y \in & \cG  \nonumber\\
  (G_xT_x)^{-1} (G_{P_y} P_y)^{-1} G_xT_x G_{P_y} P_y \in & \cG  
\end{align}
\begin{align}
  (G_yT_y)^{-1}(G_{P_{xy}}P_{xy})^{-1}G_xT_xG_{P_{xy}}P_{xy}
\in & \cG    \nonumber \\
  (G_xT_x)^{-1}(G_{P_{xy}}P_{xy})^{-1}G_yT_yG_{P_{xy}}P_{xy}\in & \cG
\end{align}
or
\begin{align}
\label{genGxyPx}
G_x(P_x(\v i))G_{P_x}^{-1}(\v i+\hat{\v x})G_x(\v i+\hat{\v x})G_{P_x}(\v i) 
 \in & \cG
\nonumber\\
G_y^{-1}(P_x(\v i)) G_{P_x}^{-1}(\v i) G_y(\v i) G_{P_x}(\v i-\hat{\v y}) 
 \in & \cG
\end{align}
\begin{align}
\label{genGxyPy}
G_y(P_y(\v i))G_{P_y}^{-1}(\v i+\hat{\v y})G_y(\v i+\hat{\v y})G_{P_y}(\v i) 
 \in & \cG
\nonumber\\
G_x^{-1}(P_y(\v i)) G_{P_y}^{-1}(\v i) G_x(\v i) G_{P_y}(\v i-\hat{\v x}) 
 \in & \cG
\end{align}
\begin{align}
\label{genGxyPxy}
G_y^{-1}(P_{xy}(\v i)) G_{P_{xy}}^{-1}(\v i) G_x(\v i) 
G_{P_{xy}}(\v i-\hat{\v x}) \in & \cG \nonumber\\
G_x^{-1}(P_{xy}(\v i)) G_{P_{xy}}^{-1}(\v i) G_y(\v i) 
G_{P_{xy}}(\v i-\hat{\v y}) \in & \cG 
\end{align}
We also have $P_{xy}P_xP_{xy}P_y^{-1}= P_yP_x P_y^{-1}P_x^{-1} =1$. Thus
\begin{align}
G_{P_{xy}} P_{xy}G_{P_x}P_x G_{P_{xy}} P_{xy} (G_{P_y}P_y)^{-1} \in & \cG
\nonumber\\
G_{P_y} P_yG_{P_x}P_x (G_{P_y} P_y)^{-1} (G_{P_x}P_x)^{-1} \in & \cG
\end{align}
which implies
\begin{align}
\label{genPxyPP}
G_{P_{xy}}(\v i) G_{P_x}(P_{xy}(\v i))
G_{P_{xy}}(P_{xy}P_x(\v i)) G_{P_y}^{-1}(\v i)  \in & \cG
\nonumber\\
G_{P_y}(\v i) G_{P_x}(P_y(\v i))
G_{P_y}^{-1}(P_x(\v i)) G_{P_x}^{-1}(\v i)  \in & \cG
\end{align}
The fact $T^2 =1$ leads to condition
\begin{equation}
\label{genTT}
 G_T^2(\v i) \in \cG
\end{equation}
and $P_x^2=P_y^2=P_{xy}^2=1$ leads to
\begin{align}
\label{genPP}
 G_{P_x}(\v i) G_{P_x}(P_x(\v i)) \in & \cG  \nonumber\\
 G_{P_y}(\v i) G_{P_y}(P_y(\v i)) \in & \cG  \nonumber\\
 G_{P_{xy}}(\v i) G_{P_{xy}}(P_{xy}(\v i)) \in & \cG  
\end{align}

The above conditions completely determine the PSG's. The solutions of the
above equations for $\cG=Z_2$, $U(1)$, and $SU(2)$ allow us to obtain PSG's
for $Z_2$, $U(1)$, and $SU(2)$ spin liquids.  However, we would like to point
out that the above conditions define the so called algebraic PSG's, which are
somewhat different from the invariant PSG defined in section \ref{psg}. More
precisely, an algebraic PSG is defined for a given IGG and a given symmetry
group SG. It is a group equipped with a projection $P$ and satisfies the
following conditions
\begin{align}
IGG \subset & PSG,\ \ \ P(PSG)=SG \\
P(gu) = & P(u),\ \ \ \hbox{for any $u\in PSG$ and $g \in IGG$} \nonumber 
\end{align}
It is clear that an invariant PSG is
always an algebraic PSG. However, some algebraic PSG's are not invariant
PSG's.  This is because a generic ansatz $u_{\v i\v j}$ that are invariant
under an algebraic PSG may be invariant under a larger invariant PSG.  If we
limit ourselves to spin liquids constructed using $u_{\v i\v j}$, then an
algebraic PSG characterizes a mean-field spin liquid only when it
is also an invariant PSG at the same time.

We would like to remark that the definition of invariant PSG can be
generalized. In section \ref{psg}, the invariant PSG is defined as a
collection of transformations that leave an ansatz $u_{\v i\v j}$ invariant.
More generally, a spin liquid is not only characterized by the two-point
correlation $(U_{\v i\v j})_{\al\bt} =\vev{\psi_{\al \v i}\psi^\dag_{\bt\v
j}}$ but also by many-point correlations such as $(U_{\v i\v j\v m\v
n})_{\al\bt\ga\la} =\vev{ \psi_{\al \v i} \psi_{\bt\v j} \psi^\dag_{\ga\v m}
\psi^\dag_{\la\v n} }$. We may define the generalized invariant PSG as a
collection of transformations that leave many-point correlation invariant. It
would be very interesting to see if the generalized invariant PSG coincide
with the algebraic PSG.

\subsection{Classification of $Z_2$ projective symmetry groups }
\label{sec:Z2cl}

We have seen that there are only two types of $Z_2$ spin liquids which have
{\em only} the translation symmetry.  However, spin liquids with more
symmetries can have more types.  In this section, we are going to construct
all (algebraic) PSG's associated with the symmetry group generated by
$T_{x,y}$, $P_{x,y,xy}$, and $T$ for the case $\cG=Z_2$. This allows us to
obtain a classification of mean-field symmetric $Z_2$ spin liquids.

We start with $Z_2$ spin liquids with only translation symmetry.  First let us
add the time reversal symmetry.  An arbitrary ansatz has the time reversal
symmetry if it satisfies
\begin{align}
\label{Tinv}
 G_T T(u_{\v i\v j}) =& u_{\v i\v j}  \nonumber\\
 T(u_{\v i\v j}) \equiv & - u_{\v i\v j}
\end{align}
For $Z_2$ spin liquid, the condition \Eq{genGxyT} becomes
\begin{align}
\label{GxyT}
&G_x^{-1}(\v i)G_T^{-1}(\v i) G_x(\v i) G_T(\v i-\hat{\v x}) = \eta_{xt}\tau^0
\nonumber\\
&G_y^{-1}(\v i)G_T^{-1}(\v i) G_y(\v i) G_T(\v i-\hat{\v y}) = \eta_{yt}\tau^0
\end{align}
where $\eta_{xt,yt}=\pm 1$.
For $Z_2$ spin liquids, $G_{x,y} \propto \tau^0$ and
the above four conditions (labeled by $\eta_{xt,yt}=\pm 1$ )
on $G_T$ can be simplified
\begin{align}
\label{Z2GxyT}
&G_T^{-1}(\v i) G_T(\v i-\hat{\v x}) = \eta_{xt}\tau^0
\nonumber\\
&G_T^{-1}(\v i) G_T(\v i-\hat{\v y}) = \eta_{yt}\tau^0
\end{align}
This leads to four types of $G_T$ labeled by $\eta_{xt,yt}=\pm 1$
\begin{align}
\label{GT}
  G_T(\v i) = & \eta_{yt}^{i_y} \eta_{xt}^{i_x} g_T 
\end{align}
where $g_T$ satisfies $g_T^2 = \pm \tau^0$.  We see that $g_T$ has two gauge
inequivalent choices $g_T=\tau^0, i\tau^3$.  Thus the symmetry group generated
by $T_{x,y}$ and $T$ has $2\times 4\times 2=16$ different 
extensions (or 16 different PSG's) if $\cG=Z_2$.  There can be (at most) 16
different mean-field $Z_2$ spin liquids which have {\em only} translation and
the time reversal symmetries.

Next let us add three types of parity symmetries.  An arbitrary ansatz has the
parity symmetries if it satisfies
\begin{align}
\label{Pinv1}
 G_{P_x} P_x(u_{\v i\v j}) =& u_{\v i\v j}  \nonumber\\
 P_x (u_{\v i\v j}) \equiv &  u_{P_x(\v i), P_x(\v j)} \nonumber\\
 P_x(\v i) = & (-i_x,i_y)
\end{align}
\begin{align}
\label{Pinv2}
 G_{P_y} P_y(u_{\v i\v j}) =& u_{\v i\v j}  \nonumber\\
 P_y (u_{\v i\v j}) \equiv &  u_{P_y(\v i), P_y(\v j)} \nonumber\\
 P_y(\v i) = & (i_x,-i_y)
\end{align}
\begin{align}
\label{Pinv3}
 G_{P_{xy}} P_{xy}(u_{\v i\v j}) =& u_{\v i\v j}  \nonumber\\
 P_{xy} (u_{\v i\v j}) \equiv &  u_{P_{xy}(\v i), P_{xy}(\v j)} \nonumber\\
 P_{xy}(\v i) = & (i_y,i_x)
\end{align}
For $Z_2$ spin liquids, \Eq{genGxyPx}, \Eq{genGxyPy}, and \Eq{genGxyPxy}
reduce to 
\begin{align}
\label{GxyPx}
G_x(P_x(\v i))G_{P_x}^{-1}(\v i+\hat{\v x})G_x(\v i+\hat{\v x})G_{P_x}(\v i) 
 =&  \eta_{xpx}\tau^0
\nonumber\\
G_y^{-1}(P_x(\v i)) G_{P_x}^{-1}(\v i) G_y(\v i) G_{P_x}(\v i-\hat{\v y}) 
 =&  \eta_{ypx}\tau^0
\end{align}
\begin{align}
\label{GxyPy}
G_y(P_y(\v i))G_{P_y}^{-1}(\v i+\hat{\v y})G_y(\v i+\hat{\v y})G_{P_y}(\v i) 
 =&  \eta_{ypy}\tau^0
\nonumber\\
G_x^{-1}(P_y(\v i)) G_{P_y}^{-1}(\v i) G_x(\v i) G_{P_y}(\v i-\hat{\v x}) 
 =&  \eta_{xpy}\tau^0
\end{align}
\begin{align}
\label{GxyPxy}
G_y^{-1}(P_{xy}(\v i)) G_{P_{xy}}^{-1}(\v i) G_x(\v i) 
G_{P_{xy}}(\v i-\hat{\v x}) =&  \eta_{xp{xy}}\tau^0 \nonumber\\
G_x^{-1}(P_{xy}(\v i)) G_{P_{xy}}^{-1}(\v i) G_y(\v i) 
G_{P_{xy}}(\v i-\hat{\v y}) =&  \eta_{yp{xy}}\tau^0 
\end{align}
where $\eta_{xpx,xpy,xp{xy}}=\pm 1$
and $\eta_{ypx,ypy,yp{xy}}=\pm 1$.

We will consider the two cases \Eq{GZ2t1} and \Eq{GZ2t2} separately.
First we assume
$G_x(\v i)=G_y(\v i)=\tau^0$. In this case,
\Eq{GxyPx}, \Eq{GxyPy}, and \Eq{GxyPxy} can be simplified
\begin{align}
\label{Z2GxyPx1}
G_{P_x}^{-1}(\v i+\hat{\v x})
G_{P_x}(\v i) 
 =&  \eta_{xpx}\tau^0
\nonumber\\
G_{P_x}^{-1}(\v i) 
G_{P_x}(\v i-\hat{\v y}) 
 =&  \eta_{ypx}\tau^0
\end{align}
\begin{align}
\label{Z2GxyPy1}
G_{P_y}^{-1}(\v i+\hat{\v y})
G_{P_y}(\v i) 
 =&  \eta_{ypy}\tau^0
\nonumber\\
G_{P_y}^{-1}(\v i) 
G_{P_y}(\v i-\hat{\v x}) 
 =&  \eta_{xpy}\tau^0
\end{align}
\begin{align}
\label{Z2GxyPxy1}
G_{P_{xy}}^{-1}(\v i) 
G_{P_{xy}}(\v i-\hat{\v x}) =&  \eta_{xp{xy}}\tau^0 \nonumber\\
G_{P_{xy}}^{-1}(\v i) 
G_{P_{xy}}(\v i-\hat{\v y}) =&  \eta_{yp{xy}}\tau^0   .
\end{align}
We find
\begin{align}
\label{GPxyxy1}
 G_{P_x}(\v i) =&  \eta_{xpx}^{i_x}\eta_{ypx}^{i_y}  g_{P_x} 
\nonumber\\
 G_{P_y}(\v i) =&  \eta_{xpy}^{i_x}\eta_{ypy}^{i_y}  g_{P_y} 
\nonumber\\
 G_{P_{xy}}(\v i) =&  \eta_{xpxy}^{i_y} \eta_{ypxy}^{i_x} g_{P_{xy}} 
\end{align}
where $g_{P_x}^2=\pm\tau^0$, $g_{P_y}^2=\pm\tau^0$, and 
$g_{P_{xy}}^2=\pm\tau^0$.
$\eta$'s and $g$'s in the above equation are not independent. 
From \Eq{genPP}, we find that
\begin{align}
&\eta_{xpxy}^{i_x} \eta_{ypxy}^{i_y}
 \eta_{xpxy}^{i_y} \eta_{ypxy}^{i_x} g_{P_{xy}} ^2 \nonumber\\
 =& \pm \eta_{xpxy}^{i_x} \eta_{ypxy}^{i_y}
 \eta_{xpxy}^{i_y} \eta_{ypxy}^{i_x} \nonumber\\
 =& \pm \tau^0
\end{align}
which requires 
$\eta_{xpxy}= \eta_{ypxy} \equiv \eta_{pxy}$.
From
\Eq{genPxyPP} we see that
\begin{align}
\label{etapxyxygg}
& \eta_{xpxy}^{i_x} \eta_{ypxy}^{i_y}\eta_{xpx}^{i_y}  
 \eta_{ypx}^{i_x} \eta_{xpxy}^{i_y}\eta_{ypxy}^{i_x} 
 \eta_{xpy}^{i_x}\eta_{ypy}^{i_y} \times \nonumber\\ 
&\ \ \ \ g_{P_{xy}} g_{P_x} g_{P_{xy}} g_{P_y}^{-1} 
=\pm \tau^0  \nonumber\\
& g_{P_y} g_{P_x} g_{P_y}^{-1} g_{P_x}^{-1} 
=\pm \tau^0 
\end{align}
We find
\begin{align}
\label{etapxyxy}
 \eta_{xpy} \eta_{xpxy} \eta_{ypx} \eta_{ypxy} =&  1 \nonumber\\
 \eta_{ypy} \eta_{ypxy} \eta_{xpx} \eta_{xpxy} =&  1 
\end{align}
and
\begin{align}
\label{gPxyxy}
g_{P_{xy}} g_{P_x} g_{P_{xy}} g_{P_y} ^{-1} =&  \pm \tau^0 
\nonumber \\
g_{P_y} g_{P_x} g_{P_y}^{-1} g_{P_x}^{-1} & =\pm \tau^0 
\end{align}
From \Eq{etapxyxy} we find 
$\eta_{xpx}= \eta_{ypy} $ and
$\eta_{xpy}= \eta_{ypx} $.
\Eq{GPxyxy1} becomes
\begin{align}
\label{GPxyxy}
 G_{P_x}(\v i) =&  \eta_{xpx}^{i_x}\eta_{xpy}^{i_y}  g_{P_x} 
\nonumber\\
 G_{P_y}(\v i) =&  \eta_{xpy}^{i_x}\eta_{xpx}^{i_y}  g_{P_y} 
\nonumber\\
 G_{P_{xy}}(\v i) =&  \eta_{pxy}^{\v i}   g_{P_{xy}} 
\end{align}
Now the three $\eta_{xpx}$, $\eta_{xpy}$, and $\eta_{pxy}$ are independent.
We note that gauge transformation $W_{\v i} = \eta_{wx}^{i_x} \eta_{wy}^{i_y}$
with $\eta_{wx,wy}=\pm$ does not change the form of $G_{x,y}$ in 
\Eq{GZ2t1} and \Eq{GZ2t2}. Thus we can use such gauge transformation to
further simplify $G_{P_{x,y,xy}}$. We find that the gauge transformation
$W_{\v i} = (-)^{i_x}$ changes $\eta_{pxy}$ to $-\eta_{pxy}$. Thus
we can always set $\eta_{pxy}=1$.
In the following we will choose the gauge in which $\eta_{pxy}=1$.

We also find that \Eq{genTPxyxy} requires 
\begin{align}
\label{etaxtyt}
 \eta_{xt} = \eta_{yt} 
\end{align}
and
\begin{align}
\label{gTPxyxy}
 g_T^{-1} g_{P_x}^{-1} g_T g_{P_x} =& \pm \tau^0 \nonumber\\
 g_T^{-1} g_{P_y}^{-1} g_T g_{P_y} =& \pm \tau^0 \nonumber\\
 g_T^{-1} g_{P_{xy}}^{-1} g_T g_{P_{xy}} =& \pm \tau^0 
\end{align}
Thus we only have two types of $G_T(\v i)$
\begin{equation}
\label{GT1}
 G_T(\v i) = \eta_t^{\v i} g_T
\end{equation}
labeled by $\eta_t=\pm 1$.

In the following, we will list all the gauge inequivalent solutions
for $g$'s from \Eq{gPxyxy} and \Eq{gTPxyxy}.
Most of them are obtained by setting $g$'s to be one of
$\tau^\mu$, $\mu=0,1,2,3$.
\begin{align}
g_{P{xy}}=& \tau^0 &g_{P_x}=&  \tau^0 &g_{P_y} =&  \tau^0 &g_T =&  \tau^0;
\label{gggg1}\\
g_{P{xy}}=&  \tau^0 &g_{P_x}=& i\tau^3 &g_{P_y} =& i\tau^3 &g_T =&  \tau^0;
\label{gggg3}
\end{align}
\begin{align}
g_{P{xy}}=& i\tau^3 &g_{P_x}=&  \tau^0 &g_{P_y} =&  \tau^0 &g_T =&  \tau^0;
\label{gggg6}\\
g_{P{xy}}=& i\tau^3 &g_{P_x}=& i\tau^3 &g_{P_y} =& i\tau^3 &g_T =&  \tau^0;
\label{gggg9}\\
g_{P{xy}}=& i\tau^3 &g_{P_x}=& i\tau^1 &g_{P_y} =& i\tau^1 &g_T =&  \tau^0;
\label{gggg12}
\end{align}
\begin{align}
g_{P{xy}}=&  \tau^0 &g_{P_x}=&  \tau^0 &g_{P_y} =&  \tau^0 &g_T =& i\tau^3;
\label{gggg2}\\
g_{P{xy}}=&  \tau^0 &g_{P_x}=& i\tau^3 &g_{P_y} =& i\tau^3 &g_T =& i\tau^3;
\label{gggg4}\\
g_{P{xy}}=&  \tau^0 &g_{P_x}=& i\tau^1 &g_{P_y} =& i\tau^1 &g_T =& i\tau^3;
\label{gggg5}
\end{align}
\begin{align}
g_{P{xy}}=& i\tau^3 &g_{P_x}=&  \tau^0 &g_{P_y} =&  \tau^0 &g_T =& i\tau^3;
\label{gggg7}\\
g_{P{xy}}=& i\tau^3 &g_{P_x}=& i\tau^3 &g_{P_y} =& i\tau^3 &g_T =& i\tau^3;
\label{gggg10}\\
g_{P{xy}}=& i\tau^3 &g_{P_x}=& i\tau^1 &g_{P_y} =& i\tau^1 &g_T =& i\tau^3;
\label{gggg15} 
\end{align}
\begin{align}
g_{P{xy}}=& i\tau^1 &g_{P_x}=&  \tau^0 &g_{P_y} =&  \tau^0 &g_T =& i\tau^3;
\label{gggg8}\\
g_{P{xy}}=& i\tau^1 &g_{P_x}=& i\tau^3 &g_{P_y} =& i\tau^3 &g_T =& i\tau^3;
\label{gggg13}\\
g_{P{xy}}=& i\tau^1 &g_{P_x}=& i\tau^1 &g_{P_y} =& i\tau^1 &g_T =& i\tau^3;
\label{gggg11}\\
g_{P{xy}}=& i\tau^1 &g_{P_x}=& i\tau^2 &g_{P_y} =& i\tau^2 &g_T =& i\tau^3;
\label{gggg14}
\end{align}
\begin{align}
g_{P{xy}}=& i\tau^{12} &g_{P_x}=& i\tau^1 &g_{P_y} =& i\tau^2 
&g_T =& i\tau^0; \label{gggg16} \\
g_{P{xy}}=& i\tau^{12} &g_{P_x}=& i\tau^1 &g_{P_y} =& i\tau^2 
&g_T =& i\tau^3; \label{gggg17} 
\end{align}
where 
\begin{equation}
 \tau^{ab} =\frac{\tau^a+\tau^b}{\sqrt{2}},\ \ \ 
 \tau^{a\bar b} =\frac{\tau^a-\tau^b}{\sqrt{2}}.
\end{equation}
The above 17 solutions, when combined with 8 choices of $\eta_t$,
$\eta_{xpx}$, and $ \eta_{xpy} $ (see \Eq{GT1} and \Eq{GPxyxy}), give us
136 different PSG's for the case $G_x(\v i)=G_y(\v i)=\tau^0$.

For $Z_2$ spin liquid
with $G_x(\v i)=(-)^{i_y}\tau^0$, $G_y(\v i)=\tau^0$,
\Eq{GxyPx}, \Eq{GxyPy}, and \Eq{GxyPxy} can be simplified as
\begin{align}
\label{Z2GxyPx2}
G_{P_x}^{-1}(\v i+\hat{\v x})
G_{P_x}(\v i) 
 =&  \eta_{xpx}\tau^0
\nonumber\\
G_{P_x}^{-1}(\v i) 
G_{P_x}(\v i-\hat{\v y}) 
 =&  \eta_{ypx}\tau^0
\end{align}
\begin{align}
\label{Z2GxyPy2}
G_{P_y}^{-1}(\v i+\hat{\v y})
G_{P_y}(\v i) 
 =&  \eta_{ypy}\tau^0
\nonumber\\
G_{P_y}^{-1}(\v i) 
G_{P_y}(\v i-\hat{\v x}) 
 =&  \eta_{xpy}\tau^0
\end{align}
\begin{align}
\label{Z2GxyPxy2}
(-)^{i_y} G_{P_{xy}}^{-1}(\v i) 
G_{P_{xy}}(\v i-\hat{\v x}) =&  \eta_{xp{xy}}\tau^0 \nonumber\\
(-)^{i_x} G_{P_{xy}}^{-1}(\v i) 
G_{P_{xy}}(\v i-\hat{\v y}) =&  \eta_{yp{xy}}\tau^0   .
\end{align}
The above equations can be solved and we get
\begin{align}
\label{GPxyxy2a}
 G_{P_x}(\v i) =&  \eta_{xpx}^{i_x}\eta_{ypx}^{i_y} g_{P_x} 
\nonumber\\
 G_{P_y}(\v i) =&  \eta_{xpy}^{i_x}\eta_{ypy}^{i_y} g_{P_y} 
\nonumber\\
 G_{P_{xy}}(\v i) =&  (-)^{i_x i_y}
  \eta_{xp{xy}}^{i_x}\eta_{yp{xy}}^{i_y} g_{P_{xy}} 
\end{align}
\Eq{genPP} and \Eq{genPxyPP} still leads to $\eta_{xpxy}=\eta_{ypxy}$ and
\Eq{etapxyxygg}. Thus 
\begin{align}
\label{GPxyxy2}
 G_{P_x}(\v i) =&  \eta_{xpx}^{i_x}\eta_{xpy}^{i_y} g_{P_x} 
\nonumber\\
 G_{P_y}(\v i) =&  \eta_{xpy}^{i_x}\eta_{xpx}^{i_y}  g_{P_y} 
\nonumber\\
 G_{P_{xy}}(\v i) =&  (-)^{i_x i_y} g_{P_{xy}} 
\end{align}
\Eq{gPxyxy} and \Eq{gTPxyxy} are still valid here, which lead to the same
choices for $g$'s.  For the case $(-)^{i_y}G_x(\v i)=G_y(\v i)=\tau^0$, the
17 choices of $g$'s in \Eq{gggg1} to \Eq{gggg17}, when combined with 8
choices of $\eta_t$, $\eta_{xpx}$, and $ \eta_{xpy} $ again give us 136
different PSG's through \Eq{GT1} and \Eq{GPxyxy2}.

Now we would like to consider which of the translation symmetric ansatz
in \Eq{Z2t1} or \Eq{Z2t2} have the parity and the time reversal symmetries.
We note that three parity symmetries also imply the $90^\circ$ rotation
symmetry. 

After two parity transformations $P_x$ and $P_y$, we find
$u_{\v m}$ in \Eq{Z2t1} or \Eq{Z2t2} satisfies
\begin{equation}
 u_{-\v m} = 
  \eta_{xpy}^{\v m}  
  \eta_{xpx}^{\v m}  
 g_{P_y} g_{P_x} u_{\v m} g_{P_x}^{-1} g_{P_y}^{-1}
\end{equation}
After the time reversal transformation, we have
\begin{equation}
 -u_{\v m} = \eta_{t}^{\v m}  g_T u_{\v m} g_T^{-1} 
\end{equation}
Thus $u_{\v m} = u_{\v m}^\mu \tau^\mu$ must satisfy
\begin{align}
\label{Z2ugeta1}
 u_{\v m}^0 = & 0,\ \ \ \hbox{if $ \eta_{xpy}^{\v m}  
  \eta_{xpx}^{\v m}  =1$ or
$\eta_{t}^{\v m} = 1$ }
\nonumber\\
 u_{\v m}^l\tau^l = & 
  \eta_{xpy}^{\v m}  
  \eta_{xpx}^{\v m}  
 g_{P_y} g_{P_x} u_{\v m}^l\tau^l g_{P_x}^{-1} g_{P_y}^{-1}
\nonumber\\
 -u_{\v m}^l\tau^l = & \eta_{t}^{\v m}  g_T u_{\v m}^l\tau^l g_T^{-1} 
\end{align}
in order to have the parity and the time reversal symmetries.  We see that
$u_{\v i\v j}=0$ if $g_T=\tau^0$ and $\eta_t=1$ and $u_{\v m}^l=0$,
$l=1,2,3$, if $\eta_t = \eta_{xpx}\eta_{xpy}$ and $g_T=\pm g_{P_y}g_{P_x}$.
There are $2\times 6 \times 4 = 48$ PSG's with $g_T=\tau^0$ and $\eta_t=1$.
There are $2\times 1 \times 2 = 4$ PSG's with $g_T\neq \tau^0$, $g_T=\pm
g_{P_y}g_{P_x}$, $\eta_t = \eta_{xpx}\eta_{xpy}$, and $\eta_t=1$.  There
are $2\times 6 \times 2 = 24$ PSG's with $g_T=\pm g_{P_y}g_{P_x}$, $\eta_t
= \eta_{xpx}\eta_{xpy}$, and $\eta_t=-1$.  
Since the ansatz that are invariant under the above PSG's have
$u_{\v m}^l=0$, those ansatz are actually invariant under larger PSG's
with IGG equal or larger than $SU(2)$. 
Thus there are at most $272 - 48 -4
-24=196$ different mean-field 
$Z_2$ spin liquids that can be constructed form $u_{\v i\v j}$.  

For ansatz of type \Eq{Z2t1} 
the parity symmetries also require that
\begin{align}
\label{Z2ugeta2}
u_{P_x(\v m)}=&\eta_{xpx}^{m_x}\eta_{xpy}^{m_y} 
g_{P_x}^{-1} u_{\v m} g_{P_x} \nonumber\\
u_{P_y(\v m)}=&\eta_{xpy}^{m_x}\eta_{xpx}^{m_y} 
g_{P_y}^{-1} u_{\v m} g_{P_y} \nonumber\\
u_{P_{xy}(\v m)}=&\eta_{pxy}^{\v m} 
g_{P_{xy}}^{-1} u_{\v m} g_{P_{xy}} 
\end{align}
For ansatz of type \Eq{Z2t2} 
the parity symmetries require that
\begin{align}
\label{Z2ugeta3}
u_{P_x(\v m)}=&\eta_{xpx}^{m_x}\eta_{xpy}^{m_y} 
g_{P_x}^{-1} u_{\v m} g_{P_x} \nonumber\\
u_{P_y(\v m)}=&\eta_{xpy}^{m_x}\eta_{xpx}^{m_y} 
g_{P_y}^{-1} u_{\v m} g_{P_y} \nonumber\\
u_{P_{xy}(\v m)}=& (-)^{m_xm_y} \eta_{pxy}^{\v m} 
g_{P_{xy}}^{-1} u_{\v m} g_{P_{xy}} 
\end{align}
For each choice of $g$'s and $\eta$'s, \Eq{Z2ugeta1}, \Eq{Z2ugeta2}, and
\Eq{Z2ugeta3} allow us to construct $Z_2$ symmetric ansatz $u_{\v i\v j}$.

\subsection{Classification of $U(1)$ projective symmetry groups}

In this section we will use PSG to classify quantum orders in $U(1)$ spin
liquids by finding the PSG with
IGG $\cG=U(1)$.  First we note that elements in the $U(1)$ IGG must have a
form $e^{i\th \v v_{\v i}\cdot \v \tau}$ where $\v v_{\v i}$ is a site
dependent vector. We can always choose a gauge such that $\v v_{\v i}$ all
point to the same direction, say, the $\tau^3$ direction.  We will call
this gauge canonical gauge.  We also find that $|\v v_{\v i}|$ must be
independent of $\v i$ in order for $u_{\v i\v j}$ to be non-zero and
invariant under the IGG.  Thus, in the canonical gauge, IGG has a form
\begin{equation}
 \cG = \{ e^{i\th\tau^3}| \th \in [0,2\pi)\}
\end{equation}
and the ansatz $u_{\v i\v j}$ has a form
\begin{equation}
\label{u03}
 u_{\v i\v j} = u_{\v i\v j}^0 \tau^0 + u_{\v i\v j}^3 \tau^3 
\end{equation}
We see that the flux through any loops is in the $\tau^3$ direction.
Due to the translation symmetry of the ansatz, the absolute value of the flux
must be translation invariant, but the sign may change as we translate the
loops. Thus, the loop operator have a form
\begin{equation}
 P_{C_{\v i}} = (\tau^1)^{n_{\v i}} P_{C_{\v i = 0}} (\tau^1)^{n_{\v i}}
\end{equation}
where $n_{\v i} =0,1$ and $C_{\v i}$ is loop with base point $\v i$.  Here
the two loops $C_{\v i}$ and $C_{\v j}$ are related by a translation and
have the same shape. Now we can choose a different gauge by making a gauge
transformation $W_{\v i}=(i\tau^1)^{n_{\v i}}$. In the new gauge we have
\begin{align}
\label{uPCcG}
 u_{\v i\v j} =& 
 (i\tau^1)^{n_{\v i}-n_{\v j}}
 u_{\v i\v j}^0 \tau^0
  + 
 (i\tau^1)^{n_{\v i}+n_{\v j}}
  u_{\v i\v j}^3 \tau^3   \nonumber\\
 P_{C_{\v i}} =&  P_{C_{\v i = 0}} \nonumber\\
 \cG =& \{ e^{i(-)^{n_{\v i}}\th\tau^3}| \th \in [0,2\pi)\}
\end{align}
Since the loop operators are uniform, we will call the new gauge uniform
gauge.  

Let us first work in the uniform gauge.  From the translation invariance of
$P_{C_{\v i}}$
\begin{align}
 P_{C_{\v i}} =& G_x(\v i) P_{C_{\v i-\hat{\v x}}}G_x^{-1}(\v i)   
 = G_x(\v i) P_{C_{\v i}}G_x^{-1}(\v i)   
 \nonumber\\
 P_{C_{\v i}} =& G_y(\v i) P_{C_{\v i-\hat{\v y}}}G_y^{-1}(\v i)  
 = G_y(\v i) P_{C_{\v i}}G_y^{-1}(\v i)   
\end{align}
we find that $G_{x,y}$ have a form
\begin{align}
\label{U1GxGy1a}
 G_x(\v i) = & g_3(\th_x(\v i))
 \nonumber\\
 G_y(\v i) = & g_3(\th_y(\v i))
\end{align}

Now we switch to the canonical gauge.
We note that a gauge transformation that keep an ansatz to have the form
in the canonical gauge \Eq{u03} must have a one the following two forms 
\begin{align}
\label{WU1a}
W_{\v i}=g_3(\th(\v i))  \\
\label{WU1b}
W_{\v i}=i\tau^1 g_3(\th(\v i))
\end{align}
if we require that $u_{\v i\v j}\neq 0$.  (More precisely, we require that
any two points on the lattice can be connected by several non-zero $u_{\v
i\v j}$'s. We will call such an ansatz connected.) Thus for spin
liquids with connected $u_{\v i\v j}$, $G_{x,y}$ must take one of the above
two forms in the canonical gauge, since $G_{x,y}$ are two special gauge
transformations.  From \Eq{U1GxGy1a}, we find that $G_{x,y}$ have a form
\begin{align}
\label{U1GxGy1c}
 G_y(\v i) = & 
(-i\tau^1)^{n_{\v i}} 
g_3(\th_y(\v i))
(i\tau^1)^{n_{\v i-\hat {\v y} } } 
 \nonumber\\
 G_x(\v i) = & 
(-i\tau^1)^{n_{\v i}} 
g_3(\th_x(\v i))
(i\tau^1)^{n_{\v i-\hat {\v x}} }
\end{align}
in the canonical gauge.  Thus $n_{\v i}$ can only be one of the following
four choices: $n_{\v i}=0$, $n_{\v i}=(1-(-)^{\v i})/2$, $n_{\v
i}=(1-(-)^{i_y})/2$, and $n_{\v i}=(1-(-)^{i_x})/2$.  In these four cases,
$G_{x,y}$ take one of the above two forms and $u_{\v i\v j}$ can be
connected.  

Let us consider those cases in turn. We will work in the canonical gauge.
When $n_{\v i}=0$, $G_{x,y}$ have a form
\begin{align}
 G_x(\v i) =  g_3(\th_x(\v i)),\ \ \
 G_y(\v i) =  g_3(\th_y(\v i))
\end{align}
Since the gauge transformation $W_{\v i} = g_3(\th_{\v i})$ keep an ansatz
and its PSG in the canonical gauge, we can use such kind of gauge
transformation to simplify $G_{x,y}$ by setting $\th_y(\v i) = 0$ and
$\th_x(i_y=0,i_x) =0$.  Now \Eq{genGxGy} takes a form
\begin{align}
 G_x(\v i)   G_x(\v i-\hat{\v y})^{-1}  
= g_3(\vphi) 
\end{align}
for a constant $\vphi$. This allows us to obtain
\begin{align}
\label{U1GxGy1}
 G_x(\v i) =  g_3(i_y \vphi+\th_x),\ \ \ 
 G_y(\v i) =  g_3(\th_y) .
\end{align}
The translation symmetric ansatz has a form
\begin{align}
\label{U1tr1}
u_{\v i,\v i+\v m} = i\rho_{\v m} g_3(-m_y i_x \vphi +\phi)
\end{align}
where $\rho_{\v m} > 0$.  The above ansatz describes particle hopping in
uniform ``magnetic field'' with $e^{i\vphi\tau^3}$ flux per plaquette.  In
this case $\vphi/\pi$ should be a rational number $\vphi/\pi = p/q$
(between $0$ and $1$) so that the ansatz can be put on a finite lattice.
Thus $\vphi/\pi$ should be viewed as a discrete label and different
rational numbers between $0$ and $1$ will gives rise to different
type of spin liquids. 

When $n_{\v i}=(1+(-)^{\v i})/2$, $G_{x,y}$ have a form
\begin{align}
 G_x(\v i) =   g_3(\th_x(\v i))i\tau^1,\ \ \
 G_y(\v i) =   g_3(\th_y(\v i))i\tau^1 
\end{align}
Again we can use the gauge transformation $W_{\v i} = g_3(\th_{\v i})$ 
to simplify $G_{x,y}$
by setting $\th_y(\v i) = 0$ and $\th_x(i_y=0,i_x) =0$.  Now \Eq{genGxGy}
takes a form
\begin{align}
 G_x(\v i) \tau^1 G_x(\v i-\hat{\v y})^{-1}\tau^1 =& g_3(\vphi)  
\end{align}
or
\begin{align}
\th_x(\v i) + 
\th_x(\v i-\hat{\v y}) = \vphi  
\end{align}
for a constant $\vphi$. This allows us to obtain
\begin{align}
 G_x(\v i) =  g_3((-)^{i_y}\phi_{tr} + \th_x)i\tau^1,\ \ \ 
 G_y(\v i) =  g_3(\th_y)i\tau^1.
\end{align}
where $\phi_{tr}\in [0,\pi)$.  A gauge transformation $W_{\v i}=g_3(
-(-)^{i_y} \phi_{tr}/2 )$ change the above to
\begin{align}
\label{U1GxGy2}
 G_x(\v i) =  g_3(\th_x)i\tau^1,\ \ \ 
 G_y(\v i) =  g_3(\th_y)i\tau^1.
\end{align}
The translation symmetric ansatz has a form
\begin{align}
\label{U1tr2}
u_{\v i,\v i+\v m} = i\rho_{\v m} 
g_3\left( (-)^{\v i} \phi_{\v m} \right)
\end{align}

When $n_{\v i}=(1+(-)^{i_x})/2$, $G_{x,y}$ have a form
\begin{align}
 G_x(\v i) =   g_3(\th_x(\v i))i\tau^1,\ \ \
 G_y(\v i) =   g_3(\th_y(\v i)) 
\end{align}
After using a gauge transformation $W_{\v i} = g_3(\th_{\v i})$ to simplify
$G_{x,y}$
by setting $\th_y(\v i) = 0$,  \Eq{genGxGy} takes a form
\begin{align}
 G_x(\v i) G_x(\v i-\hat{\v y})^{-1} =&  g_3(\vphi)  
\end{align}
or
\begin{align}
\th_x(\v i) - 
\th_x(\v i-\hat{\v y}) = \vphi  
\end{align}
for a constant $\vphi$. This allows us to obtain
\begin{align}
 G_x(\v i) =  g_3(i_y\vphi + \th_x)i\tau^1,\ \ \ 
 G_y(\v i) =  g_3(\th_y).
\end{align}
where $\vphi\in [0,\pi)$.  A gauge transformation $W_{\v i}=g_3( -i_y
\vphi/2 )$ change the above to
\begin{align}
\label{U1GxGy3}
 G_x(\v i) =  g_3(\th_x)i\tau^1,\ \ \ 
 G_y(\v i) =  g_3(\th_y).
\end{align}
The translation symmetric ansatz has a form
\begin{align}
\label{U1tr3}
u_{\v i,\v i+\v m} = i\rho_{\v m} 
g_3\left( (-)^{i_x} \phi_{\v m} \right)
\end{align}

To summarize, \Eq{U1GxGy1}, \Eq{U1GxGy2} and \Eq{U1GxGy3} are the most
general translation PSG's that allow non-zero $u_{\v i\v j}$.  \Eq{U1tr1},
\Eq{U1tr2}, and \Eq{U1tr3} are the most general translation symmetric
mean-field ansatz for $U(1)$ spin liquids.

Next, we would like to include more symmetries. 
We first consider the translation PSG in
\Eq{U1GxGy1}.  When $\vphi=0$ the
translation PSG has a form
\begin{align}
\label{U1AGxGy}
 G_x(\v i) = g_3(\th_x), \ \ \ \
 G_y(\v i) = g_3(\th_y).
\end{align}
The corresponding spin liquids will be called type U1A spin liquids.  When
$\vphi=\pi$ the translation PSG has a form
\begin{align}
\label{U1BGxGy}
 G_x(\v i) = (-)^{i_y}g_3(\th_x), \ \ \ \
 G_y(\v i) =          g_3(\th_y).
\end{align}
and the corresponding spin liquids will be called type U1B spin liquids.
For other value of $\vphi$, we will call the corresponding spin liquids 
type U1$^m_n$ spin liquids, where $m/n = \vphi/\pi$ mod $1$.
The translation PSG's \Eq{U1GxGy2} and \Eq{U1GxGy3} will correspond to type
U1C and type U1D spin liquids respectively. 

Let us first consider the type U1A spin liquids.  To add the time reversal
symmetry, we note that, for type U1A spin liquid, the condition
\Eq{genGxyT} becomes
\begin{align}
\label{U1AGxyT}
&G_T^{-1}(\v i)  G_T(\v i-\hat{\v x}) \in U(1)
\nonumber\\
&G_T^{-1}(\v i)  G_T(\v i-\hat{\v y}) \in U(1) 
\end{align}
This leads to two types of $G_T$ 
\begin{equation}
\label{U1AGT1}
 G_T = g_3(\v i \cdot \v \vphi_t+\th_t),\ 
     i g_3(\v i \cdot \v \vphi_t+\th_t)\tau^1
\end{equation}
Since $T^2=1$ and $G_T^2 \in U(1)$, the above becomes
\begin{equation}
\label{U1AGT}
 G_T = \eta_{xt}^{i_x}\eta_{yt}^{i_y} g_3(\th_t),\ 
      i g_3(\v i \cdot \v \vphi_t+ \th_t)\tau^1
\end{equation}

To add three types of parity symmetries, we note that, for type U1A spin
liquids, \Eq{genGxyPx}, \Eq{genGxyPy}, and \Eq{genGxyPxy} reduce to 
\begin{align}
\label{U1AGxyPx}
G_{P_x}^{-1}(\v i+\hat{\v x})G_{P_x}(\v i) 
 \in & U(1)
\nonumber\\
G_{P_x}^{-1}(\v i) G_{P_x}(\v i-\hat{\v y}) 
 \in & U(1)
\end{align}
\begin{align}
\label{U1AGxyPy}
G_{P_y}^{-1}(\v i+\hat{\v y})G_{P_y}(\v i) 
 \in & U(1)
\nonumber\\
 G_{P_y}^{-1}(\v i)  G_{P_y}(\v i-\hat{\v x}) 
 \in & U(1)
\end{align}
\begin{align}
\label{U1AGxyPxy}
G_{P_{xy}}^{-1}(\v i) G_{P_{xy}}(\v i-\hat{\v y}) \in & U(1)  .
\end{align}
We find that $G_{P_x,P_y,P_{xy}}$ can have the following forms
\begin{align}
\label{U1AGPxPyPxy1}
 G_{P_x} =& g_3(\v i \cdot \v \vphi_{px}+\th_{px}),\ 
          i g_3(\v i \cdot \v \vphi_{px}+\th_{px})\tau^1 \nonumber\\
 G_{P_y} =& g_3(\v i \cdot \v \vphi_{py}+\th_{py}),\ 
          i g_3(\v i \cdot \v \vphi_{py}+\th_{py})\tau^1 \nonumber\\
 G_{P_{xy}} =& g_3(\v i \cdot \v \vphi_{pxy}+\th_{pxy}),\ 
             i g_3(\v i \cdot \v \vphi_{pxy}+\th_{pxy})\tau^1 
\end{align}
Note that the gauge transformation $W_i=g_3(\v i\cdot \v \th)$ does not
change $G_{x,y}$.  Thus, we can use it to simplify $G_{P_x,P_y}$ and get
\begin{align}
\label{U1AGPxPyPxy2}
 G_{P_x} =& g_3(i_y \vphi_{px}+\th_{px}),\ 
          i g_3(i_x \vphi_{px}+\th_{px})\tau^1 \nonumber\\
 G_{P_y} =& g_3(i_x \vphi_{py}+\th_{py}),\ 
          i g_3(i_y \vphi_{py}+\th_{py})\tau^1 \\
 G_{P_{xy}} =& g_3(\v i \cdot \v \vphi_{pxy}+\th_{pxy}),\ 
             i g_3(\v i \cdot \v \vphi_{pxy}+\th_{pxy})\tau^1 
\nonumber 
\end{align}
From the condition \Eq{genPP}, we find that
\begin{align}
\label{U1AGPxPyPxy3}
 G_{P_x} =& \eta_{ypx}^{i_y} g_3(\th_{px}),\ 
          i \eta_{xpx}^{i_x} g_3(\th_{px})\tau^1 \nonumber\\
 G_{P_y} =& \eta_{xpy}^{i_x} g_3(\th_{py}),\ 
          i \eta_{ypy}^{i_y} g_3(\th_{py})\tau^1  \\
 G_{P_{xy}} =& g_3((i_x-i_y) \vphi_{pxy}+\th_{pxy}),\ 
             i g_3((i_x+i_y) \vphi_{pxy}+\th_{pxy})\tau^1 
	     \nonumber 
\end{align}

$G_{P_x,P_y,P_{xy},T}$ should also satisfy \Eq{genPxyPP} and
\Eq{genTPxyxy}.  We find $G_{P_{xy}}$ can be obtained from $G_{P_x,P_y}$
through \Eq{genPxyPP} and $G_T$ from $G_{P_x,P_y, P_{xy}}$ through
\Eq{genTPxyxy}. This leads to the following 24 sets of solutions 
\begin{align}
 G_{P_x} =& \eta_{ypx}^{i_y} g_3(\th_{px}),\ \ \ \
 G_{P_y} = \eta_{ypx}^{i_x} g_3(\th_{py})  \nonumber\\
 G_{P_{xy}} =& g_3(\th_{pxy}), \ \ g_3(\th_{pxy})i\tau^1
\nonumber\\
 G_T =& \eta_{t}^{\v i} g_3(\th_t)|_{\eta_t=-1},\ \
        \eta_{t}^{\v i} g_3(\th_t)i\tau^1
\end{align}
and
\begin{align}
 G_{P_x} =& \eta_{xpx}^{i_x} g_3(\th_{px})i\tau^1,\ \ \ \
 G_{P_y} = \eta_{xpx}^{i_y} g_3(\th_{py})i\tau^1  \nonumber\\
 G_{P_{xy}} =& g_3(\th_{pxy}),\ g_3(\th_{pxy})i\tau^1  \nonumber \\
 G_T =& \eta_{t}^{\v i} g_3(\th_t)|_{\eta_t=-1},\ \ 
        \eta_{t}^{\v i} g_3(\th_t)i\tau^1
\end{align}
When combined with the type U1A translation PSG \Eq{U1AGxGy}, the above 24
sets of solutions give us 24 different PSG's. 
A labeling scheme of the above PSG's is given below \Eq{U1Apsg2}.

Now let us consider the form of ansatz that is invariant under the above
PSG's. The translation symmetry requires that
\begin{align}
 u_{\v i,\v i+\v m} = u_{\v m} = u^0_{\v m} \tau^0 + u^3_{\v m} \tau^3
\end{align}
The $180^\circ$ rotation symmetry requires that, 
for
$
G_{P_x} = \eta_{ypx}^{i_y} g_3(\th_{px}),
G_{P_y} = \eta_{ypx}^{i_x} g_3(\th_{py}) 
$,
\begin{align}
u_{\v m} = \eta_{ypx}^{\v m} u_{-\v m}  
         = \eta_{ypx}^{\v m} u_{\v m}^\dag  
\end{align}
and for
$
G_{P_x} = \eta_{xpx}^{i_x} g_3(\th_{px})i\tau^1,
G_{P_y} = \eta_{xpx}^{i_y} g_3(\th_{py})i\tau^1 
$,
\begin{align}
u_{\v m} = \eta_{xpx}^{\v m} u_{-\v m}  
         = \eta_{xpx}^{\v m} u_{\v m}^\dag  
\end{align}
The time reversal symmetry requires that, 
for
$ G_T = \eta_{t}^{\v i} g_3(\th_t)|_{\eta_t=-1}$
\begin{align}
u_{\v m} = -(-)^{\v m} u_{\v m}  ,
\end{align}
for
$ G_T = \eta_{t}^{\v i} g_3(\th_t)i\tau^1$
\begin{align}
u_{\v m} = 
-\eta_t^{\v m} u^0_{\v m}\tau^0  + \eta_t^{\v m} u^3_{\v m}\tau^3  ,
\end{align}

We find the following 8 sets of ansatz that give rise to $U(1)$ symmetric
spin liquids:
U1A$00[0,1]1$ and U1A$11[0,1]1$,
\begin{align}
\label{U1Aan1}
 u_{\v i,\v i+\v m} =& u^3_{\v m} \tau^3 .
\end{align}
U1A$0n[0,1]x$ and U1A$x1[0,1]x$,
\begin{align}
\label{U1Aan2}
 u_{\v i,\v i+\v m} =& u^0_{\v m} \tau^0 + u^3_{\v m} \tau^3 \nonumber\\
 u^0_{\v m}=& 0,\ \ \ \hbox{if $\v m= even$} \nonumber\\
 u^3_{\v m}=& 0,\ \ \ \hbox{if $\v m= odd$} .
\end{align}
Other 16 PSG's lead to $SU(2)$ spin liquids and can be dropped.

To obtain the PSG's for type U1B symmetric spin liquids, we would like to
first prove a general theorem.  Given a PSG generated by $G_{x,y,T}$ and
$G_{P_x,P_y,P_{xy}}$, the following generators 
\begin{align}
\label{newPSG}
\t G_x(\v i) =&   (-)^{i_y} G_x(\v i) , 
&\t G_y(\v i) =&    G_y(\v i) , 
\nonumber \\
\t G_{P_x}(\v i) =&   G_{P_x}(\v i),  
&\t G_{P_y}(\v i) =&  G_{P_y}(\v i), 
\nonumber\\
\t G_{P_{xy}}(\v i) =&   (-)^{i_xi_y} G_{P_{xy}}(\v i), 
&\t G_T(\v i) =&   G_T(\v i)  .
\end{align}
generate a new PSG.  The new PSG has the same IGG and is an
extension of the same symmetry group as the original PSG.  If an ansatz
$u_{\v i\v j}$ is described by a PSG $(G_{x,y,T},\ G_{P_x,P_y,P_{xy}})$, a
new ansatz described by PSG $(\t G_{x,y,T},\ \t G_{P_x,P_y,P_{xy}})$ can be
constructed
\begin{align}
\label{utoup}
\t u_{\v i\v j} = & (-)^{(j_y-i_y)i_x} u_{\v i\v j}, \ \ \ \
\hbox{for $(j_x-i_x)(j_y-i_y)=even$} \nonumber\\
\t u_{\v i\v j} = & 0,  \ \ \ \
\hbox{for $(j_x-i_x)(j_y-i_y)=odd$} 
\end{align}
The new ansatz
$\t u_{\v i\v j}$
has the same symmetry and the same gauge structure as $u_{\v i\v j}$.

To obtain the above result, we note that the following ansatz
\begin{align}
\label{u00}
u_{\v i\v j}=& (-)^{(j_y-i_y)i_x} \tau^0
, \ \ \ \hbox{for $ (j_x-i_x)(j_y-i_y)=even$ } \nonumber\\
u_{\v i\v j}=&  0
, \ \ \ \hbox{for $ (j_x-i_x)(j_y-i_y)=odd$ } 
\end{align}
has all the translation, parity, and the time reversal symmetries and has an
$SU(2)$ invariant gauge group.  The PSG of the ansatz has a subgroup
\begin{align}
 G_x(\v i) =&   (-)^{i_y}  \tau^0, 
& G_y(\v i) =&     \tau^0, 
\nonumber \\
 G_{P_x}(\v i) =&   \tau^0,  
& G_{P_y}(\v i) =&  \tau^0, 
\nonumber\\
 G_{P_{xy}}(\v i) =&   (-)^{i_xi_y}\tau^0 
\end{align}
The above properties of $\t u_{\v i\v j}$ can be obtained after realizing
that $\t u_{\v i\v j}$ can be obtained by combining $u_{\v i\v j}$ with the
ansatz in \Eq{u00}.  We see that the mapping has a meaning of adding
$\pi$-flux to each plaquette.

Using the mapping \Eq{newPSG} and the results for the type U1A symmetric
spin liquids, we find that the type U1B symmetric spin liquids are also
classified by 24 PSG's. $G_{P_x,P_y,P_{xy},T}$ of those PSG's are given by
\begin{align}
 G_{P_x} =& \eta_{ypx}^{i_y} g_3(\th_{px}),\ \ \ \
 G_{P_y} = \eta_{ypx}^{i_x} g_3(\th_{py}),  \nonumber\\
(-)^{i_xi_y} G_{P_{xy}} =& g_3(\th_{pxy}), \ \ g_3(\th_{pxy})i\tau^1 
\nonumber \\
 G_T =& \eta_{t}^{\v i} g_3(\th_t)|_{\eta_t=-1},\ \
        \eta_{t}^{\v i} g_3(\th_t)i\tau^1,
\end{align}
and
\begin{align}
 G_{P_x} =& \eta_{xpx}^{i_x} g_3(\th_{px})i\tau^1,\ \ \ \
 G_{P_y} = \eta_{xpx}^{i_y} g_3(\th_{py})i\tau^1  ,\nonumber\\
(-)^{i_xi_y} G_{P_{xy}} =& g_3(\th_{pxy}), \ \ g_3(\th_{pxy})i\tau^1 ,
\nonumber \\
 G_T =& \eta_{t}^{\v i} g_3(\th_t)|_{\eta_t=-1},\ \ 
        \eta_{t}^{\v i} g_3(\th_t)i\tau^1,
\end{align}
A labeling scheme of the above PSG's is given below \Eq{U1Bpsg2}. 

Next we consider the form of ansatz that is invariant under the above
PSG's. The translation symmetry requires that
\begin{align}
 u_{\v i,\v i+\v m} = (-)^{i_xm_y} u_{\v m} = 
(-)^{i_xm_y}(u^0_{\v m} \tau^0 + u^3_{\v m} \tau^3)
\end{align}
The $180^\circ$ rotation symmetry requires that, 
for
$
G_{P_x} = \eta_{ypx}^{i_y} g_3(\th_{px}),
G_{P_y} = \eta_{ypx}^{i_x} g_3(\th_{py}) 
$,
\begin{align}
u_{\v m} = \eta_{ypx}^{\v m} u_{-\v m}  
         = \eta_{ypx}^{\v m} (-)^{m_xm_y} u_{\v m}^\dag  
\end{align}
and for
$
G_{P_x} = \eta_{xpx}^{i_x} g_3(\th_{px})i\tau^1,
G_{P_y} = \eta_{xpx}^{i_y} g_3(\th_{py})i\tau^1 
$,
\begin{align}
u_{\v m} = \eta_{xpx}^{\v m} u_{-\v m}  
         = \eta_{xpx}^{\v m} (-)^{m_xm_y} u_{\v m}^\dag  
\end{align}
The time reversal symmetry requires that, 
for
$ G_T = \eta_{t}^{\v i} g_3(\th_t)|_{\eta_t=-1}$
\begin{align}
u_{\v m} = -(-)^{\v m} u_{\v m}  ,
\end{align}
for
$ G_T = \eta_{t}^{\v i} g_3(\th_t)i\tau^1$
\begin{align}
u_{\v m} = 
-\eta_t^{\v m} u^0_{\v m}\tau^0  + \eta_t^{\v m} u^3_{\v m}\tau^3  ,
\end{align}

We find the following 8 sets of ansatz that give rise to $U(1)$ symmetric
spin liquids:
U1B$00[0,1]1$ and U1B$11[0,1]1$,
\begin{align}
\label{U1Ban1}
 u_{\v i,\v i+\v m} =& u^3_{\v m} \tau^3 . \nonumber\\
 u^3_{\v m}=& 0,\ \ \ \hbox{if $m_x= odd$ and $m_y=odd$} .
\end{align}
U1B$0n[0,1]x$ and U1B$x1[0,1]x$,
\begin{align}
\label{U1Ban2}
 u_{\v i,\v i+\v m} =& u^0_{\v m} \tau^0 + u^3_{\v m} \tau^3 \nonumber\\
 u^0_{\v m}=& 0,\ \ \ \hbox{if $\v m= even$} \nonumber\\
 u^3_{\v m}=& 0,\ \ \ \hbox{if $m_x= odd$ or $m_y=odd$ } .
\end{align}
Other 16 PSG's lead to $SU(2)$ spin liquids and can be dropped.

Next we consider the type U1C spin liquids.  To add the time reversal
symmetry, we note that the condition \Eq{genGxyT} becomes
\begin{align}
\label{U1CGxyT}
&\tau^1G_T^{-1}(\v i) \tau^1 G_T(\v i-\hat{\v x}) \in U(1)
\nonumber\\
&\tau^1G_T^{-1}(\v i) \tau^1 G_T(\v i-\hat{\v y}) \in U(1) 
\end{align}
This leads to the following $G_T$ 
\begin{equation}
\label{U1CGT1}
 G_T = 
g_3((-)^{\v i}\phi(\v i)),\ \ 
g_3((-)^{\v i}\phi(\v i))
i\tau^1.
\end{equation}
where $\phi(\v i)$ satisfies
\begin{align}
 \phi(\v i + \hat{\v x}) - \phi(\v i) = & (-)^{\v i} \vphi_1 \hbox{ mod } 2\pi
\nonumber\\
 \phi(\v i + \hat{\v y}) - \phi(\v i) = & (-)^{\v i} \vphi_2 \hbox{ mod } 2\pi
\end{align}
The solution exist only for two cases where
$\vphi_1 - \vphi_2 = 0 \hbox{ mod } \pi$:
\begin{align}
 \phi(\v i + \hat{\v x}) - \phi(\v i) = & -(-)^{\v i} 2\th_t \hbox{ mod } 2\pi
\nonumber\\
 \phi(\v i + \hat{\v y}) - \phi(\v i) = & -(-)^{\v i} 2\th_t \hbox{ mod } 2\pi
\end{align}
and
\begin{align}
 \phi(\v i + \hat{\v x}) - \phi(\v i) = & -(-)^{\v i} (2\th_t +\pi) \hbox{ mod } 2\pi
\nonumber\\
 \phi(\v i + \hat{\v y}) - \phi(\v i) = & -(-)^{\v i} 2\th_t \hbox{ mod } 2\pi
\end{align}
The two solutions are given by
\begin{align}
 \phi(\v i) = & \vphi_t + (-)^{\v i} \th_t   \nonumber\\
 \phi(\v i) = & \vphi_t + (-)^{\v i} \th_t + i_x \pi 
\end{align}
Thus $G_T$ can take the following four forms
\begin{align}
\label{U1CGT2}
 G_T = 
 & g_3( (-)^{\v i} \vphi_t + \th_t ),          \\
 & g_3( i_x \pi + (-)^{\v i} \vphi_t + \th_t ), \nonumber\\
 & g_3( (-)^{\v i} \vphi_t + \th_t )i\tau^1,          \nonumber \\
 & g_3( i_x \pi + (-)^{\v i} \vphi_t + \th_t ) i\tau^1  . \nonumber 
\end{align}
Since $T^2=1$ and $G_T^2 \in U(1)$, the above becomes
\begin{align}
\label{U1CGT}
 G_T =& \eta_{xt}^{i_x} \eta_{yt}^{i_y} g_3(\th_t), \nonumber\\
 & g_3( (-)^{\v i} \vphi_t + \th_t )i\tau^1,          \nonumber \\
 & g_3( i_x \pi + (-)^{\v i} \vphi_t + \th_t ) i\tau^1  . 
\end{align}
where $\eta_{xt,yt}=\pm 1$.

To add three types of parity symmetries, we note that, for type U1C spin
liquids, \Eq{genGxyPx}, \Eq{genGxyPy}, and \Eq{genGxyPxy} reduce to 
\begin{align}
\label{U1CGxyPx}
\tau^1 G_{P_x}^{-1}(\v i+\hat{\v x})\tau^1 G_{P_x}(\v i) 
 \in & U(1),
\nonumber\\
\tau^1G_{P_x}^{-1}(\v i) \tau^1  G_{P_x}(\v i-\hat{\v y}) 
 \in & U(1),
\end{align}
\begin{align}
\label{U1CGxyPy}
\tau^1 G_{P_y}^{-1}(\v i+\hat{\v y})\tau^1 G_{P_y}(\v i) 
 \in & U(1),
\nonumber\\
\tau^1 G_{P_y}^{-1}(\v i) \tau^1  G_{P_y}(\v i-\hat{\v x}) 
 \in & U(1),
\end{align}
\begin{align}
\label{U1CGxyPxy}
\tau^1 G_{P_{xy}}^{-1}(\v i) \tau^1  
G_{P_{xy}}(\v i-\hat{\v x}) \in & U(1) ,\nonumber\\
\tau^1 G_{P_{xy}}^{-1}(\v i) \tau^1 
G_{P_{xy}}(\v i-\hat{\v y}) \in & U(1)  .
\end{align}
After a calculation similar to that for $G_t$, we find that
$G_{P_x,P_y,P_{xy}}$ can have the following forms
\begin{align}
 G_{P_x} =
 & g_3( (-)^{\v i} \vphi_{px} + \th_{px} ),          \nonumber \\
 & g_3( i_x \pi + (-)^{\v i} \vphi_{px} + \th_{px} ), \nonumber\\
 & g_3( (-)^{\v i} \vphi_{px} + \th_{px} )i\tau^1,          \nonumber \\
 & g_3( i_x \pi + (-)^{\v i} \vphi_{px} + \th_{px} ) i\tau^1  ;
\end{align}
\begin{align}
 G_{P_y} =
 & g_3( (-)^{\v i} \vphi_{py} + \th_{py} ),          \nonumber \\
 & g_3( i_x \pi + (-)^{\v i} \vphi_{py} + \th_{py} ), \nonumber\\
 & g_3( (-)^{\v i} \vphi_{py} + \th_{py} )i\tau^1,          \nonumber \\
 & g_3( i_x \pi + (-)^{\v i} \vphi_{py} + \th_{py} ) i\tau^1  ; 
\end{align}
\begin{align}
 G_{P_{xy}} =
 & g_3( (-)^{\v i} \vphi_{pxy} + \th_{pxy} ),          \nonumber \\
 & g_3( i_x \pi + (-)^{\v i} \vphi_{pxy} + \th_{pxy} ), \nonumber\\
 & g_3( (-)^{\v i} \vphi_{pxy} + \th_{pxy} )i\tau^1,          \nonumber \\
 & g_3( i_x \pi + (-)^{\v i} \vphi_{pxy} + \th_{pxy} ) i\tau^1  ; 
\end{align}
From the condition \Eq{genPP}, we find that
\begin{align}
 G_{P_x} =
 & \eta_{xpx}^{i_x} \eta_{ypx}^{i_y} g_3(\th_{px}), \nonumber\\
 & g_3( (-)^{\v i} \vphi_{px} + \th_{px} )i\tau^1,          \nonumber \\
 & g_3( i_x \pi + (-)^{\v i} \vphi_{px} + \th_{px} ) i\tau^1  ;
\end{align}
\begin{align}
 G_{P_y} =
 & \eta_{xpy}^{i_x} \eta_{ypy}^{i_y} g_3(\th_{px}), \nonumber\\
 & g_3( (-)^{\v i} \vphi_{py} + \th_{py} )i\tau^1,          \nonumber \\
 & g_3( i_x \pi + (-)^{\v i} \vphi_{py} + \th_{py} ) i\tau^1  ; 
\end{align}
\begin{align}
 G_{P_{xy}} =
 & g_3( \th_{pxy} ),          \nonumber \\
 & g_3( i_x \pi + (-)^{\v i}\frac{\pi}{4} + \th_{pxy} ), \nonumber\\
 & g_3( (-)^{\v i} \vphi_{pxy} + \th_{pxy} )i\tau^1 ;
\end{align}
The first condition in \Eq{genPxyPP} requires  $G_{Px,Py}$ should have the
same number of $\tau^1$.  The second condition in \Eq{genPxyPP} further
requires that $\vphi_{px} = \vphi_{py} $ mod $\pi/2$.  We note that
$G_{x,y}$ in \Eq{U1GxGy2} are invariant under gauge transformation $W_{\v
i}=g_3((-)^{\v i}\phi )$. Using such a gauge transformation, we can set
$\vphi_{px} =0$ and $\vphi_{py} =0$ mod $\pi/2$.  This leads to
\begin{align}
 G_{P_x} =
 & \eta_{xpx}^{i_x} \eta_{ypx}^{i_y} g_3(\th_{px}), \nonumber\\
 & \eta_{xpx}^{i_x} g_3( \th_{px} )i\tau^1;
\end{align}
\begin{align}
 G_{P_y} =
 & \eta_{xpy}^{i_x} \eta_{ypy}^{i_y} g_3(\th_{px}), \nonumber\\
 & \eta_{xpx}^{i_y} \eta_{pp}^{\v i} g_3( \th_{py} )i\tau^1;
\end{align}
\begin{align}
 G_{P_{xy}} =
 & g_3( \th_{pxy} ),          \nonumber \\
 & g_3( i_x \pi + (-)^{\v i}\frac{\pi}{4} + \th_{pxy} ), \nonumber\\
 & g_3( (-)^{\v i} \vphi_{pxy} + \th_{pxy} )i\tau^1 ;
\end{align}

We find $G_{P_{xy}}$ can be determined from $G_{P_x,P_y}$
through \Eq{genPxyPP} and $G_T$ from $G_{P_x,P_y, P_{xy}}$ through
\Eq{genTPxyxy}. Thus from \Eq{genPxyPP} and \Eq{genTPxyxy},
we find the following 60 sets of solutions
for $G_{P_x,P_y,P_{xy},T}$:
\begin{align}
 G_{P_x} =& \eta_{xpx}^{i_x}  \eta_{ypx}^{i_y}   g_3(\th_{px}),\ \ \ \
 G_{P_y} = \eta_{ypx}^{i_x}   \eta_{xpx}^{i_y}   g_3(\th_{py}) , \nonumber\\
 G_{P_{xy}} =
 & \eta_{pxy}^{i_x}g_3( \eta_{pxy}^{\v i}\frac{\pi}{4} + \th_{pxy} ), \nonumber\\
G_T = &\eta_t^{\v i} g_3(\th_t)|_{\eta_t=-1}, \ \ 
       \eta_{pxy}^{i_x} g_3(\th_t)i\tau^1 .
\end{align}

\begin{align}
 G_{P_x} =&\eta_{xpx}^{i_x}                   g_3(\th_{px})i\tau^1,\ \ \ \
 G_{P_y} = \eta_{xpx}^{i_y} \eta_{pxy}^{\v i} g_3(\th_{py})i\tau^1  ,
\nonumber\\
 G_{P_{xy}} =
 & \eta_{pxy}^{i_x}g_3( \eta_{pxy}^{\v i}\frac{\pi}{4} + \th_{pxy} ), \nonumber\\
G_T = &\eta_t^{\v i} g_3(\th_t)|_{\eta_t=-1}, \ \ 
       \eta_{pxy}^{i_x} \eta_t^{\v i} g_3(\th_t)i\tau^1 .
\end{align}

\begin{align}
 G_{P_x} =& \eta_{xpx}^{i_x}  \eta_{ypx}^{i_y}   g_3(\th_{px}),\ \ \ \
 G_{P_y} = \eta_{ypx}^{i_x}   \eta_{xpx}^{i_y}   g_3(\th_{py}),  \nonumber\\
 G_{P_{xy}} =
 & g_3( \th_{pxy} )i\tau^1,  \nonumber\\
G_T = &\eta_t^{\v i} g_3(\th_t)|_{\eta_t=-1} .
\end{align}

\begin{align}
 G_{P_x} =& \eta_{xpx}^{i_x}  \eta_{ypx}^{i_y}   g_3(\th_{px}),\ \ \ \
 G_{P_y} = \eta_{ypx}^{i_x}   \eta_{xpx}^{i_y}   g_3(\th_{py}),  \nonumber\\
 G_{P_{xy}} =
 & g_3( \eta_{pxy}^{\v i}\frac{\pi}{4} + \th_{pxy} )i\tau^1,  \nonumber\\
G_T = & \eta_{pxy}^{i_x} \eta_t^{\v i} g_3(\th_t)i\tau^1 .
\end{align}

\begin{align}
 G_{P_x} =&\eta_{xpx}^{i_x}                   g_3(\th_{px})i\tau^1,\ \ \ \
 G_{P_y} = \eta_{xpx}^{i_y} \eta_{pxy}^{\v i} g_3(\th_{py})i\tau^1,  
\nonumber\\
 G_{P_{xy}} =
 & g_3( \eta_{pxy}^{\v i} \frac{\pi}{4} + \th_{pxy} )i\tau^1 , \nonumber\\
G_T = &\eta_t^{\v i} g_3(\th_t)|_{\eta_t=-1}, \ \ 
       \eta_t^{\v i} \eta_{pxy}^{i_x} g_3(\th_t)i\tau^1 .
\end{align}
When combined with the type U1C translation PSG \Eq{U1GxGy2}, the above 60
sets of solutions give us 60 different type U1C PSG's.
A labeling scheme of the above PSG's is given below \Eq{PSGU1Cd}. 

Now let us consider the form of ansatz that is invariant under the above
type U1C PSG's. The translation symmetry requires that
\begin{align}
 u_{\v i,\v i+\v m} = u^0_{\v m} \tau^0 + (-)^{\v i}u^3_{\v m} \tau^3
\end{align}

For PSG's
U1C$[00,nn][0,n,1]n$, U1C$11[0,1]n$, and U1C$x1[n,x]n$,
the $180^\circ$ rotation symmetry requires that, 
\begin{align*}
u^0_{\v m} = -u^0_{\v m} ,\ \ \
u^3_{\v m} = (-)^{\v m}u^3_{\v m} .
\end{align*}
The time reversal symmetry requires that, 
\begin{align*}
u^0_{\v m} = -(-)^{\v m}u^0_{\v m} ,\ \ \
u^3_{\v m} = -(-)^{\v m}u^3_{\v m} .
\end{align*}
The ansatz have a form
\begin{align}
\label{U1Can1}
 u_{\v i,\v i+\v m} =& (-)^{\v m} u^3_{\v m} \tau^3 \nonumber\\
 u^3_{\v m}=& 0,\ \ \ \hbox{if $\v m= even$} .
\end{align}
which describe $SU(2)$ spin liquids.

For PSG's
\begin{align}
\label{U1Clb2}
&\text{U1C}[n0,0n][0,n,1]n, && 
 \text{U1C}11[n,x]n,    \nonumber\\
&\text{U1C}x1 [0,1]n, 
\end{align}
the $180^\circ$ rotation symmetry requires that 
\begin{align*}
u^0_{\v m} = -(-)^{\v m}u^0_{\v m} ,\ \ \
u^3_{\v m} = u^3_{\v m} .
\end{align*}
The time reversal symmetry requires that 
\begin{align*}
u^0_{\v m} = -(-)^{\v m}u^0_{\v m} ,\ \ \
u^3_{\v m} = -(-)^{\v m}u^3_{\v m} .
\end{align*}
The ansatz have a form
\begin{align}
\label{U1Can2}
 u_{\v i,\v i+\v m} =& u^0_{\v m}\tau^0 + (-)^{\v m} u^3_{\v m} \tau^3 
 \nonumber\\
 u^{0,3}_{\v m}=& 0,\ \ \ \hbox{if $\v m= even$} .
\end{align}

For PSG's
\begin{align}
\label{U1Clb3}
&\text{U1C}[00,nn][0,1]1, && 
 \text{U1C}11[0,1]1,
\end{align}
the $180^\circ$ rotation symmetry requires that, 
\begin{align*}
u^0_{\v m} =          -u^0_{\v m} ,\ \ \
u^3_{\v m} = (-)^{\v m}u^3_{\v m} .
\end{align*}
The time reversal symmetry requires that, 
\begin{align*}
u^0_{\v m} = -u^0_{\v m} ,\ \ \
u^3_{\v m} =  u^3_{\v m} .
\end{align*}
The ansatz have a form
\begin{align}
\label{U1Can3}
 u_{\v i,\v i+\v m} =& (-)^{\v m} u^3_{\v m} \tau^3 \nonumber\\
 u^3_{\v m}=& 0,\ \ \ \hbox{if $\v m= odd$} .
\end{align}
The ansatz gives rise to $U(1)\times U(1)$ spin liquids since $u_{\v i\v j}$
only connect points within two different sublattices.

For PSG's
\begin{align}
\label{U1Clb4}
&\text{U1C}[n0,0n][0,1]1, && 
 \text{U1C}x1 [0,1] 1, 
\end{align}
the $180^\circ$ rotation symmetry requires that 
\begin{align*}
u^0_{\v m} = -(-)^{\v m}u^0_{\v m} ,\ \ \
u^3_{\v m} =            u^3_{\v m} .
\end{align*}
The time reversal symmetry requires that 
\begin{align*}
u^0_{\v m} = -u^0_{\v m} ,\ \ \
u^3_{\v m} =  u^3_{\v m} .
\end{align*}
The ansatz have a form
\begin{align}
\label{U1Can4}
 u_{\v i,\v i+\v m} =& (-)^{\v m} u^3_{\v m} \tau^3 .
\end{align}

For PSG's
\begin{align}
\label{U1Clb5}
&\text{U1C}[00,nn]1x, && 
 \text{U1C}11[0,1]x,   
\end{align}
the $180^\circ$ rotation symmetry requires that, 
\begin{align*}
u^0_{\v m} =          -u^0_{\v m} ,\ \ \
u^3_{\v m} = (-)^{\v m}u^3_{\v m} .
\end{align*}
The time reversal symmetry requires that, 
\begin{align*}
u^0_{\v m} = -(-)^{\v m}u^0_{\v m} ,\ \ \
u^3_{\v m} =  (-)^{\v m}u^3_{\v m} .
\end{align*}
The ansatz have a form
\begin{align}
\label{U1Can5}
 u_{\v i,\v i+\v m} =& (-)^{\v m} u^3_{\v m} \tau^3 \nonumber\\
 u^3_{\v m}=& 0,\ \ \ \hbox{if $\v m= odd$} .
\end{align}
The ansatz gives rise to $U(1)\times U(1)$ spin liquids since $u_{\v i\v j}$
only connect points within two different sublattices.

For PSG's
\begin{align}
\label{U1Clb6}
&\text{U1C}[n0,0n]1x, && 
 \text{U1C}x1 [0,1] x    ,
\end{align}
the $180^\circ$ rotation symmetry requires that 
\begin{align*}
u^0_{\v m} = -(-)^{\v m}u^0_{\v m} ,\ \ \
u^3_{\v m} =            u^3_{\v m} .
\end{align*}
The time reversal symmetry requires that 
\begin{align*}
u^0_{\v m} = -(-)^{\v m}u^0_{\v m} ,\ \ \
u^3_{\v m} =  (-)^{\v m}u^3_{\v m} .
\end{align*}
The ansatz have a form
\begin{align}
\label{U1Can6}
 u_{\v i,\v i+\v m} =& (-)^{\v m} u^3_{\v m} \tau^3 \nonumber\\
 u^0_{\v m}=& 0,\ \ \ \hbox{if $\v m= even$} \nonumber\\
 u^3_{\v m}=& 0,\ \ \ \hbox{if $\v m= odd$} .
\end{align}

For PSG's
\begin{align}
\label{U1Clb7}
&\text{U1C}[00,nn][n,x]1, && 
 \text{U1C}x1[n,x] 1,   
\end{align}
the $180^\circ$ rotation symmetry requires that, 
\begin{align*}
u^0_{\v m} = -u^0_{\v m} ,\ \ \
u^3_{\v m} = (-)^{\v m}u^3_{\v m} .
\end{align*}
The time reversal symmetry requires that, 
\begin{align*}
u^0_{\v m} = -(-)^{m_x}u^0_{\v m} ,\ \ \
u^3_{\v m} = (-)^{m_x}u^3_{\v m} .
\end{align*}
The ansatz have a form
\begin{align}
\label{U1Can7}
 u_{\v i,\v i+\v m} =& (-)^{\v m} u^3_{\v m} \tau^3 \nonumber\\
 u^3_{\v m}=& 0,\ \ \ \hbox{if $m_x=odd$ or $m_y=odd$} .
\end{align}
The ansatz gives rise to $(U(1))^4$ spin liquids since $u_{\v i\v j}$
only connect points within four different sublattices.

For PSG's
\begin{align}
\label{U1Clb8}
&\text{U1C}[n0, 0n][n,x]1, && 
 \text{U1C} 11[n,x]1,    
\end{align}
the $180^\circ$ rotation symmetry requires that, 
\begin{align*}
u^0_{\v m} = -(-)^{\v m}u^0_{\v m} ,\ \ \
u^3_{\v m} =            u^3_{\v m} .
\end{align*}
The time reversal symmetry requires that, 
\begin{align*}
u^0_{\v m} = -(-)^{m_x}u^0_{\v m} ,\ \ \
u^3_{\v m} = (-)^{m_x}u^3_{\v m} .
\end{align*}
The ansatz have a form
\begin{align}
\label{U1Can8}
 u_{\v i,\v i+\v m} =& (-)^{\v m} u^3_{\v m} \tau^3 \nonumber\\
 u^0_{\v m}=& 0,\ \ \ \hbox{if $m_x=even$ or $m_y=odd$} \nonumber\\
 u^3_{\v m}=& 0,\ \ \ \hbox{if $m_x=odd$} .
\end{align}

For PSG's
\begin{align}
\label{U1Clb9}
&\text{U1C}[00,nn]xx, && 
 \text{U1C}x1[n,x]x,   
\end{align}
the $180^\circ$ rotation symmetry requires that, 
\begin{align*}
u^0_{\v m} = -u^0_{\v m} ,\ \ \
u^3_{\v m} = (-)^{\v m}u^3_{\v m} .
\end{align*}
The time reversal symmetry requires that, 
\begin{align*}
u^0_{\v m} = -(-)^{m_y}u^0_{\v m} ,\ \ \
u^3_{\v m} = (-)^{m_y}u^3_{\v m} .
\end{align*}
The ansatz have a form
\begin{align}
\label{U1Can9}
 u_{\v i,\v i+\v m} =& (-)^{\v m} u^3_{\v m} \tau^3 \nonumber\\
 u^3_{\v m}=& 0,\ \ \ \hbox{if $m_x=odd$ or $m_y=odd$} .
\end{align}
The ansatz gives rise to $(U(1))^4$ spin liquids since $u_{\v i\v j}$
only connect points within four different sublattices.

For PSG's
\begin{align}
\label{U1Clb10}
&\text{U1C}[n0, 0n]xx, && 
 \text{U1C} 11[n,x]x,    
\end{align}
the $180^\circ$ rotation symmetry requires that, 
\begin{align*}
u^0_{\v m} = -(-)^{\v m}u^0_{\v m} ,\ \ \
u^3_{\v m} =            u^3_{\v m} .
\end{align*}
The time reversal symmetry requires that, 
\begin{align*}
u^0_{\v m} = -(-)^{m_y}u^0_{\v m} ,\ \ \
u^3_{\v m} = (-)^{m_y}u^3_{\v m} .
\end{align*}
The ansatz have a form
\begin{align}
\label{U1Can10}
 u_{\v i,\v i+\v m} =& (-)^{\v m} u^3_{\v m} \tau^3 \nonumber\\
 u^0_{\v m}=& 0,\ \ \ \hbox{if $m_x=odd$ or $m_y=even$} \nonumber\\
 u^3_{\v m}=& 0,\ \ \ \hbox{if $m_y=odd$} .
\end{align}

The type U1D spin liquids always break the parity generated by $P_{xy}$ and
\Eq{genGxyPxy} cannot be satisfied. Thus there is no type U1D symmetric
spin liquid.

Last, let us consider the type U1$^m_n$ spin liquids. Instead of finding a
classification of U1$^m_n$ spin liquids, here, we will just consider the
following example:
\begin{align}
\label{U1mnS}
u_{\v i,\v i+\hat{\v x}} = \chi \tau^3,\ \ \
u_{\v i,\v i+\hat{\v y}} = \chi g_3( \frac{m\pi}{n} i_x)\tau^3 .
\end{align}
One can check that the above ansatz describes 
a symmetric spin liquid. Its PSG is given by
\begin{align}
\label{U1mnPSG}
 G_x =&   g_3(-\frac{m\pi}{n} i_y +\th_x),\ \ \ 
 G_y =  g_3(\th_y) ,  \\
 G_{P_x} =&  i(-)^{\v i}\tau^1 g_3(\th_{px}), \ \ \
 G_{P_y} = i(-)^{\v i}\tau^1 g_3(\th_{py}), \nonumber\\
 G_{P_{xy}} =&  i(-)^{\v i}\tau^1 g_3(\frac{m\pi}{n} i_xi_y+\th_{pxy}),\ \ \
 G_T = (-)^{\v i} g_3(\th_t) . \nonumber 
\end{align}
The form of $G_{x,y}$ tells us that \Eq{U1mnS} indeed describes a U1$^m_n$
spin liquid. Using the labeling scheme for the U1[A,B,C] PSG's, we can label
the above PSG by U1$^m_n xxxn$.  From the above example, we see that there
are infinite different spin liquids of type U1$^m_n$ , at least one for each
rational number $m/n$ between $0$ and $1$.

In summary, we find 8 type U1A, 8 type U1B, and 30 type U1C
symmetric $U(1)$ spin liquids. But there is an infinite number of type U1$^m_n$
spin liquids.

\subsection{Classification of $SU(2)$ projective symmetry groups}

In this section we will use PSG to classify quantum orders in mean-field
symmetric $SU(2)$ spin liquids.  We need to find the 
extensions of the symmetry transformations 
when IGG $\cG=SU(2)$.  First, we assume
that, for a $SU(2)$ spin liquid, we can always choose a gauge such that
$u_{\v i\v j}$ has a form
\begin{equation}
\label{u0}
 u_{\v i\v j} = u_{\v i\v j}^0 \tau^0  .
\end{equation}
We will call this gauge canonical gauge.  In the canonical gauge, IGG has a
form $ \cG = SU(2) $. Here we will only consider spin liquids described by 
non-zero $u_{\v i\v j}$. In this case the gauge transformations that keep
$u_{\v i\v j}$ to have the form in the canonical gauge are given by
\begin{equation}
 W_{\v i} = \eta(\v i) g
\end{equation}
where $\eta(\v i) =\pm 1$ for each $\v i$ and $g\in SU(2)$. 
The gauge transformations $G_{x,y}$
associated with the translation also take the above form:
\begin{align}
 G_x(\v i) =& \eta_x(\v i) g_x   \nonumber\\
 G_y(\v i) =& \eta_y(\v i) g_y   
\end{align}

Note that gauge transformation $W_{\v i} = \eta(\v i)\tau^0$ still keep
$u_{\v i\v j}$ in the canonical gauge.  So we can use such gauge
transformation to simplify $G_x(\v i)$ and $G_y(\v i)$ (see \Eq{WGUWU}) and
get
\begin{align}
 G_x(\v i) =& g_x   \nonumber\\
 G_y(\v i) =& \eta_y(\v i) g_y   
\end{align}
with $\eta_y(i_x=0,i_y)=1$.  Now \Eq{genGxGy} takes a form
\begin{align}
 \eta_y(\v i-\hat{\v x})  \eta_y(\v i) \in SU(2)
\end{align}
We find that there are only two different PSG's for translation symmetric
ansatz
\begin{align}
\label{SU2GxGyA}
 G_x(\v i) =& g_x   & G_y(\v i) =& g_y   \\
\label{SU2GxGyB}
 G_x(\v i) =& g_x   & G_y(\v i) =& (-)^{i_x} g_y 
\end{align}
The two PSG's lead to the following two translation symmetric ansatz
\begin{align}
\label{SU2uA}
 u_{\v i,\v i+\v m} =&  u_{\v m}^0\tau^0  \\
\label{SU2uB}
 u_{\v i,\v i+\v m} =&  (-)^{m_x i_y} u_{\v m}^0\tau^0  
\end{align}

Next we will consider the case $ G_x(\v i) = g_x$ and $G_y(\v i) = g_y$ and
add more symmetries. First let us add the three parities $P_{x,y,xy}$.
\Eq{genGxyPx}, \Eq{genGxyPy}, and \Eq{genGxyPxy} can be simplified
\begin{align}
\label{SU2GxyPx1}
G_{P_x}^{-1}(\v i+\hat{\v x})
G_{P_x}(\v i) 
 \in & \cG
\nonumber\\
G_{P_x}^{-1}(\v i) 
G_{P_x}(\v i-\hat{\v y}) 
 \in & \cG
\end{align}
\begin{align}
\label{SU2GxyPy1}
G_{P_y}^{-1}(\v i+\hat{\v y})
G_{P_y}(\v i) 
 \in & \cG
\nonumber\\
G_{P_y}^{-1}(\v i) 
G_{P_y}(\v i-\hat{\v x}) 
 \in & \cG
\end{align}
\begin{align}
\label{SU2GxyPxy1}
G_{P_{xy}}^{-1}(\v i) 
G_{P_{xy}}(\v i-\hat{\v x}) \in & \cG \nonumber\\
G_{P_{xy}}^{-1}(\v i) 
G_{P_{xy}}(\v i-\hat{\v y}) \in & \cG   .
\end{align}
We find
\begin{align}
\label{SU2GPxyxy1}
 G_{P_x}(\v i) =&  \eta_{xpx}^{i_x}\eta_{ypx}^{i_y}  g_{P_x} 
\nonumber\\
 G_{P_y}(\v i) =&  \eta_{xpy}^{i_x}\eta_{ypy}^{i_y}  g_{P_y} 
\nonumber\\
 G_{P_{xy}}(\v i) =&  \eta_{xpxy}^{i_y} \eta_{ypxy}^{i_x} g_{P_{xy}} 
\end{align}
where $g_{P_x}\in SU(2)$, $g_{P_y}\in SU(2)$, and $g_{P_{xy}}\in SU(2)$.
$\eta$'s in the above equation are not independent.  From \Eq{genPP}, we
find that
\begin{align}
&\eta_{xpxy}^{i_x} \eta_{ypxy}^{i_y}
 \eta_{xpxy}^{i_y} \eta_{ypxy}^{i_x} g_{P_{xy}}^2 \in  \cG
\end{align}
which requires 
$\eta_{xpxy}= \eta_{ypxy} \equiv \eta_{pxy}$.
From
\Eq{genPxyPP} we see that
\begin{align}
\label{SU2etapxyxygg}
& \eta_{xpxy}^{i_x} \eta_{ypxy}^{i_y}\eta_{xpx}^{i_y}  
 \eta_{ypx}^{i_x} \eta_{xpxy}^{i_y}\eta_{ypxy}^{i_x} 
 \eta_{xpy}^{i_x}\eta_{ypy}^{i_y} \nonumber\\ 
&\ \ g_{P_{xy}} g_{P_x} g_{P_{xy}} g_{P_y}^{-1} 
\in \cG 
\end{align}
We find
\begin{align}
\label{SU2etapxyxy}
 \eta_{xpy} \eta_{xpxy} \eta_{ypx} \eta_{ypxy} =&  1 \nonumber\\
 \eta_{ypy} \eta_{ypxy} \eta_{xpx} \eta_{xpxy} =&  1 
\end{align}
When combined with $\eta_{xpxy}= \eta_{ypxy}$, we see that
$\eta_{xpx}= \eta_{ypy} $ and
$\eta_{xpy}= \eta_{ypx} $.
\Eq{SU2GPxyxy1} becomes
\begin{align}
\label{SU2GPxyxy}
 G_{P_x}(\v i) =&  \eta_{xpx}^{i_x}\eta_{xpy}^{i_y}  g_{P_x} 
\nonumber\\
 G_{P_y}(\v i) =&  \eta_{xpy}^{i_x}\eta_{xpx}^{i_y}  g_{P_y} 
\nonumber\\
 G_{P_{xy}}(\v i) =&  \eta_{pxy}^{\v i}   g_{P_{xy}} 
\end{align}
Now the three $ \eta_{xpx}$, $ \eta_{xpy}$, and $ \eta_{pxy}$ are independent.

Similarly from \Eq{genGxyT} and \Eq{genTPxyxy}
we find there are only two types of $G_T$:
\begin{equation}
\label{SU2GT2}
 G_T(\v i) = \eta_t^{\v i} g_T
\end{equation}
labeled by $\eta_t=\pm1$. We note that $\eta_t=1$ implies $u_{\v i\v j}=0$
for the $SU(2)$ spin liquids. Thus we can only choose $\eta_t=-1$
\begin{equation}
\label{SU2GT1}
 G_T(\v i) = (-)^{\v i} g_T
\end{equation}

\Eq{SU2GxGyA}, \Eq{SU2GPxyxy} and \Eq{SU2GT1} give us PSG's for symmetric
$SU(2)$ spin liquids. They are labeled by $\eta_{xpx,xpy,pxy}$.  Again we can
use the gauge transformation $W_{\v i} = (-)^{i_x}$ to set $\eta_{pxy}=1$.
Thus there are only four PSG's labeled by $\eta_{xpx,xpy}$

Now we consider the case
$ G_x(\v i) = g_x$ and $G_y(\v i) = (-)^{i_x} g_y$.
\Eq{genGxyPx}, \Eq{genGxyPy}, and \Eq{genGxyPxy} have the form
\begin{align}
\label{SU2GxyPx2}
G_{P_x}^{-1}(\v i+\hat{\v x})
G_{P_x}(\v i) 
 \in & \cG
\nonumber\\
G_{P_x}^{-1}(\v i) 
G_{P_x}(\v i-\hat{\v y}) 
 \in & \cG
\end{align}
\begin{align}
\label{SU2GxyPy2}
G_{P_y}^{-1}(\v i+\hat{\v y})
G_{P_y}(\v i) 
 \in & \cG
\nonumber\\
G_{P_y}^{-1}(\v i) 
G_{P_y}(\v i-\hat{\v x}) 
 \in & \cG
\end{align}
\begin{align}
\label{SU2GxyPxy2}
(-)^{i_y}G_{P_{xy}}^{-1}(\v i) 
G_{P_{xy}}(\v i-\hat{\v x}) \in & \cG \nonumber\\
(-)^{i_x}G_{P_{xy}}^{-1}(\v i) 
G_{P_{xy}}(\v i-\hat{\v y}) \in & \cG   .
\end{align}
We find
\begin{align}
\label{SU2GPxyxy2}
 G_{P_x}(\v i) =&  \eta_{xpx}^{i_x}\eta_{ypx}^{i_y}  g_{P_x} 
\nonumber\\
 G_{P_y}(\v i) =&  \eta_{xpy}^{i_x}\eta_{ypy}^{i_y}  g_{P_y} 
\nonumber\\
 G_{P_{xy}}(\v i) =&  (-)^{i_xi_y}\eta_{xpxy}^{i_y} \eta_{ypxy}^{i_x} g_{P_{xy}} 
\end{align}
where $g_{P_x}\in SU(2)$, $g_{P_y}\in SU(2)$, and 
$g_{P_{xy}}\in SU(2)$.
$\eta$'s in the above equation are not independent. 
We find that
$\eta_{xpxy}= \eta_{ypxy} \equiv \eta_{pxy}$,
$\eta_{xpx}= \eta_{ypy} $ and
$\eta_{xpy}= \eta_{ypx} $.
After setting $\eta_{pxy}=1$ through gauge transformation 
$W_{\v i}=(-)^{i_x}$,
\Eq{SU2GPxyxy2} becomes
\begin{align}
\label{SU2GPxyxy3}
 G_{P_x}(\v i) =&  \eta_{xpx}^{i_x}\eta_{xpy}^{i_y}  g_{P_x} 
\nonumber\\
 G_{P_y}(\v i) =&  \eta_{xpy}^{i_x}\eta_{xpx}^{i_y}  g_{P_y} 
\nonumber\\
 G_{P_{xy}}(\v i) =&  (-)^{i_xi_y}  g_{P_{xy}} 
\end{align}
and the two $ \eta_{xpx}$ and $ \eta_{xpy}$ are
independent.  The gauge transformation associated with the time reversal
transformation is still given by \Eq{SU2GT1}.  \Eq{SU2GxGyB},
\Eq{SU2GPxyxy3} and \Eq{SU2GT1} give us PSG's for symmetric $SU(2)$ spin
liquids. 

We see there are total of $2\times 2^2=8$ different $SU(2)$ PSG's. 
A labeling scheme of those 8 $SU(2)$ is given below \Eq{SU2Bpsg}.
Those $SU(2)$ PSG's are algebraic PSG's. In the following we will see which
of them lead to symmetric $SU(2)$ spin liquids.

First we consider which of the translation symmetric ansatz
in \Eq{SU2uA} have the parity and the time reversal symmetries.
After two parity transformations $P_x$ and $P_y$, we find
$u_{\v m}$ in \Eq{SU2uA} satisfies
\begin{equation}
 u_{-\v m} = 
  (\eta_{xpy}\eta_{xpx})^{\v m}  
 g_{P_y} g_{P_x} u_{\v m} g_{P_x}^{-1} g_{P_y}^{-1}
\end{equation}
or
\begin{align}
 -u^0_{\v m} = (\eta_{xpy}\eta_{xpx})^{\v m}  u^0_{\v m}
\end{align}
After the time reversal transformation, we have
\begin{equation}
 -u_{\v m} = (-)^{\v m}  g_T u_{\v m} g_T^{-1} 
\end{equation}
Thus the SU2A$n0$ and SU2A$0n$ symmetric ansatz have a form
\begin{align}
\label{SU2Au}
u_{\v m} = & u_{\v m}^0 \tau^0    \nonumber\\
 u_{\v m}^0 = & 0,\ \ \ \hbox{if $\v m = even$ }
\end{align}
The other two PSG's SU2A$[00,nn]$ leads to vanishing 
$u_{\v i\v j}$ and should be dropped.

For the translation symmetric ansatz
in \Eq{SU2uB} 
the $180^\circ$ rotation symmetry requires that
\begin{align}
 -(-)^{m_xm_y} u^0_{\v m} = (\eta_{xpy}\eta_{xpx})^{\v m}  u^0_{\v m}
\end{align}
The time reversal transformation requires that
\begin{equation}
 -u^0_{\v m} = (-)^{\v m}  u^0_{\v m} 
\end{equation}
Thus the SU2B$n0$ and SU2B$0n$ symmetric ansatz have a form
\begin{align}
\label{SU2Bu}
u_{\v m} = & (-)^{i_ym_x} u_{\v m}^0 \tau^0    \nonumber\\
 u_{\v m}^0 = & 0,\ \ \ \hbox{if $\v m = even$ }
\end{align}
The other two PSG's SU2B$[00,nn]$ leads to vanishing 
$u_{\v i\v j}$ and are dropped.

We see that only 4 of the 8 $SU(2)$ PSG's leads to symmetric $SU(2)$
ansatz. Thus there are only 4 $SU(2)$ symmetric spin liquids at mean-field
level.

For ansatz of type \Eq{SU2uA} 
the parity symmetries also require that
\begin{align}
\label{SU2ugeta2}
u_{P_x(\v m)}=&\eta_{xpx}^{m_x}\eta_{xpy}^{m_y} 
 u_{\v m} \nonumber\\
u_{P_y(\v m)}=&\eta_{xpy}^{m_x}\eta_{xpx}^{m_y} 
u_{\v m}  \nonumber\\
u_{P_{xy}(\v m)}=&\eta_{pxy}^{\v m} 
u_{\v m}  
\end{align}
For ansatz of type \Eq{SU2uB} 
the parity symmetries require that
\begin{align}
\label{SU2ugeta3}
u_{P_x(\v m)}=&\eta_{xpx}^{m_x}\eta_{xpy}^{m_y} 
 u_{\v m} \nonumber\\
u_{P_y(\v m)}=&\eta_{xpy}^{m_x}\eta_{xpx}^{m_y} 
 u_{\v m}  \nonumber\\
u_{P_{xy}(\v m)}=& (-)^{m_xm_y} \eta_{pxy}^{\v m} 
 u_{\v m}  
\end{align}
For each choice of $\eta$'s, \Eq{SU2Au}, \Eq{SU2Bu}, \Eq{SU2ugeta2}, and
\Eq{SU2ugeta3} allow us to construct ansatz $u_{\v i\v j}$ for symmetric
$SU(2)$ spin liquids.

\section{Symmetric perturbations  around symmetric spin liquids}

\subsection{Construction of symmetric perturbations}

Let us consider a perturbation $\del u_{\v i\v j}$ around  a general
mean-field ansatz $u_{\v i\v j}$.  We would like to find all the symmetric
perturbations that do not change the symmetries of the original ansatz.  
Let $PSG_0$ be the PSG of the ansatz $u_{\v i\v j}$.
Clearly, the PSG of the perturbed ansatz $u_{\v i\v j}+\del u_{\v i\v j}$,
$PSG_1$, is a subgroup of $PSG_0$. If we require the two ansatz $u_{\v i\v j}$
and $u_{\v i\v j}+\del u_{\v i\v j}$ to have the same symmetry, then the two
$PSG_0$ and $PSG_1$ must satisfy
\begin{equation}
\label{SG0SG1}
 PSG_0/IGG_0 = PSG_1/IGG_1
\end{equation}
where $IGG_{0,1}$ are the IGG of $PSG_{0,1}$. We see that the low energy
gauge group of the perturbed ansatz is equal or less than the low energy gauge
group of the original ansatz.

In the next a few subsections, we will use the following steps to find
symmetric perturbations.  We first choose $IGG_1$ to be $IGG_0$ or a subgroup
of $IGG_0$. Second, we find all the gauge inequivalent subgroup of $PSG_0$:
$PSG_1 \in PSG_0$, which has $IGG_0$ as its IGG and satisfies \Eq{SG0SG1}.
Last, we find all the ansatz that are invariant under $PSG_1$.

\subsection{ Symmetric perturbations around the $Z_2$-linear 
state Z2A$003z$}

First let us apply the above approach to find all the symmetric ansatz near the
the $Z_2$-linear state \Eq{Z2lE}.
Here by symmetric ansatz we mean the ansatz with the translation, the
time reversal and the three parity symmetries.
The IGG that leaves
the $Z_2$-linear ansatz \Eq{Z2lE} invariant is $\cG = Z_2$.
The PSG of ansatz \Eq{Z2lE} is given by
\begin{align}
\label{PSGZ2lE}
 G_x(\v i) =&  \tau^0, & G_y(\v i) =&  \tau^0  \nonumber \\
 G_{P_x}(\v i) =&  \tau^0 & G_{P_y}(\v i) =&  \tau^0
\nonumber\\
 G_{P_{xy}}(\v i) =&  \tau^3 & G_T(\v i) =&  (-)^{\v i} \tau^0 
\end{align}
This is one of the PSG of the $Z_2$ spin liquids labeled by Z2A$003z$.
We note that the IGG for any ansatz $u_{\v i\v j}$ is at least $Z_2$.
Thus the above $Z_2$ PSG is already minimal, \ie non of its subgroup can
be regarded as the PSG of some {\em symmetric} ansatz.
Therefore,
all the symmetric perturbations around the $Z_2$-linear state are 
invariant under the above PSG \Eq{PSGZ2lE}.
From \Eq{Z2ugeta1}  and \Eq{Z2ugeta2}, we find that the most general
symmetric spin liquid with PSG 
Z2A$\tau^0_+\tau^0_+\tau^3\tau^3_-$ have the form
\begin{align}
\label{Z2lg1}
 u_{\v i, \v i +\v m}  =&   u^l_{\v m} \tau^l|_{l=1,2,3}  \nonumber\\
 u^{1,2}_{P_{xy}(\v m)} =&  - u^{1,2}_{\v m}  \nonumber\\
  u^{3}_{P_{xy}(\v m)} =&   u^{3}_{\v m}  \nonumber\\
  u^{1,2,3}_{P_x(\v m)} =&    u^{1,2,3}_{\v m}  \nonumber\\
  u^{1,2,3}_{P_y(\v m)} =&    u^{1,2,3}_{\v m}  \nonumber\\
 u_{\v m} =&  0,\ \ \ \hbox{for $\v m = even$}
\end{align}

\subsection{ Symmetric perturbations around the $U(1)$-linear 
state U1C$n01n$}
\label{U1pert}

The above analysis can also be used to obtain all the symmetric
perturbations around the $U(1)$-linear ansatz 
in \Eq{U1lA}.
The invariant gauge group is $\cG=\{ e^{i\th \tau^3} \}$.

The ansatz is invariant under
translations by $\hat x$ and $\hat y$ followed by gauge transformation 
$i\tau^{\th_x}$ and
$i\tau^{\th_y}$, 
where $i\tau^\th\equiv i(\cos\th\tau^1 + \sin\th\tau^2)$. 
The ansatz also has the time reversal and the three parity symmetries.
The PSG of the ansatz is generated by
\begin{align}
\label{PSGU1lin}
 G_x(\v i) =&  i\tau^{\th_x}, & G_y(\v i) =&  i\tau^{\th_y}  \nonumber \\
 G_{P_x}(\v i) =&  (-)^{i_x}g_3(\th_{px}) & G_{P_y}(\v i) &=
(-)^{i_y}g_3(\th_{py})
\nonumber\\
 G_{P_{xy}}(\v i) =&  \tau^{\th_{pxy}} & G_T(\v i) =&  (-)^{\v i}g_3(\th_T) 
\end{align}
where $ g_a(\th) = e^{i \th \tau^a}$ and
$\th_{x,y,px,py,pxy,T}$ can take any values.
We see that the ansatz \Eq{U1lA} is labeled by 
U1C$n0 1n$.

First let us consider the symmetric perturbations that do not break the $U(1)$
gauge structure. 
Since the perturbed ansatz are required to invariant under the same IGG
and have the same symmetry as the original $U(1)$-linear ansatz \Eq{U1lA},
the perturbations must be invariant under the original PSG
\Eq{PSGU1lin}. The translation symmetry require the perturbations
to have a form
\begin{equation}
 \del u_{\v i,\v i+\v m} = 
\del u^0_{\v m}\tau^0 +
(-)^{\v i} \del u^3_{\v m}\tau^3 
\end{equation}
The $180^\circ$ rotation symmetry $P_xP_y$ requires that
\begin{align*}
 \del u^0_{-\v m} =& \del u^0_{\v m} (-)^{\v m}  \nonumber\\ 
 \del u^3_{-\v m} =& \del u^3_{\v m} (-)^{\v m} 
\end{align*}
and the time reversal symmetry requires
\begin{align*}
 -\del u^0_{\v m} =& \del u^0_{\v m} (-)^{\v m}  \nonumber\\ 
 -\del u^3_{\v m} =& \del u^3_{\v m} (-)^{\v m} 
\end{align*}
Thus the symmetric ansatz with PSG \Eq{PSGU1lin} are given by
\begin{align}
\label{U1glin}
 u_{\v i, \v i +\v m}  =&  
 u^0_{\v m}\tau^0 +
(-)^{\v i}  u^3_{\v m}\tau^3 
\nonumber\\
 u^{0,3}_{\v m} =&  0,\ \ \ \hbox{for $\v m = even$}  \nonumber\\
  u^{0}_{P_{xy}(\v m)} =&   u^{0}_{\v m}  \nonumber\\
  u^{3}_{P_{xy}(\v m)} =&  - u^{3}_{\v m}  \nonumber\\
  u^{0,3}_{P_x(\v m)} =&   (-)^{m_x} u^{0,3}_{\v m}  \nonumber\\
  u^{0,3}_{P_y(\v m)} =&   (-)^{m_y} u^{0,3}_{\v m}  
\end{align}
The above represent most general 
ansatz around the $U(1)$-linear state
that do not break any symmetries and do not change the quantum order in the
state.

Next we consider the symmetric perturbations that break the $U(1)$
gauge structure down to a $Z_2$ gauge structure. 
The IGG becomes
$\cG=Z_2$ for the perturbed ansatz.
We first need to find subgroups of \Eq{PSGU1lin} which have the reduced IGG
and the same symmetries. The elements in new PSG have the following form
\begin{align}
 G_x(\v i) =&  \pm i\tau^{\th_x}, & 
 G_y(\v i) =&  \pm i\tau^{\th_y}  
\nonumber \\
 G_{P_x}(\v i) =&  \pm (-)^{i_x}g_3(\th_{px}) & 
 G_{P_y}(\v i) =&  \pm (-)^{i_y}g_3(\th_{py})
\nonumber\\
 G_{P_{xy}}(\v i) =&   \pm i\tau^{\th_{pxy}} & 
 G_T(\v i)        =&  \pm (-)^{\v i}g_3(\th_T) 
\end{align}
where $\th_{x,y,px,py,pxy,T}$ each takes a fixed value.
The $\pm$ signs are independent from each other and come from the $Z_2$ IGG.
Different choices of
$\th_{x,y,px,py,pxy,T}$ give us different subgroups which lead to different
classes of $Z_2$ symmetric perturbations. 

To obtain the consistent choices of
$\th_{x,y,px,py,pxy,T}$, 
we note that $G_{x,y}$, $G_{P_x,P_y,P_{xy}}$ and $G_T$ must satisfy
\Eq{genGxGy}, \Eq{genGxyT}, \Eq{genGxyPx}, \Eq{genGxyPy}, \Eq{genGxyPxy}, 
\Eq{genTPxyxy}, and \Eq{genPxyPP}, with $\cG=\{ \pm \tau^0 \}$.
Those equations reduce to 
\begin{equation}
\label{Z2U1GxGy}
 \tau^{\th_x} \tau^{\th_y} \tau^{\th_x} \tau^{\th_y} =\pm \tau^0
\end{equation}
\begin{align}
\label{Z2U1GxyT}
 \tau^{\th_x} g_3^{-1}(\th_T) \tau^{\th_x} g_3(\th_T) = & \pm \tau^0
\nonumber\\
 \tau^{\th_y} g_3^{-1}(\th_T) \tau^{\th_y} g_3(\th_T) = & \pm \tau^0
\end{align}
\begin{align}
\label{Z2U1GxyPx}
 \tau^{\th_x} g_3^{-1}(\th_{px}) \tau^{\th_x} g_3(\th_{px}) = & \pm \tau^0
\nonumber\\
 \tau^{\th_y} g_3^{-1}(\th_{px}) \tau^{\th_y} g_3(\th_{px}) = & \pm \tau^0
\end{align}
\begin{align}
\label{Z2U1GxyPy}
 \tau^{\th_y} g_3^{-1}(\th_{py}) \tau^{\th_y} g_3(\th_{py}) = & \pm \tau^0
\nonumber\\
 \tau^{\th_x} g_3^{-1}(\th_{py}) \tau^{\th_x} g_3(\th_{py}) = & \pm \tau^0
\end{align}
\begin{align}
\label{Z2U1GxyPxy}
 \tau^{\th_y} \tau^{\th_{pxy}} \tau^{\th_x} \tau^{\th_{pxy}} = & \pm \tau^0
\nonumber\\
 \tau^{\th_x} \tau^{\th_{pxy}} \tau^{\th_y} \tau^{\th_{pxy}} = & \pm \tau^0
\end{align}
\begin{align}
\label{Z2U1TPxyxy}
g_3^{-1}(\th_T) g_3^{-1}(\th_{px}) g_3(\th_T) g_3(\th_{px}) =&\pm \tau^0
\nonumber\\
g_3^{-1}(\th_T) g_3^{-1}(\th_{py}) g_3(\th_T) g_3(\th_{py}) =&\pm \tau^0
\nonumber\\
g_3^{-1}(\th_T) \tau^{\th_{pxy}} g_3(\th_T) \tau^{\th_{pxy}} =&\pm \tau^0
\end{align}
\begin{align}
\label{Z2U1PxyPP}
\tau^{\th_{pxy}} g_3(\th_{px})\tau^{\th_{pxy}} g_3^{-1}(\th_{py}) 
=& \pm \tau^0  \nonumber\\
g_3(\th_{py})g_3(\th_{px}) 
g_3^{-1}(\th_{py})g_3^{-1}(\th_{px}) 
=& \pm \tau^0  
\end{align}
Since $P_x^2=P_y^2=P_{xy}^2=T^2=1$, we also have
\begin{align}
\label{Z2U1pxpyT}
 g_3^2(\th_{px}) =& \pm \tau^0, & g_3^2(\th_{py}) =& \pm \tau^0, 
\nonumber\\
 g_3^2(\th_T) =& \pm \tau^0 .
\end{align}

We can choose a gauge to make $\th_x=0$. \Eq{Z2U1GxGy} has two solutions
\begin{align}
 G_x =& i\tau^1, & G_y =&i\tau^1  \\
 G_x =& i\tau^1, & G_y =&i\tau^2  
\end{align}
When $G_x = i\tau^1$, and $ G_y =i\tau^1$, 
we find the following 8 solutions for
\Eq{Z2U1GxGy}, \Eq{Z2U1GxyT}, \Eq{Z2U1GxyPx}, \Eq{Z2U1GxyPy}, \Eq{Z2U1GxyPxy}, 
\Eq{Z2U1TPxyxy}, and \Eq{Z2U1PxyPP}, with $\cG=\{ \pm \tau^0 \}$.
\begin{align}
 G_x(\v i) =&  i\tau^1,  \ \ \ \
 G_y(\v i) = i\tau^1, 
\nonumber\\
 (-)^{i_x} G_{P_x}(\v i) &=
 (-)^{i_y} G_{P_y}(\v i) = \tau^0,\ i\tau^3,
\nonumber \\
 G_{P_{xy}}(\v i) =&   i\tau^1,\ i\tau^2\ \ \ \
 (-)^{\v i}G_T(\v i) = \tau^0,\ i\tau^3; 
\end{align}
We can make a gauge transformation $W_{\v i} = (i\tau^1)^{\v i}$
(see \Eq{WGUWU}) to change the above to
\begin{align}
 G_x(\v i) =&  \tau^0,  \ \ \ \
 G_y(\v i) = \tau^0, 
\nonumber\\
  G_{P_x}(\v i) &=
  G_{P_y}(\v i) = \tau^0,\ i(-)^{\v i}\tau^3,
\nonumber \\
 G_{P_{xy}}(\v i) =&   i\tau^1,\ i(-)^{\v i}\tau^2\ \ \ \
 G_T(\v i) = (-)^{\v i}\tau^0,\ i\tau^3;  \nonumber\\
 \cG =&  \{ \pm \tau^0 \}
\end{align}
We can use a gauge transformation $W_{\v i}=(-)^{i_x}$ to change
$G_{P_{xy}}(\v i) =  i(-)^{\v i}\tau^2$ to
$G_{P_{xy}}(\v i) =  i\tau^2$ without affecting other $G$'s.
Now we see that 
$G_{P_{xy}}(\v i) =  i(-)^{\v i}\tau^2$ and
$G_{P_{xy}}(\v i) =  i\tau^1$ give rise to gauge equivalent PSG's.
Thus we only have 4 different PSG's
\begin{align}
\label{Z2U1psg}
 G_x(\v i) =&  \tau^0,  \ \ \ \
 G_y(\v i) = \tau^0, 
\nonumber\\
  G_{P_x}(\v i) &=
  G_{P_y}(\v i) = \tau^0,\ i(-)^{\v i}\tau^3,
\nonumber \\
 G_{P_{xy}}(\v i) =&   i\tau^1\ \ \ \
 G_T(\v i) = (-)^{\v i}\tau^0,\ i\tau^3;  \nonumber\\
 \cG =&  \{ \pm \tau^0 \}
\end{align}

When $G_x = i\tau^1$, and $ G_y =i\tau^2$, 
we find the following 4 solutions for
\Eq{Z2U1GxGy}, \Eq{Z2U1GxyT}, \Eq{Z2U1GxyPx}, \Eq{Z2U1GxyPy}, \Eq{Z2U1GxyPxy}, 
\Eq{Z2U1TPxyxy}, and \Eq{Z2U1PxyPP}, with $\cG=\{ \pm \tau^0 \}$.
\begin{align}
 G_x(\v i) =&  i\tau^1,  \ \ \ \
 G_y(\v i) = i\tau^2, 
\nonumber\\
 (-)^{i_x} G_{P_x}(\v i) &=
 (-)^{i_y} G_{P_y}(\v i) = \tau^0,\ i\tau^3,
\nonumber \\
 G_{P_{xy}}(\v i) =&   i\tau^{12}\ \ \ \
 (-)^{\v i}G_T(\v i) = \tau^0,\ i\tau^3; 
\end{align}
We can make a gauge transformation 
$W_{\v i} = (i\tau^1)^{i_x} (i\tau^2)^{i_y} $
(see \Eq{WGUWU}) to change the above to
\begin{align}
\label{Z2U1psg1}
 G_x(\v i) =&  (-)^{i_y} \tau^0,  \ \ \ \
 G_y(\v i) = \tau^0, 
\nonumber\\
  G_{P_x}(\v i) &=
  G_{P_y}(\v i) = \tau^0,\ i(-)^{\v i}\tau^3,
\nonumber \\
 G_{P_{xy}}(\v i) =&   (-)^{i_xi_y} i\tau^1  \ \ \ \
 G_T(\v i) = (-)^{\v i}\tau^0,\ i\tau^3;  \nonumber\\
 \cG =&  \{ \pm \tau^0 \}
\end{align}
We note the 4 PSG's in \Eq{Z2U1psg1} can be obtained from the 4 PSG's
in \Eq{Z2U1psg} through the transformation \Eq{utoup}.

We find that all the symmetric spin liquids around the $U(1)$-linear
state \Eq{U1lA} that break the $U(1)$ gauge structure to a $Z_2$ gauge
structure can be divided into eight classes. 
Using PSG's in \Eq{Z2U1psg}, we find
translation symmetry requires the ansatz to have a form
\begin{equation}
u_{\v i,\v i+\v m} =
u^\mu_{\v m}\tau^\mu 
\end{equation}
The $180^\circ$ rotation (generated by $P_xP_y$) symmetry
requires that
\begin{align*}
u_{\v m} =& u_{-\v m}  
\end{align*}
which implies $u^0_{\v m}=0$.
The time reversal $T$ symmetry
requires that
\begin{align*}
u_{\v m} =& -(-)^{\v m} u_{\v m} 
\end{align*}
or
\begin{align*}
u_{\v m} =& -\tau^3 u_{\v m}\tau^3  
\end{align*}
The four ansatz for PSG's in \Eq{Z2U1psg} are given by
Z2A$[\tau^0_+\tau^0_+, \tau^3_-\tau^3_-]\tau^1\tau^0_-$:
\begin{align}
\label{Z2U1g1}
  u_{\v i,\v i+\v m} = &  u^l_{\v m}\tau^l 
\\
 u^{1,2,3}_{\v m} =&  0,\ \ \ \hbox{for $\v m = even$}  \nonumber
\end{align}
and 
Z2A$[\tau^0_+\tau^0_+, \tau^3_-\tau^3_-]\tau^1\tau^3_+$:
\begin{align}
\label{Z2U1g2}
  u_{\v i,\v i+\v m} = & 
  u^1_{\v m}\tau^1 + 
  u^2_{\v m}\tau^2 
\end{align}

Using PSG's in \Eq{Z2U1psg1}, we find
translation symmetry requires the ansatz to have a form
\begin{equation}
u_{\v i,\v i+\v m} = (-)^{i_xm_y}
u^\mu_{\v m}\tau^\mu 
\end{equation}
the $180^\circ$ rotation symmetry requires that
\begin{align*}
(-)^{i_xm_y}u_{\v m}=(-)^{(i_x+m_x)m_y}u_{\v m}^\dag   
\end{align*}
or
\begin{align*}
 u^0_{\v m} =&  0,\ \ \ \hbox{for $m_x=even$ or $m_y=even$} \nonumber\\
 u^{1,2,3}_{\v m} =&  0,\ \ \ \hbox{for $m_x=odd$ and $m_y=odd$}
\end{align*}
The time reversal $T$ symmetry
requires that
\begin{align*}
u_{\v m} =& -(-)^{\v m} u_{\v m} 
\end{align*}
or
\begin{align*}
u_{\v m} =& -\tau^3 u_{\v m}\tau^3  
\end{align*}
The four ansatz for PSG's in \Eq{Z2U1psg1} are given by
Z2B$[\tau^0_+\tau^0_+, \tau^3_-\tau^3_-]\tau^1\tau^0_-$:
\begin{align}
\label{Z2U1g3}
  u_{\v i,\v i+\v m} = &  (-)^{i_xm_y} u^l_{\v m}\tau^l 
\\
 u^{1,2,3}_{\v m} =&  0,\ \ \ \hbox{for $\v m = even$}  \nonumber
\end{align}
and
Z2B$[\tau^0_+\tau^0_+, \tau^3_-\tau^3_-]\tau^1\tau^3_+$:
\begin{align}
\label{Z2U1g4}
  u_{\v i,\v i+\v m} = & 
  (-)^{i_xm_y}(
  u^1_{\v m}\tau^1 + 
  u^2_{\v m}\tau^2 )
\\
 u^{1,2}_{\v m} =&  0,\ \ \ \hbox{for $m_x=odd$ and $m_y=odd$}
\nonumber 
\end{align}

The eight different $Z_2$ spin liquids have different quantum orders.
They can transform into each other via the $U(1)$-linear spin liquids. 
without any change of symmetries.
Those transitions are continuous transitions without broken symmetries.

\subsection{ Symmetric perturbations around the $SU(2)$-gapless 
state SU2A$n0$}

In this subsection,
we would like to consider the symmetric perturbations around
the $SU(2)$-gapless ansatz \Eq{SU2gl},
which describes a SU2A$n0$ spin liquid.
The invariant gauge group is $\cG=SU(2)$.
The PSG of the ansatz is generated by
\begin{align}
\label{PSGSU2gl}
 G_x(\v i) =&  g_x, & G_y(\v i) =&  g_y  \nonumber \\
 G_{P_x}(\v i) =&  (-)^{i_x} g_{px}  & G_{P_y}(\v i) =&  (-)^{i_y} g_{py}
\nonumber\\
 G_{P_{xy}}(\v i) =&  g_{pxy} & G_T(\v i) =&  (-)^{\v i}g_T 
\end{align}
where $g_{x,y,px,py,pxy,T} \in SU(2)$.
Thus the $SU(2)$-gapless state is labeled by SU2A$\tau^0_-\tau^0_+$.

First let us consider the symmetric perturbations that do not break the $SU(2)$
gauge structure. To have the $SU(2)$ gauge structure,
the perturbations must be invariant under the gauge
transformations in $\cG$ and satisfy $\del u_{\v i\v j}\propto \tau^0$.
To have the symmetries, 
the perturbations must be invariant under PSG
in \Eq{PSGSU2gl}. The translation symmetry require the perturbations
to have a form
\begin{equation}
 \del u_{\v i,\v i+\v m} = \del u^0_{\v m}\tau^0 
\end{equation}
The $180^\circ$ rotation symmetry $P_xP_y$ and the time reversal symmetry $T$
require that
\begin{align*}
 \del u^0_{-\v m} =& \del u^0_{\v m} (-)^{\v m}  \nonumber\\
 -\del u^0_{\v m} =& \del u^0_{\v m} (-)^{\v m} 
\end{align*}
Thus the symmetric ansatz with PSG \Eq{PSGSU2gl} are given by
\begin{align}
\label{SU2glg}
 u_{\v i, \v i +\v m}  =&   u^0_{\v m}\tau^0 
\nonumber\\
 u^{0}_{\v m} =&  0,\ \ \ \hbox{for $\v m = even$}  \nonumber\\
  u^{0}_{P_{xy}(\v m)} =&   u^{0}_{\v m}  \nonumber\\
  u^{0}_{P_x(\v m)} =&   (-)^{m_x} u^{0}_{\v m}  \nonumber\\
  u^{0}_{P_y(\v m)} =&   (-)^{m_y} u^{0}_{\v m}  
\end{align}
The above represent most general ansatz around the $SU(2)$-gapless state that
do not break any symmetries and do not change the quantum order in the state.
It describes the most general $SU(2)$-gapless state with quantum order
SU2A$\tau^0_-\tau^0_+$.

Next we consider the symmetric perturbations that break the $SU(2)$
gauge structure down to a $U(1)$ gauge structure. 
The invariant gauge group becomes $\cG=U(1)$ for the perturbed ansatz.
We first need to find subgroups of \Eq{PSGSU2gl} by
choosing a fixed value for each $g_{x,y,px,py,pxy,T}$.
We choose $g_{x,y,px,py,pxy,T}$ in such a way that
\Eq{genGxGy}, \Eq{genGxyT}, \Eq{genGxyPx}, \Eq{genGxyPy}, \Eq{genGxyPxy}, 
\Eq{genTPxyxy}, and \Eq{genPxyPP} can be satisfied 
when we limit $\cG$ to a $U(1)$ subgroup. (Those equations are always
satisfied when $\cG = SU(2)$).
Since the original invariant gauge group is formed by constant gauge
transformations, its $U(1)$ subgroup is also formed by constant gauge
transformations. We can choose a gauge such that the $U(1)$ invariant gauge 
group
is given by $\cG=\{ g_3(\th)| \th\in [0,2\pi) \}$.

To obtain the consistent choices of
$g_{x,y,px,py,pxy,T}$, 
we note that $G_{x,y}$, $G_{P_x,P_y,P_{xy}}$ and $G_T$ must satisfy
\Eq{genGxGy}, \Eq{genGxyT}, \Eq{genGxyPx}, \Eq{genGxyPy}, \Eq{genGxyPxy}, 
\Eq{genTPxyxy}, and \Eq{genPxyPP}, with $\cG=\{ g_3(\th)| \th\in [0,2\pi) \}$.
Those equations reduce to 
\begin{equation}
\label{U1SU2glGxGy}
 g_x g_y g_x^{-1} g_y^{-1} \in U(1)
\end{equation}
\begin{align}
\label{U1SU2glGxyT}
 g_x^{-1} g_T^{-1}g_xg_T \in & U(1)
&
 g_y^{-1} g_T^{-1} g_yg_T \in & U(1)
\end{align}
\begin{align}
\label{U1SU2glGxyPx}
 g_x g_{px}^{-1} g_x g_{px} \in & U(1)
&
 g_y^{-1} g_{px}^{-1}g_y g_{px} \in & U(1)
\end{align}
\begin{align}
\label{U1SU2glGxyPy}
 g_y g_{py}^{-1} g_y g_{py} \in & U(1)
&
 g_x^{-1} g_{py}^{-1} g_x g_{py} \in & U(1)
\end{align}
\begin{align}
\label{U1SU2glGxyPxy}
 g_y^{-1} g_{pxy}^{-1} g_x g_{pxy} \in & U(1)
&
 g_x^{-1} g_{pxy}^{-1} g_y g_{pxy} \in & U(1)
\end{align}
\begin{align}
\label{U1SU2glTPxyxy}
g_T^{-1} g_{px}^{-1} g_T g_{px} \in & U(1)
&
g_T^{-1} g_{py}^{-1} g_T g_{py} \in & U(1)
\nonumber\\
g_T^{-1} g_{pxy}^{-1} g_T g_{pxy} \in & U(1)
\end{align}
\begin{align}
\label{U1SU2glPxyPP}
g_{pxy} g_{px}g_{pxy} g_{py}^{-1}
\in & U(1)&
g_{py}g_{px} 
g_{py}^{-1}g_{px}^{-1}
\in & U(1)
\end{align}
Since $P_x^2=P_y^2=P_{xy}^2=T^2=1$, we also have
\begin{align}
\label{U1SU2glpxpyT}
 g_{px}^2 \in & U(1) , & g_{py}^2 \in & U(1) , 
\nonumber\\
 g_{pxy}^2 \in & U(1), & g_T^2 \in & U(1)  .
\end{align}

Solving the above equations,
we find 16 different PSG's with $U(1)$ invariant gauge group.
The following are their generators and their labels.\\
U1A$\tau^0_-\tau^0_+[\tau^0,\tau^1][\tau^0_-,\tau^1_-]$ 
( which is gauge equivalent to 
U1A$\tau^0_+\tau^0_+[\tau^0,\tau^1][\tau^0_-,\tau^1_+]$): 
\begin{align}
\label{U1SU2glPSG1}
& G_x(\v i)  = g_3(\th_x), &&G_y(\v i) = g_3(\th_y)  \\
& (-)^{i_x}G_{P_x}(\v i) =  g_3(\th_{px})
&& (-)^{i_y} G_{P_y}(\v i) = g_3(\th_{py})
\nonumber\\
& G_{P_{xy}}(\v i) =  g_3(\th_{pxy}),\ i\tau^{\th_{pxy}}
&& (-)^{\v i} G_T(\v i) = g_3(\th_T),\ i\tau^{\th_T}  \nonumber 
\end{align}
U1A$\tau^1_-\tau^1_+[\tau^0,\tau^1][\tau^0_-,\tau^1_-]$: 
\begin{align}
\label{U1SU2glPSG2}
& G_x(\v i)  = g_3(\th_x), && G_y(\v i) = g_3(\th_y)  \\
& (-)^{i_x}G_{P_x}(\v i) =   i\tau^{\th_{px}}
&& (-)^{i_y} G_{P_y}(\v i) =  i\tau^{\th_{py}}
\nonumber\\
& G_{P_{xy}}(\v i) =  g_3(\th_{pxy}),\ i\tau^{\th_{pxy}}
&& (-)^{\v i} G_T(\v i) = g_3(\th_T),\ i\tau^{\th_T}  \nonumber 
\end{align}
U1C$\tau^0_-\tau^0_+[\tau^0_+,\tau^1_+][\tau^0_-,\tau^1_-]$: 
\begin{align}
\label{U1SU2glPSG3}
& G_x(\v i)  = i\tau^{\th_x}, &&G_y(\v i) = i\tau^{\th_y}  \\
& (-)^{i_x}G_{P_x}(\v i) =  g_3(\th_{px})
&& (-)^{i_y} G_{P_y}(\v i) = g_3(\th_{py})
\nonumber\\
& G_{P_{xy}}(\v i) =  g_3(\th_{pxy}),\ i\tau^{\th_{pxy}}
&& (-)^{\v i} G_T(\v i) = g_3(\th_T),\ i\tau^{\th_T}  \nonumber 
\end{align}
U1C$\tau^1_-\tau^1[\tau^0_+,\tau^1_+][\tau^0_-,\tau^1_-]$: 
\begin{align}
\label{U1SU2glPSG4}
& G_x(\v i)  = i\tau^{\th_x}, && G_y(\v i) = i\tau^{\th_y}  \\
& (-)^{i_x}G_{P_x}(\v i) =   i\tau^{\th_{px}}
&& (-)^{i_y} G_{P_y}(\v i) =  i\tau^{\th_{py}}
\nonumber\\
& G_{P_{xy}}(\v i) =  g_3(\th_{pxy}),\ i\tau^{\th_{pxy}}
&& (-)^{\v i} G_T(\v i) = g_3(\th_T),\ i\tau^{\th_T}  \nonumber 
\end{align}

The $180^\circ$ rotation symmetry $P_xP_y$ and the time reversal symmetry $T$
require that
\begin{align*}
 u_{-\v i,-\v i-\v m} =&  u_{\v i,\v i+\v m} (-)^{\v m}  
= u_{-\v i-\v m,-\v i}^\dag   \nonumber\\
-  u_{\v i,\v i+\v m} =& g_T  u_{\v i,\v i+\v m}g_T^{-1} (-)^{\v m}  
\end{align*}
We find that all the symmetric spin liquids around the $SU(2)$-gapless
state \Eq{SU2gl} that break the $SU(2)$ gauge structure to a $U(1)$ gauge
structure can be divided into 12 classes. They are given by
(using abbreviated notation)
 U1C$n0[0,1]n$:
\begin{align}
\label{U1SU2gl1}
  u_{\v i,\v i+\v m} &=
 u^0_{\v m}\tau^0 
+ (-)^{\v i}  u^3_{\v m}\tau^3 
\nonumber\\
 u^{0,3}_{\v m} =&  0,\ \ \ \hbox{for $\v m = even$}  
\\
 G_x(\v i) =&  i\tau^{\th_x},  \ \ \ \
 G_y(\v i) = i\tau^{\th_y}, 
\nonumber\\
 (-)^{i_x} G_{P_x}(\v i) =&  g_3(\th_{px}), \ \ \ \
 (-)^{i_y} G_{P_y}(\v i) = g_3(\th_{py}) ,
\nonumber \\
 G_{P_{xy}}(\v i) =&  g_3(\th_{pxy}),\ i\tau^{\th_{pxy}} \ \ \ \
 (-)^{\v i}G_T(\v i) =  g_3(\th_T); \nonumber 
\end{align}
 U1C$n0[0,1]x$:
\begin{align}
\label{U1SU2gl2}
  u_{\v i,\v i+\v m} &=
 u^0_{\v m}\tau^0 
+ (-)^{\v i}  u^3_{\v m}\tau^3 
\nonumber\\
 u^{0}_{\v m} =&  0,\ \ \ \hbox{for $\v m = even$}  \nonumber\\
 u^{3}_{\v m} =&  0,\ \ \ \hbox{for $\v m = odd$}  
\\
 G_x(\v i) =&  i\tau^{\th_x},  \ \ \ \
 G_y(\v i) = i\tau^{\th_y}, 
\nonumber\\
 (-)^{i_x} G_{P_x}(\v i) =&  g_3(\th_{px}), \ \ \ \
 (-)^{i_y} G_{P_y}(\v i) = g_3(\th_{py}) ,
\nonumber \\
 G_{P_{xy}}(\v i) =&  g_3(\th_{pxy}),\ i\tau^{\th_{pxy}} \ \ \ \
 (-)^{\v i}G_T(\v i) =  i\tau^{\th_T}; \nonumber 
\end{align}
 U1C$x1[0,1]n$:
\begin{align}
\label{U1SU2gl3}
  u_{\v i,\v i+\v m} &=
 u^0_{\v m}\tau^0 
+ (-)^{\v i}  u^3_{\v m}\tau^3 
\nonumber\\
 u^{0,3}_{\v m} =&  0,\ \ \ \hbox{for $\v m = even$}  
\\
 G_x(\v i) =&  i\tau^{\th_x},  \ \ \ \
 G_y(\v i) = i\tau^{\th_y}, 
\nonumber\\
 (-)^{i_x} G_{P_x}(\v i) =&  i\tau^{\th_{px}}, \ \ \ \
 (-)^{i_y} G_{P_y}(\v i) = i\tau^{\th_{py}} ,
\nonumber \\
 G_{P_{xy}}(\v i) =&  g_3(\th_{pxy}),\ i\tau^{\th_{pxy}} \ \ \ \
 (-)^{\v i}G_T(\v i) =  g_3(\th_T); \nonumber 
\end{align}
 U1C$x1[0,1]x$:
\begin{align}
\label{U1SU2gl4}
  u_{\v i,\v i+\v m} &=
 u^0_{\v m}\tau^0 
+ (-)^{\v i}  u^3_{\v m}\tau^3 
\nonumber\\
 u^{0}_{\v m} =&  0,\ \ \ \hbox{for $\v m = even$}  \nonumber\\
 u^{3}_{\v m} =&  0,\ \ \ \hbox{for $\v m = odd$}  
\\
 G_x(\v i) =&  i\tau^{\th_x},  \ \ \ \
 G_y(\v i) = i\tau^{\th_y}, 
\nonumber\\
 (-)^{i_x} G_{P_x}(\v i) =&  i\tau^{\th_{px}}, \ \ \ \
 (-)^{i_y} G_{P_y}(\v i) = i\tau^{\th_{py}} ,
\nonumber \\
 G_{P_{xy}}(\v i) =&  g_3(\th_{pxy}),\ i\tau^{\th_{pxy}} \ \ \ \
 (-)^{\v i}G_T(\v i) =  i\tau^{\th_T}; \nonumber 
\end{align}
 U1A$n0[0,1]x$ (which is gauge equivalent to U1A$00[0,1]1$):
\begin{align}
\label{U1SU2gl5}
  u_{\v i,\v i+\v m} &=
 u^0_{\v m}\tau^0 
+  u^3_{\v m}\tau^3 
\nonumber\\
 u^{0}_{\v m} =&  0,\ \ \ \hbox{for $\v m = even$}  \nonumber\\
 u^{3}_{\v m} =&  0,\ \ \ \hbox{for $\v m = odd$}  
\\
 G_x(\v i) =&  g_3(\th_x),  \ \ \ \
 G_y(\v i) = g_3(\th_y), 
\nonumber\\
 (-)^{i_x} G_{P_x}(\v i) =&  g_3(\th_{px}), \ \ \ \
 (-)^{i_y} G_{P_y}(\v i) = g_3(\th_{py}) ,
\nonumber \\
 G_{P_{xy}}(\v i) =&  g_3(\th_{pxy}),\ i\tau^{\th_{pxy}} \ \ \ \
 (-)^{\v i}G_T(\v i) =  i\tau^{\th_T}; \nonumber 
\end{align}
 U1A$x1[0,1]x$:
\begin{align}
\label{U1SU2gl6}
  u_{\v i,\v i+\v m} &=
 u^0_{\v m}\tau^0 
+   u^3_{\v m}\tau^3 
\nonumber\\
 u^{0}_{\v m} =&  0,\ \ \ \hbox{for $\v m = even$}  \nonumber\\
 u^{3}_{\v m} =&  0,\ \ \ \hbox{for $\v m = odd$}  \\
 G_x(\v i) =&  g_3(\th_x),  \ \ \ \
 G_y(\v i) = g_3(\th_y), 
\nonumber\\
 (-)^{i_x} G_{P_x}(\v i) =&  i\tau^{\th_{px}}, \ \ \ \
 (-)^{i_y} G_{P_y}(\v i) = i\tau^{\th_{py}} ,
\nonumber \\
 G_{P_{xy}}(\v i) =&  g_3(\th_{pxy}),\ i\tau^{\th_{pxy}} \ \ \ \
 (-)^{\v i}G_T(\v i) =  i\tau^{\th_T}; \nonumber 
\end{align}
The 12 different $U(1)$ spin liquids have different quantum orders.

We can use a gauge transformation $W_{\v i}=(i\tau^1)^{\v i}$ to make
ansatz \Eq{U1SU2gl1} to \Eq{U1SU2gl4} translation invariant. We get
U1C$[n0, x1][0,1]n$:
\begin{align}
\label{U1SU2gl1tr}
 u_{\v i,\v i+\v m} &=
 u^0_{\v m}(i\tau^1)^{\v m} 
+   u^3_{\v m} (i\tau^1)^{\v m}\tau^3 
\nonumber\\
 u^{0,3}_{\v m} =&  0,\ \ \ \hbox{for $\v m = even$}  \\
 G_x(\v i) =&  \tau^{0},  \ \ \ \
 G_y(\v i) = \tau^{0}, 
\nonumber\\
 G_{P_x}(\v i) =&  G_{P_y}(\v i) = \tau^0 ,\ i\tau^1
\nonumber \\
 G_{P_{xy}}(\v i) =&  \tau^0,\ i\tau^1 \ \ \ \
 G_T(\v i) = (-)^{\v i} \tau^0; \nonumber 
\end{align}
and U1C$[n0, x1][0,1]x$:
\begin{align}
\label{U1SU2gl2tr}
 u_{\v i,\v i+\v m} &=
u^0_{\v m}(i\tau^1)^{\v m} 
+   u^3_{\v m} (i\tau^1)^{\v m}\tau^3 
\nonumber\\
 u^{0}_{\v m} =&  0,\ \ \ \hbox{for $\v m = even$}  \nonumber\\
 u^{3}_{\v m} =&  0,\ \ \ \hbox{for $\v m = odd$}  \\
 G_x(\v i) =&  \tau^{0},  \ \ \ \
 G_y(\v i) = \tau^{0}, 
\nonumber\\
 G_{P_x}(\v i) =&  G_{P_y}(\v i) = \tau^0 ,\ i\tau^1
\nonumber \\
 G_{P_{xy}}(\v i) =&  \tau^0,\ i\tau^1 \ \ \ \
 G_T(\v i) =  i\tau^{2}; \nonumber 
\end{align}

Now we consider the symmetric perturbations that break the $SU(2)$
gauge structure down to a $Z_2$ gauge structure. 
The invariant gauge group becomes $\cG=Z_2$ for the perturbed ansatz.
We choose a fixed value for each $g_{x,y,px,py,pxy,T}$
in \Eq{PSGSU2gl} such that
\Eq{genGxGy}, \Eq{genGxyT}, \Eq{genGxyPx}, \Eq{genGxyPy}, \Eq{genGxyPxy}, 
\Eq{genTPxyxy}, and \Eq{genPxyPP} can be satisfied 
when we limit $\cG$ to a $Z_2$ subgroup. 
Since the original invariant gauge group is formed by constant gauge
transformations, its $Z_2$ subgroup is also formed by constant gauge
transformations, which 
is given by $\cG=\{ \pm\tau^0 \}$.

To obtain the consistent choices of
$g_{x,y,px,py,pxy,T}$, 
we note that $G_{x,y}$, $G_{P_x,P_y,P_{xy}}$ and $G_T$ must satisfy
\Eq{genGxGy}, \Eq{genGxyT}, \Eq{genGxyPx}, \Eq{genGxyPy}, \Eq{genGxyPxy}, 
\Eq{genTPxyxy}, and \Eq{genPxyPP}, with $\cG=\{ \pm\tau^0 \}$.
Those equations reduce to 
\begin{equation}
\label{Z2SU2glGxGy}
 g_x g_y g_x^{-1} g_y^{-1} = \pm\tau^0
\end{equation}
\begin{align}
\label{Z2SU2glGxyT}
 g_x^{-1} g_T^{-1}g_xg_T =& \pm\tau^0
&
 g_y^{-1} g_T^{-1} g_yg_T =& \pm\tau^0
\end{align}
\begin{align}
\label{Z2SU2glGxyPx}
 g_x g_{px}^{-1} g_x g_{px} =& \pm\tau^0
&
 g_y^{-1} g_{px}^{-1}g_y g_{px} =& \pm\tau^0
\end{align}
\begin{align}
\label{Z2SU2glGxyPy}
 g_y g_{py}^{-1} g_y g_{py} =& \pm\tau^0
&
 g_x^{-1} g_{py}^{-1} g_x g_{py} =& \pm\tau^0
\end{align}
\begin{align}
\label{Z2SU2glGxyPxy}
 g_y^{-1} g_{pxy}^{-1} g_x g_{pxy} =& \pm\tau^0
&
 g_x^{-1} g_{pxy}^{-1} g_y g_{pxy} =& \pm\tau^0
\end{align}
\begin{align}
\label{Z2SU2glTPxyxy}
g_T^{-1} g_{px}^{-1} g_T g_{px} =& \pm\tau^0
&
g_T^{-1} g_{py}^{-1} g_T g_{py} =& \pm\tau^0
\nonumber\\
g_T^{-1} g_{pxy}^{-1} g_T g_{pxy} =& \pm\tau^0
\end{align}
\begin{align}
\label{Z2SU2glPxyPP}
g_{pxy} g_{px}g_{pxy} g_{py}^{-1}
=& \pm\tau^0&
g_{py}g_{px} 
g_{py}^{-1}g_{px}^{-1}
=& \pm\tau^0
\end{align}
Since $P_x^2=P_y^2=P_{xy}^2=T^2=1$, we also have
\begin{align}
\label{Z2SU2glpxpyT}
 g_{px}^2 =& \pm\tau^0 & g_{py}^2 =& \pm\tau^0, 
\nonumber\\
 g_{pxy}^2 =& \pm\tau^0& g_T^2 =& \pm\tau^0.
\end{align}

From \Eq{Z2SU2glPxyPP}, we see that $g_{px}p_{py}= \pm g_{py}g_{px}$.
$g_{px,py}$ has the following 5 gauge inequivalent choices
\begin{align}
 g_{px} =& \tau^0,\ i\tau^3 & g_{py} =& \tau^0,\ i\tau^3  \\
 g_{px} =& i\tau^1 & g_{py} =& i\tau^2  
\end{align}
Also, according to \Eq{Z2SU2glPxyPP}, $g_{px}= \pm g_{pxy} g_{py}g_{pxy}$.
This requires $ g_{px} =  g_{py} $ if $ [g_{px},  g_{py}]=0 $.
Similarly, we also have $ g_{x} =  g_{y} $ if $ [g_{x},  g_{y}]=0 $.
Many solutions can be obtained by simply choosing each of
$g_{x,y,T}$ and $g_{px,py,pxy}$ to have one of the four values:
$(\tau^0,i\tau^{1,2,3})$. 
We find the following 65 solutions
\begin{align}
 G_x =&  G_y = \tau^0, & (-)^{\v i}G_T=&\tau^0,  \nonumber \\
 (-)^{i_x}G_{P_x}  =& (-)^{i_y}G_{P_y}=\tau^0,\ i\tau^3  &  
 G_{P_{xy}} =& \tau^0,\ i\tau^3 ;\label{1ggg}  
 \end{align}
 \begin{align}
 G_x =&  G_y = \tau^0, & (-)^{\v i}G_T=&\tau^0,  \nonumber \\
 (-)^{i_x}G_{P_x}  =& (-)^{i_y}G_{P_y}=i\tau^1  &  
 G_{P_{xy}} =& i\tau^2 ;\label{2ggg}  
 \end{align}
 \begin{align}
 G_x =&  G_y = \tau^0, & (-)^{\v i}G_T=&\tau^0,  \nonumber \\
 (-)^{i_x}G_{P_x}  =& i\tau^1, & (-)^{i_y}G_{P_y}=&i\tau^2  \nonumber\\
 G_{P_{xy}} =& i\tau^{12} ;\label{2aggg}  
\end{align}
\begin{align}
 G_x =&  G_y = \tau^0, & (-)^{\v i}G_T=&i\tau^3,  \nonumber \\
 (-)^{i_x}G_{P_x}  =& (-)^{i_y}G_{P_y}=\tau^0,\ i\tau^3  &  
 G_{P_{xy}} =& \tau^0,\ i\tau^3 ;\label{3ggg}  
 \end{align}
 \begin{align}
 G_x =&  G_y = \tau^0, & (-)^{\v i}G_T=&i\tau^3,  \nonumber \\
 (-)^{i_x}G_{P_x}  =& (-)^{i_y}G_{P_y}=\tau^0,\ i\tau^{1,2,3}  &  
 G_{P_{xy}} =& i\tau^1 ;\label{4ggg}
 \end{align}
 \begin{align}
 G_x =&  G_y = \tau^0, & (-)^{\v i}G_T=&i\tau^3,  \nonumber \\
 (-)^{i_x}G_{P_x}  =& (-)^{i_y}G_{P_y}=i\tau^1  &  
 G_{P_{xy}} =& \tau^0,\ i\tau^{1,2,3};\label{5ggg}
 \end{align}
 \begin{align}
 G_x =&  G_y = \tau^0, & (-)^{\v i}G_T=&\tau^3,  \nonumber \\
 (-)^{i_x}G_{P_x}  =& i\tau^1, & (-)^{i_y}G_{P_y}=&i\tau^2  \nonumber\\
 G_{P_{xy}} =& i\tau^{12} ;\label{5aggg}  
\end{align}
\begin{align}
 G_x =&  G_y = i\tau^3, & (-)^{\v i}G_T=&\tau^0,  \nonumber \\
 (-)^{i_x}G_{P_x}  =& (-)^{i_y}G_{P_y}=\tau^0,\ i\tau^3  &  
 G_{P_{xy}} =& \tau^0,\ i\tau^3 ;\label{6ggg}  
 \end{align}
 \begin{align}
 G_x =&  G_y = i\tau^3, & (-)^{\v i}G_T=&\tau^0,  \nonumber \\
 (-)^{i_x}G_{P_x}  =& (-)^{i_y}G_{P_y}=\tau^0, i\tau^{1,2,3}  &  
 G_{P_{xy}} =& i\tau^1 ;\label{7ggg}
 \end{align}
 \begin{align}
 G_x =&  G_y = i\tau^3, & (-)^{\v i}G_T=&\tau^0,  \nonumber \\
 (-)^{i_x}G_{P_x}  =& (-)^{i_y}G_{P_y}=i\tau^1  &  
 G_{P_{xy}} =& \tau^0,\ i\tau^{1,2,3};\label{8ggg}  
\end{align}
\begin{align}
 G_x =&  G_y = i\tau^3, & (-)^{\v i}G_T=&i\tau^3,  \nonumber \\
 (-)^{i_x}G_{P_x}  =& (-)^{i_y}G_{P_y}=\tau^0,\ i\tau^3  &  
 G_{P_{xy}} =& \tau^0,\ i\tau^3 ;\label{9ggg}  
 \end{align}
 \begin{align}
 G_x =&  G_y = i\tau^3, & (-)^{\v i}G_T=&i\tau^3,  \nonumber \\
 (-)^{i_x}G_{P_x}  =& (-)^{i_y}G_{P_y}=\tau^0,\ i\tau^{1,2,3}  &  
 G_{P_{xy}} =& i\tau^1 ;\label{10ggg}  
 \end{align}
 \begin{align}
 G_x =&  G_y = i\tau^3, & (-)^{\v i}G_T=&i\tau^3,  \nonumber \\
 (-)^{i_x}G_{P_x}  =& (-)^{i_y}G_{P_y}=i\tau^1  &  
 G_{P_{xy}} =& \tau^0,\ i\tau^{1,2,3};\label{11ggg}  
\end{align}
\begin{align}
 G_x =&  G_y = i\tau^3, & (-)^{\v i}G_T=&i\tau^1,  \nonumber \\
 (-)^{i_x}G_{P_x}  =& (-)^{i_y}G_{P_y}=\tau^0,\ i\tau^{1,2,3}  &  
 G_{P_{xy}} =& \tau^0,
i\tau^{1,2,3}  ;\label{12ggg}  
\end{align}
\begin{align}
 G_x =& i\tau^1, & G_y =& i\tau^2, \nonumber\\
 G_{P_{xy}} =& i\tau^{12} , & (-)^{\v i}G_T=&\tau^0, i\tau^3,  \nonumber \\
 (-)^{i_x}G_{P_x}  =& i\tau^2, & (-)^{i_y}G_{P_y}=&i\tau^1  ;\label{13ggg}  
\end{align}
\begin{align}
 G_x =& i\tau^1, & G_y =& i\tau^2, \nonumber\\
 G_{P_{xy}} =& i\tau^{12} , & (-)^{\v i}G_T=&\tau^0, i\tau^3,  \nonumber \\
 (-)^{i_x}G_{P_x}  =& i\tau^1, & (-)^{i_y}G_{P_y}=&i\tau^2  ;\label{14ggg}  
\end{align}
\begin{align}
 G_x =& i\tau^1, & G_y =& i\tau^2, \nonumber\\
 G_{P_{xy}} =& i\tau^{12} , & (-)^{\v i}G_T=&\tau^0,\ i\tau^3,  \nonumber \\
 (-)^{i_x}G_{P_x}  =& \tau^0, & (-)^{i_y}G_{P_y}=&\tau^0  ;\label{16ggg}  
\end{align}

Using a gauge transformation $W_{\v i}=(i\tau^3)^{\v i}(\pm)^{i_x}$,
we can change \Eq{6ggg} -- \Eq{12ggg} to standard $Z_2$ form
\Eq{GZ2t1} or \Eq{GZ2t2}. We choose the $\pm$ to remove the $(-)^{\v i}$
factor in $G_{P_{xy}}$. We get
\begin{align}
 G_x =&  G_y = \tau^0, & (-)^{\v i}G_T=&\tau^0,  \nonumber \\
 G_{P_x}  =& G_{P_y}=\tau^0,\ i\tau^3  &  
 G_{P_{xy}} =& \tau^0,\ i\tau^3 ;\label{6gggt}  
 \end{align}
 \begin{align}
 G_x =&  G_y = \tau^0, & (-)^{\v i}G_T=&\tau^0,  \nonumber \\
 G_{P_x}  =& G_{P_y}=\tau^{0,3}, (-)^{\v i}i\tau^{1,2}  &  
 G_{P_{xy}} =& i\tau^1 ;\label{7gggt}
 \end{align}
 \begin{align}
 G_x =&  G_y = \tau^0, & (-)^{\v i}G_T=&\tau^0,  \nonumber \\
 G_{P_x}  =& G_{P_y}=(-)^{\v i} i\tau^1  &  
 G_{P_{xy}} =& \tau^{0},\ i\tau^{1,2,3};\label{8gggt}  
\end{align}
\begin{align}
 G_x =&  G_y = \tau^0, & (-)^{\v i}G_T=&i\tau^3,  \nonumber \\
 G_{P_x}  =& G_{P_y}=\tau^0,\ i\tau^3  &  
 G_{P_{xy}} =& \tau^0,\ i\tau^3 ;\label{9gggt}  
 \end{align}
 \begin{align}
 G_x =&  G_y = \tau^0, & (-)^{\v i}G_T=&i\tau^3,  \nonumber \\
 G_{P_x}  =& G_{P_y}=\tau^{0,3},\ (-)^{\v i}i\tau^{1,2}  &  
 G_{P_{xy}} =& i\tau^1 ;\label{10gggt}  
 \end{align}
 \begin{align}
 G_x =&  G_y = \tau^0, & (-)^{\v i}G_T=&i\tau^3,  \nonumber \\
 G_{P_x}  =& G_{P_y}=(-)^{\v i}i\tau^1  &  
 G_{P_{xy}} =& \tau^{0},\ i\tau^{1,2,3};\label{11gggt}  
\end{align}
\begin{align}
 G_x =&  G_y = \tau^0, & G_T=&i\tau^1,  \nonumber \\
 G_{P_x}  =& G_{P_y}=\tau^{0,3},(-)^{\v i}i\tau^{1,2}  &  
 G_{P_{xy}} =& \tau^{0},\ i\tau^{1,2,3}  ;\label{12gggt}  
\end{align}

Using a gauge transformation $W_{\v i}=(i\tau^1)^{i_x}(i\tau^2)^{i_y}$,
we can change \Eq{13ggg} -- \Eq{16ggg} to more standard forms (see
\Eq{GZ2t1} and \Eq{GZ2t2}):
\begin{align}
 G_x =& (-)^{i_y} \tau^0, & G_y =& \tau^0, \nonumber\\
 G_{P_{xy}} =& (-)^{i_xi_y} i\tau^{12} , & 
 G_T=&(-)^{\v i}\tau^0,\ i\tau^3,  \nonumber \\
 (-)^{i_x}G_{P_x}  =& i\tau^2, & (-)^{i_y}G_{P_y}=&i\tau^1  ;\label{13gggt}  
\end{align}
\begin{align}
 G_x =& (-)^{i_y} \tau^0, & G_y =& \tau^0, \nonumber\\
 G_{P_{xy}} =& (-)^{i_xi_y} i\tau^{12} , & 
 G_T=&(-)^{\v i} \tau^0,\ i\tau^3,  \nonumber \\
 (-)^{i_y}G_{P_x}  =& i\tau^1, & (-)^{i_x}G_{P_y}=&i\tau^2  ;\label{14gggt}  
\end{align}
\begin{align}
 G_x =& (-)^{i_y}\tau^0, & G_y =& \tau^0, \nonumber\\
 G_{P_{xy}} =& (-)^{i_xi_y} i\tau^{12} , & 
 G_T=&(-)^{\v i} \tau^0,\ i\tau^3,  \nonumber \\
 G_{P_x}  =& \tau^0, & G_{P_y}=&\tau^0  ;\label{16gggt}  
\end{align}

Now we can list all PSG's that describe the $Z_2$ spin liquids in the
neighborhood of the $SU(2)$-gapless state using the notation
Z2A$...$ or Z2B$...$.
We find that the 65 PSG's obtained before lead to 58 gauge inequivalent
PSG's.

In the following, we will list all the 58 PSG's. We will also 
construct ansatz for those PSG's.
First let us consider PSG of form Z2A$...$.
For those PSG the ansatz can be written as
\begin{equation}
 u_{\v i,\v i+\v m} = u_{\v m}
\end{equation}
In the following we consider the constraint imposed by the $180^\circ$
rotation symmetry and the time reversal symmetry.

For PSG
\begin{align}
&Z2A[\tau^0_-\tau^0_+, \tau^3_-\tau^3_+][\tau^0,\tau^3]\tau^0_- 
\nonumber\\
&Z2A\tau^3_-\tau^3_+\tau^1\tau^0_- 
\nonumber\\
\end{align}
(here we have used the notation $[a,b][c,d]$ to represent four combinations
$ac,ad,bc,bd$),
the $180^\circ$ rotation symmetry generated by $P_xP_y$ requires that
\begin{align*}
 u_{-\v m} =&  u_{\v m} (-)^{\v m}  = u_{\v m}^\dag   
\end{align*}
The time reversal symmetry $T$ requires that
\begin{align*}
- u_{\v m} =& u_{\v m} (-)^{\v m}  
\end{align*}
The above two equations require that $u_{\v i\v j}\propto \tau^0$, which
describe $SU(2)$ spin liquids.

For PSG
\begin{align}
&Z2A\tau^3_-\tau^3_- \tau^{0,1,2,3}\tau^0_- 
\nonumber\\
&Z2A[\tau^0_+\tau^0_+,\tau^3_+\tau^3_+] \tau^{0,1,3}\tau^0_- 
\end{align}
the $180^\circ$ rotation symmetry 
and the time reversal symmetry require that
\begin{align*}
 u_{-\v m} =&  u_{\v m}  = u_{\v m}^\dag   
\nonumber\\
- u_{\v m} =& u_{\v m} (-)^{\v m}  
\end{align*}
The above two equations give us
\begin{align}
\label{Z2SU2gl1}
 u_{\v i,\v i+\v m} =&  
 u^1_{\v m}\tau^1 +
 u^2_{\v m}\tau^2 + u^3_{\v m}\tau^3 
\nonumber\\
 u^{1,2,3}_{\v m} =&  0,\ \ \ \hbox{for $\v m = even$}  
\end{align}

For PSG 
\begin{equation}
Z2A\tau^1_-\tau^2_+\tau^{12} \tau^0_- ,
\end{equation}
the $180^\circ$ rotation and
the time reversal symmetries require that
\begin{align*}
 u_{-\v m} =&  \tau^3 u_{\v m}\tau^3 (-)^{\v m}   = u_{\v m}^\dag   
\nonumber\\
- u_{\v m} =& u_{\v m} (-)^{\v m}  
\end{align*}
We find
\begin{align}
\label{Z2SU2gl2}
 u_{\v i,\v i+\v m} =&  
 u^0_{\v m}\tau^0 +
 u^1_{\v m}\tau^1 +
 u^2_{\v m}\tau^2   
\nonumber\\
 u^{0,1,2}_{\v m} =&  0,\ \ \ \hbox{for $\v m = even$}  
\end{align}

For PSG
\begin{align}
&Z2A[\tau^0_-\tau^0_+, \tau^3_-\tau^3_+]\tau^{0,1,3}\tau^3_- 
\nonumber\\
&Z2A\tau^1_-\tau^1_+\tau^{0,1,2,3}\tau^3_- 
\end{align}
the $180^\circ$ rotation and the time reversal symmetries require that
\begin{align*}
 u_{-\v m} =&  u_{\v m} (-)^{\v m}  = u_{\v m}^\dag   
\nonumber\\
- u_{\v m} =& \tau^3 u_{\v m}\tau^3  (-)^{\v m}  
\end{align*}
The ansatz has a form
\begin{align}
\label{Z2SU2gl3}
 u_{\v i,\v i+\v m} =&  
 u^0_{\v m}\tau^0 +
 u^1_{\v m}\tau^1 +
 u^2_{\v m}\tau^2  
\nonumber\\
 u^{0}_{\v m} =&  0,\ \ \ \hbox{for $\v m = even$}  
\nonumber\\
 u^{1,2}_{\v m} =&  0,\ \ \ \hbox{for $\v m = odd$}  
\end{align}

For PSG's
\begin{align}
&Z2A\tau^3_-\tau^3_- \tau^3\tau^3_- 
\nonumber\\
&Z2A\tau^1_-\tau^1_- \tau^{0,1,2,3}\tau^3_- 
\nonumber\\
&Z2A[\tau^0_+\tau^0_+,\tau^3_+\tau^3_+]\tau^{0,1,3}\tau^3_- 
\end{align}
the $180^\circ$ rotation and the time reversal symmetries require that
\begin{align*}
 u_{-\v m} =&  u_{\v m} = u_{\v m}^\dag   
\nonumber\\
- u_{\v m} =& \tau^3 u_{\v m}\tau^3  (-)^{\v m}  
\end{align*}
The ansatz has a form
\begin{align}
\label{Z2SU2gl4}
 u_{\v i,\v i+\v m} =&  
 u^1_{\v m}\tau^1 +
 u^2_{\v m}\tau^2 +
 u^3_{\v m}\tau^3  
\nonumber\\
 u^{3}_{\v m} =&  0,\ \ \ \hbox{for $\v m = even$}  
\nonumber\\
 u^{1,2}_{\v m} =&  0,\ \ \ \hbox{for $\v m = odd$}  
\end{align}

For PSG
Z2A$\tau^1_-\tau^2_+\tau^{12}\tau^3_-$,
the $180^\circ$ rotation and
the time reversal symmetries require that
\begin{align*}
 u_{-\v m} =&  \tau^3 u_{\v m}\tau^3 (-)^{\v m}   = u_{\v m}^\dag   
\nonumber\\
- u_{\v m} =& \tau^3 u_{\v m}\tau^3  (-)^{\v m}  
\end{align*}
We find $u_{\v i\v j} \propto \tau^0$. The spin liquid constructed 
from $u_{\v i\v j}$ is a $SU(2)$ spin liquid.

For PSG's
\begin{align}
&Z2A\tau^3_-\tau^3_- \tau^{0,1,3}\tau^3_+ 
\nonumber\\
&Z2A\tau^1_-\tau^1_- \tau^{0,1,2,3}\tau^3_+ 
\nonumber\\
&Z2A\tau^0_+\tau^0_+ \tau^{0,1,3}\tau^3_+ 
\nonumber\\
&Z2A\tau^1_+\tau^1_+ \tau^{0,1,2,3}\tau^3_+ 
\end{align}
the $180^\circ$ rotation and the time reversal symmetries require that
\begin{align*}
 u_{-\v m} =&  u_{\v m} = u_{\v m}^\dag   
\nonumber\\
- u_{\v m} =& \tau^3 u_{\v m}\tau^3  
\end{align*}
The ansatz has a form
\begin{align}
\label{Z2SU2gl4a}
 u_{\v i,\v i+\v m} =&  u^1_{\v m}\tau^1 + u^1_{\v m}\tau^2 
\end{align}

There are six PSG's of form Z2B$...$.
The first two are
\begin{align}
Z2B[\tau^1_-\tau^2_+,\tau^1_+\tau^2_- ] \tau^{12}\tau^0_-.
\end{align}
The $180^\circ$ rotation symmetry requires that
\begin{align*}
(-)^{i_xm_y}u_{\v m}=(-)^{\v m}(-)^{(i_x+m_x)m_y}\tau^3 u_{\v m}^\dag\tau^3    
\end{align*}
or
\begin{align*}
 u^{0,1,2}_{\v m} =&  0,\ \ \ \hbox{for $m_x=even$ and $m_y=even$} \nonumber\\
 u^{3}_{\v m} =&  0,\ \ \ \hbox{for $m_x=odd$ or $m_y=odd$} 
\end{align*}
The time reversal symmetry requires that
\begin{align*}
- u_{\v m} =& u_{\v m} (-)^{\v m}  
\end{align*}
We find
\begin{align}
\label{Z2SU2gl5}
 u_{\v i,\v i+\v m} =&  
(-)^{m_yi_x} (
 u^0_{\v m}\tau^0 +
 u^1_{\v m}\tau^1 +
 u^2_{\v m}\tau^2 )  
\nonumber\\
 u^{0,1,2}_{\v m} =&  0,\ \ \ \hbox{for $\v m = even$}  
\end{align}

For PSG
\begin{align}
Z2B\tau^0_+\tau^0_+\tau^3\tau^0_-
\end{align}
the $180^\circ$ rotation symmetry requires that
\begin{align*}
(-)^{i_xm_y}u_{\v m}=(-)^{(i_x+m_x)m_y}u_{\v m}^\dag   
\end{align*}
or
\begin{align*}
 u^0_{\v m} =&  0,\ \ \ \hbox{for $m_x=even$ or $m_y=even$} \nonumber\\
 u^{1,2,3}_{\v m} =&  0,\ \ \ \hbox{for $m_x=odd$ and $m_y=odd$}
\end{align*}
The time reversal symmetry requires that
\begin{align*}
- u_{\v m} =& u_{\v m} (-)^{\v m}  
\end{align*}
The above two equations give us
\begin{align}
\label{Z2SU2gl6}
 u_{\v i,\v i+\v m} =&  
(-)^{i_xm_y}( u^1_{\v m}\tau^1 +
 u^2_{\v m}\tau^2 + u^3_{\v m}\tau^3 )
\nonumber\\
 u^{1,2,3}_{\v m} =&  0,\ \ \ \hbox{for $\v m = even$}  
\end{align}

For PSG 
\begin{align}
Z2B[\tau^1_-\tau^2_+,\tau^1_+\tau^2_- ] \tau^{12}\tau^3_+
\end{align}
the $180^\circ$ rotation symmetry requires that
\begin{align*}
(-)^{i_xm_y}u_{\v m}=(-)^{\v m}(-)^{(i_x+m_x)m_y}\tau^3 u_{\v m}^\dag\tau^3    
\end{align*}
or
\begin{align*}
 u^{0,1,2}_{\v m} =&  0,\ \ \ \hbox{for $m_x=even$ and $m_y=even$} \nonumber\\
 u^{3}_{\v m} =&  0,\ \ \ \hbox{for $m_x=odd$ or $m_y=odd$} 
\end{align*}
The time reversal symmetry requires that
\begin{align*}
- u_{\v m} =& \tau^3 u_{\v m}\tau^3
\end{align*}
\begin{align}
\label{Z2SU2gl7}
 u_{\v i,\v i+\v m} =&  
(-)^{i_xm_y}( 
 u^1_{\v m}\tau^1 +
 u^2_{\v m}\tau^2 )
\\
 u^{1,2}_{\v m} =&  0,\ \ \ \hbox{for $m_x=even$ and $m_y=even$} \nonumber
\end{align}

For PSG
\begin{align}
Z2B\tau^0_+\tau^0_+\tau^3\tau^3_+
\end{align}
the $180^\circ$ rotation symmetry requires that
\begin{align*}
(-)^{i_xm_y}u_{\v m}=(-)^{(i_x+m_x)m_y}u_{\v m}^\dag   
\end{align*}
or
\begin{align*}
 u^0_{\v m} =&  0,\ \ \ \hbox{for $m_x=even$ or $m_y=even$} \nonumber\\
 u^{1,2,3}_{\v m} =&  0,\ \ \ \hbox{for $m_x=odd$ and $m_y=odd$}
\end{align*}
The time reversal symmetry requires that
\begin{align*}
- u_{\v m} =& \tau^3 u_{\v m}\tau^3  
\end{align*}
The ansatz has a form
\begin{align}
\label{Z2SU2gl8}
 u_{\v i,\v i+\v m} =&  (-)^{i_xm_y}(u^1_{\v m}\tau^1 + u^1_{\v m}\tau^2 )
 \\
 u^{1,2}_{\v m} =&  0,\ \ \ \hbox{for $m_x=odd$ and $m_y=odd$} \nonumber 
\end{align}

We find that there are 52 different $Z_2$ spin liquids in the neighborhood
of the $SU(2)$-gapless state \Eq{SU2gl}.  Those $Z_2$ spin liquids can be
constructed through $u_{\v i\v j}$.

\subsection{ Symmetric perturbations around the $SU(2)$-linear 
state SU2B$n0$}

In this subsection,
we would like to consider the symmetric perturbations around
the $SU(2)$-linear ansatz \Eq{SU2lA}, 
which describes a SU2B$n0$ spin liquid.
The invariant gauge group is $\cG=SU(2)$.
The PSG of the ansatz is generated by
\begin{align}
\label{PSGSU2lin}
 G_x(\v i) =&  (-)^{i_y} g_x, & G_y(\v i) =&  g_y  \nonumber \\
 G_{P_x}(\v i) =&  (-)^{i_x} g_{px}  & G_{P_y}(\v i) =&  (-)^{i_y} g_{py}
\nonumber\\
 G_{P_{xy}}(\v i) =&  (-)^{i_xi_y} g_{pxy} & G_T(\v i) =&  (-)^{\v i}g_T 
\end{align}
where $g_{x,y,px,py,pxy,T} \in SU(2)$.

First let us consider the symmetric perturbations that do not break the $SU(2)$
gauge structure. To have the $SU(2)$ gauge structure and the symmetries, 
the perturbations must be invariant under PSG
in \Eq{PSGSU2lin}. The translation symmetry require the perturbations
to have a form
\begin{equation}
 \del u_{\v i,\v i+\v m} = (-)^{i_xm_y} \del u^0_{\v m}\tau^0 
\end{equation}
The $180^\circ$ rotation symmetry $P_xP_y$ and the time reversal symmetry $T$
require that
\begin{align*}
(-)^{(i_x+m_x)m_y} 
(\del u^0_{\v m})^\dag =& (-)^{i_xm_y}\del u^0_{\v m} (-)^{\v m}
\nonumber\\
 -\del u^0_{\v m} =& \del u^0_{\v m} (-)^{\v m} 
\end{align*}
Thus the symmetric ansatz with PSG \Eq{PSGSU2lin} are given by
\begin{align}
\label{SU2lin}
 u_{\v i, \v i +\v m}  =&   (-)^{i_xm_y} u^0_{\v m}\tau^0 
\\
 u^{0}_{\v m} =&  0,\ \ \ \hbox{for $\v m = even$}  \nonumber
\end{align}
The above represent most general ansatz around the $SU(2)$-linear state
that do not break any symmetries and do not change the quantum order in the
state.

To obtain other symmetric perturbations around the $SU(2)$-linear state
SU2B$n 0$, we can use the mapping \Eq{newPSG} and \Eq{utoup}.  We
first note that the PSG's that describe the spin liquids around the
$SU(2)$-linear state can be obtained from those around the $SU(2)$-gapless
state SU2A$n 0$. This is because mapping described by \Eq{newPSG} maps
the SU2A$n0$ PSG \Eq{PSGSU2gl} to the SU2B$n0$ PSG
\Eq{PSGSU2lin}.  Using this results, we can obtain all the PSG's that
describe the spin liquids in the neighborhood of the $SU(2)$-linear state.
The SU2A$n 0$ PSG for the $SU(2)$-gapless state have 16 different
subgroups with IGG=$U(1)$ (see \Eq{U1SU2glPSG1} - \Eq{U1SU2glPSG4}) and 58
subgroups with IGG=$Z_2$ (see \Eq{1ggg} - \Eq{16gggt}).  Therefore,
the SU2B$n 0$ PSG of the $SU(2)$-linear state also have 16 different
subgroups with IGG=$U(1)$  and 58 subgroups with IGG=$Z_2$ (see \Eq{1ggg} -
\Eq{16gggt}).

The 16 subgroups with IGG=$U(1)$ can be obtained from \Eq{U1SU2glPSG1} -
\Eq{U1SU2glPSG4} through the mapping \Eq{newPSG}.
They are
U1B$\tau^0_-\tau^0_+[\tau^0,\tau^1][\tau^0_-,\tau^1_-]$ 
( which is gauge equivalent to 
U1B$\tau^0_+\tau^0_+[\tau^0,\tau^1][\tau^0_-,\tau^1_+]$): 
\begin{align}
\label{U1SU2lPSG1}
& G_x(\v i)  = (-)^{i_y} g_3(\th_x), &&G_y(\v i) = g_3(\th_y)\\
& (-)^{i_x}G_{P_x}(\v i) =  g_3(\th_{px})
&& (-)^{\v i} G_T(\v i) = g_3(\th_T),\ i\tau^{\th_T}  
\nonumber\\
& (-)^{i_y} G_{P_y}(\v i) = g_3(\th_{py})
&& (-)^{i_xi_y}G_{P_{xy}}(\v i) =  g_3(\th_{pxy}),\ i\tau^{\th_{pxy}}
\nonumber 
\end{align}
U1B$\tau^1_-\tau^1_+[\tau^0,\tau^1][\tau^0_-,\tau^1_-]$: 
\begin{align}
\label{U1SU2lPSG2}
& G_x(\v i)  = (-)^{i_y} g_3(\th_x), && G_y(\v i) = g_3(\th_y) \\
& (-)^{i_x}G_{P_x}(\v i) =   i\tau^{\th_{px}}
&& (-)^{\v i} G_T(\v i) = g_3(\th_T),\ i\tau^{\th_T}  
\nonumber\\
& (-)^{i_y} G_{P_y}(\v i) =  i\tau^{\th_{py}}
&& (-)^{i_xi_y}G_{P_{xy}}(\v i) =  g_3(\th_{pxy}),\ i\tau^{\th_{pxy}}
\nonumber 
\end{align}
U1C$\tau^0_-\tau^0_+\tau^0_-[\tau^0_-,\tau^1_+]$,
U1C$\tau^0_-\tau^0_+\tau^1_-\tau^1_+$ and
U1C$\tau^0_-\tau^0_+\tau^1\tau^0_-$:
\begin{align}
\label{U1SU2lPSG3}
& G_x(\v i)  = i(-)^{i_y} \tau^{\th_x}, &&G_y(\v i) = i\tau^{\th_y} \\
& (-)^{i_x}G_{P_x}(\v i) =  g_3(\th_{px})
&& (-)^{\v i} G_T(\v i) = g_3(\th_T),\ i\tau^{\th_T}  
\nonumber\\
& (-)^{i_y} G_{P_y}(\v i) = g_3(\th_{py})
&& (-)^{i_xi_y}G_{P_{xy}}(\v i) =  g_3(\th_{pxy}),\ i\tau^{\th_{pxy}}
\nonumber 
\end{align}
U1C$\tau^1_+\tau^1_+[\tau^0_-,\tau^1_-][\tau^0_-,\tau^1_-]$:
\begin{align}
\label{U1SU2lPSG4}
& G_x(\v i)  = i(-)^{i_y} \tau^{\th_x}, && G_y(\v i) = i\tau^{\th_y}\\
& (-)^{i_x}G_{P_x}(\v i) =   i\tau^{\th_{px}}
&& (-)^{\v i} G_T(\v i) = g_3(\th_T),\ i\tau^{\th_T}  
\nonumber\\
& (-)^{i_y} G_{P_y}(\v i) =  i\tau^{\th_{py}}
&& (-)^{i_xi_y}G_{P_{xy}}(\v i) =  g_3(\th_{pxy}),\ i\tau^{\th_{pxy}}
\nonumber 
\end{align}
The labels for the last two equations are obtained by making a gauge
transformation to put $G_{x,y}$ in a more standard form (see
\Eq{U1SU2lPSG3a} and \Eq{U1SU2lPSG4a}).

After obtaining the PSG's, we can construct the ansatz which are invariant
under those PSG's. We note that for the above PSG's
the time reversal symmetry $T$ requires that
\begin{align*}
-  u_{\v i,\v i+\v m} =& g_T  u_{\v i,\v i+\v m}g_T^{-1} (-)^{\v m}  
\nonumber\\
g_T =& \tau^0,\ i\tau^1
\end{align*}
and the $180^\circ$ rotation symmetry $P_xP_y$ requires that
\begin{align}
\label{u180}
 u_{-\v i,-\v i-\v m} =& u_{\v i,\v i+\v m} (-)^{\v m}  
= u_{-\v i-\v m,-\v i}^\dag   
\end{align}
When $G_x(\v i) = i(-)^{i_y}\tau^{\th_x}$, $G_y(\v i) = i\tau^{\th_y}$, 
$u_{\v i\v j}$ has a form 
$u_{\v i,\v i+\v m} =(-)^{i_xm_y}( u^0_{\v m}\tau^0
+(-)^{\v i}u^3_{\v m}\tau^3 )$.
\Eq{u180} reduces to
\begin{align*}
u^0_{\v m} =& 0,\ \ \ \hbox{if $m_x=even$ and $m_y = even$} \nonumber\\ 
u^3_{\v m} =& 0,\ \ \ \hbox{if $m_x= odd$ and $m_y = odd$}  
\end{align*}
When $G_x(\v i) = (-)^{i_y}g_3(\th_x)$, $G_y(\v i) = g_3(\th_y)$, 
$u_{\v i\v j}$ has a form 
$u_{\v i,\v i+\v m} =(-)^{i_xm_y}( u^0_{\v m}\tau^0
+u^3_{\v m}\tau^3 )$.
\Eq{u180} reduces to
\begin{align*}
u^0_{\v m} =& 0,\ \ \ \hbox{if $m_x=even$ and $m_y = even$} \nonumber\\ 
u^3_{\v m} =& 0,\ \ \ \hbox{if $m_x= odd$ or $m_y = odd$}  
\end{align*}
Using the above results,
we find that all the symmetric spin liquids around the $SU(2)$-linear
state \Eq{SU2lA} that break the $SU(2)$ gauge structure to a $U(1)$ gauge
structure can be divided into 12 classes. They are given by
U1C$n0[n, 1]n$:
\begin{align}
\label{U1SU2lin1}
 u_{\v i,\v i+\v m} &=
(-)^{i_xm_y}( u^0_{\v m}\tau^0 
+ (-)^{\v i} u^3_{\v m}\tau^3 )
\\
u^{0,3}_{\v m} =& 0,\ \ \ \hbox{if $\v m= even$} \nonumber\\ 
 G_x(\v i) =&  i(-)^{i_y}\tau^{\th_x},  \ \ \ \ G_y(\v i) = i\tau^{\th_y}, 
\nonumber\\
 (-)^{i_x} G_{P_x}(\v i) =&  g_3(\th_{px}), \ \ \ \
 (-)^{i_y} G_{P_y}(\v i) = g_3(\th_{py}) ,
\nonumber \\
(-)^{i_xi_y} G_{P_{xy}}(\v i) =&  g_3(\th_{pxy}),\ i\tau^{\th_{pxy}} \ \ \ \
 (-)^{\v i}G_T(\v i) =  g_3(\th_T); \nonumber 
\end{align}
U1C$n0[n,x]1$:
\begin{align}
\label{U1SU2lin2}
 u_{\v i,\v i+\v m} &=
(-)^{i_xm_y}( u^0_{\v m}\tau^0 
+ (-)^{\v i} u^3_{\v m}\tau^3 )
\\
u^0_{\v m} =& 0,\ \ \ \hbox{if $\v m= even$} \nonumber\\ 
u^3_{\v m} =& 0,\ \ \ \hbox{if $m_x=odd$ or $m_y = odd$}  \nonumber\\
 G_x(\v i) =&  i(-)^{i_y}\tau^{\th_x},  \ \ \ \ G_y(\v i) = i\tau^{\th_y}, 
\nonumber\\
 (-)^{i_x} G_{P_x}(\v i) =&  g_3(\th_{px}), \ \ \ \
 (-)^{i_y} G_{P_y}(\v i) = g_3(\th_{py}) ,
\nonumber \\
(-)^{i_xi_y} G_{P_{xy}}(\v i) =&  g_3(\th_{pxy}),\ i\tau^{\th_{pxy}} \ \ \ \
 (-)^{\v i}G_T(\v i) =  i\tau^{\th_T}; \nonumber 
\end{align}
U1C$11[n,x]n$:
\begin{align}
\label{U1SU2lin3}
  u_{\v i,\v i+\v m} &=
(-)^{i_xm_y}( u^0_{\v m}\tau^0 
+ (-)^{\v i} u^3_{\v m}\tau^3 )
\\
u^{0,3}_{\v m} =& 0,\ \ \ \hbox{if $\v m= even$} \nonumber\\ 
 G_x(\v i) =&  i(-)^{i_y}\tau^{\th_x},  \ \ \ \
 G_y(\v i) = i\tau^{\th_y}, 
\nonumber\\
 (-)^{i_x} G_{P_x}(\v i) =&  i\tau^{\th_{px}}, \ \ \ \
 (-)^{i_y} G_{P_y}(\v i) = i\tau^{\th_{py}} ,
\nonumber \\
(-)^{i_xi_y} G_{P_{xy}}(\v i) =&  g_3(\th_{pxy}),\ i\tau^{\th_{pxy}} \ \ \ \
 (-)^{\v i}G_T(\v i) =  g_3(\th_T); \nonumber 
\end{align}
U1C$11[n,x]x$:
\begin{align}
\label{U1SU2lin4}
u_{\v i,\v i+\v m} &=
(-)^{i_xm_y}( u^0_{\v m}\tau^0 
+ (-)^{\v i} u^3_{\v m}\tau^3 )
\\
u^0_{\v m} =& 0,\ \ \ \hbox{if $\v m= even$} \nonumber\\ 
u^3_{\v m} =& 0,\ \ \ \hbox{if $m_x=odd$ or $m_y = odd$}  \nonumber\\
 G_x(\v i) =&  i(-)^{i_y}\tau^{\th_x},  \ \ \ \
 G_y(\v i) = i\tau^{\th_y}, 
\nonumber\\
 (-)^{i_x} G_{P_x}(\v i) =&  i\tau^{\th_{px}}, \ \ \ \
 (-)^{i_y} G_{P_y}(\v i) = i\tau^{\th_{py}} ,
\nonumber \\
(-)^{i_xi_y} G_{P_{xy}}(\v i) =&  g_3(\th_{pxy}),\ i\tau^{\th_{pxy}} \ \ \ \
 (-)^{\v i}G_T(\v i) =  i\tau^{\th_T}; \nonumber 
\end{align}
U1B$00[0,1]1$:
\begin{align}
\label{U1SU2lin5}
  u_{\v i,\v i+\v m} &=
(-)^{i_xm_y}( u^0_{\v m}\tau^0 
+ u^3_{\v m}\tau^3 )
\\
u^0_{\v m} =& 0,\ \ \ \hbox{if $\v m= even$} \nonumber\\ 
u^3_{\v m} =& 0,\ \ \ \hbox{if $m_x=odd$ or $m_y = odd$}  \nonumber\\
 G_x(\v i) =&  (-)^{i_y}g_3(\th_x),  \ \ \ \
 G_y(\v i) = g_3(\th_y), 
\nonumber\\
 (-)^{i_x} G_{P_x}(\v i) =&  g_3(\th_{px}), \ \ \ \
 (-)^{i_y} G_{P_y}(\v i) = g_3(\th_{py}) ,
\nonumber \\
(-)^{i_xi_y} G_{P_{xy}}(\v i) =&  g_3(\th_{pxy}),\ i\tau^{\th_{pxy}} \ \ \ \
 (-)^{\v i}G_T(\v i) =  i\tau^{\th_T}; \nonumber 
\end{align}
U1B$x1[0,1]x$:
\begin{align}
\label{U1SU2lin6}
u_{\v i,\v i+\v m} &=
(-)^{i_xm_y}( u^0_{\v m}\tau^0 
+  u^3_{\v m}\tau^3 )
\\
u^0_{\v m} =& 0,\ \ \ \hbox{if $\v m= even$} \nonumber\\ 
u^3_{\v m} =& 0,\ \ \ \hbox{if $m_x=odd$ or $m_y = odd$}  \nonumber\\
 G_x(\v i) =&  (-)^{i_y}g_3(\th_x),  \ \ \ \
 G_y(\v i) = g_3(\th_y), 
\nonumber\\
 (-)^{i_x} G_{P_x}(\v i) =&  i\tau^{\th_{px}}, \ \ \ \
 (-)^{i_y} G_{P_y}(\v i) = i\tau^{\th_{py}} ,
\nonumber \\
(-)^{i_xi_y} G_{P_{xy}}(\v i) =&  g_3(\th_{pxy}),\ i\tau^{\th_{pxy}} \ \ \ \
 (-)^{\v i}G_T(\v i) =  i\tau^{\th_T}; \nonumber 
\end{align}

Using the gauge transformation $W_{\v i} = g_3((-)^{i_y}\pi/4)$ we can change
\Eq{U1SU2lPSG3} and
\Eq{U1SU2lPSG4} to
\begin{align}
\label{U1SU2lPSG3a}
 G_x  =& i \tau^{\th_x}, \ \ \ G_y = i\tau^{\th_y}, \nonumber \\
G_{P_x} =&  (-)^{i_x} g_3(\th_{px}),
\ \ \  
G_{P_y} = (-)^{i_y} g_3(\th_{py}),
\nonumber\\
 G_{P_{xy}} =& 
(-)^{i_xi_y}g_3(((-)^{i_y}-(-)^{i_x})\frac{\pi}{4} +\th_{pxy}),
\nonumber\\
&
(-)^{i_xi_y}g_3(((-)^{i_y}+(-)^{i_x})\frac{\pi}{4} +\th_{pxy})i\tau^1,
\nonumber\\
  G_T =& (-)^{\v i}g_3(\th_T),\ (-)^{i_x}g_3(\th_T)i\tau^1  
\end{align}
\begin{align}
\label{U1SU2lPSG4a}
 G_x  = &i \tau^{\th_x}, \ \ \ G_y = i\tau^{\th_y},\nonumber \\
G_{P_x} =&   (-)^{\v i} i\tau^{\th_{px}},
\ \ \  G_{P_y} =  i\tau^{\th_{py}}
\nonumber\\
 G_{P_{xy}} =&   
(-)^{i_xi_y}g_3(((-)^{i_y}-(-)^{i_x})\frac{\pi}{4} +\th_{pxy}),
\nonumber\\
&
(-)^{i_xi_y}g_3(((-)^{i_y}+(-)^{i_x})\frac{\pi}{4} +\th_{pxy})i\tau^1,
\nonumber\\
  G_T =& (-)^{\v i}g_3(\th_T),\ (-)^{i_x}g_3(\th_T)i\tau^1  
\end{align}
We note that
\begin{align}
&
(-)^{i_xi_y}g_3(((-)^{i_y}-(-)^{i_x})\frac{\pi}{4} +\th_{pxy})
\nonumber\\
=& (-)^{i_x}g_3((-)^{\v i}\frac{\pi}{4} + \th_{pxy}^\prime)
\end{align}
and
\begin{align}
&
(-)^{i_xi_y}g_3(((-)^{i_y}+(-)^{i_x})\frac{\pi}{4} +\th_{pxy})
\nonumber\\
=& g_3( (-)^{\v i}\frac{\pi}{4}+\th_{pxy}^\prime)
\end{align}
Thus the above two sets of PSG's are labeled by
U1C$\tau^0_-\tau^0_+[\tau^0_-,\tau^1_-][\tau^0_-,\tau^1_+]$
and
U1C$\tau^1_+\tau^1_+[\tau^0_-,\tau^1_-][\tau^0_-,\tau^1_-]$
respectively (see \Eq{PSGU1Ca} - \Eq{PSGU1Cd}).
We also note that the PSG
U1C$\tau^0_-\tau^0_+\tau^1_-\tau^0_-$
is gauge equivalent to
U1C$\tau^0_-\tau^0_+\tau^1_+\tau^0_-$.
Using the gauge transformation $W_{\v i} = g_3(-(-)^{\v i}\pi/8)$ we can change
\Eq{U1SU2lPSG3a} and
\Eq{U1SU2lPSG4a} to
\begin{align}
\label{U1SU2lPSG3aa}
 G_x  =& i \tau^{\th_x}, \ \ \ G_y = i\tau^{\th_y} \nonumber \\
G_{P_x} =&  (-)^{i_x} g_3(\th_{px})
\ \ \  
G_{P_y} = (-)^{i_y} g_3(\th_{py})
\nonumber\\
 G_{P_{xy}} =& 
(-)^{i_xi_y}g_3(((-)^{i_y}-(-)^{i_x})\frac{\pi}{4} +\th_{pxy}),
g_3(
\th_{pxy})i\tau^1,
\nonumber\\
  G_T =& (-)^{\v i}g_3(\th_T),\ (-)^{i_y}g_3((-)^{\v i}\pi/4)i\tau^{\th_T}  
\end{align}
\begin{align}
\label{U1SU2lPSG4aa}
 G_x  = &i \tau^{\th_x}, \ \ \ G_y = i\tau^{\th_y}\nonumber \\
G_{P_x} =&   
ig_3(\th_{px}+(-)^{\v i}\frac{\pi}{4})\tau^1
\ \ \  G_{P_y} =  
ig_3(\th_{py}-(-)^{\v i}\frac{\pi}{4})\tau^1
\nonumber\\
 G_{P_{xy}} =&   
(-)^{i_xi_y}g_3(((-)^{i_y}-(-)^{i_x})\frac{\pi}{4} +\th_{pxy}),
g_3(
\th_{pxy})i\tau^1,
\nonumber\\
 G_T =& (-)^{\v i} g_3(\th_T),\ (-)^{i_y}g_3((-)^{\v i}\pi/4)i\tau^{\th_T}  
\end{align}
Then we can use
the gauge transformation $W_{\v i} = (i\tau^1)^{\v i}$ to change
\Eq{U1SU2lPSG3aa} and
\Eq{U1SU2lPSG4aa} to
\begin{align}
\label{U1SU2lPSG3b}
 G_x  =& g_3((-)^{\v i} \th_x), \ \ \ 
  G_y  = g_3((-)^{\v i} \th_y)
\nonumber \\
       G_{P_x} =& g_3((-)^{\v i}\th_{px})
\ \ \  G_{P_y} =  g_3((-)^{\v i}\th_{py})
\nonumber\\
G_{P_{xy}} =&  
(-)^{i_xi_y}g_3(((-)^{i_x}-(-)^{i_y})\frac{\pi}{4} +(-)^{\v i}\th_{pxy}),
\nonumber\\
&
g_3((-)^{\v i}\th_{pxy})i\tau^1,
\nonumber\\
 G_T =& (-)^{\v i} g_3((-)^{\v i} \th_T), 
 (-)^{i_y} g_3((-)^{\v i} \th_T)i\tau^{1\bar 2}  
\end{align}
\begin{align}
\label{U1SU2lPSG4b}
 G_x  =& g_3((-)^{\v i} \th_x), \ \ \ 
  G_y  = g_3((-)^{\v i} \th_y), \ \ \ 
\nonumber \\
    G_{P_x} =&   (-)^{i_x} g_3((-)^{\v i}\th_{px})i\tau^{1\bar 2} 
\ \ \  G_{P_y} = (-)^{i_y} g_3((-)^{\v i}\th_{py})i\tau^{12}
\nonumber\\
G_{P_{xy}} =&  
(-)^{i_xi_y}g_3(((-)^{i_x}-(-)^{i_y})\frac{\pi}{4} +(-)^{\v i}\th_{pxy}),
\nonumber\\
&
g_3((-)^{\v i}\th_{pxy})i\tau^1,
\nonumber\\
G_T =& (-)^{\v i}g_3((-)^{\v i} \th_T), 
(-)^{i_y} g_3((-)^{\v i} \th_T)i\tau^{1\bar 2}
\end{align}

We can use a gauge transformation
 $W_{\v i}=(i\tau^1)^{\v i}g_3(-(-)^{\v i}\pi/8) g_3((-)^{i_y}\pi/4)$ 
to simplify the
ansatz \Eq{U1SU2lin1} - \Eq{U1SU2lin4}.
After the gauge transformation, the IGG is given by
$\{ g_3((-)^{\v i}\th) \}$. The ansatz has a form
$ u_{\v i,\v i+\v m} = u^1_{\v m}\tau^1 +  u^2_{\v m}\tau^2  $ for
$\v m =odd$ and
$ u_{\v i,\v i+\v m} = u^0_{\v m}\tau^0 +  u^3_{\v m}\tau^3  $ for
$\v m =even$.
We find the ansatz \Eq{U1SU2lin1} - \Eq{U1SU2lin4} become
\begin{align}
\label{U1SU2lin1tr}
u_{\v i,\v i+\v m} =&  
u^1_{\v m}\tau^1
+u^2_{\v m}\tau^2 
\nonumber \\
u_{\v m} =& 0,\ \ \ \hbox{for $\v m = even$}  \\
 G_x  =& g_3((-)^{\v i} \th_x), \ \ \ 
  G_y  = g_3((-)^{\v i} \th_y)
\nonumber \\
       G_{P_x} =& g_3((-)^{\v i}\th_{px})
\ \ \  G_{P_y} =  g_3((-)^{\v i}\th_{py})
\nonumber\\
 G_{P_{xy}} =&  
(-)^{i_xi_y}g_3(((-)^{i_x}-(-)^{i_y})\frac{\pi}{4} +(-)^{\v i}\th_{pxy}),
\nonumber\\
&
g_3((-)^{\v i}\th_{pxy})i\tau^1,
\nonumber\\
  G_T =& (-)^{\v i}g_3((-)^{\v i} \th_T)  \nonumber
\end{align}
\begin{align}
\label{U1SU2lin2tr}
u_{\v i,\v i+\v m} =&   
  u^1_{\v m} \tau^1 
+ u^2_{\v m} \tau^2 
+ u^3_{\v m} \tau^3
\nonumber \\
u^1_{\v m} \neq & 0,\ \ \ \hbox{for $m_x = even$ and $m_y=odd$} \nonumber  \\
u^2_{\v m} \neq & 0,\ \ \ \hbox{for $m_x = odd$ and $m_y=even$} \nonumber  \\
u^3_{\v m} \neq & 0,\ \ \ \hbox{for $m_x = even$ and $m_y=even$}  \\
 G_x  =& g_3((-)^{\v i} \th_x), \ \ \ 
  G_y  = g_3((-)^{\v i} \th_y)
\nonumber \\
       G_{P_x} =& g_3((-)^{\v i}\th_{px})
\ \ \  G_{P_y} =  g_3((-)^{\v i}\th_{py})
\nonumber\\
 G_{P_{xy}} =&  
(-)^{i_xi_y}g_3(((-)^{i_x}-(-)^{i_y})\frac{\pi}{4} +(-)^{\v i}\th_{pxy}),
\nonumber\\
&
g_3((-)^{\v i}\th_{pxy})i\tau^{12},
\nonumber\\
  G_T =& (-)^{i_y}g_3((-)^{\v i} \th_T)i\tau^1  \nonumber
\end{align}
\begin{align}
\label{U1SU2lin3tr}
u_{\v i,\v i+\v m} =&  
  u^1_{\v m}\tau^1 
+ u^2_{\v m}\tau^2 
\nonumber \\
u^{1,2}_{\v m} = & 0,\ \ \ \hbox{for $\v m = even$} \\
 G_x  =& g_3((-)^{\v i} \th_x), \ \ \ 
 G_y  =  g_3((-)^{\v i} \th_y)
\nonumber \\
       G_{P_x} =& (-)^{i_x} g_3((-)^{\v i}\th_{px})i\tau^1,
\nonumber\\
       G_{P_y} =& (-)^{i_y} g_3((-)^{\v i}\th_{py})i\tau^2
\nonumber\\
 G_{P_{xy}} =&  
(-)^{i_xi_y}g_3(((-)^{i_x}-(-)^{i_y})\frac{\pi}{4} +(-)^{\v i}\th_{pxy}),
\nonumber\\
&
g_3((-)^{\v i}\th_{pxy})i\tau^{12},
\nonumber\\
  G_T =& (-)^{\v i}g_3((-)^{\v i} \th_T) \nonumber
\end{align}
\begin{align}
\label{U1SU2lin4tr}
u_{\v i,\v i+\v m} =&  u^1_{\v m}\tau^1 
+  u^2_{\v m}\tau^2
+  u^3_{\v m}\tau^3
\nonumber \\
u^1_{\v m} \neq & 0,\ \ \ \hbox{for $m_x = even$ and $m_y=odd$}\nonumber \\
u^2_{\v m} \neq & 0,\ \ \ \hbox{for $m_x = odd$ and $m_y=even$}\nonumber \\
u^3_{\v m} \neq & 0,\ \ \ \hbox{for $m_x = even$ and $m_y = even$}  \\
 G_x  =& g_3((-)^{\v i} \th_x), \ \ \ 
  G_y  = g_3((-)^{\v i} \th_y)
\nonumber \\
       G_{P_x} =& (-)^{i_x} g_3((-)^{\v i}\th_{px})i\tau^1,
\nonumber\\
       G_{P_y} =& (-)^{i_y} g_3((-)^{\v i}\th_{py})i\tau^2
\nonumber\\
 G_{P_{xy}} =&  
(-)^{i_xi_y}g_3(((-)^{i_x}-(-)^{i_y})\frac{\pi}{4} +(-)^{\v i}\th_{pxy}),
\nonumber\\
&
g_3((-)^{\v i}\th_{pxy})i\tau^{12},
\nonumber\\
  G_T =& (-)^{i_y}g_3((-)^{\v i} \th_T) i\tau^1 \nonumber
\end{align}
In \Eq{U1SU2lin2tr}, \Eq{U1SU2lin3tr} 
and \Eq{U1SU2lin4tr} we have made additional gauge
transformation $(\tau^{1\bar 2},\tau^{12}) -> (\tau^1,\tau^2)$.

Using the mapping \Eq{newPSG}, we can obtain all the PSG's for the 
$Z_2$ symmetric
spin liquids near the $SU(2)$-linear state from the 58 $Z_2$ PSG's obtained
in the last subsection for the $SU(2)$-gapless state.
We note that, under the mapping \Eq{newPSG}, 
a $Z_2$ PSG labeled by
Z2A$abcd$ will be mapped into a PSG labeled by Z2B$abcd$ and
a $Z_2$ PSG labeled by
Z2B$abcd$ will be mapped into a PSG labeled by Z2A$abcd$.
In the following, we will list all the 58 $Z_2$ PSG's for the spin liquids
near the $SU(2)$ linear spin liquid. We will also 
construct ansatz for those PSG's.
First let us consider PSG's of form Z2B$...$.
For those PSG's the ansatz can be written as
\begin{equation}
 u_{\v i,\v i+\v m} = (-)^{i_x m_y} u_{\v m}
\end{equation}
In the following we consider the constraint imposed by the $180^\circ$
rotation symmetry and the time reversal symmetry.

For PSG
\begin{align}
&Z2B[\tau^0_-\tau^0_+, \tau^3_-\tau^3_+][\tau^0,\tau^3]\tau^0_- 
\nonumber\\
&Z2B\tau^3_-\tau^3_+\tau^1\tau^0_- 
\end{align}
the $180^\circ$ rotation symmetry generated by $P_xP_y$ requires that
\begin{align*}
(-)^{i_xm_y}u_{\v m}=(-)^{\v m}(-)^{(i_x+m_x)m_y}u_{\v m}^\dag   
\end{align*}
or
\begin{align*}
 u^0_{\v m} =&  0,\ \ \ \hbox{for $m_x=even$ and $m_y=even$} \nonumber\\
 u^{1,2,3}_{\v m} =&  0,\ \ \ \hbox{for $m_x=odd$ or $m_y=odd$}
\end{align*}
The time reversal symmetry $T$ requires that
\begin{align*}
- u_{\v m} =& u_{\v m} (-)^{\v m}  
\end{align*}
The above two equations give us
\begin{align}
\label{Z2SU2lin0}
 u_{\v i,\v i+\v m} =&  (-)^{i_xm_y} u^\mu_{\v m}\tau^\mu 
\nonumber\\
 u^0_{\v m} =&  0,\ \ \ \hbox{for $\v m=even$ } \\
 u^{1,2,3}_{\v m} =&  0,\ \ \ \hbox{for $m_x=odd$ or $m_y=odd$}  \nonumber 
\end{align}

For PSG
\begin{align}
&Z2B\tau^3_-\tau^3_- \tau^{0,1,2,3}\tau^0_- 
\nonumber\\
&Z2B[\tau^0_+\tau^0_+,\tau^3_+\tau^3_+] \tau^{0,1,3}\tau^0_- 
\end{align}
the $180^\circ$ rotation symmetry requires that
\begin{align*}
(-)^{i_xm_y}u_{\v m}=(-)^{(i_x+m_x)m_y}u_{\v m}^\dag   
\end{align*}
or
\begin{align*}
 u^0_{\v m} =&  0,\ \ \ \hbox{for $m_x=even$ or $m_y=even$} \nonumber\\
 u^{1,2,3}_{\v m} =&  0,\ \ \ \hbox{for $m_x=odd$ and $m_y=odd$}
\end{align*}
The time reversal symmetry requires that
\begin{align*}
- u_{\v m} =& u_{\v m} (-)^{\v m}  
\end{align*}
The above two equations give us
\begin{align}
\label{Z2SU2lin1}
 u_{\v i,\v i+\v m} =&  
(-)^{i_xm_y}( u^1_{\v m}\tau^1 +
 u^2_{\v m}\tau^2 + u^3_{\v m}\tau^3 )
\nonumber\\
 u^{1,2,3}_{\v m} =&  0,\ \ \ \hbox{for $\v m = even$}  
\end{align}

For PSG 
\begin{equation}
Z2B\tau^1_-\tau^2_+\tau^{12} \tau^0_- ,
\end{equation}
the $180^\circ$ rotation symmetry requires that
\begin{align*}
(-)^{i_xm_y}u_{\v m}=(-)^{\v m}(-)^{(i_x+m_x)m_y}\tau^3 u_{\v m}^\dag\tau^3    
\end{align*}
or
\begin{align*}
 u^{0,1,2}_{\v m} =&  0,\ \ \ \hbox{for $m_x=even$ and $m_y=even$} \nonumber\\
 u^{3}_{\v m} =&  0,\ \ \ \hbox{for $m_x=odd$ or $m_y=odd$} 
\end{align*}
The time reversal symmetry requires that
\begin{align*}
- u_{\v m} =& u_{\v m} (-)^{\v m}  
\end{align*}
We find
\begin{align}
\label{Z2SU2lin2}
 u_{\v i,\v i+\v m} =&  
(-)^{i_xm_y}( u^0_{\v m}\tau^0 +
 u^1_{\v m}\tau^1 +
 u^2_{\v m}\tau^2   )
\nonumber\\
 u^{0,1,2}_{\v m} =&  0,\ \ \ \hbox{for $\v m = even$}  
\end{align}

For PSG
\begin{align}
&Z2B[\tau^0_-\tau^0_+, \tau^3_-\tau^3_+]\tau^{0,1,3}\tau^3_- 
\nonumber\\
&Z2B\tau^1_-\tau^1_+\tau^{0,1,2,3}\tau^3_- 
\end{align}
the $180^\circ$ rotation symmetry requires that
\begin{align*}
(-)^{i_xm_y}u_{\v m}=(-)^{\v m}(-)^{(i_x+m_x)m_y}u_{\v m}^\dag   
\end{align*}
or
\begin{align*}
 u^0_{\v m} =&  0,\ \ \ \hbox{for $m_x=even$ and $m_y=even$} \nonumber\\
 u^{1,2,3}_{\v m} =&  0,\ \ \ \hbox{for $m_x=odd$ or $m_y=odd$}
\end{align*}
The time reversal symmetry requires that
\begin{align*}
- u_{\v m} =& \tau^3 u_{\v m}\tau^3  (-)^{\v m}  
\end{align*}
The ansatz has a form
\begin{align}
\label{Z2SU2lin3}
 u_{\v i,\v i+\v m} =&  
(-)^{i_xm_y}( u^0_{\v m}\tau^0 +
 u^1_{\v m}\tau^1 +
 u^2_{\v m}\tau^2  )
\\
 u^{0}_{\v m} =&  0,\ \ \ \hbox{for $\v m = even$}  
\nonumber\\
 u^{1,2}_{\v m} =&  0,\ \ \ \hbox{for $m_x=odd$ or $m_y=odd$}  \nonumber 
\end{align}

For PSG's
\begin{align}
&Z2B\tau^3_-\tau^3_- \tau^3\tau^3_- 
\nonumber\\
&Z2B\tau^1_-\tau^1_- \tau^{0,1,2,3}\tau^3_- 
\nonumber\\
&Z2B[\tau^0_+\tau^0_+,\tau^3_+\tau^3_+]\tau^{0,1,3}\tau^3_- 
\end{align}
the $180^\circ$ rotation symmetry requires that
\begin{align*}
(-)^{i_xm_y}u_{\v m}=(-)^{(i_x+m_x)m_y}u_{\v m}^\dag   
\end{align*}
or
\begin{align*}
 u^0_{\v m} =&  0,\ \ \ \hbox{for $m_x=even$ or $m_y=even$} \nonumber\\
 u^{1,2,3}_{\v m} =&  0,\ \ \ \hbox{for $m_x=odd$ and $m_y=odd$}
\end{align*}
The time reversal symmetry requires that
\begin{align*}
- u_{\v m} =& \tau^3 u_{\v m}\tau^3  (-)^{\v m}  
\end{align*}
The ansatz has a form
\begin{align}
\label{Z2SU2lin4}
 u_{\v i,\v i+\v m} =&  
(-)^{i_xm_y}( u^1_{\v m}\tau^1 +
 u^2_{\v m}\tau^2 +
 u^3_{\v m}\tau^3  )
\\
 u^{3}_{\v m} =&  0,\ \ \ \hbox{for $\v m = even$}  
\nonumber\\
 u^{1,2}_{\v m} =&  0,\ \ \ \hbox{for $m_x=odd$ or $m_y=odd$} \nonumber
\end{align}

For PSG
\begin{align}
Z2B\tau^1_-\tau^2_+\tau^{12}\tau^3_-,
\end{align}
the $180^\circ$ rotation symmetry requires that
\begin{align*}
(-)^{i_xm_y}u_{\v m}=(-)^{\v m}(-)^{(i_x+m_x)m_y}\tau^3 u_{\v m}^\dag\tau^3    
\end{align*}
or
\begin{align*}
 u^{0,1,2}_{\v m} =&  0,\ \ \ \hbox{for $m_x=even$ and $m_y=even$} \nonumber\\
 u^{3}_{\v m} =&  0,\ \ \ \hbox{for $m_x=odd$ or $m_y=odd$} 
\end{align*}
The time reversal symmetry requires that
\begin{align*}
- u_{\v m} =& \tau^3 u_{\v m}\tau^3  (-)^{\v m}  
\end{align*}
We get
\begin{align}
\label{Z2SU2lin4b}
 u_{\v i,\v i+\v m} =&  
(-)^{i_xm_y}( u^0_{\v m}\tau^0 +
 u^1_{\v m}\tau^1 +
 u^2_{\v m}\tau^2 )
\\
 u^{0}_{\v m} =&  0,\ \ \ \hbox{for $\v m = even$}  
\nonumber\\
 u^{1,2}_{\v m} =&  0,\ \ \ \hbox{for $m_x=even$ or $m_y=even$} \nonumber
\end{align}

For PSG's
\begin{align}
&Z2B\tau^3_-\tau^3_- \tau^{0,1,3}\tau^3_+ 
\nonumber\\
&Z2B\tau^1_-\tau^1_- \tau^{0,1,2,3}\tau^3_+ 
\nonumber\\
&Z2B\tau^0_+\tau^0_+ \tau^{0,1,3}\tau^3_+ 
\nonumber\\
&Z2B\tau^1_+\tau^1_+ \tau^{0,1,2,3}\tau^3_+ 
\end{align}
the $180^\circ$ rotation symmetry requires that
\begin{align*}
(-)^{i_xm_y}u_{\v m}=(-)^{(i_x+m_x)m_y}u_{\v m}^\dag   
\end{align*}
or
\begin{align*}
 u^0_{\v m} =&  0,\ \ \ \hbox{for $m_x=even$ or $m_y=even$} \nonumber\\
 u^{1,2,3}_{\v m} =&  0,\ \ \ \hbox{for $m_x=odd$ and $m_y=odd$}
\end{align*}
The time reversal symmetry requires that
\begin{align*}
- u_{\v m} =& \tau^3 u_{\v m}\tau^3  
\end{align*}
The ansatz has a form
\begin{align}
\label{Z2SU2lin4a}
 u_{\v i,\v i+\v m} =&  (-)^{i_xm_y}(u^1_{\v m}\tau^1 + u^1_{\v m}\tau^2 )
 \\
 u^{1,2}_{\v m} =&  0,\ \ \ \hbox{for $m_x=odd$ and $m_y=odd$} \nonumber 
\end{align}

There are six PSG's of form Z2A$...$ whose ansatz have a form
\begin{equation}
 u_{\v i,\v i+\v m} = u_{\v m}
\end{equation}
The first two are
\begin{align}
Z2A[\tau^1_-\tau^2_+,\tau^1_+\tau^2_- ] \tau^{12}\tau^0_-.
\end{align}
Their ansatz have a form \Eq{Z2SU2gl2}.
For PSG
\begin{align}
Z2A\tau^0_+\tau^0_+\tau^3\tau^0_-
\end{align}
the ansatz have a form \Eq{Z2SU2gl1}.
For 
\begin{align}
Z2A[\tau^1_-\tau^2_+,\tau^1_+\tau^2_- ] \tau^{12}\tau^3_+
\end{align}
the $180^\circ$ rotation and
the time reversal symmetries require that
\begin{align*}
 u_{-\v m} =&  \tau^3 u_{\v m}\tau^3 (-)^{\v m}   = u_{\v m}^\dag   
\nonumber\\
- u_{\v m} =& \tau^3 u_{\v m}\tau^3  
\end{align*}
which gives us
\begin{align}
\label{Z2SU2lin7}
 u_{\v i,\v i+\v m} =&  
 u^1_{\v m}\tau^1 +
 u^2_{\v m}\tau^2   
\nonumber\\
 u^{1,2}_{\v m} =&  0,\ \ \ \hbox{for $\v m = even$}  
\end{align}
For PSG
\begin{align}
Z2A\tau^0_+\tau^0_+\tau^3\tau^3_+
\end{align}
the ansatz has a form \Eq{Z2SU2gl4a}.

In summary, we find that 
there are 12 classes of perturbations around the $SU(2)$-linear spin liquid
that break the $SU(2)$ gauge structure
down to a $U(1)$ gauge structure, and
there are 58 classes of perturbations that break the $SU(2)$ gauge structure
down to a $Z_2$ gauge structure.
The resulting $U(1)$ and $Z_2$ spin liquids can be
constructed through $u_{\v i\v j}$.

\bibliographystyle{aip}
\bibliography{/home/wen/bib/wencross,/home/wen/bib/publst,/home/wen/bib/htc,/home/wen/bib/fqh,/home/wen/bib/misc}

\end{document}